# Measurement of the $t\bar{t}Z$ Production Cross Section in the Final State with Three Charged Leptons using 36.1 fb$^{-1}$ of $pp$ Collisions at 13 TeV at the ATLAS Detector

Dissertation

zur Erlangung des mathematisch-naturwissenschaftlichen Doktorgrades
„Doctor rerum naturalium"
der Georg-August-Universität Göttingen

im Promotionsprogramm ProPhys
der Georg-August University School of Science (GAUSS)

vorgelegt von

Nils-Arne Rosien

aus Dannenberg (Elbe)

Göttingen, 2017

Betreuungsausschuss

Prof. Dr. Arnulf Quadt
Prof. Dr. Ariane Frey

Mitglieder der Prüfungskommission:

Referent: Prof. Dr. Arnulf Quadt
II. Physikalisches Institut, Georg-August-Universität Göttingen

Korreferent: Prof. Dr. Stan Lai
II. Physikalisches Institut, Georg-August-Universität Göttingen

Weitere Mitglieder der Prüfungskommission:

Prof. Dr. Baida Achkar
II. Physikalisches Institut, Georg-August-Universität Göttingen

Prof. Laura Covi, PhD
Institut für Theoretische Physik, Georg-August-Universität Göttingen

Prof. Dr. Ariane Frey
II. Physikalisches Institut, Georg-August-Universität Göttingen

Prof. Dr. Steffen Schumann
Institut für Theoretische Physik, Georg-August-Universität Göttingen

Tag der mündlichen Prüfung: 22.01.2018

Referenz: II.Physik-UniGö-Diss-2017/04

„Also lautet ein Beschluß,
Daß der Mensch was lernen muß. -
Nicht allein das Abc
Bringt den Menschen in die Höh';
Nicht allein in Schreiben, Lesen
Übt sich ein vernünftig Wesen;
Nicht allein in Rechnungssachen
Soll der Mensch sich Mühe machen,
Sondern auch der Weisheit Lehren
Muß man mit Vergnügen hören."

*-Wilhelm Busch: Max und Moritz - Vierter Streich*

# Measurement of the $t\bar{t}Z$ Production Cross Section in the Final State with Three Charged Leptons using 36.1 fb$^{-1}$ of $pp$ Collisions at 13 TeV at the ATLAS Detector

## Abstract

A measurement of the production cross section for a top quark pair in association with a $Z$ boson ($t\bar{t}Z$) is presented in this PhD thesis. Final states with exactly three charged leptons (electrons or muons) are used, taking into account the decay of the top quark pair in the lepton+jets channel and the decay of the $Z$ boson into two charged leptons. The dataset used for this analysis corresponds to 36.1 fb$^{-1}$ of proton-proton collisions at a centre-of-mass energy of 13 TeV, recorded during 2015 and 2016 by the ATLAS detector at the Large Hadron Collider.

The result of a profile likelihood fit to the event yields in four signal enriched regions and two background enriched regions is $\sigma_{t\bar{t}Z} = 966^{+114}_{-102}(\text{stat.})^{+115}_{-114}(\text{syst.})$ fb. The observed (expected) significance is 7.2 (6.4) standard deviations from the background-only hypothesis. Within the experimental uncertainties, the result is in good agreement with the Standard Model prediction. This result is compared with two other $t\bar{t}Z$ analysis channels, using the same dataset but different lepton multiplicities. The analysis presented here is found to be the most sensitive one in terms of observed significance. The result of a combined fit of all three analysis channels is discussed. Two feasibility studies of possible future $t\bar{t}Z$ analysis techniques are demonstrated.

# Messung des Wirkungsquerschnitts der $t\bar{t}Z$-Produktion im Endzustand mit drei geladenen Leptonen in 36.1 fb$^{-1}$ Daten aus $pp$-Kollisionen bei 13 TeV am ATLAS-Detektor

## Zusammenfassung

In dieser Dissertation wird eine Messung des Wirkungsquerschnitts für die Produktion von Top-Quark-Paaren, zusammen mit einem $Z$-Boson ($t\bar{t}Z$), präsentiert. Endzustände mit exakt drei geladenen Leptonen (Elektronen oder Myonen) werden verwendet, was den Zerfall des Top-Quark-Paares im Lepton+Jets-Kanal und den Zerfall des $Z$-Bosons in geladene Leptonen berücksichtigt. Der Datensatz für diese Analyse entspricht einer integrierten Luminosität von 36.1 fb$^{-1}$ aus Proton-Proton-Kollisionen bei einer Schwerpunktsenergie von 13 TeV, aufgezeichnet in 2015 und 2016 vom ATLAS-Detektor am LHC.

Das Ergebnis eines Profile-Likelihood-Fits an die Anzahl der Ereignisse in vier Regionen, die mit Signalereignissen angereichert sind, und zwei Regionen, die mit Untergrundereignissen angereichert sind, ist $\sigma_{t\bar{t}Z} = 966^{+114}_{-102}(\text{stat.})^{+115}_{-114}(\text{syst.})$ fb. Die beobachtete (erwartete) Signifikanz beträgt 7.2 (6.4) Standardabweichungen von der Nullhypothese. Innerhalb der Messunsicherheiten zeigt das Resultat eine gute Übereinstimmung mit der Vorhersage des Standardmodells.

Dieses Resultat wird mit zwei anderen $t\bar{t}Z$-Analysekanälen verglichen, die denselben Datensatz verwenden, jedoch eine andere Anzahl von geladenen Leptonen voraussetzen. Die hier vorgestellte Analyse stellt sich dabei, gemessen an der beobachteten Signifikanz, als die sensitivste von den dreien heraus. Des Weiteren wird das Resultat eines kombinierten Fits unter Verwendung dieser drei Analysekanäle diskutiert. Zudem werden zwei Machbarkeitsstudien von möglichen zukünftigen $t\bar{t}Z$-Analysetechniken diskutiert.

# Contents

Contents

# Introduction

The epigraph of this thesis begins with the words "*So it is decided that man has to learn something*" which I personally interpret as a call to the innate curiosity of every person to expand their knowledge about the world. Uncovering the secrets of the universe has always been one of the greatest driving factors of society. It is in the nature of humankind to explore the unknown and to ask what lies beyond. One of the most important questions that scientists ask is about the elementary rules of nature.

One attempt to answer this question is the Standard Model of particle physics. It contains all particles and interactions that are supposed to be fundamental. At the Large Hadron Collider (LHC) at CERN, the properties of the Standard Model, as well as possible extensions, are studied. The top quark, discovered in 1995 [1, 2] is one of the elementary particles within the Standard Model. Over the years, studies of its properties have formed a distinct field of research. The ATLAS experiment, located at the LHC, has a large focus on this field, called top quark physics.

One of the properties of the top quark, that is of particular interest for this thesis, is its coupling to the $Z$ boson. This coupling can for instance give access to the third component of the weak isospin of the top quark. An observable sensitive to this coupling is the production cross section of a top quark pair in association with a $Z$ boson, called $t\bar{t}Z$. Measuring this cross section is the effort of an analysis performed by the ATLAS collaboration [3]. The project presented in this PhD thesis is a part of this analysis[1].

---

1. To separate the content of this PhD project from the collaborative effort of the analysis group, the ATLAS measurement, with all of its analysis channels, implemented methods and other contributions from $\sim 15$ different researchers, is referred to as the "overall" analysis in this thesis.

## 1. Introduction

The PhD project focuses on the $t\bar{t}Z$ decay signature with exactly three leptons (called the *trilepton channel*) which is the most significant channel of this analysis. The dataset used for this analysis was taken during 2015 and 2016 in proton-proton collisions at a centre-of-mass energy of $\sqrt{s} = 13$ TeV, corresponding to an integrated luminosity of $\int \mathcal{L} \mathrm{d}t = 36.1$ fb$^{-1}$. The $t\bar{t}Z$ cross section in the trilepton channel is extracted from a profile likelihood fit to the event yields in four signal regions and two additional control regions. The control regions are used to determine the correct normalisation of the dominating backgrounds.

The Standard Model of particle physics with the included particles and their interactions will be discussed in Chapter 2. This chapter will put a special focus on the top quark and its properties since it is the most important particle for this analysis. The $t\bar{t}Z$ process will be discussed in Chapter 3. Both its theoretical background and the latest $t\bar{t}Z$ analyses are presented. The different analysis strategies and results of both the latest ATLAS and CMS $t\bar{t}Z$ cross section measurements are compared.

Chapter 4 presents the LHC and the ATLAS detector as the experimental setup of this analysis. To be able to conduct the analysis, physics objects such as electrons, jets and missing transverse momentum need to be defined. The reconstruction of these objects from raw detector data is discussed in Chapter 5. The dataset used for this analysis, as well as the modelling of the signal and background processes are presented in Chapter 6. The signal and control regions for the trilepton channel are presented in Chapter 7. This chapter also discusses the different signal and background contributions to each channel and motivates the selection criteria. It also compares the expectation from simulated events to data before performing a fit.

Systematic uncertainties play a large role in this analysis. Not only do they have a big influence on the total uncertainty. To constrain them, they are also included in the profile likelihood fit as nuisance parameters. Systematic uncertainties are presented and discussed in detail in Chapter 8.

Since the analysis presented in this thesis is part of a more comprehensive cross section measurement of the $t\bar{t}Z$ process, and a similar process called $t\bar{t}W$, the other channels of the overall analysis are presented in Chapter 9. The strategy of the analysis presented in this thesis, including the principle of a profile likelihood fit and a test of the fitting framework using an Asimov fit, is presented in Chapter 10. The results of this fit to data are shown in Chapter 11. This chapter also discusses the behaviour of the profile likelihood fit itself and compares the results of the fit in the trilepton channel with the other analysis channels. In addition, the result of a combined fit in all channels is presented. Chapter 12 presents two feasibility studies for future modifications of the $t\bar{t}Z$ analysis. It discusses an alternative approach on $b$-tagging, called *continuous b-tagging*, and a simple application of reconstruction information of the $t\bar{t}Z$ process, obtained by a framework called *KLFitter* [4].

## The Top Quark in the Standard Model of Elementary Particle Physics

The foundation of this thesis is the Standard Model of particle physics, describing all known elementary particles and their interactions on a microscopic scale. On the other hand, the most important elementary particle for this thesis is the top quark, together with its properties. This thesis will test if the top quark actually behaves as predicted by the Standard Model in terms of the cross section of the $t\bar{t}Z$ process (see Chapter 3). This chapter will present the Standard Model of particle physics with all of its particles, as well as a short overview of the electroweak and strong interactions. It will also discuss the current limitations of the Standard Model and possible extensions to it. The second part of this chapter gives an overview of the top quark in terms of its production processes, decay channels and other properties.

For the rest of this thesis, natural units are used, where the speed of light in vacuum $c$ and the reduced Planck constant $\hbar$ are set to $c = \hbar = 1$.

## 2.1. The Standard Model of elementary particle physics

The current knowledge about all elementary particles and their fundamental interactions is condensed into the Standard Model of elementary particle physics, theoretically developed in the 1960s and 1970s [5–18]. It includes three generations of spin-1/2 fermions, divided into quarks and leptons. Four types of spin-1 gauge bosons are included as the mediators of the electromagnetic, weak (unified into the electroweak) and strong interactions. The spin-0 Higgs boson is included as a consequence of the Higgs field, granting mass to all of the massive particles mentioned above [9–12]. The Higgs boson was discovered in 2012 by the ATLAS and CMS collaborations [19, 20] and is the elementary particle in the Standard Model that was most recently discovered. All particles described by the Standard Model are listed in Figure 2.1 together with their masses.

Figure 2.1.: Elementary particles included in the Standard Model together with their masses: up quark ($u$), down quark ($d$), charm quark ($c$), strange quark ($s$), top quark ($t$), bottom quark ($b$), electron ($e$), muon ($\mu$), tau lepton ($\tau$), the neutrinos associated with the charged leptons ($\nu$), photon ($\gamma$), $W$ boson ($W$), $Z$ boson ($Z$), gluon ($g$) and the Higgs boson ($H$) [21].

It is ambiguous to talk about the "current knowledge" of particle physics, since there are many open questions in terms of experimental observations, as well as theoretical concepts and limitations, see Section 2.1.5. However, theories explaining these open questions are not part of the Standard Model yet, since there are not yet any indications for the new phenomena that they predict. Probing the Standard Model and testing possible extensions to it is one of the main purposes of experimental particle physics.

## 2.1.1. Gauge bosons and fundamental interactions

Local gauge invariance is one of the theoretical foundations of the Standard Model. Imposing this requirement on Standard Model Lagrangians yields spin-1 particles, called *gauge bosons*. Gauge bosons are the mediator particles of the interactions within the Standard Model, described by quantum field theories. At a low energy scale, the photon is the mediator particle of quantum electrodynamics (QED), based on the local U(1) gauge symmetry. The photon couples to the electric charge of a particle and is massless.

Although its existence was hypothesised a long time, Einstein's interpretation of the photoelectric effect [22] and Planck's interpretation of black-body radiation [23] at the beginning of the $20^{th}$ can be marked as discoveries of the photon.

The $W$ and $Z$ bosons, together with the photon, are the gauge bosons of the unified theory of *electroweak interaction*. Electroweak unification takes place at high energy scales of $\mathcal{O}(100\,\text{GeV})$. At lower energy scales, this interaction manifests into the electromagnetic and weak interaction. The electroweak interaction is based on a $\text{SU}(2) \times \text{U}(1)$ gauge symmetry. The important charges for this interaction are the third component of the weak isospin $T^3$ and the weak hypercharge $Y$, see Section 2.1.3. The $W$ and $Z$ bosons are massive (see Figure 2.1), and the $W$ boson carries an electric charge of $\pm 1e$, where $e$ is the elementary charge. The $W$ and $Z$ bosons were predicted by the theory of electroweak interaction by Glashow, Salam and Weinberg [5–8]. First experimental hints of the existence of the $Z$ boson were found with the discovery of weak neutral currents in 1973 by the Gargamelle experiment [24, 25]. Both bosons were directly discovered in 1983 by the UA1 and UA2 experiments [26–29], located at the Super Proton Synchrotron (SPS) at CERN. The electroweak interaction is described in Section 2.1.3.

The massless gluons are the mediators of the strong interaction, based on a $\text{SU}(3)$ gauge symmetry and described by the theory of *quantum chromodynamics* (QCD), see Section 2.1.4. The gluons couple to the colour charge and carry colour charge themselves. Based on the different colour charge combinations, eight different gluons exist. They were discovered in the late 1970s at the electron-positron colliders DORIS and PETRA at DESY in events containing three jets [30–35].

Theories that describe the unification of the electroweak interaction with the strong interaction are not yet experimentally verified. Furthermore, describing gravity as a quantum field theory is theoretically challenging. Therefore, gravity and further unifications of fundamental forces are not yet included in the Standard Model, see Section 2.1.5.

## 2.1.2. Fermions

The elementary particles of the Standard Model classified as fermions are spin-1/2 particles. They can be divided into the strongly interacting quarks and into leptons, the latter do not carry any colour charge. Both leptons and quarks can be categorised into three fermion generations, sorted by their masses. All leptons and quarks of one mass generation respectively, together with their weak isospin, electric charge and weak hypercharge, are listed in Table 2.1 (the colour charge of quarks is not shown). These properties are identical for each fermion generation.

| Lepton | $T$ | $T^3$ | $Q$ | $Y$ |
|---|---|---|---|---|
| $\nu_L$ | 1/2 | 1/2 | 0 | −1 |
| $\ell_L^-$ | 1/2 | −1/2 | −1 | −1 |
| | | | | |
| $\ell_R^-$ | 0 | 0 | −1 | −2 |

| Quark | $T$ | $T^3$ | $Q$ | $Y$ |
|---|---|---|---|---|
| $u_L$ | 1/2 | 1/2 | 2/3 | 1/3 |
| $d_L$ | 1/2 | −1/2 | −1/3 | 1/3 |
| $u_R$ | 0 | 0 | 2/3 | 4/3 |
| $d_R$ | 0 | 0 | −1/3 | −2/3 |

Table 2.1.: Leptons (left) and quarks (right) of one mass generation, respectively, listed together with some of their basic properties: the weak isospin $T$ and its third component $T^3$, electric charge $Q$ and weak hypercharge $Y$. Additionally, quarks carry colour charge. The electric charge is given as multiples of the elementary charge $e$.

Fermions are grouped into left-handed chiral isospin doublets and right-handed chiral isospin singlets, since the weak interaction is maximally parity violating [36] and therefore only couples to left-handed fermions and their right-handed antiparticles. The weak hypercharge $Y$, the third component of the weak isospin $T^3$ and the electric charge $Q$ are related via the Gell-Mann-Nishijima formula [37,38]

$$Q = T^3 + \frac{Y}{2}.$$ (2.1)

Leptons are grouped into charged leptons and *neutrinos*. For left-handed leptons, neutrinos ($T^3 = 1/2$) are the weak isospin partners of the charged leptons ($T^3 = -1/2$). In the Standard Model, neutrinos are massless and only carry weak isospin and weak hypercharge. Therefore, they only occur as left-handed particles and right-handed antiparticles, see Section 2.1.3. Other than the neutrinos, charged leptons carry mass and electric charge. Therefore, they can also interact via the electromagnetic interaction and also occur as right-handed weak isospin singlets.

The charged lepton of the first generation is the *electron* which was directly discovered in 1896 by Thomson in cathode ray experiments [39]. The charged lepton of the second generation is the *muon*, often denoted by the Greek letter $\mu$. It was discovered in 1936 in experiments with cosmic radiation [40,41]. The *tau* lepton, denoted by the letter $\tau$, is the charged lepton of the third generation. It was discovered in the mid 1970s at the SPEAR collider at SLAC in $e^+e^- \to \tau^+\tau^- \to e^\pm\mu^\mp + 4\nu$ events [42] (where $\nu$ stands for a neutrino of any flavour).

The neutrinos of the Standard Model are named after their corresponding isospin partners: *electron-neutrino* ($\nu_e$), *muon-neutrino* ($\nu_\mu$) and *tau-neutrino* ($\nu_\tau$). The electron-neutrino was first hypothesised in 1930 by Pauli to describe energy and spin conservation in the radioactive $\beta$ decay [43]. It was discovered in 1956 by Cowan and Reines [44] in the reaction $\bar{\nu}_e + p^+ \rightarrow n^0 + e^+$, using electron-antineutrinos ($\bar{\nu}_e$) from beta decays, produced in a nuclear reactor. The muon-neutrino was discovered in 1962 at the Alternating Gradient Synchrotron (AGS) in pion decays [45]. The tau-neutrino was discovered in 2000 at the DONUT experiment in the decays of mesons containing charm quarks [46].

Quarks are all massive and carry colour charge in addition to the charges shown in Table 2.1. Therefore, together with gluons, they are the only elementary particles in the Standard Model that interact via the strong interaction. Quarks, forming bound stated via the strong interaction, are the constituents of hadronic matter, such as baryons (three quarks) and mesons (quark-antiquark)[1]. Left-handed quarks are organised in weak isospin doublets, with an *up-type quark* with $T^3 = 1/2$ and a *down-type* quark with $T^3 = -1/2$. Right-handed quarks are organised in weak isospin singlets, respectively.

The quarks of the first generation, the *up* and *down* quarks (denoted by $u$ and $d$), are the basic constituents of protons ($uud$) and neutrons ($udd$), as well as of many other hadronic bound states. The down-type quark of the second generation is called the *strange* quark ($s$). The existence of these three quarks was predicted by Gell-Mann and Zweig in 1964 [13,14,47]. This prediction was confirmed in 1968 by deep inelastic scattering experiments at SLAC [48–50]. These three quarks are often called *light flavour quarks*.

The *charm* quark ($c$) is the up-type quark of the second generation. It was proposed in 1970 as part of the GIM mechanism, to explain the suppression of flavour changing decays of mesons containing a strange quark, such as $K_L \rightarrow \mu^+\mu^-$ [51,52]. The discovery of the $J/\psi$ meson (quark content $c\bar{c}$) in 1974 also marked the discovery of the charm quark. It was almost simultaneously discovered by experiments located at the AGS and SPEAR accelerators. At the AGS, a fixed-target experiment was used, measuring the invariant mass spectrum of the electron-positron pairs from the process $p + Be \rightarrow J/\psi + X \rightarrow e^+e^- + X$ [53]. At SPEAR, the processes $e^+e^- \rightarrow J/\psi \rightarrow e^+e^-$ and $e^+e^- \rightarrow J/\psi \rightarrow \mu^+\mu^-$ were used to discover the $J/\psi$ meson [54].

---

1. Other known hadronic bound states are tetraquarks (two quarks, two antiquarks) and pentaquarks (four quarks, one antiquark).

## 2. The top quark in the Standard Model of elementary particle physics

The down-type quark of the third generation is the *bottom* quark ($b$) and its up-type isospin partner is the *top* quark ($t$). Both were hypothesised in 1973 by Kobayashi and Maskawa to explain CP violation in the weak interaction [55]. The bottom quark was discovered in 1977 at Fermilab by measuring the invariant mass spectrum of $\mu^+\mu^-$ pairs produced in fixed-target collisions [56]. The top quark was discovered in 1995 by the CDF and DØ collaborations at the Tevatron in proton-antiproton collisions [1, 2]. It is the heaviest particle of the Standard Model with a mass of $m_t \approx 173$ GeV. It will be discussed in further detail in Section 2.2.

### 2.1.3. The electroweak interaction

The theory of the electroweak interaction [5–8] is a unification of the electromagnetic interaction (U(1) gauge symmetry) and the weak interaction (SU(2) gauge symmetry). Therefore, the electroweak interaction makes use of an SU(2)×U(1) gauge symmetry. The resulting gauge bosons of this theory are the massive $W$ and $Z$ bosons, as well as the massless photon.

For the electroweak interaction, the Dirac Lagrangian

$$\mathcal{L} = \underbrace{i\bar{\psi}\gamma^\mu D_\mu \psi}_{\text{kinetic term}} - \underbrace{m\bar{\psi}\psi}_{\text{mass term}} \tag{2.2}$$

for a fermion doublet $\psi$ needs to be invariant under local SU(2)×U(1) gauge transformations of the kind

$$\psi \rightarrow \underbrace{\exp\left(i\frac{g'}{2}Y\,\alpha(x)\right)}_{\text{U(1) part}} \underbrace{\exp\left(ig\,\vec{\tau}\cdot\vec{\Lambda}(x)\right)}_{\text{SU(2) part}} \psi \; . \tag{2.3}$$

It can be divided into a part for the local U(1) transformation (as for QED) and for the local SU(2) transformation. Note that while the U(1) and SU(2) parts are mathematically identical to the respective terms in the electromagnetic and weak interaction, they do not represent these interactions. The gamma matrices are denoted by $\gamma^\mu$. The factors $g'$ and $g$ are the coupling strengths corresponding to the U(1) and SU(2), respectively. The Pauli matrices $\vec{\tau}$ are the generators of the SU(2) group. The weak hypercharge is denoted by $Y$, thus belonging to the U(1) part of the local gauge transformation. The U(1) part represents a local phase change $\alpha(x)$ and the SU(2) part represents a local rotation of the weak isospin according to $\vec{\Lambda}(x)$.

To ensure gauge invariance under this SU(2)×U(1) transformation, the covariant derivative $D_\mu$ in Equation (2.2) needs to be chosen as

$$D_\mu = \partial_\mu + i \frac{g'}{2} Y B_\mu + i g \vec{\tau} \cdot \vec{W}_\mu \ . \tag{2.4}$$

This includes four new massless gauge fields: one gauge field $B_\mu$ for the U(1) transformation and three gauge fields $\vec{W}_\mu = (W^1_\mu, W^2_\mu, W^3_\mu)$ for the SU(2) transformation. However, measurements show that the electroweak predictions include one massless gauge boson (the photon) and three massive gauge bosons ($Z$, $W^+$ and $W^-$ bosons). The masses of these bosons are included in the Lagrangian via electroweak symmetry breaking, described by the Higgs mechanism [9–12], which is not discussed here. The resulting fields are

$$
\begin{aligned}
A_\mu &= B_\mu \cos\theta_W + W^{(3)}_\mu \sin\theta_W \ , \\
Z_\mu &= -B_\mu \sin\theta_W + W^{(3)}_\mu \cos\theta_W \ , \\
W^\pm_\mu &= \frac{1}{\sqrt{2}} \left( W^{(1)}_\mu \mp i W^{(2)}_\mu \right) \ ,
\end{aligned}
\tag{2.5}
$$

where $A_\mu$ is the massless photon field and $W^\pm_\mu$ and $Z_\mu$ are the fields for the massive $W$ and $Z$ gauge bosons. The angle $\theta_W$ is the Weinberg angle with $\sin^2\theta_W \approx 0.23$ [21], defined as $\tan\theta_W = g'/g$. The couplings of the photon ($e$), the $Z$ boson ($g_Z$) and the $W$ boson ($g_W$) are related via

$$e = g_W \sin\theta_W = g_Z \sin\theta_W \cos\theta_W \ . \tag{2.6}$$

The resulting mass terms for the gauge bosons in the Lagrangian are

$$
\begin{aligned}
\mathcal{L}_{\text{boson mass}} &= \left| \left( -i\frac{g}{2}\vec{\tau}\cdot\vec{W}_\mu - i\frac{g'}{2}B_\mu \right)\phi \right|^2 \\
&= \underbrace{\left(\frac{1}{2}vg\right)^2 W^+_\mu W^{-\mu}}_{W \text{ mass term}} + \underbrace{\frac{1}{8}v^2 \left( W^{(3)}_\mu \ , \ B_\mu \right) \begin{pmatrix} g^2 & -gg' \\ -gg' & g'^2 \end{pmatrix} \begin{pmatrix} W^{\mu(3)} \\ B^\mu \end{pmatrix}}_{Z \text{ mass term}} ,
\end{aligned}
\tag{2.7}
$$

where $\phi$ is the scalar Higgs doublet and $v \approx 246$ GeV [21] is its vacuum expectation value. The masses for the $W$ and $Z$ bosons can therefore be calculated as

$$M_W = \frac{1}{2}vg \qquad \text{and} \qquad M_Z = \frac{1}{2}v\sqrt{g^2 + g'^2} \ . \tag{2.8}$$

## 2. The top quark in the Standard Model of elementary particle physics

The electroweak symmetry breaking also introduces mass terms for fermions in terms of *Yukawa couplings*:

$$\mathcal{L}_{\text{Yukawa}} = -g_i \left[ \bar{L}_i \phi R_i + (\bar{L}_i \phi R_i)^\dagger \right] \ , \tag{2.9}$$

where $L_i$ are the left-handed fermions in weak isospin doublets, $R_i$ are the corresponding right-handed fermions with the same quark or gluon flavour and $g_i$ is the Yukawa coupling of the fermions to the Higgs field. The index $i$ runs over all fermions. It is obvious from this expression, that neutrinos are massless within the Standard Model, since they do not exist as weak isospin singlets. The masses of the fermions can be determined via their Yukawa coupling as

$$m_i = g_i \frac{v}{\sqrt{2}} \ . \tag{2.10}$$

The covariant derivative in Equation (2.4) not only includes gauge bosons to the Lagrangian, but also their interactions with fermions via the kinetic term, see Equation (2.2). The part of the electroweak Lagrangian describing the interactions of the gauge bosons with the fermions is

$$\begin{aligned}
\mathcal{L}_{\text{EW}}^{\text{fermion-gauge}} = &- \underbrace{\frac{g_W}{2\sqrt{2}} \sum_i \bar{\psi}_i \gamma^\mu (1 - \gamma^5) \, (T^+ W_\mu^+ + T^- W_\mu^-) \, \psi_i}_{\text{fermion-}W \text{ interaction}} \\
&- \underbrace{e \sum_i Q_i \, \bar{\psi}_i \, \gamma^\mu \, \psi_i \, A_\mu}_{\text{fermion-}\gamma \text{ interaction}} \\
&- \underbrace{\frac{g_Z}{2} \sum_i \bar{\psi}_i \, \gamma^\mu (C_V^i - C_A^i \gamma^5) \, \psi_i \, Z_\mu}_{\text{fermion-}Z \text{ interaction}} \ .
\end{aligned} \tag{2.11}$$

The index $i$ runs over all fermions. The electric charge of the fermions is denoted by $Q_i$. The weak isospin raising and lowering operators $T^+$ and $T^-$ describe the coupling of up-type isospin particles to down-type isospin particles via $W$ exchange, such as charged leptons to their corresponding neutrinos or up-type quarks to down-type quarks. While leptons can only couple to their weak isospin partners via an interaction with a $W$ boson, quarks are able to couple to quarks of other generations via this interaction. This is possible because the weak eigenstates of quarks are not the same as their mass eigenstates. The flavour changing charged currents across quark generations are quantified by the CKM matrix [55, 57]. Complex phases in the CKM matrix enable CP violation.

The part of the Lagrangian describing the interaction of the $W$ boson to fermions has a "vector - axial vector" structure with the term $\gamma^\mu (1 - \gamma^5)$ (with the matrix $\gamma^5 = i\gamma^0\gamma^1\gamma^2\gamma^3$). This means that the $W$ boson can only couple to left-handed par-ticles or right-handed antiparticles. The Lagrangian part for the interaction of photons with fermions however is purely vectorial. The photon can therefore couple to both left- and right-handed fermions. The part of the Lagrangian describing the interaction of the $Z$ boson with fermions however has a more complicated interaction, described by the vector coupling $C_V$ and axial vector coupling $C_A$ in the $\gamma^\mu(C_V^i - C_A^i\gamma^5)$ term:

$$
\begin{aligned}
C_V^i &= T_i^3 - 2Q_i \sin^2\theta_W\,, \\
C_A^i &= T_i^3\,.
\end{aligned}
\tag{2.12}
$$

Therefore, measurements of observables sensitive to the interaction of the $Z$ boson to a specific fermion can probe the third component of the weak isospin $T^3$ of that fermion, see also Chapter 3 in case of the top quark.

### 2.1.4. The strong interaction

The strong interaction, described by the theory of quantum chromodynamics (QCD), is based on the non-abelian SU(3) gauge symmetry and is mediated by eight massless gauge bosons, called gluons [13–18]. The corresponding elementary charge of the strong interaction is the colour charge, which can be red ($R$), green ($G$) or blue ($B$), or the corresponding anticolour charge. The only elementary particles that carry colour charge in the Standard Model are quarks and gluons. Quarks carry a colour charge while antiquarks carry the corresponding anticolour charge ($\bar{R}$, $\bar{G}$ or $\bar{B}$). Gluons both carry colour and anticolour charges and can be arranged in a colour octet in bracket notation:

$$
\begin{aligned}
&|G\bar{B}\rangle\,, &&|R\bar{B}\rangle\,, &&-|G\bar{R}\rangle\,, \\
&|R\bar{G}\rangle\,, &&-|B\bar{R}\rangle\,, &&|B\bar{G}\rangle\,, \\
&\tfrac{1}{\sqrt{2}}(|G\bar{G}\rangle - |R\bar{R}\rangle), \\
&\tfrac{1}{\sqrt{6}}(|R\bar{R}\rangle + |G\bar{G}\rangle - 2\,|B\bar{B}\rangle)\,.
\end{aligned}
\tag{2.13}
$$

Because gluons carry colour charges themselves, gluon self interactions are possible in QCD.

Isolated particles must not have any colour charge in total, the phenomenon of which is called *colour confinement* [58]. Therefore, isolated quarks or gluons do not exist. Quarks and gluons need to form colourless hadronic states such as mesons (colour-anticolour) or baryons (red, green, blue). The latter colour combination is colourless because the combination of red, green and blue yields white, which is analogous to colours in visible light.

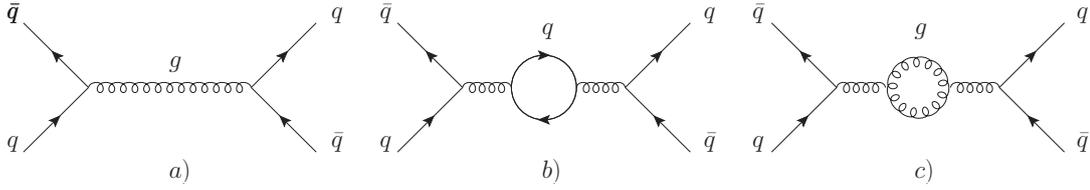

Figure 2.2.: Generic Feynman diagrams for quark-antiquark scattering via gluon exchange. a) leading order process, b) example for vacuum polarisation via a quark loop, c) example for vacuum polarisation via a gluon loop due to gluon self coupling.

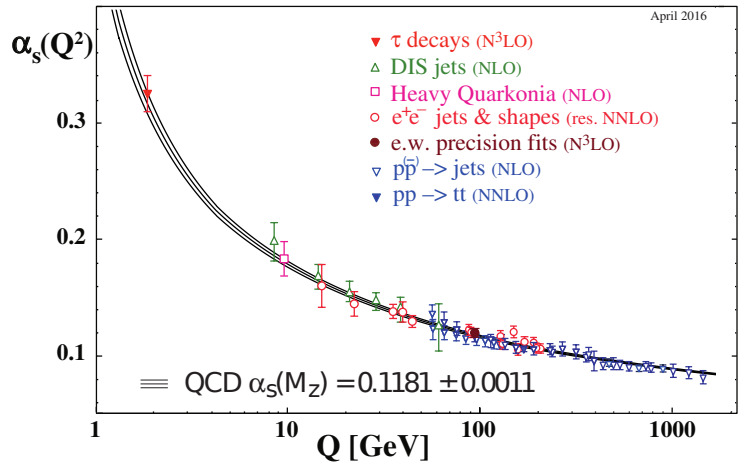

Figure 2.3.: Theoretical prediction of the strong coupling constant $\alpha_S$ depending on the energy scale $Q$ (represented by the solid lines), compared to results of $\alpha_S$ measurements (represented by the markers) [21].

Due to the gluon self coupling, the dependence of the strong coupling constant $\alpha_S$ on the energy scale is different than for the electromagnetic coupling constant. In Figure 2.2, generic Feynman diagrams for quark-antiquark scattering via gluon exchange are shown. For the vacuum polarisation, not only quark loops, but also gluon loops can occur. The boson self coupling makes the difference when compared to QED vacuum polarisations in this case, since only fermion loops can occur in QED vacuum polarisation. Figure 2.3 shows the dependence of $\alpha_S$ on the energy scale $Q$ together with the results of various measurements of this variable. This shows that for higher energy scales, and therefore smaller length scales, $\alpha_S$ decreases, which is called *asymptotic freedom* [17,18].

In case two quarks have momenta in different directions, forcing them to fly further apart, confinement and the fact that the force from strong interaction between them increases with distance, cause a process called *hadronisation*. It will be discussed here using a simplified picture in the style of the string model [59], see Figure 2.4. While the distance between two quarks gets larger, the potential energy induced by the strong force increases, which is represented by colour "tubes". Once this energy is high enough, the colour tubes break up and a new quark-antiquark pair is produced due to the famous relation $E = mc^2$ between energy and mass. The resulting quarks form *colour strings* (hence the name of the model) with colour tubes between them, as it is energetically most favourable. This process is repeated until the kinetic energies of the quarks with opposite momentum vectors are not sufficient to break up the colour strings and form new quark-antiquark pairs again. These hadrons form sprays of particles, called *jets*, which can be detected, for example, in the ATLAS hadronic calorimeter (see Section 4.2.4) and are reconstructed objects used for the analysis presented in this thesis (see Section 5.4).

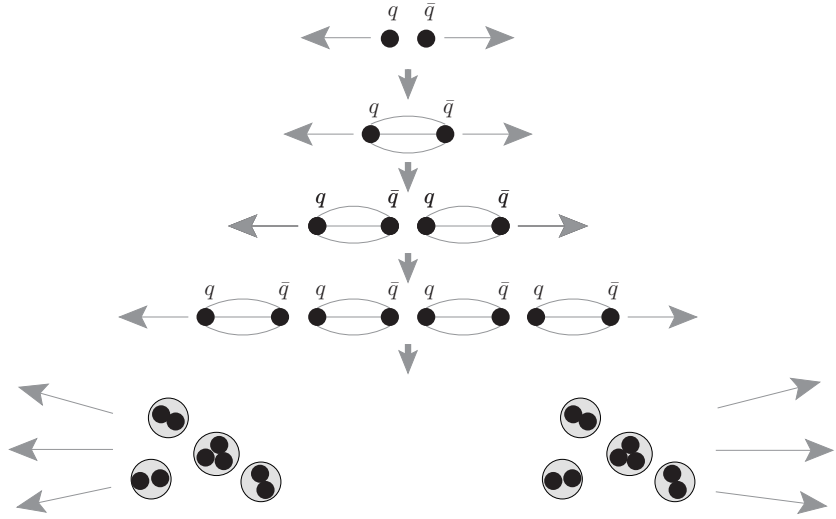

Figure 2.4.: Simplified model of hadronisation. The light grey lines between the quarks represent the colour field from the strong interaction.

### 2.1.5. Limitations of the Standard Model

Over the years, the Standard Model of particle physics has been proven to be very robust. Most measurements show good agreement with Standard Model predictions and attempts to find new physics at particle colliders could only push the exclusion limits to higher energy scales or lower cross section values, so far [60, 61]. However, there are strong indications that the Standard Model is not yet complete. These indications come from theoretical physics, as well as astrophysical and cosmological observations. Finding hints for physics beyond the Standard Model (also called *BSM physics*) is one of the main purposes of measurements at the LHC. This is due to the fact that collider experiments provide a well controllable environment with the possibility to reproduce experimental observations, which is often not the case for astrophysical or cosmological experiments.

Even measurements of Standard Model properties also probe indirectly BSM physics, since deviations from these properties could be the indicator of new physics processes. Therefore, it is not only important for these kinds of measurements to measure a certain nominal value, but also to carefully evaluate the sources of experimental uncertainties and to compare the results to actual Standard Model predictions.

One of the most obvious limitations of the Standard Model is the fact that it does not include gravity, which is approximately 37 orders of magnitude weaker than the weak interaction at the scale of quarks. On much larger scales, gravity describes the interaction between masses via the theory of general relativity. In order to have a theory that describes gravity together with the other fundamental forces, the Standard Model needs to be expanded. However, describing gravity in terms of quantum field theory has been proven to be mathematically challenging [62]. No evidence of the effects of gravity has been found yet at collider experiments such as the LHC [63].

A similar theoretical consideration for the limitations of the Standard Model is that it describes the unification of the electromagnetic and the weak interactions into the electroweak interaction but not the unification of the electroweak interaction with the strong interaction into a *Grand Unified Theory* (GUT) [64]. Several attempts have been made to describe GUT theoretically but none of them could be proven experimentally so far. The energy scale, at which grand unification is expected, is at $E_{\mathrm{GUT}} \approx 10^{15} - 10^{16}$ GeV, which is still far from being accessible by current collider experiments ($E_{\mathrm{LHC}} \approx m_t \approx m_{\mathrm{Higgs}} \approx m_W \approx m_Z \approx 10^2$ GeV). The next logical step after GUT would be a *theory of everything* (ToE) which would unify GUT with gravity. This is even harder to prove experimentally at collider experiments, since the ToE would first manifest at the Planck scale ($E_{\mathrm{Planck}} \approx 10^{19}$ GeV).

Another theoretical limitation of the Standard Model is the low mass of the Higgs boson of $m_H \approx 125$ GeV in comparison with the Planck scale. In the Standard Model, this can only be understood by a fine-tuning of the parameters for quantum loop corrections to the Higgs boson mass. However, supersymmetric extensions of the Standard Model (SUSY) [65] can easily explain the low Higgs boson mass by introducing bosonic partners for each Standard Model fermion and fermionic partners for each Standard Model boson. Newly introduced loop corrections for the SUSY partners naturally cancel the Standard Model loop contributions in a way that the Higgs boson mass at the measured energy scale does not require arbitrary fine tuning. However, no direct experimental evidence of SUSY has been found yet [61].

Recent observations from the Planck telescope [66] indicate that matter made up of ordinary particles described by the Standard Model only accounts for 4.9% of the total mass-energy of the universe. The other contributions come from dark energy (68.3%) and the so-called *dark matter* (26.8%). Dark matter was first used to explain observations of the motion of galaxy clusters [67, 68] and the rotation curves of galaxies [69]. These observations indicate that the visible matter makes up only a small fraction of the actual mass of these objects. The missing mass is explained by dark matter. In many theoretical descriptions, dark matter therefore only interacts via gravity to explain these kind of observations and a very weak kind of interaction, possibly not described by the Standard Model, responsible for its production. Candidates for dark matter particles with these properties are called *weakly interacting massive particles* (WIMPs) [70]. WIMPs are not described by the Standard Model. However, they are predicted by theories of SUSY, universal extra dimensions [71] and little Higgs models [72, 73]. No reliable evidence of direct dark matter detection has been found yet [74, 75].

As it was first theoretically predicted in 1957 [76, 77], neutrinos can change their lepton flavour due to a process called *neutrino oscillation*. This was first observed in the flux of electron neutrinos from the sun [78]. Compared to the theoretical prediction of solar electron neutrino production, less electron neutrinos were detected. This was interpreted as the oscillation of electron neutrinos into other flavours. Neutrino oscillation implies that neutrinos have a mass, which is not predicted by the Standard Model [79, 80].

| | | $m_t$ [GeV] |
|---|---|---|
| Tevatron ($\sqrt{s} = 1.96$ TeV) and LHC ($\sqrt{s} = 7$ TeV) | [82] | $173.34 \pm 0.27$(stat.) $\pm 0.71$(syst.) |
| ATLAS 7+8 TeV (l+jets and dil.) | [83] | $172.51 \pm 0.27$(stat.) $\pm 0.42$(syst.) |
| CMS 7+8 TeV | [84] | $172.44 \pm 0.13$(stat.) $\pm 0.47$(syst.) |
| CMS 13 TeV (l+jets) | [85] | $172.25 \pm 0.08$(stat.+JSF) $\pm 0.62$(syst.) |

Table 2.2.: Combinations of top quark mass measurements prior to Run-II of the LHC [82–84], as well as the CMS measurement at $\sqrt{s} = 13$ TeV [85].

## 2.2. The top quark

Since its discovery in 1995 by the CDF and DØ collaborations [1, 2] at the Tevatron, studies of the top quark have formed a distinct field of research in elementary particle physics, with hundreds of scientists all around the world dedicating their work to this particle. The reasons for this excitement are manifold. Measuring the properties of the top quark is a reliable test of the Standard Model. Since it has a lifetime of approximately $0.5 \times 10^{-24}$ s, which is much shorter than the hadronisation time of approximately $10^{-23}$ s, it is possible to study the decay of the top quark before it forms a hadronic bound state [81]. This means that the top quark properties can be described using perturbative QCD without dealing with low-energy QCD effects. All other quarks hadronise before decaying and can therefore only be studied in hadronic bound states.

Several measurements of the top quark mass have been performed to this day. Combined results prior to Run-II of the LHC, as well as the CMS measurement[2] at $\sqrt{s} = 13$ TeV, are listed in Table 2.2. The top quark is the heaviest particle of the Standard Model, with a Yukawa coupling to the Higgs field of approximately one. This raises the question of whether the top quark plays a special role in electroweak symmetry breaking.

Up to this day, the only experimental facilities capable of producing top quarks were and are the Tevatron proton-antiproton collider at Fermilab and the LHC at CERN, see Chapter 4. During Run-I of the LHC, roughly 6 million top quark pairs have been produced at the ATLAS detector. During the Run-II data taking periods of the ATLAS detector for the years 2015, 2016 and 2017, already roughly 75 million top quark pairs have been produced [86] with similar numbers for CMS. Therefore, the LHC can be considered to be a top quark factory.

---

2. No ATLAS measurement has been performed at $\sqrt{s} = 13$ TeV up to this date.

The year 2017 was an exciting one for top quark physics. Many interesting measurements were published in this field, using sophisticated analysis techniques on the datasets from both Run-I and Run-II. Due to the sheer amount of interesting results, not all of them can be discussed here. The search for the top quark pair production in association with a Higgs boson ($t\bar{t}H$) was performed by both the ATLAS [87,88] and the CMS [89,90] collaborations at $\sqrt{s} = 13$ TeV. This process allows the Yukawa coupling of the top quark to the Higgs field to be tested. The ATLAS collaboration claimed an observation of this process for the Higgs decay modes into $b\bar{b}$, $\gamma\gamma$ and $ZZ^* \to 4\ell$ with an observed (expected) significance of $4.2\sigma$ ($3.8\sigma$). The CMS collaboration achieved an expected (observed) significance of $3.3\sigma$ ($2.5\sigma$) for the Higgs decay channels of $WW^*$, $ZZ^*$ and $\tau\tau$ into multilepton final states. In addition, CMS conducted a search of the associated production of a single top quark and a Higgs boson [91] at $\sqrt{s} = 13$ TeV, which is also sensitive to the top quark Yukawa coupling. The measurement of the production cross section of a top quark pair in association with a photon ($t\bar{t}\gamma$) at $\sqrt{s} = 8$ TeV was conducted at both ATLAS and CMS [92,93]. The ATLAS collaboration also published a measurement of the direct top quark decay width at $\sqrt{s} = 8$ TeV [94], while the CMS collaboration published a search for the production of events containing four top quarks [95]. A combination of $t\bar{t}$ charge asymmetry measurements from the ATLAS and CMS collaborations was also published [96]. The CMS collaboration published a measurement of the $t\bar{t}Z$ and $t\bar{t}W$ cross sections [97], similar to the analysis presented in this thesis, see Chapter 3. Both experiments performed measurements of the production of a single top quark in association with a $Z$ boson ($tZ$) [98–100]. One of the $tZ$ measurements from CMS is a search at $\sqrt{s} = 8$ TeV and the other CMS measurement claims evidence for this process at $\sqrt{s} = 13$ TeV with an observed (expected) significance of $3.7\sigma$ ($3.1\sigma$). The corresponding ATLAS measurement yields an observed (expected) significance of $4.2\sigma$ ($5.4\sigma$) at $\sqrt{s} = 13$ TeV. Similar to $t\bar{t}Z$, this process is also sensitive to the $t\bar{t}Z$ vertex (see Chapter 3). Further interesting top quark measurements published in the year 2017 are discussed below. All of these results are in good agreement with the Standard Model predictions.

### 2.2.1. Top quark decay

The top quark decays approximately 100% of the time into a $W$ boson and a bottom quark due to $|V_{tb}| \approx 1$, where $V_{tb}$ is the CKM matrix element, describing the coupling of a top quark to a bottom quark and a $W$ boson. The $W$ boson further decays in approximately 33% of all cases into a charged lepton (electron, muon or tau lepton) and the corresponding neutrino of the same generation. In approximately 67% of all cases, it decays into an up-type and a down-type quark [21]. Therefore, the decay signature of each top quark is either one jet, a charged lepton and a neutrino, or it has a signature of three jets. Jets from bottom quarks (also called *b-jets*) can be identified via the relatively long lifetime of bottom flavoured mesons with a method called *b-tagging*, see Section 5.5 in this thesis. Neutrinos cannot be detected by current detectors for collider experiments. However, the contribution of neutrino momentum can be identified via the momentum conservation law with an observable called *missing transverse momentum* ($E_{\mathrm{T}}^{\mathrm{miss}}$), see Section 5.6.

For top quark pairs ($t\bar{t}$ pairs), the combination of these two decay modes leads to three different $t\bar{t}$ decay channels. The *all-jets channel* (also called the *fully hadronic channel*) is the case where both top quarks decay into three jets each, resulting in two $b$-jets and four non-$b$-jets. For the next decay channel, both $W$ bosons from the top quarks decay leptonically, resulting in two charged leptons, two $b$-jets and missing transverse momentum, which is called the *dilepton channel*. The third decay channel is the *lepton+jets* channel (also called the *semileptonic* channel) in which one $W$ boson from the $t\bar{t}$ pair decays into quarks and the other one decays into a charged lepton and a neutrino. The lepton+jets channel therefore contains two $b$-jets, two non-$b$-jets, one charged lepton and missing transverse momentum. All $t\bar{t}$ decay channels with their approximate branching fractions and their decay signatures are listed in Table 2.3. In the case of the analysis presented in this thesis, tau leptons are not reconstructed. Therefore, their decay products contribute to the other $t\bar{t}$ decay channels. In this case, the $t\bar{t}$ decay channels are the all-jets channel (or fully hadronic channel), lepton-plus jets channel (or semilepton channel) for $e$+jets and $\mu$+jets, as well as the dilepton channel for the electron-electron, electron-muon (both permutations) and muon-muon signatures.

For many $t\bar{t}$ measurements at the LHC, using the current Run-II dataset, statistical uncertainties no longer play a large role. Therefore, the $e\mu$ dilepton channel can be considered as the "golden channel", since the contribution of background events is low [101]. However, the process of $t\bar{t}$ production in association with a $Z$ boson ($t\bar{t}Z$), which is discussed in this thesis, has a cross section three orders of magnitude lower that the $t\bar{t}$ cross section at $\sqrt{s} = 13$ TeV ($\sigma_{t\bar{t}} \approx 800$ pb versus $\sigma_{t\bar{t}Z} \approx 0.8$ pb). Therefore, the golden channel of the $t\bar{t}Z$ analysis is the one where the $t\bar{t}$ pair decays semileptonically and the $Z$ boson decays into charged leptons. In this case, the $Z$ boson is reconstructed easily. In addition, the physics backgrounds can be relatively well controlled and the branching fraction of the $t\bar{t}$ pair is relatively high. This channel is called the *trilepton channel*, which is the main topic of this thesis. The dileptonic decay of the $t\bar{t}$ pair is used in the *tetralepton channel*, where the $Z$ boson is required to decay into charged leptons. This channel has a high signal purity but statistical uncertainties are a limiting factor due to the small branching fraction of the dileptonic $t\bar{t}$ decay. The all-jets decay of the $t\bar{t}$ pair is used in the so-called $2\ell$OSSF channel, where the $Z$ boson is required to decay into charged leptons. While the branching ratio of the fully hadronic $t\bar{t}$ decay is high, this $t\bar{t}Z$ channel suffers from high background contributions from $t\bar{t}$+jets and $Z$+jets. These channels are discussed in more detail over the course of this thesis, whereas the decay channels of the $Z$ boson into quarks and neutrinos are not taken into account for current $t\bar{t}Z$ analyses at ATLAS and CMS.

| Decay channel | Approximate branching fraction | Decay signature |
|---|---|---|
| All-jets | 46% | 2 $b$-jets, 4 non-$b$-jets |
| $e$+jets | 15% | 2 $b$-jets, 2 non-$b$-jets, $E_\mathrm{T}^\mathrm{miss}$, 1 electron |
| $\mu$+jets | 15% | 2 $b$-jets, 2 non-$b$-jets, $E_\mathrm{T}^\mathrm{miss}$, 1 muon |
| $\tau$+jets | 15% | 2 $b$-jets, 2 non-$b$-jets, $E_\mathrm{T}^\mathrm{miss}$, 1 tau lepton* |
| $ee$ dilepton | 1% | 2 $b$-jets, $E_\mathrm{T}^\mathrm{miss}$, 2 electrons |
| $e\mu$ dilepton | 2% | 2 $b$-jets, $E_\mathrm{T}^\mathrm{miss}$, 1 electron, 1 muon |
| $\mu\mu$ dilepton | 1% | 2 $b$-jets, $E_\mathrm{T}^\mathrm{miss}$, 2 muons |
| $e\tau$ dilepton | 2% | 2 $b$-jets, $E_\mathrm{T}^\mathrm{miss}$, 1 electron, 1 tau lepton* |
| $\mu\tau$ dilepton | 2% | 2 $b$-jets, $E_\mathrm{T}^\mathrm{miss}$, 1 muon, 1 tau lepton* |
| $\tau\tau$ dilepton | 1% | 2 $b$-jets, $E_\mathrm{T}^\mathrm{miss}$, 2 tau leptons* |

Table 2.3.: Top quark pair decay branching fractions. The missing transverse momentum is denoted by $E_\mathrm{T}^\mathrm{miss}$.
*) If the tau lepton is not reconstructed, the channels containing this lepton contribute to the other channels via the hadronic and leptonic tau lepton decay modes. This is the case for the analysis presented in this thesis.

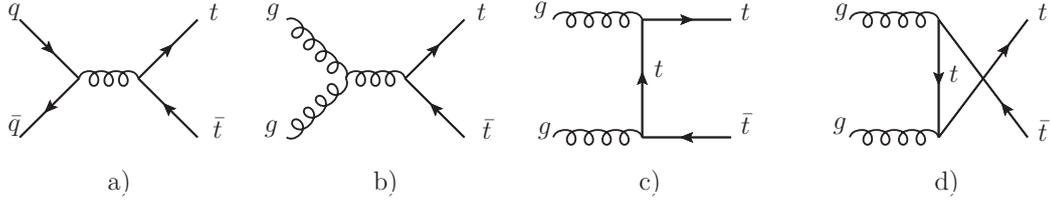

Figure 2.5.: Leading order Feynman diagrams for the top quark pair production in hadronic collisions. The quark-antiquark annihilation (a) and the gluon-gluon fusion processes (b, c and d) are shown.

## 2.2.2. Top quark production

At hadron colliders, the top quark can be produced via two main processes: top quark pair ($t\bar{t}$) production and the production of single top quarks. The dominating process of the two is the $t\bar{t}$ pair production. The corresponding leading order Feynman diagrams are shown in Figure 2.5. The strong interaction is, by far, the most dominant contribution to the $t\bar{t}$ production, while electroweak top quark pair production can be neglected at hadron colliders. At the LHC, both gluon-gluon fusion and quark-antiquark annihilation processes can produce $t\bar{t}$ pairs. The ratio between these two contributions depends on the centre-of-mass energy of the collision, due to the different momentum fractions of gluons and quarks with respect to the colliding protons, expressed by parton distribution functions (PDFs). For the Tevatron, the quark-antiquark annihilation was the dominant production mechanism with a contribution to the $t\bar{t}$ production of approximately 85% for proton-antiproton collisions at $\sqrt{s} = 1.96$ TeV. For proton-proton collisions at $\sqrt{s} = 13$ TeV at the LHC, gluon-gluon fusion dominates, with approximately 90% of all $t\bar{t}$ pairs being produced this way. For a centre-of-mass energy of 13 TeV, the prediction of the $t\bar{t}$ cross section is $\sigma_{t\bar{t}} = 823^{+40}_{-46}$ pb, calculated at next-to-next-to leading order (NNLO) accuracy in $\alpha_s$ and including next-to-next-to-leading logarithm (NNLL) soft gluon terms [86]. It has been measured in several analyses at $\sqrt{s} = 13$ TeV by the ATLAS and CMS collaborations. The results are listed in Table 2.4.

The differential partonic cross sections of top quark pair production via quark-antiquark annihilation and gluon-gluon fusion at leading order are given by [108]

$$
\begin{aligned}
\frac{\mathrm{d}\hat{\sigma}_{q\bar{q} \to t\bar{t}}}{\mathrm{d}z} &= \frac{\pi\alpha_s^2}{9s}\,\beta\left(2 - (1 - z^2)\beta\right), \\
\frac{\mathrm{d}\hat{\sigma}_{gg \to t\bar{t}}}{\mathrm{d}z} &= \frac{\pi\alpha_s^2}{96s}\,\beta\,\frac{7 + 9z^2\beta^2}{(1 - z^2\beta^2)^2}\left(1 + 2\beta^2 - 2z^2\beta^2 - 2\beta^4 + 2z^2\beta^2 - z^4\beta^4\right),
\end{aligned}
\tag{2.14}
$$

| Measurement | | $\int \mathcal{L} \mathrm{d}t$ | $\sigma_{t\bar{t}} \pm$ (stat.) $\pm$ (syst.) $\pm$ (lumi) |
|---|---|---|---|
| ATLAS dilepton $e\mu$ | [102] | 3.2 fb$^{-1}$ | $818 \pm \phantom{0}8 \pm \phantom{0}27 \pm 19$ pb |
| ATLAS dilepton $ee/\mu\mu$ | [103] | 85 pb$^{-1}$ | $749 \pm 57 \pm \phantom{0}79 \pm 74$ pb |
| ATLAS $\ell$+jets | [103] | 85 pb$^{-1}$ | $817 \pm 13 \pm 103 \pm 88$ pb |
| CMS dilepton $e\mu$ | [104] | 43 pb$^{-1}$ | $746 \pm 58 \pm \phantom{0}53 \pm 36$ pb |
| CMS dilepton $e\mu$ | [105] | 2.2 fb$^{-1}$ | $815 \pm \phantom{0}9 \pm \phantom{0}38 \pm 19$ pb |
| CMS $\ell$+jets | [106] | 2.2 fb$^{-1}$ | $888 \pm \phantom{0}2 \pm \phantom{0}26 \pm 20$ pb |
| CMS all-jets | [107] | 2.53 fb$^{-1}$ | $834 \pm 25 \pm 118 \pm 23$ pb |
| Exact NNLO QCD + NNLL | [86] | — | $823^{+40}_{-46}$ pb |

Table 2.4.: ATLAS and CMS measurements of the top quark pair production cross section at $\sqrt{s} = 13$ TeV, compared to the NNLO QCD theory prediction.

where $s$ is the squared centre-of-mass energy, $m_t$ is the top quark mass and $z = \cos\theta$ is the cosine of the angle between the initial parton and the final top quark. The expression $\beta = \sqrt{1 - 4m_t^2/s}$ describes the velocity of the top quark in the centre-of-mass system of the colliding partons. These cross sections need to be convoluted with PDFs to account for the different momentum fractions of the interacting partons from the colliding protons. The differential cross section is usually measured with respect to the transverse momentum of the top quark, the transverse momentum of the $t\bar{t}$ pair or the invariant mass of the $t\bar{t}$ pair. The latter distribution, measured by the ATLAS and CMS collaborations at $\sqrt{s} = 8$ TeV, is shown in Figure 2.6.

Single top quark production is mediated by the charged-current electroweak interaction. It can be divided into three different channels: $s$-channel, $t$-channel and $W$-boson associated production (also called $Wt$ channel). All production channels have different cross sections, which are all lower than for $t\bar{t}$ production, due to the different coupling strengths of the electroweak and strong interactions. The leading order Feynman diagrams of these processes are shown in Figure 2.7. Single top quark production was first observed in 2009 by the CDF and DØ collaborations at the Tevatron for the combined $s$- and $t$-channel production [109–111]. It was not possible to observe top quark production via the $Wt$ channel at the Tevatron due to its small cross section in proton-antiproton collisions at $\sqrt{s} = 1.96$ TeV.

Single top quark production can be used to study the coupling of the $W$ boson to the top quark in its production [112, 113]. This approach is complementary to studies of the top quark $Wtb$ decay vertex. Combining single top quark measurements with, for example, measurements of the helicity of the $W$ boson from the top quark decay [114] can constrain anomalous couplings from effective field theories at the $Wtb$ vertex, see for example Reference [115]. The parameter $|V_{tb}|^2$ is also determined in many single top measurements due to its relation to the $Wtb$ vertex, where $V_{tb}$ is the CKM matrix element for the interaction between the top and bottom quark.

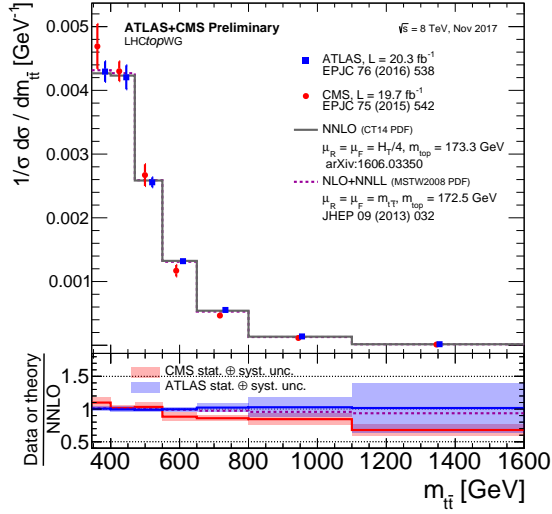

Figure 2.6.: Normalised differential $t\bar{t}$ cross sections at $\sqrt{s} = 8$ TeV as a function of the invariant mass of the $t\bar{t}$ system, measured by the ATLAS and CMS collaborations and compared with theoretical predictions. *This figure was provided by the LHCtopWG.*

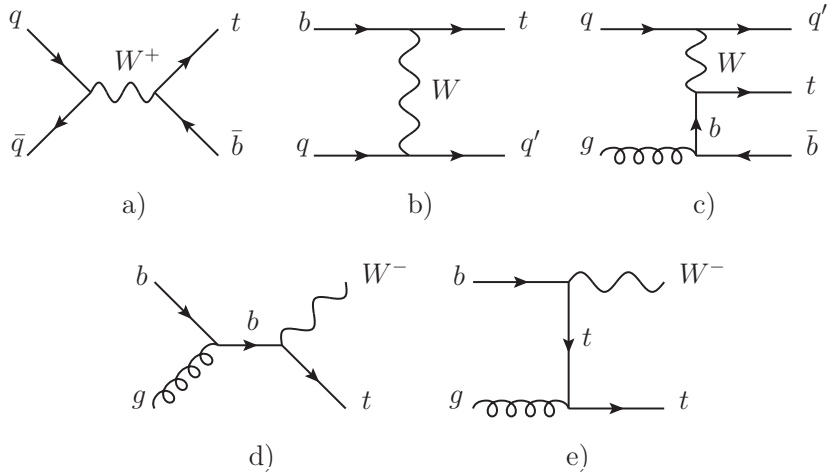

Figure 2.7.: Leading order Feynman diagrams for single top quark production in the *s*-channel (a), *t*-channel $2 \to 2$ (b), *t*-channel $2 \to 3$ (c) and *W*-associated production (d and e).

|  |  | $t$-channel cross section [pb] |
|---|---|---|
| ATLAS | [122] | $247 \pm 6(\text{stat.}) \pm 5(\text{lumi}) \pm 45(\text{syst.})$ |
| CMS | [123] | $238 \pm 13(\text{stat.}) \pm 5(\text{lumi}) \pm 12(\text{exp.}) \pm 26(\text{theo.})$ |
| Theory | [119, 120] | $217^{+6.6}_{-4.6}(\text{scale}) \pm 6.2(\text{PDF}+\alpha_S)$ |

Table 2.5.: Latest $t$-channel single top quark cross section results from the ATLAS and CMS collaborations, using data taken at $\sqrt{s} = 13$ TeV. The results are compared with the theoretical prediction from NLO QCD calculations.

The $t$-channel process has the highest cross section amongst the single top quark production processes an the LHC. Theoretically, it can be described by the $2 \to 2$ process (Figure 2.7b) or the $2 \to 3$ process (Figure 2.7c). The description of the $2 \to 2$ processes needs to take into account bottom quarks in the colliding protons, which is called *five-flavour scheme* (for the $u$, $d$, $c$, $s$ and $b$ quarks), whereas the $2 \to 3$ process can be described by a four-flavour scheme [116–118]. The theoretical prediction for the $t$-channel cross sections for single top and antitop quark production at $\sqrt{s} = 13$ TeV are $\sigma_{t\text{-channel}}(tq) = 136.0^{+5.4}_{-4.6}$ pb and $\sigma_{t\text{-channel}}(\bar{t}q) = 81.0^{+4.1}_{-3.6}$ pb, respectively, calculated at NLO QCD in the five-flavour scheme [119, 120]. This process was measured both during Run-I and Run-II by the ATLAS and CMS collaborations, see Table 2.5 for results with data taken at $\sqrt{s} = 13$ TeV. It was also measured at the Tevatron [121].

The theoretical cross section of the single top quark production, associated with a $W$ boson (Figure 2.7d and e, also called the $Wt$ channel), at $\sqrt{s} = 13$ TeV is $\sigma_{Wt\text{-channel}} = 71.7 \pm 1.8(\text{scale}) \pm 3.4(\text{PDF})$ pb, calculated at approximate next-to-next-to-leading order (aNNLO) QCD. The uncertainties correspond to QCD scale choices and PDF uncertainties [124]. It is the single top quark production process with the second highest cross section at the LHC. This process was measured during Run-I and Run-II of the LHC by the ATLAS and CMS collaborations, see Table 2.6 for the results using data taken at $\sqrt{s} = 13$ TeV.

The $s$-channel single top quark cross section at $\sqrt{s} = 13$ TeV is $\sigma_{s\text{-channel}} = 11.17 \pm 0.18(\text{scale}) \pm 0.38(\text{PDF})$ pb, also calculated at aNNLO QCD [124]. While this process had the second highest contribution to the single top quark production at the Tevatron, it has the smallest cross section at the LHC. The $s$-channel production was first observed as a separate process in 2014 by the CDF and DØ collaborations at the Teva-tron [127]. Due to the large background contributions for this process at the LHC and the relatively small cross section, identifying this process is a difficult task under these conditions. Evidence for the $s$-channel single top production at the LHC was found by the ATLAS collaboration at $\sqrt{s} = 8$ TeV [128]. However, during Run-II of the LHC, the contribution of the background processes is even larger. Therefore, the $s$-channel single top production has not been observed yet at $\sqrt{s} = 13$ TeV, since a much larger dataset is needed to reduce statistical uncertainties [129].

| | | $Wt$ channel cross section [pb] |
|---|---|---|
| ATLAS | [125] | $94 \phantom{.1} \pm 10 \ (\text{stat.}) \pm 2 \phantom{.1} (\text{lumi})^{+28}_{-22} \phantom{00} (\text{syst.})$ |
| CMS | [126] | $63.1 \pm 1.8 (\text{stat.}) \pm 2.1 (\text{lumi}) \pm 6.0 (\text{syst.})$ |
| Theory | [124] | $71.7 \pm 1.8 (\text{scale}) \pm 3.4 (\text{PDF})$ |

Table 2.6.: Latest $Wt$-channel single top quark cross section results from the ATLAS and CMS collaborations, using data taken at $\sqrt{s} = 13$ TeV. The results are compared with the theoretical prediction from aNNLO QCD calculations.

# The $t\bar{t}Z$ Process

The production of a top quark pair in association with a $Z$ boson is called $t\bar{t}Z$. The aim of this thesis is to measure the cross section of this process in the so-called *trilepton* channel, where the $Z$ boson decays into charged leptons and the top-quark pair decays semileptonically. The $t\bar{t}Z$ process is usually measured together with the $t\bar{t}$ production in association with a $W$ boson, called $t\bar{t}W$, due to the similar decay signatures and background processes. When discussed together, both $t\bar{t}Z$ and $t\bar{t}W$ processes are summarised under the name $t\bar{t}V$. The leading order Feynman diagrams of both of these processes are shown in Figure 3.1. While the $t\bar{t}W$ process can only occur via the radiation of a $W$ boson from the initial quarks in the Standard Model (*initial state radiation*, ISR), the $t\bar{t}Z$ process can both occur via $Z$ boson radiation from final state quarks (*final state radiation*, FSR) and ISR. Therefore, the $t\bar{t}Z$ process gives access to the coupling of the $Z$ boson to the top quark (called $t\bar{t}Z$ *coupling*) via final state radiation. Due to the dominance of up quarks over down quarks in the parton distribution functions (PDFs) of the colliding protons, the $t\bar{t}W$ production which results in a positively charged $W$ boson dominates over the production of a negatively charged $W$ boson. The $t\bar{t}Z$ process has an inevitable overlap between the $Z$ boson and an off-shell photon ($\gamma^*$), which is small for on-shell $Z$ production but can have none-negligible contributions in the off-shell $Z$ regime[1].

---

[1]. To avoid divergences for low virtual boson masses, the Monte Carlo samples used in this thesis have a minimum mass cut of 5 GeV applied.

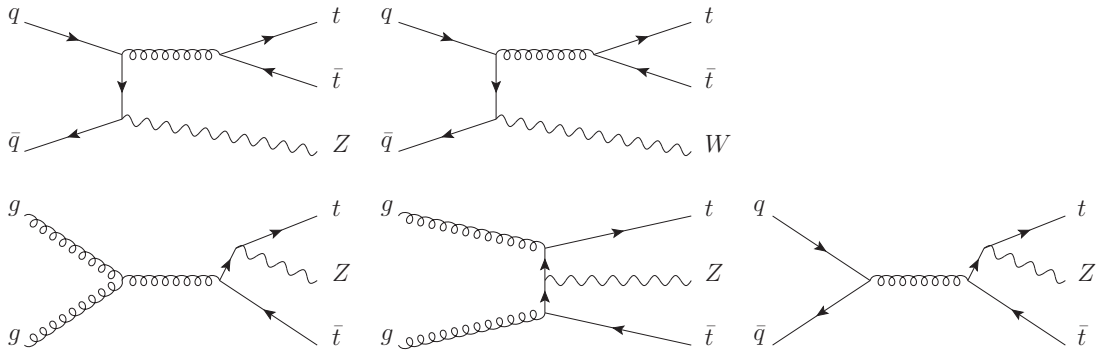

Figure 3.1.: Leading order Feynman diagrams of the $t\bar{t}Z$ and $t\bar{t}W$ processes. The upper two diagrams show $t\bar{t}Z$ production from initial state radiation (left) and $t\bar{t}W$ production (right). The lower diagrams show $t\bar{t}Z$ production from final state radiation.

The $t\bar{t}Z$ and $t\bar{t}W$ cross sections have been measured by the ATLAS and CMS collaborations during Run-I of the LHC [130–133]. The Run-I ATLAS $t\bar{t}Z$ measurement yields an observed significance of $4.2\sigma$ and the most sensitive Run-I CMS $t\bar{t}Z$ measurement [133] yields an observed significance of $6.4\sigma$, corresponding to the deviation from the background-only hypothesis. During Run-II of the LHC, the ATLAS collaboration has conducted a measurement using data taken during 2015 [134] and the CMS collaboration conducted measurements using data taken during 2016 [97, 135]. The Run-II measurements are further discussed in Section 3.2. This thesis presents a first measurement of the $t\bar{t}Z$ and $t\bar{t}W$ cross sections with the full dataset taken during 2015 and 2016 with the ATLAS detector. The focus of this thesis and of the PhD topic is on the trilepton channel sensitive to the $t\bar{t}Z$ process.

Analysing the $t\bar{t}Z$ process is the only way to reliably access the $t\bar{t}Z$ coupling, since the strong interaction heavily dominates the process $q\bar{q} \to t\bar{t}$ and the electroweak production via virtual $Z$ bosons is suppressed. In the future, it will be possible to measure the $t\bar{t}Z$ coupling at electron-positron colliders in $t\bar{t}$ production via the exchange of a virtual $Z$ boson. However, no electron-positron collider has been built so far with a centre-of-mass energy high enough to produce $t\bar{t}$ pairs. Lepton colliders fulfilling this requirement, such as the International Linear Collider (ILC) [136–140] or the Compact Linear Collider (CLIC) [141–143], are still in the planning phase.

The $t\bar{t}Z$ and $t\bar{t}W$ processes are important backgrounds for other analyses at the LHC, such as $t\bar{t}H$ measurements[2] (see for example Reference [88]) and searches for parti-cles from supersymmetric theories in multilepton final states (see for example References [144–146]). The $t\bar{t}V$ processes can also be used to study electroweak symmetry breaking via the interactions of the $W$ and $Z$ bosons. Due to the production of the $t\bar{t}W$ process via initial state $W$ boson radiation only, this process is a potential candidate to probe PDFs. Both the $t\bar{t}Z$ and $t\bar{t}W$ processes are sensitive to physics beyond the Standard Model, such as vector-like quarks [147,148], exotic Higgs models [149] or technicolour [150–154].

---

2. The top quark pair production in association with a Higgs boson is called $t\bar{t}H$.

## 3.1. The $t\bar{t}Z$ process in theory

The cross sections of the $t\bar{t}Z$, $t\bar{t}W$ and $t\bar{t}H$ processes, depending on the centre-of-mass energy $\sqrt{s}$ at proton-proton colliders, such as the LHC, are shown in Figure 3.2. The $t\bar{t}W$ cross section is separated for the $t\bar{t}W^+$ and $t\bar{t}W^-$ processes. The right hand side of the figure shows the ratio between these cross sections and the $t\bar{t}$ production cross section, depending on $\sqrt{s}$. Both $t\bar{t}Z$ and $t\bar{t}W$ cross sections increase with $\sqrt{s}$. However, the ratio between the $t\bar{t}Z$ and $t\bar{t}$ cross sections increases with $\sqrt{s}$, while the $t\bar{t}W$ to $t\bar{t}$ ratio decreases. The reason for the decrease of the ratio for $t\bar{t}W$ is that this process only occurs via quark-antiquark annihilation due to the ISR coupling of the $W$ boson. With increasing $\sqrt{s}$, the PDFs change in a way such that gluon-gluon fusion becomes even more dominant for the $t\bar{t}$ production. Therefore, the $t\bar{t}W$ to $t\bar{t}$ cross section ratio and also the $t\bar{t}Z$ contribution from ISR decrease, while the FSR contributions increase for $t\bar{t}Z$.

The NLO QCD predictions of the $t\bar{t}Z$ and $t\bar{t}W$ cross sections at a centre-of-mass energy of 13 TeV, used for Monte Carlo normalisation in the analysis presented in this thesis, are $\sigma_{t\bar{t}Z} = 0.84 \pm 0.10$ pb and $\sigma_{t\bar{t}W} = 0.60 \pm 0.08$ pb [155]. They are calculated with MadGraph5_aMC@NLO [156] interfaced with Pythia8 [157], using the `NNPDF3.0NLO` PDF [158] and the `A14` Monte Carlo tunes [159] (see Section 6.2.1). NLO electroweak corrections are applied. The errors include PDF uncertainties and the uncertainties from the renormalisation and factorisation scale choices.

According to Equation (2.11), the $t\bar{t}Z$ coupling is described by the following electroweak Lagrangian:

$$\mathcal{L}_{t\bar{t}Z} \propto \bar{u}(p_t) \left[ \gamma^\mu (C_V^t - C_A^t \gamma_5) \right] v(p_{\bar{t}}) Z_\mu \,, \tag{3.1}$$

where $\bar{u}(p_t)$ and $v(p_{\bar{t}})$ are the Dirac spinors of the top quarks and $Z_\mu$ is the gauge field corresponding to the $Z$ boson. According to Equation (2.12), the vector and axial vector

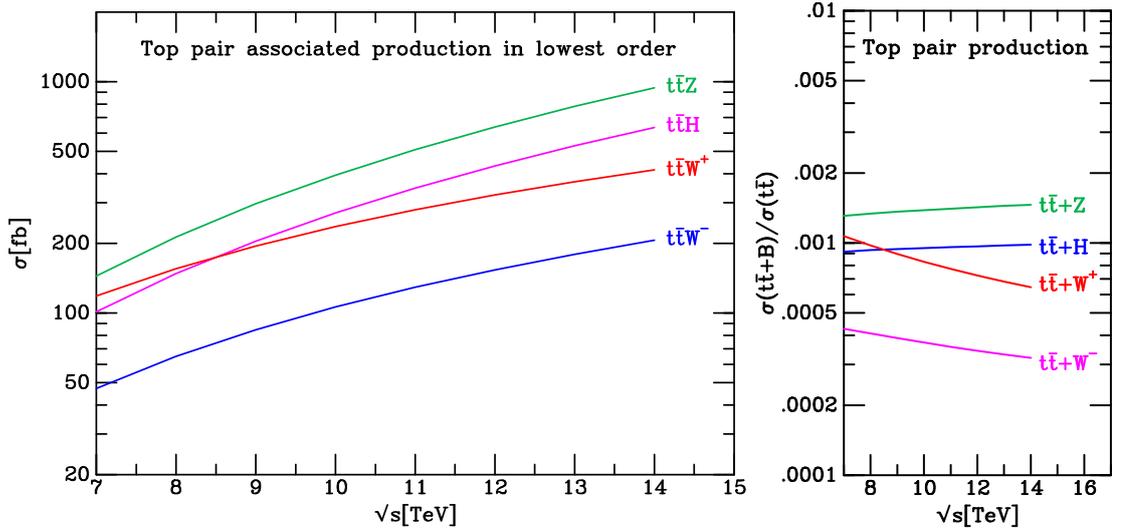

Figure 3.2.: Left: Leading order cross sections for top quark pair production processes in association with bosons at $pp$ colliders as a function of the centre-of-mass energy [160].
Right: Ratio between the leading order cross sections for these processes and the leading order $t\bar{t}$ cross section at $pp$ colliders as a function of the centre-of-mass energy [161].

couplings for the top quark in the Standard Model are

$$C_{\mathrm{V}}^t = C_{\mathrm{V}}^{t,\mathrm{SM}} = T_t^3 - 2Q_t \sin^2 \theta_W \,,$$
$$C_{\mathrm{A}}^t = C_{\mathrm{A}}^{t,\mathrm{SM}} = T_t^3 \,. \tag{3.2}$$

The weak mixing angle is denoted as $\theta_W$ and $Q_t = 2/3e$ is the electric charge of the top quark. The third component of the weak isospin of the top quark is denoted as $T_t^3$ with a Standard Model value of $T_t^3 = 1/2$. Since the third component of the top quark's weak isospin influences the $t\bar{t}Z$ Lagrangian, the $t\bar{t}Z$ process is sensitive to $T_t^3$ in the final state radiation process.

If any kind of physics beyond the Standard Model exists at a higher mass scale $\Lambda$, it is possible to quantify deviations from the Standard Model expectations at the much lower energy scale of the LHC. This can be done with an effective field theory by introducing higher dimensional operators to the $t\bar{t}Z$ Lagrangian [162,163]. A simple expansion using higher dimensional operators is shown in [164]:

$$
\begin{aligned}
C_{\text{V}}^t &= C_{\text{V}}^{t,\text{SM}} + \frac{1}{2}\left(\frac{v^2}{\Lambda^2}\right)\text{Re}\left[C_{\phi q}^{(3,33)} - C_{\phi q}^{(1,33)} - C_{\phi u}^{33}\right], \\
C_{\text{A}}^t &= C_{\text{A}}^{t,\text{SM}} - \frac{1}{2}\left(\frac{v^2}{\Lambda^2}\right)\text{Re}\left[C_{\phi q}^{(3,33)} - C_{\phi q}^{(1,33)} + C_{\phi u}^{33}\right],
\end{aligned}
\tag{3.3}
$$

where $v \approx 246$ GeV [21] is the vacuum expectation value of the Higgs field. The higher dimensional operators are defined as

$$
\begin{aligned}
C_{\phi q}^{(3,33)} &= i\,(\phi^\dagger \tau^a D_\mu \phi)\,(\bar{t}_L \gamma^\mu \tau_a t_L), \\
C_{\phi q}^{(1,33)} &= i\,(\phi^\dagger D_\mu \phi)\,(\bar{t}_L \gamma^\mu t_L), \\
C_{\phi u}^{33} &= i\,(\phi^\dagger D_\mu \phi)\,(\bar{t}_R \gamma^\mu t_R),
\end{aligned}
\tag{3.4}
$$

where $t_L$ is the left-handed quark doublet of the third generation, $t_R$ is the right-handed top quark singlet and $\phi$ is the Higgs boson doublet. The covariant derivative including the gauge fields is denoted by $D_\mu$, and $\tau_a$ are the Pauli spin matrices. This example shows how an effective field theory can introduce higher dimensional operators to the $t\bar{t}Z$ Lagrangian. Many other extensions are possible.

Anomalous couplings can be studied through the $t\bar{t}Z$ Lagrangian as demonstrated in [165]. In this study, the invariant mass of the decay products of the $Z$ boson[3] and their angular separation are found to be sensitive to new physics at the $t\bar{t}Z$ vertex. In addition, the invariant mass of the $t\bar{t}$ pair and the transverse momentum of the $Z$ boson are identified as observables to investigate possible deviations from the expected $t\bar{t}Z$ coupling at hadron colliders, as well.

## 3.2. Recent $t\bar{t}Z$ measurements

Apart from the analysis described in this thesis, several previous measurements of the $t\bar{t}Z$ cross section have been conducted by the ATLAS and CMS collaborations. The most recent ones are discussed in this section. In Section 11.3, their results will be compared with the result of the analysis presented in this thesis. Although $t\bar{t}Z$ cross section measurements have also been performed during Run-I of the LHC, this section will only focus on measurements using data taken at a centre-of-mass energy of $\sqrt{s} = 13$ TeV. All of these analyses also determine the $t\bar{t}W$ cross section in separate fits in addition to the $t\bar{t}Z$ cross section. The two measurements discussed in this section also perform two-dimensional fits of the $t\bar{t}Z$ and $t\bar{t}W$ cross sections. The measurements presented here differ in terms of integrated luminosity of the dataset, as well as in the choices of signal regions, analysis techniques and the detector used for data taking. Their results are listed in Table 3.1.

---

3. The easiest choices are electron or muon pairs as they provide the cleanest signature in the detector.

| Experiment | $\int \mathcal{L} \mathrm{d}t$ [fb$^{-1}$] | $\sigma_{t\bar{t}Z}$ [pb] | $\sigma_{t\bar{t}W}$ [pb] |
|---|---|---|---|
| ATLAS [134] | 3.2 | $0.92 \pm 0.29(\mathrm{stat.}) \pm 0.10(\mathrm{syst.})$ | $1.50 \pm 0.72(\mathrm{stat.}) \pm 0.33(\mathrm{syst.})$ |
| CMS [135] | 12.9 | $0.70^{+0.16}_{-0.15}(\mathrm{stat.})^{+0.14}_{-0.12}(\mathrm{syst.})$ | $0.98^{+0.23}_{-0.22}(\mathrm{stat.})^{+0.22}_{-0.18}(\mathrm{syst.})$ |
| CMS [97] | 35.9 | $0.99^{+0.09}_{-0.08}(\mathrm{stat.})^{+0.12}_{-0.10}(\mathrm{syst.})$ | $0.77^{+0.12}_{-0.11}(\mathrm{stat.})^{+0.13}_{-0.12}(\mathrm{syst.})$ |
| NLO QCD +EWK [155] | — | $0.84 \pm 0.10$ | $0.60 \pm 0.08$ |

Table 3.1.: Results of the recent $t\bar{t}Z$ and $t\bar{t}W$ measurements conducted at the LHC at a centre-of-mass energy of $\sqrt{s} = 13$ TeV compared to the NLO QCD and electroweak predictions used for the Monte Carlo normalisation in this thesis. The numbers show the cross sections obtained in separate fits for the $t\bar{t}Z$ and $t\bar{t}W$ processes. The result of the analysis presented in this thesis is not included here.

Some concepts like missing transverse momentum (momentum imbalance of reconstructed particles, induced mostly by neutrinos) and $b$-tagging (identification of jets induced by bottom quarks) are explained in later chapters of this thesis, for example in Chapter 5. However, the definitions might vary between the analyses, especially for the CMS measurement. These differences cannot be covered in detail in this section. The cited papers discuss the different concepts in more detail.

### 3.2.1. ATLAS measurement with $\int \mathcal{L} \mathrm{d}t = 3.2$ fb$^{-1}$ of data

This earlier ATLAS measurement uses a dataset corresponding to an integrated luminosity of 3.2 fb$^{-1}$, taken at a centre-of-mass energy of $\sqrt{s} = 13$ TeV during the year 2015 [134]. For the $t\bar{t}Z$ cross section measurement, two channels are defined based on the number of required leptons. Only electrons and muons are considered as "leptons" in the object reconstruction. The channel requiring exactly three leptons is called the *trilepton channel* and the channel requiring exactly four leptons is called the *tetralepton channel*. Both channels are sensitive to $t\bar{t}Z$, with the $Z$ boson decaying into an electron or a muon pair. The trilepton channel takes into account the semileptonic decay of the top quark pair while the tetralepton channel is sensitive to its dileptonic decay channel. One multilepton channel is used for the $t\bar{t}W$ measurement exclusively, which requires exactly two muons with the same electric charge, called 2$\mu$SS channel. It takes into account the leptonic $W$ boson decay together with the semileptonic decay of the top quark pair. One signal region of the trilepton channel is also partially sensitive to the $t\bar{t}W$ process, where the top quark pair decays into the dilepton channel and the $W$ boson decays leptonically.

The $WZ$ and $ZZ$ processes are the most dominant backgrounds estimated with Monte Carlo simulations in the trilepton and tetralepton channels, respectively. Their normalisations are constrained by adding dedicated control regions to the fit and by including their normalisation factors as free fit parameters. Events with non-prompt leptons being reconstructed as leptons from the hard process, called *fake leptons*, are another important background source. Since this background is not well modelled in Monte Carlo simulations, it is estimated using a data driven method for the 2$\mu$SS and trilepton channels. It makes use of looser lepton definitions to calculate efficiencies for loose real and fake leptons to end up in the tight selection. This method is also called the *matrix method* [166, 167]. For the tetralepton channel, this background is estimated by scaling fake lepton events from simulation to data in dedicated control regions. Both fake lepton estimation techniques are similar to the ones described in Section 6.3.

The 2$\mu$SS channel has the highest sensitivity to the $t\bar{t}W$ process, compared to other dilepton selections with the same electric charge ($ee$ and $e\mu$), due to the large contribution of events containing electrons with misidentified charge for these selections. This was demonstrated in the ATLAS $t\bar{t}V$ measurement at $\sqrt{s} = 8$ TeV [131]. Therefore, the $ee$ and $e\mu$ channels are not taken into account for this analysis. The two muons are required to have a transverse momentum of $p_\mathrm{T} > 25$ GeV. The other requirements for the 2$\mu$SS channel are missing transverse momentum of $E_\mathrm{T}^\mathrm{miss} > 40$ GeV, a total scalar sum of the transverse momenta of leptons and jets of $H_\mathrm{T} > 240$ GeV and at least two $b$-tagged jets (according to the 77% $b$-tagging efficiency of the MV2c20 $b$-tagging algorithm [168]). To avoid overlap with the other analysis channels, additional leptons with $p_\mathrm{T} > 7$ GeV are vetoed. The main background sources are events containing fake leptons from the $t\bar{t}$ process.

The trilepton channel has the highest sensitivity to the $t\bar{t}Z$ process of all analysis channels. It is divided into four different signal regions, three being sensitive to $t\bar{t}Z$ and one being sensitive to both $t\bar{t}Z$ and $t\bar{t}W$. All of these regions require exactly three charged leptons (electrons or muons), with a leading lepton transverse momentum of $p_\mathrm{T} > 25$ GeV. The other two leptons are required to have $p_\mathrm{T} > 20$ GeV. The total elec-tric charge of the three leptons is required to be $\pm1e$. The signal regions are required to have one lepton pair with the same flavour and opposite charge (called *opposite-sign same-flavour* or *OSSF lepton pair*). For the three signal regions sensitive to $t\bar{t}Z$, the invariant mass of the OSSF lepton pair $m_{\ell\ell}$ is required to be $|m_{\ell\ell} - m_Z| < 10$ GeV, where $m_Z$ is the mass of the $Z$ boson. This invariant mass range is called the *Z-window*. This cut allows sensitivity to on-shell $Z$ bosons from the $t\bar{t}Z$ process. For the trilepton channel sensitive to $t\bar{t}W$, OSSF lepton pairs which fulfil the $Z$-window requirement are vetoed to reject on-shell $Z$ bosons. The signal regions are further divided by jet and $b$-jet multiplicities. The $t\bar{t}Z$ signal regions require at least four jets of which exactly one is $b$-tagged, exactly three jets of which at least two are $b$-tagged or at least four jets of which at least two are $b$-tagged. The trilepton region sensitive to $t\bar{t}W$ is required to have between two and four jets of which at least two are $b$-tagged. The main backgrounds in that region are $Z$+jets events with additional fake leptons and $WZ$+jets events.

The tetralepton channel requires exactly four leptons (electrons or muons), of which two pairs of leptons with opposite electric charge have to be formed, which are required to have an invariant mass of at least 10 GeV. At least one of them is required to be an OSSF lepton pair. The OSSF pair with the invariant mass closest to the $Z$ boson is called the $Z_1$ pair and the other lepton pair is called the $Z_2$ pair. Four signal regions are constructed, based on the flavour composition of the $Z_2$ pair (opposite and same flavour) and the $b$-tag multiplicity (one or at least two $b$-tags). For events with exactly one $b$-tag, an additional requirement on the scalar sum of the lowest two lepton transverse momenta is imposed. This value is required to be at least 25 GeV if the $Z_2$ leptons have the same flavour and at least 35 GeV if the $Z_2$ leptons have different flavour. The dominating background process in this channel is $ZZ$ in association with additional jets.

The resulting cross sections for this analysis are obtained by a profile likelihood fit to data, using the event yields in the signal and control regions. Both one- and two-dimensional fits of the $t\bar{t}Z$ and $t\bar{t}W$ cross sections are performed. The one-dimensional fit fixes one signal strength at the standard model expectation and considers the other one as the parameter of interest. The $WZ$ and $ZZ$ normalisation factors are included as free parameters in the $t\bar{t}Z$ fit. The resulting cross sections from the one-dimensional fit are $\sigma_{t\bar{t}Z} = 0.92\pm0.29(\text{stat.})\pm0.10(\text{syst.})$ pb and $\sigma_{t\bar{t}W} = 1.50\pm0.72(\text{stat.})\pm0.33(\text{syst.})$ pb. The corresponding observed (expected) significances are $3.9\sigma$ $(3.4\sigma)$ and $2.2\sigma$ $(1.0\sigma)$ for the $t\bar{t}Z$ and $t\bar{t}W$ processes, respectively. The two-dimensional fit varies the $t\bar{t}Z$ and $t\bar{t}W$ signal strengths simultaneously. The result is shown in Figure 3.3. Within the uncertainties, the results are in agreement with the Standard Model predictions. Due to the relatively small dataset used in this analysis, the statistical uncertainty of the dataset dominates the results.

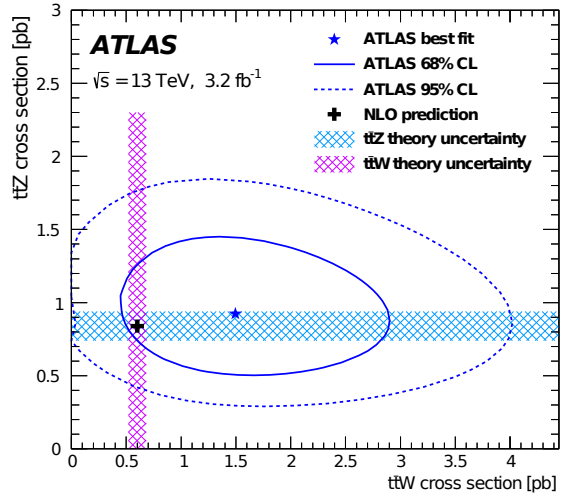

Figure 3.3.: Two-dimensional fit result of the $t\bar{t}Z$ and $t\bar{t}W$ cross sections from the ATLAS analysis using data corresponding to an integrated luminosity of 3.2 fb$^{-1}$ [134].

The difference between the 2015 analysis and the one presented in this thesis is not only an increase in data statistics. A channel requiring exactly two leptons with an opposite electric charge and same lepton flavour is added to the $t\bar{t}Z$ measurement. This channel was already used for an ATLAS $t\bar{t}V$ analysis for data taken at $\sqrt{s} = 8$ TeV [131]. For the $t\bar{t}W$ measurement, electrons are included in the definition of the corresponding dilepton channel and the signal region definitions are completely revised, for example by including events with one $b$-tagged jet, see Section 9.3. The inclusion of electrons in the $t\bar{t}W$ analysis is possible due to new techniques applied for the suppression of electrons with misidentified charge, which is the main background for the $e\mu$ and $ee$ regions of this channel. For the trilepton channel, the cuts for the signal regions are modified. This includes a split of the region sensitive to $t\bar{t}W$ and $t\bar{t}Z$ with off-shell $Z$ bosons into two separate regions, sensitive to one of the respective processes. More details on the different channels are discussed in the later parts of this thesis. Plenty of technical details have been changed between the two analyses which cannot be compared one-by-one in this section. The discussion of the setup of the analysis using 2015 and 2016 data, especially in the trilepton channel, is the topic of this thesis and will be presented in the later chapters.

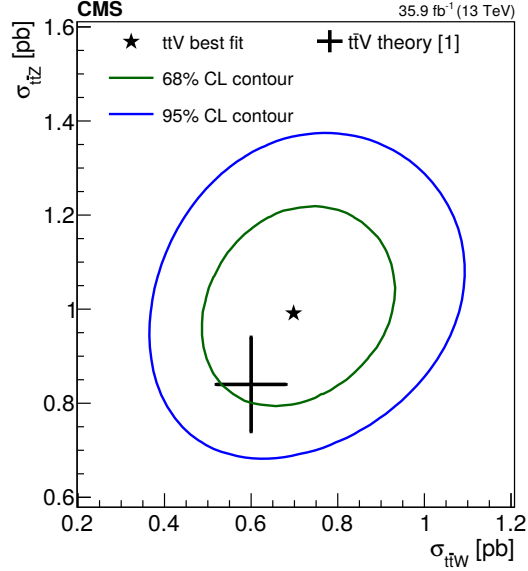

Figure 3.4.: Result of the two-dimensional fit for the $t\bar{t}Z$ and $t\bar{t}W$ cross sections from the current CMS measurement, using data corresponding to an integrated luminosity of 35.9 fb$^{-1}$ [97].

### 3.2.2. CMS measurement with $\int \mathcal{L} \mathrm{d}t = 35.9$ fb$^{-1}$ of data

The most recent CMS measurement of the $t\bar{t}Z$ and $t\bar{t}W$ cross sections uses data taken during 2016 at a centre-of-mass energy of $\sqrt{s} = 13$ TeV, corresponding to an integrated luminosity of $\int \mathcal{L} \mathrm{d}t = 35.9$ fb$^{-1}$ [97]. The $t\bar{t}W$ cross section is measured in a channel requiring exactly two leptons with the same electric charge (electrons and/or muons), called $2\ell$SS channel. The $t\bar{t}Z$ cross section is measured in trilepton and tetralepton channels. One- and two-dimensional profile likelihood fits are performed to determine the $t\bar{t}Z$ and $t\bar{t}W$ cross sections. The resulting values for the one-dimensional fits are $\sigma_{t\bar{t}Z} = 0.99^{+0.09}_{-0.08}(\text{stat.})^{+0.12}_{-0.10}(\text{syst.})$ pb and $\sigma_{t\bar{t}W} = 0.77^{+0.12}_{-0.11}(\text{stat.})^{+0.13}_{-0.12}(\text{syst.})$ pb. The observed (expected) significance of the $t\bar{t}W$ measurement is $5.3\sigma$ ($4.5\sigma$), while the significance of the $t\bar{t}Z$ measurement is only quoted to be "in excess of 5 standard deviations". The result of the two-dimensional fit is shown in Figure 3.4. The contribution from fake leptons from hadronic processes is determined using a data driven method. Events containing electrons with misidentified charge are estimated from Monte Carlo events matched to data in a dedicated control region.

The 2ℓSS channel additionally requires both leptons to have a transverse momentum of $p_T > 25$ GeV, while in the electron-electron channel, the highest lepton $p_T$ is required to be higher than 40 GeV, due to the trigger requirements. The invariant mass of the lepton pair has to be above 12 GeV and a missing transverse momentum of $E_T^{miss} > 30$ GeV is required. To veto events with $Z$ bosons decaying into an electron pair with misidentified charge, the invariant mass of the lepton pair is additionally required to fulfil $|m_{\ell\ell} - m_Z| > 15$ GeV. Events must contain at least two jets with at least one of them passing the $b$-tagging requirements. The main backgrounds in this channel are processes with fake leptons from hadronic processes and leptons with misidentified charge. To be able to reject those backgrounds efficiently, a Boosted Decision Tree (BDT) is trained, based upon the number of jets, number of $b$-jets, scalar sum of jet $p_T$, $E_T^{miss}$, leading and trailing lepton $p_T$, the invariant mass using $E_T^{miss}$ and lepton $p_T$, transverse mass, leading and subleading jet $p_T$, as well as $\Delta R$ between the trailing lepton and the nearest jet. Cuts on the BDT classifier output, as well as different jet and $b$-jet multiplicities are used to refine the signal region definitions. The regions are further divided by the charges of the lepton pair to take into account the imbalance between the $t\bar{t}W^+$ and $t\bar{t}W^-$ production. A total of twenty signal regions are defined for the 2ℓSS channel.

The trilepton channel requires the leading lepton $p_T$ to be higher than 40 GeV, the second leading lepton $p_T$ to be higher than 20 GeV and the lowest lepton $p_T$ to be higher than 10 GeV. At least one OSSF lepton pair with $|m_{\ell\ell} - m_Z| < 10$ GeV is expected, to allow sensitivity to on-shell $Z$ bosons from the $t\bar{t}Z$ process. Nine regions are defined by dividing the trilepton channel into bins of two, three, or at least four jets and zero, one, or at least two $b$-tags. Due to their low signal-to-background ratio, the regions with exactly two jets are used to constrain background uncertainties.

The tetralepton channel additionally requires missing transverse momentum and an OSSF lepton pair out of the four leptons. The leading lepton transverse momentum cut is set to $p_T > 40$ GeV and to $p_T > 10$ GeV for the other leptons. Only events with two lepton pairs with an opposite electric charge pass the selection, of which one is an OSSF pair with $|m_{\ell\ell} - m_Z| < 20$ GeV. Events with an additional OSSF pair are vetoed to reduce the $ZZ$ background. All possible lepton pairings must have an invariant mass above 12 GeV. The channel is divided into different regions in terms of jet and $b$-tag multiplicities, with a minimum requirement of zero $b$-tags and two jets.

The CMS measurement using a fraction of the 2016 dataset, corresponding to an integrated luminosity of 12.9 fb$^{-1}$ [135], is also shown in Table 3.1. This analysis uses a very similar setup as the one discussed above and is superseded by it. Some small differences in the channel definitions are, for example, the separation of the trilepton channels into different regions and the minimum $p_T$ requirement for the leading lepton ($p_T > 20$ GeV) in the tetralepton channel.

Due to the different detectors, as well as the different object reconstruction techniques, triggers, etc., there are plenty of fundamental differences between the ATLAS and CMS measurements which are not discussed here. However, different analysis techniques and signal regions between these two analyses can be compared. First, note that the CMS measurement uses a dataset with approximately the same integrated luminosity as the ATLAS measurement that is discussed in this thesis and ten times more integrated luminosity than the ATLAS measurement discussed above. Second, the CMS measurement makes use of the full 2ℓSS channel, including also events with one $b$-tag and dividing the channel by lepton charges, as it is done for the ATLAS 2ℓSS channel presented in this thesis, see Section 9.3. However, the 2ℓSS channel in the CMS measurement makes use of a multivariate analysis to further separate between signal and background, which is done in neither of the two ATLAS measurements. The trilepton channel in the CMS analysis does not include signal regions sensitive to $tt̄Z$ with off-shell $Z$ bosons or to the $tt̄W$ process. The cuts on the transverse momenta of the leptons and the signal region definitions are also different. The definitions of the tetralepton channels also differ between the ATLAS and CMS measurements.

Compared to the other measurements at $\sqrt{s} = 13$ TeV, the latest ATLAS $tt̄V$ analysis, of which this thesis describes a part, is the only one including the 2ℓOSSF channel. This channel requires exactly two charged leptons, which are required to have the same flavour and opposite electric charge, see Section 9.1. While all ATLAS and CMS measurements use profile likelihood fits to data, also allowing background uncertainties to be constrained, only the ATLAS measurements choose the $WZ$ and $ZZ$ normalisations as free floating parameters with dedicated control regions. For the CMS measurement, a similar approach is studied for the $WZ$ background, but it is concluded that there is no need of such a method, since the background modelling shows good agreement with data in a dedicated control region.

Experimental Setup

This chapter introduces the experimental setup of this analysis, which is the ATLAS detector at the *Large Hadron Collider* (LHC). The experimental data for this analysis were taken during periods in 2015 and 2016 from proton-proton collisions at a centre-of-mass energy of $\sqrt{s} = 13$ TeV. Section 4.1 introduces the LHC machine and shortly explains the other experiments located at this collider. Section 4.2 gives a comprehensive overview of the components of the ATLAS detector during the Run-II period of the LHC.

## 4.1. The Large Hadron Collider

The Large Hadron Collider is a particle accelerator and collider for protons and heavy ions. It is located at the European Organisation for Nuclear Research (CERN) close to Geneva, between 45 and 170 metres below the ground in the former tunnel of the *Large Electron-Positron collider* (LEP). The tunnel has a circumference of 26.7 kilometres. To bend the trajectory of the particles accordingly via the Lorentz force, 1232 super-conducting dipole magnets are used. Particles are accelerated using a superconducting cavity system. The tunnel has eight arcs and eight straight sections. Four of those straight sections are used as the locations of the four main experiments, ATLAS, CMS, ALICE and LHCb (see below). The following text will focus on the proton-proton operation of the LHC. If not stated otherwise, the following information is taken from [169].

## 4. Experimental setup

The main goals of the LHC are and were the discovery of the Higgs boson (discovered in 2012 by the ATLAS and CMS collaborations [19,20]), searches for processes predicted by physics beyond the Standard Model (BSM) theories, as well as studies of the properties of the top quark, the Higgs boson and other Standard Model particles and processes. For all of these studies, an energy regime higher than at any particle accelerator built before is necessary. The initially planned maximum centre-of-mass energy was $\sqrt{s} = 14$ TeV, but due to an incident with the cooling system of the superconducting dipoles at the beginning of the LHC's operation [170], the centre-of-mass energy was reduced and then raised over time, see Table 4.1. During Run-I, the centre-of-mass energies were set to 7 TeV from 2010 to 2011, and to 8 TeV in 2012. For Run-II, the centre-of-mass energy is currently set to 13 TeV, with the potential to run with the initially planned $\sqrt{s} = 14$ TeV in the future.

To be able to search for rare physics processes and to investigate their properties, those processes have to be measured at a reliable rate

$$\frac{\mathrm{d}N_{\mathrm{process}}}{\mathrm{d}t} = \mathcal{L} \cdot \sigma_{\mathrm{process}} \, . \tag{4.1}$$

The cross section of the process is denoted as $\sigma_{\mathrm{process}}$ and $\mathcal{L}$ is the *instantaneous luminosity*, which is a way to quantify the number of useful collision events. The instantaneous luminosity integrated over the time, usually over the whole run period of one year, is called the *integrated luminosity* $\int \mathcal{L} \mathrm{d}t$. Both quantities are shown in Table 4.1 for all run periods of the LHC, as they are delivered to the ATLAS detector. Also taking the ATLAS detector as an example, Figure 4.1 shows that over the course of both Run-I and Run-II, the performance in terms of integrated luminosity improves. This high performance comes with the price of a large number of interactions from the current bunch crossing or another bunch crossing shortly before or after it, called *pileup* (see Section 6.1). Table 4.1 shows the average pileup $\langle \mu \rangle$ at the ATLAS detector. Pileup is discussed in more detail in Section 6.1. The two multi-purpose detectors ATLAS and CMS are placed at the interaction points where the beam conditions are best suited for maximum instantaneous luminosity (see below).

For relativistic energies, the energy loss of a particle per revolution $\Delta E$ in the ring due to synchrotron radiation is proportional to

$$\Delta E \propto \frac{1}{R} \cdot \left(\frac{E}{m}\right)^4 \, , \tag{4.2}$$

| | Year | $\sqrt{s}$ [TeV] | Peak $\mathcal{L}$ [$10^{33}$ cm$^{-2}$s$^{-1}$] | $\int\mathcal{L}dt$ | Maximum #bunches | Minimum bunch spacing | Average $\langle\mu\rangle$ |
|---|---|---|---|---|---|---|---|
| | 2010 | 7 TeV | $0.2\times10^{33}$ | 48.1 pb$^{-1}$ | 368 | 50 ns (one fill), otherwise 150 ns | 9.1 for 7 TeV combined |
| Run-I | 2011 | 7 TeV | $3.65\times10^{33}$ | 5.46 fb$^{-1}$ | 1380 | 25 ns (one fill), otherwise 50 ns | |
| | 2012 | 8 TeV | $7.73\times10^{33}$ | 22.8 fb$^{-1}$ | 1380 | 25 ns (one fill), otherwise 50 ns | 20.7 |
| | 2013 | | No proton-proton collisions for physics | | | | |
| | 2014 | | No proton-proton collisions for physics | | | | |
| Run-II | 2015 | 13 TeV | $5.0\times10^{33}$ | 4.2 fb$^{-1}$ | 2244 | 25 ns | 13.5 |
| | 2016 | 13 TeV | $13.8\times10^{33}$ | 38.5 fb$^{-1}$ | 2220 | 25 ns | 24.9 |
| | 2017 | 13 TeV | $20.6\times10^{33}$ | 47.1 fb$^{-1}$ | 2556 | 25 ns | 38.3 |

Table 4.1.: The LHC performance for the proton-proton runs over the years [171–173]. The luminosities and the pileup refer to specifications for ATLAS.

where $R$ is the ring radius, $E$ is the energy of the particle and $m$ is its mass. Because $R$ is given due to the usage of the former LEP tunnel, heavier particles are needed to go to higher energies without the limitation of high beam energy losses. Therefore, protons are used in the LHC. However, the higher mass and kinetic energy of the protons require for much stronger magnets than the ones used for LEP. The peak dipole field for the LHC dipoles is 8.33 T for the maximum planned centre-of-mass energy of $\sqrt{s} = 14$ TeV. The coils of the LHC dipole magnets are made out of NbTi superconductor cables, cooled down to 1.9 K using suprafluid helium. In addition to the dipole magnets, used for keeping the protons in a circular orbit, quadrupole, sextupole and octupole magnets are used for beam focussing, defocussing and corrections.

To bring the proton beams up to the required energies and intensities, they have to go through a chain of pre-accelerators, see Figure 4.2. First, electrons are stripped off from hydrogen atoms. The resulting protons are then accelerated to 50 MeV using the linear accelerator LINAC 2. The Proton Synchrotron Booster accelerates the beam to 1.4 GeV which is then forwarded to the Proton Synchrotron (PS), where it is accelerated to 25 GeV. The last step in the accelerator chain is the Super Proton Synchrotron (SPS), which accelerates the beam to 450 GeV and injects it into the LHC, where the beam is finally accelerated to the desired energy.

Beams in the LHC are accelerated using a 400 MHz superconducting cavity system which also serves for beam capturing, storage and correction of longitudinal injection errors. To be able to circulate proton beams of one single fill for a long time, a high vacuum has to be achieved in the beam pipe. The vacuum at the interaction regions around the experiments is required to be below an equivalent hydrogen gas density of $10^{13}$ $H_2$ m$^{-3}$. Due to the high required peak luminosity of $\sim 10^{34}$ cm$^{-2}$s$^{-1}$ and the structure of parton distribution functions at the required centre-of-mass energy regime of $\sqrt{s} = 7 - 14$ TeV, an antiproton beam, as for the Tevatron, was not the desired design when the LHC was decided to be built. Producing that many antiprotons in order to match the luminosity requirements would be unrealistic. Conversely, gluon-gluon fusion becomes more dominant at the LHC energy regime for many important physics processes such as $t\bar{t}$, while quark-antiquark fusion contributes less. Therefore, the valence antiquarks from the antiproton are not as important as for the Tevatron energy range. This lead to the choice of building the LHC as a proton-proton collider. The peak number of bunches rotating in the LHC and the minimum bunch spacing are shown in Table 4.1.

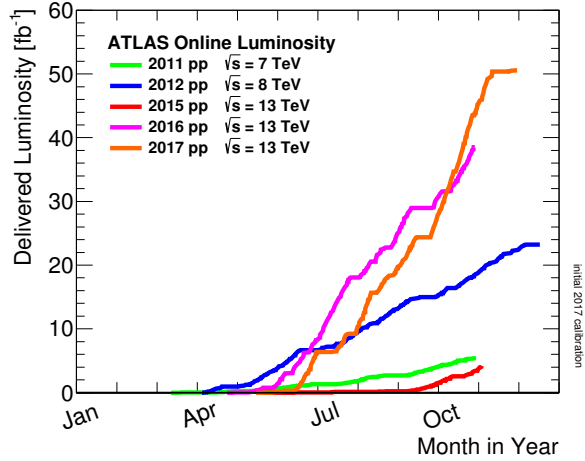

Figure 4.1.: Integrated luminosity delivered to ATLAS versus time for 2011-2017. *This figure is provided by the ATLAS Luminosity Working Group.*

Because two counter-rotating beams of particles with the same electric charge are travelling in the LHC, a separate beam pipe with a separate vacuum chamber and dipole magnetic field has to be used for each beam. Since the diameter of the LEP tunnel is too small for twice as much equipment, the LHC dipole magnets use the *twin bore magnet* design, which consists of two sets of magnet coils and beam channels within the same mechanical structure, sharing one cryostat, see Figure 4.3.

In addition to the ATLAS experiment [174] which will be comprehensively described in Section 4.2, the following experiments are located at the LHC:

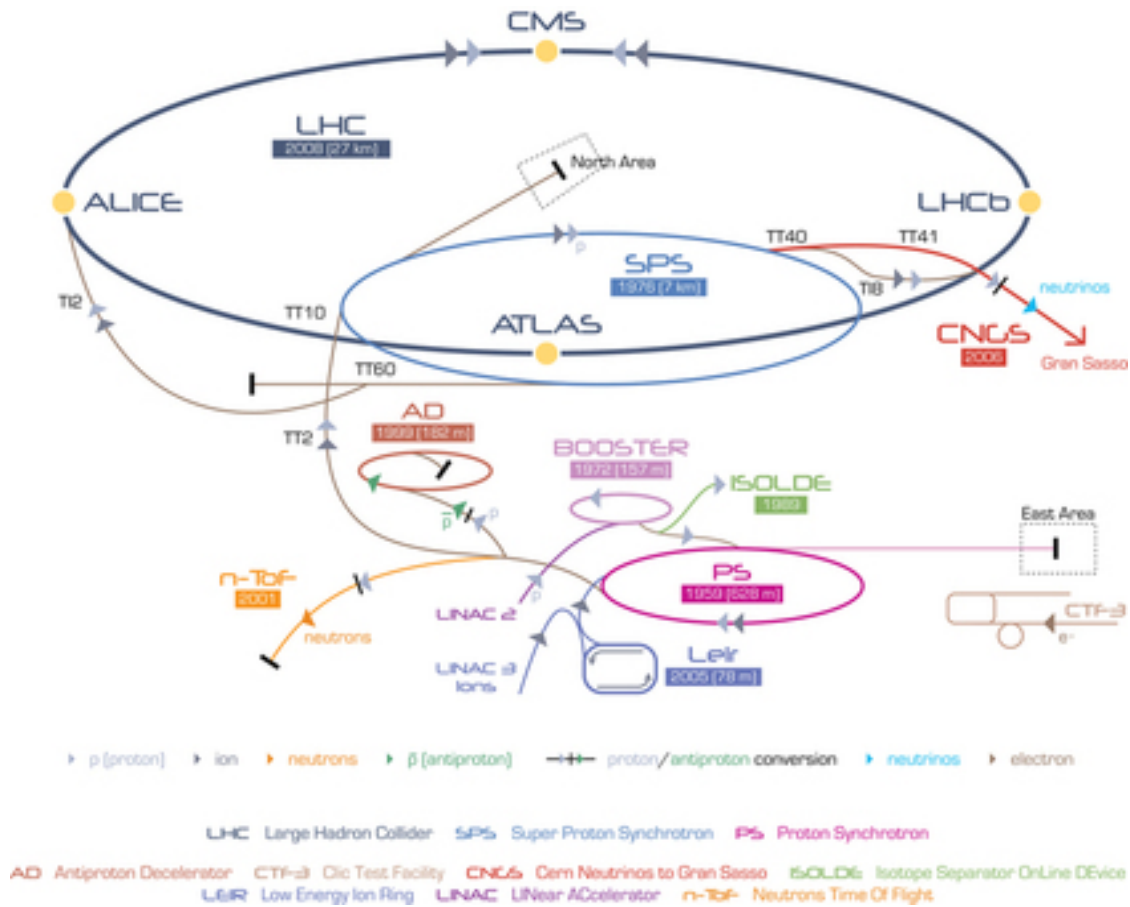

Figure 4.2.: Layout of the CERN accelerator complex. The labels show the circumference and the year of the first operation of each accelerator © CERN.

**Compact Muon Solenoid (CMS)**  Apart from ATLAS, the CMS detector [175] is the other multi-purpose experiment at the LHC. Similar to ATLAS, it covers almost $4\pi$ of the interaction point and consists of a magnet system, trackers, calorimeters and a muon spectrometer. However, there are some crucial design differences between these two detectors. The CMS detector does not rely on an additional toroidal magnetic field for muon spectroscopy. Instead, the whole magnetic field is generated by a superconducting solenoid, which produces a magnetic field of 3.8 T, and its return steel yoke makes up the majority of the detector's mass of 14,000 tonnes. Contrary to the ATLAS detector, both the tracking system and the calorimeter system are immersed in the solenoid.

The first part of the detector (described from the interaction point outwards) is the tracker, which consists of a silicon pixel detector, followed by a silicon microstrip detector.

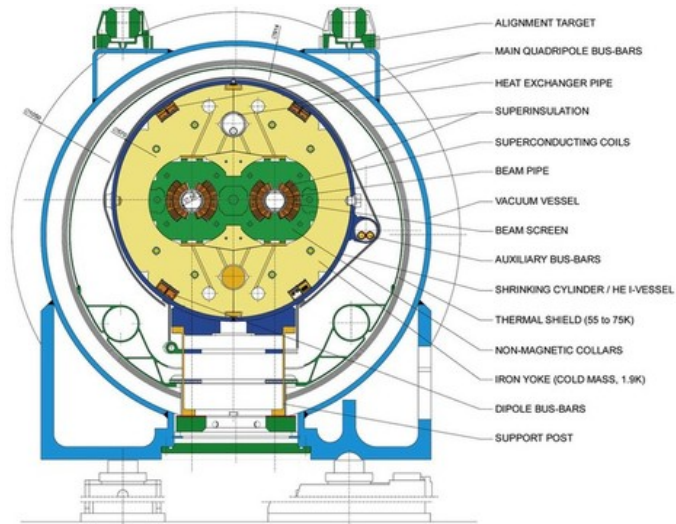

Figure 4.3.: Cross-section drawing of an LHC dipole magnet © CERN.

This part of the detector identifies the path of charged particles, which are bent in the magnetic field of the solenoid. Next is the electromagnetic calorimeter, made out of lead tungstate crystals, which measures the energy of electromagnetically interacting particles by scintillation signals. The subsequent detector part is the hadronic calorimeter, which measures the energy of hadrons. It consists of sandwich structures of metals (brass and steel) together with scintillator materials. The muon detectors are located outside the solenoid system. They consist of drift tubes, cathode strip chambers and resistive plate chambers.

All detector parts are placed in the barrel region, as well as in the end-cap regions of the detector. The detector has the overall dimensions of 21 metres in length and 15 metres in diameter. Figure 4.4 shows a cutaway drawing of the CMS detector.

**A Large Ion Collider Experiment (ALICE)**  The ALICE detector [176] is used to collect data from the heavy ion runs at the LHC. One focus of research for the experiment is the investigation of the quark-gluon plasma. To fulfil this task, the ALICE detector is designed with a different concept in mind than the ATLAS and CMS detectors, since the type and charge of the different produced hadrons need to be identified. The ALICE detector consists of 18 types of sub-detectors, which will not be discussed in detail. A tracking system, surrounded by a magnetic field of 0.5 T, consists of a silicon *Inner Tracking System*, gas filled *Time Projection Chambers* and a transition radiation detector. The *Time-Of-Flight detector* uses multi-gap resistive plate chambers to mea-sure particle velocities over a certain distance in the detector. The *High Momentum Particle Identification Detector* uses Cherenkov radiation to identify particle types. A calorimeter system is used for the determination of particle energies. A dedicated muon spectrometer, placed in the forward region of the ALICE detector, is used to study the dimuon decay signatures of heavy flavour hadrons. The detector has the overall dimensions of 26 metres in length and 16 metres in diameter. Further remnants of the collisions are detected using additional detectors placed 110 meters away from the "core" ALICE detector. Figure 4.5 shows a cutaway drawing of the ALICE detector.

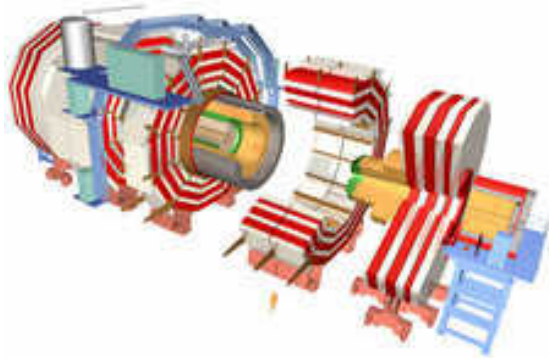

Figure 4.4.: Cutaway drawing of the CMS detector © CERN.

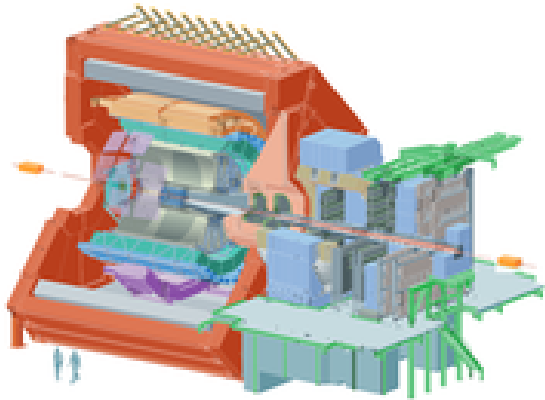

Figure 4.5.: Cutaway drawing of the ALICE detector © CERN.

**Large Hadron Collider beauty (LHCb)**  The LHCb detector [177] is built for heavy flavour physics measurements to study CP violation, rare Standard Model processes and to search for possible BSM processes. Since those events are usually boosted along the beam axis, the detector covers mostly one forward region, rather than the symmetrical

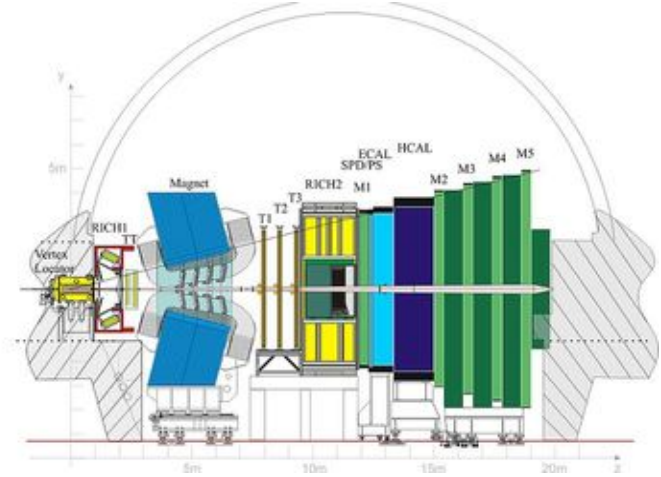

Figure 4.6.: Schematic view of the LHCb detector © CERN.

design of the ATLAS, CMS and ALICE detectors. The instantaneous luminosity delivered to LHCb is lower than for ATLAS or CMS to reduce the pileup contribution, as well as detector occupancy and radiation damage. The following LHCb subdetector systems are (from the interaction point outwards) as follows:

The *vertex locator* (VELO) is located closely to the interaction point and is used to identify primary and secondary decay vertices, for instance to determine heavy flavour meson lifetimes. It is made out of semiconductor detectors and allows a vertex resolution of 50 $\mu$m. After the VELO, the first *ring-imaging Cherenkov detector* (RICH-1) is positioned. It is designed for particle identification via Cherenkov radiation. The second system (RICH-2) is placed after the tracking system (see below) to cover a wider momentum range. The tracking system, placed around the LHCb dipole magnet, measures the tracks of charged particles bent in the magnetic field of the dipole. The momenta of these particles are measured from the curvature of the tracks. Silicon strip detectors and straw-tube detectors are used for the tracking system. Following after the RICH-2 system, the calorimeter system is placed, consisting of electromagnetic and hadronic calorimeters. The muon system is used to identify muons. One layer of the muon system is positioned between RICH-2 and the calorimeter systems and the other layers are positioned after the calorimeter system.

The LHCb detector has a length of 21 metres, a width of 13 metres and a height of 10 metres. The detector weights 5600 tons. Figure 4.6 shows a schematic view of the LHCb detector.

**Smaller experiments at the LHC**    Apart from ATLAS, CMS, ALICE and LHCb, some smaller experiments are located at the LHC, positioned close to the interaction points of the large four experiments.

The *Large Hadron Collider forward* experiment (LHCf) [178] is the smallest experiment at the LHC with only two detectors. Each of them has a size of $30 \times 10 \times 80$ centimetres and a mass of 40 kg. Both detectors are located 140 metres away from the ATLAS interaction point in the forward direction. The setup allows the measurement of particles scattered at almost zero degrees along the beam pipe to study the behaviour of cosmic rays in laboratory conditions.

The *TOTal cross section, Elastic scattering and diffraction dissociation Measurement at the LHC* (TOTEM) [179] detects particles in the forward regions produced at the CMS interaction point. It is used to study the proton-proton interaction cross section and the proton structure itself. The detector components are spread almost 220 metres in both directions in the forward regions of the interaction point, consisting of four particle telescopes and 26 roman pot detectors.

The *Monopole and Exotics Detector at the LHC* (MoEDAL) [180] searches for magnetic monopoles and highly ionising *Stable Massive Particles* (SMPs). It consists of 400 modules made from plastic nuclear-track detectors placed around the LHCb interaction point. If magnetic monopoles or SMPs are produced in the proton-proton collisions, they would break the long-chain plastic molecules in all sheets of a module, leaving holes in the sheets that trace back to the interaction point. The data are evaluated by extracting the detector modules during an LHC shutdown and investigating them off-site in a dedicated laboratory.

## 4.2. The ATLAS detector

The ATLAS detector (*A Toroidal LHC ApparatuS*) is one of the four large experiments at the LHC with a length of 44 metres, a diameter of 25 metres and 7000 tonnes of weight. It is the detector that is used to take the data for the analysis presented in this thesis and will also be assumed for the detector simulation of the Monte Carlo samples. Together with CMS, it is one of the two multi-purpose experiments at the LHC, aiming at top-quark and Higgs boson measurements, tests of the Standard Model and searches for BSM physics. It is also capable of taking data during the heavy ion runs. If not stated otherwise, the following information is taken from Reference [174]. The ATLAS detector is designed in several layers, placed in the cylindrical barrel around the beam pipe, as well as in the end-caps at both ends of the barrel. The detector is forward-backward symmetric. The inner detector is the part closest to the interaction point. It measures the tracks of charged particles, which are bent by the solenoidal magnetic field, and therefore allows the measurement of the momentum of these particles. It also serves for vertex identification. A calorimeter system measures the deposited energy of electromagnetically and hadronically interacting particles. The muon spectrometer is the outermost part of the detector, detecting and measuring the properties of the muons escaping the other parts of the detector. The magnetic field inside ATLAS is provided by a 2 T solenoidal magnet system, wrapped around the inner detector, and an air-core toroidal magnet system for muon spectrometry.

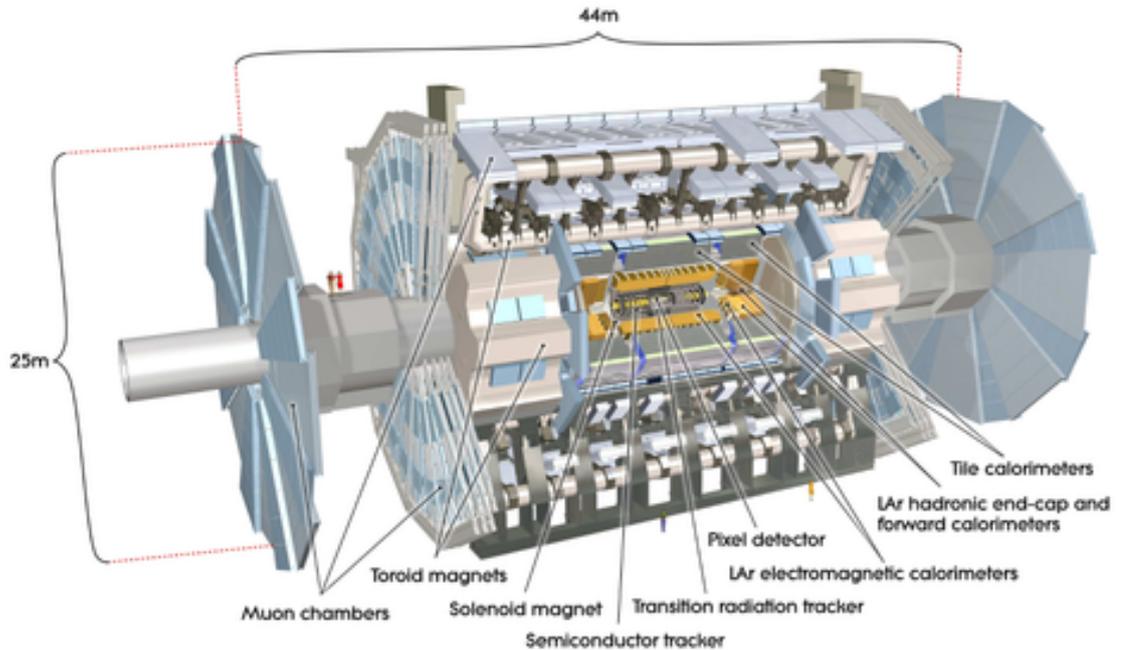

Figure 4.7.: Cutaway view of the ATLAS detector © CERN.

The high instantaneous luminosity and particle energies, as well as the requirements for high precision measurements, impose high standards on the design of the ATLAS detector. Up to this day, it has demonstrated a remarkable performance. The ATLAS experiment just celebrated its $25^{th}$ anniversary which was marked by the publication of the letter of intent in 1992 [181]. Figure 4.7 shows a cutaway view of the detector and its subsystems, which will be further explained in the following sections.

### 4.2.1. The ATLAS coordinate system

The ATLAS coordinate system is needed for the definitions of the variables used in this analysis, as well as for the discussion of object definitions and further descriptions of the detector components. The origin of the coordinate system is the nominal interaction point. The $z$-coordinate is defined as the right-handed beam direction and the $x - y$ plane is defined transversely to it, with the $x$ axis pointing towards the centre of the LHC ring and the $y$ axis pointing upwards. The angle $\phi$ is defined around the beam axis and the angle $\theta$ is defined as the angle from the beam axis. The pseudorapidity is defined as

$$\eta = -\ln \tan \left( \frac{\theta}{2} \right),$$

(4.3)

where the difference in pseudorapidity is invariant under boosts in the $z$-direction. The distance between two objects in the $\eta - \phi$ plane is expressed as

$$\Delta R = \sqrt{\Delta \eta^2 + \Delta \phi^2}.$$

(4.4)

### 4.2.2. Magnet system and magnetic field

The magnetic field of the ATLAS magnet system (see Figure 4.8) bends the tracks of charged particles via the Lorentz force and therefore allows the determination of their momentum and charge. It consists of one solenoid, one barrel toroid and two end-cap toroids (one in each end-cap). The whole magnetic system has a length of 26 metres, a diameter of 22 metres and stores an energy of around 1.6 GJ. The volume in which the total magnetic field is over 50 mT fills approximately 12,000 m$^3$.

The solenoid is located in the barrel section between the inner detector and the electromagnetic calorimeter. It provides an axial field of 2 T for the track measurements in the inner detector. It is designed in a way to reduce the amount of dead material between the inner detector and the calorimeters.

The toroids are installed within the muon spectrometer. They serve for momentum and charge identification measurements of muons in that sub-detector. The barrel toroid produces a magnetic field of around 0.5 T. The toroids located in the end-cap regions produce a magnetic field of around 1 T.

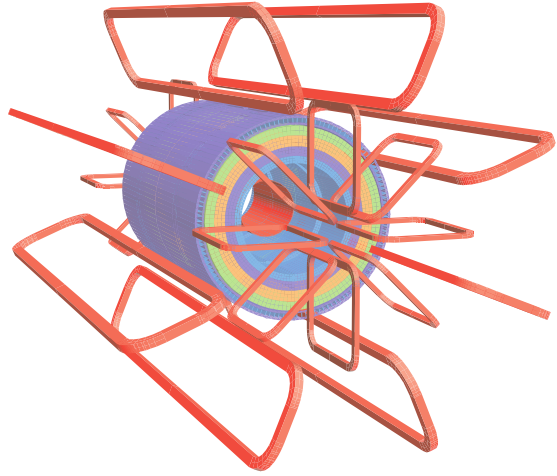

Figure 4.8.: ATLAS magnet system (displayed in red). The toroid system lies outside and the solenoid system lies inside the calorimeter system © CERN.

### 4.2.3. Inner detector

The inner detector (ID, see Figure 4.9) serves to identify the origin of vertices for particle reconstruction and *b*-tagging (see Chapter 5). It also serves for momentum measurements using tracks of electrically charged particles bent in the 2 T magnetic field provided by the solenoid, see Section 4.2.2. It has to deal with high track densities, as well as with high doses of radiation which requires both fine detector granularity and radiation hardness.

The transverse momentum resolution of a tracking detector can be determined in general via the *Glückstern formula* [182]

$$\frac{\sigma(p_{\mathrm{T}})}{p_{\mathrm{T}}} = \frac{\sigma_x \cdot p_{\mathrm{T}}}{0.3 \cdot B \cdot L^2} \sqrt{\frac{720}{N+4}}\,, \tag{4.5}$$

where $\sigma_x$ is the spatial resolution orthogonal to the tracks, N is the number of identified track points, $L$ is the length of the track and $B$ corresponds to the solenoidal magnetic field inside the inner detector. The transverse momentum resolution of the ATLAS inner detector is

$$\frac{\sigma(p_{\mathrm{T}})}{p_{\mathrm{T}}} \approx 0.05\%\, p_{\mathrm{T}}(\mathrm{GeV}) \,\oplus\, 1\%\,. \tag{4.6}$$

The intrinsic spatial accuracies of the inner detector sub-systems are shown in Table 4.2. The ID consists of the following detector layers, listed from the interaction point outwards.

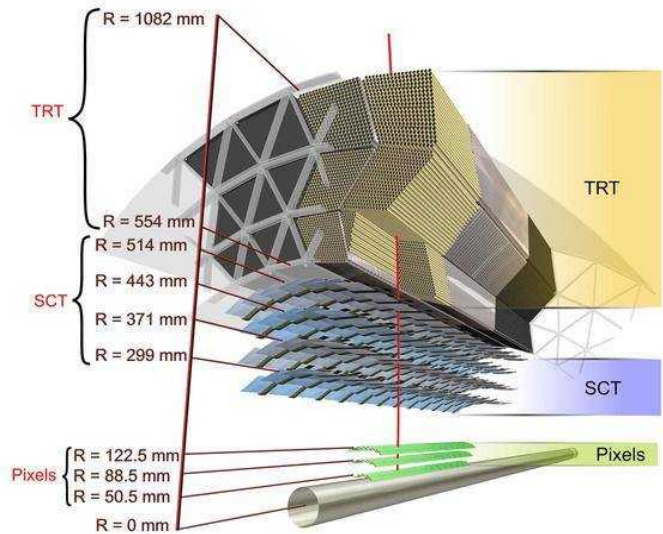

Figure 4.9.: The ATLAS inner detector in the barrel region. The figure shows the Pixel detector layout during Run-I. An additional pixel layer, called *IBL*, was added for Run-II © CERN.

### The Pixel detector and the semiconductor tracker

Both the Pixel detector and the semiconductor tracker (SCT) have sensors made out of silicon and work with the same principle: in most of the modules, the silicon semiconductor material is n-doped on one side and p-doped on the other side to create a p-n junction. A reverse bias voltage is applied to extend the depletion depth. Once an electrically charged particle passing the p-n junction creates electron-hole pairs via ionisation, the produced charges drift towards the electrodes and are collected there. The difference between the Pixel detector and the SCT is that the Pixel detector consists of silicon pixel sensors while the SCT consists of strip sensors.

The ATLAS Pixel detector [186] covers a pseudorapidity range of $|\eta| < 2.5$. Since it sits closest to the beam pipe, it is designed to withstand a large amount of radiation. The barrel region of the ATLAS Pixel detector consists of the IBL as the "new" innermost layer (see below) and three more layers of the "initial" Pixel detector initially installed for Run-I. In the end-cap regions, three Pixel detector discs are placed on each side. The "initial" Pixel detector consists of 1744 silicon pixel sensors of 250 $\mu$m thickness, with 46,080 readout channels per sensor and a nominal pixel size of $50 \times 400$ $\mu$m$^2$.

|  | Intrinsic accuracy ($\mu$m) |
|---|---|
| **IBL** | 10 ($R - \phi$) 75 ($z$) |
| **Pixel detector (without IBL)** | |
| Barrel | 10 ($R - \phi$) 115 ($z$) |
| Disks | 10 ($R - \phi$) 115 ($R$) |
| **SCT** | |
| Barrel | 17 ($R - \phi$) 580 ($z$) |
| Disks | 17 ($R - \phi$) 580 ($R$) |
| **TRT** | 130 |

Table 4.2.: Intrinsic measurement accuracies of the inner detector sub-systems. For the Pixel detector and the SCT, the single-module accuracies are shown. For the TRT, the drift-time accuracy of a single straw is shown [174, 183–185].

The Insertable $B$-Layer (IBL) [183,184] is an additional layer of the Pixel detector which was added during the long shutdown 1 (LS1) in 2014. It is located between the innermost layer ($B$-layer) of the "initial" Pixel detector and the beam pipe. It has an average radius of only 33 mm. For its installation during the upgrade, the beam pipe had to be replaced by a smaller one. The IBL allows for the improvement of tracking and vertex identification, as well as for the compensation of radiation damage in the other layers of the Pixel detector accumulated during Run-I. The improvement of tracking and vertex identification is achieved with the inclusion of the IBL by adding more track points, which are also closer to the interaction point than for the "initial" Pixel detector. Since during the operation of the LHC, radiation damage is causing some of the "initial" Pixel detector's modules to fail, the new IBL modules should balance out this effect. The sensors of the IBL in the more forward regions have the electrodes passing through the bulk of the silicon sensor, allowing a lower bias voltage, which improves the performance after irradiation of the sensor. These are called 3D-silicon pixel sensors. In the more central region, the electrodes are attached to the surfaces of the sensors, as in the "initial" Pixel detector, which is also known as the planar design. However, these sensors use an n-in-n technology, for better performance after irradiation. The IBL has a pseudorapidity range of $|\eta| < 3$. Figure 4.10 shows the Pixel detector layout including the IBL.

The semiconductor tracker (SCT) is the next part of the inner detector. It consists of 4088 silicon-strip detector modules with a total of 15,392 silicon sensors. The modules are arranged in four barrel layers and nine discs on each end-cap side. They are arranged in a way that tracks usually are identified within four layers of the SCT. The modules contain 285 $\mu$m thick n-type strip sensors with p-type implants. The SCT covers a pseudorapidity range of $|\eta| < 2.5$.

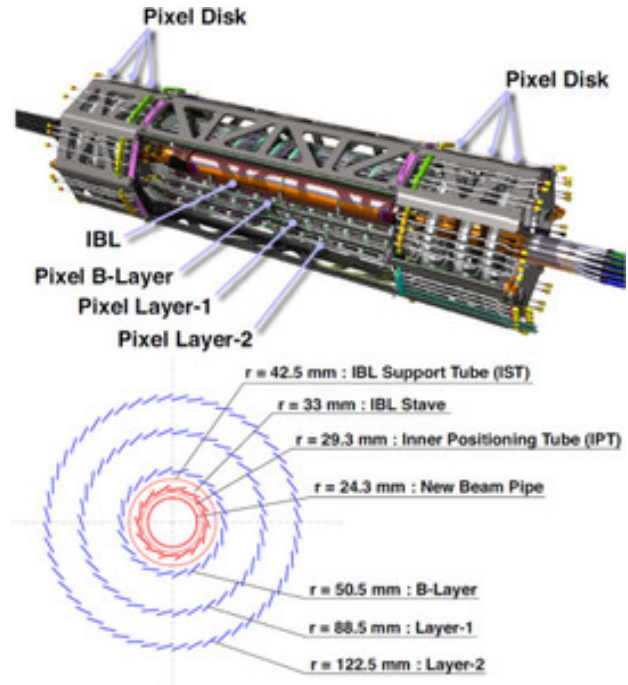

Figure 4.10.: Schematic view of the ATLAS Pixel detector, including the IBL © CERN.

**Transition Radiation Tracker**

The outermost part of the inner detector is the Transition Radiation Tracker (TRT). It is a straw-tube tracker, consisting of 300,000 proportional drift tubes with a diameter of 4 mm, filled with a $Xe/CO_2/O_2$ gas mixture. The anode is made of gold-plated tungsten. Charged particles passing through the gas leave a trail of ions. The produced electrons are then accelerated towards the anode due to the applied voltage, creating short, localised gas amplification avalanches close to the anode, where the charges are collected. Since this effect depends on the ratio $E/m$ between the energy and the mass of the particle, the TRT can be used for particle identification, for instance to differentiate between electrons and charged pions [187].

The TRT is divided into a barrel region and two end-cap regions. The barrel region covers a pseudorapidity range of $|\eta| < 1$ and consists of 1.5 m long straw tubes parallel to the beam axis. The end-cap TRTs on both sides consist of 0.4 m long straws placed perpendicular to the beam axis and cover the range of $1 < |\eta| < 2$. The TRT provides around 30 two-dimensional space points per charged particle track and a resolution of $\sim 130\ \mu$m.

### 4.2.4. Calorimetry

Calorimeters measure the energy of showers produced by particles passing through them. This concept can be described most easily for the electromagnetic showers. After the radiation length $X_0$, which is a material constant, high-energetic electrons lose on average all but $1/e$ of their energy via bremsstrahlung in the electromagnetic calorimeter[1]. The photons from the bremsstrahlung however split up into electron-positron pairs. In each instance of bremsstrahlung or pair production, the particles in the final state each carry half of the initial particle's energy. The two processes of bremsstrahlung and electron-positron pair production are repeated, until a critical energy $E_c$ is surpassed when electrons mostly lose their energy via ionisation processes.

The produced cascade of electrons, positrons and photons is called an electromagnetic shower. The length of the shower is logarithmically dependent on the energy of the incoming particle. Therefore, the energy of the incoming particle is determined by the structure of the shower. Electromagnetic calorimeters are built in a way that they cause these showers and detect them, for example via scintillation processes or ionisation. Hadronic showering works in a similar way, but due to the different nature of QCD, those processes cannot be described with such a simple example. For hadronic showers, the nuclear interaction length $\lambda$ takes over the role of the radiation length $X_0$.

The ATLAS calorimeter system (see Figure 4.11, left) is placed outside of the solenoid and is made out of sampling calorimeters. Sampling calorimeters work on the principle that absorbers induce the particle showers while active material is used to detect the shower particles. The calorimeter system is made out of different sections for the energy measurement of electromagnetically and hadronically interacting particles. The calorimeter system is also able to locate the clusters of deposited energy. It covers a pseudorapidity range of $|\eta| < 4.9$.

The calorimeter system is optimised in a way that electromagnetically and hadronically interacting particles deposit all of their energy in it, except for muons (see Section 4.2.5), shielding the muon spectrometer from additional particles as a side effect. This means that most energy from electrons and photons is deposited in the electromagnetic calorimeter, which has a thickness of more than 22 radiation lengths $X_0$ in the barrel and more than 24 $X_0$ in the end-caps. For hadronic interactions, the calorimeter system has a thickness of 11 $\lambda$ at $\eta = 0$, so that almost all energy from hadrons is expected to be deposited in the calorimeter system.

---

1. $e \approx 2.71828$ is Euler's number.

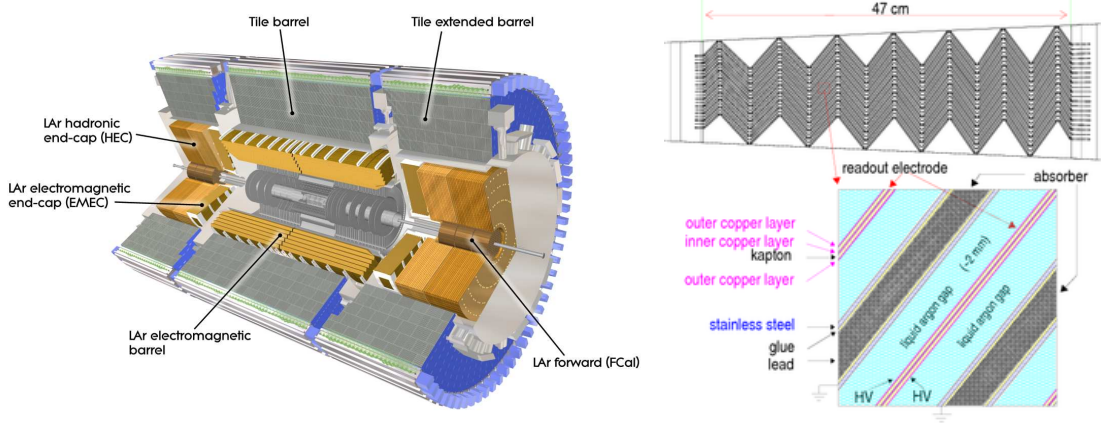

Figure 4.11.: Left: Cutaway view of the ATLAS calorimeter system © CERN.
Right: Accordion structure of the LAr electromagnetic calorimeter © CERN.

The energy resolution of the calorimeter system can be parametrised as

$$\frac{\sigma(E)}{E} = \frac{a}{\sqrt{E(\text{GeV})}} \oplus b\,. \tag{4.7}$$

For the electromagnetic calorimeter, the corresponding parameters are the stochastic term $a \approx 10\%$ and the constant $b \approx 0.7\%$, reflecting local non-uniformities in the calorimeter response. For the hadronic calorimeter, the parameters are $a \approx 50\%$ and $b \approx 3\%$ in the barrel and end-cap parts and $a \approx 100\%$ and $b \approx 10\%$ in the forward region of the detector system.

**LAr electromagnetic calorimeter**

The electromagnetic calorimeter consists of a part ($|\eta| < 1.475$) in the barrel region and two parts each in both of the end-cap areas ($1.375 < |\eta| < 2.5$ and $2.5 < |\eta| < 3.2$). To avoid as much dead material in the detector as possible, the electromagnetic calorimeter is housed in the same vacuum vessel as the central solenoid. The electromagnetic calorimeter is made out of liquid argon (LAr) with copper-kapton electrodes as the active material and lead absorber plates.

Shower particles ionise the argon atoms. Between the electrodes and the absorbers, a high voltage is applied. This causes the charges to drift towards the electrodes and the absorbers respectively. The electrodes are used to collect the charges drifting towards them to detect the shower particles.

To allow complete $\phi$ symmetry without any azimuthal cracks, the calorimeter layers are arranged in an accordion shape (see Figure 4.11, right). One part of the LAr forward calorimeters is also used for electromagnetic calorimetry, see below.

**Hadronic calorimeters**

The part of the ATLAS calorimeter system used for measuring the energy from hadronic showers is located after the electromagnetic calorimeter (from the interaction point outwards). It is divided into three different parts: a tile calorimeter for the barrel part, LAr end-cap calorimeters (HECs) at both sides and LAr forward calorimeters (FCal) at both sides to cover the high-$|\eta|$ regime.

The tile calorimeter consists of a central barrel region ($|\eta| < 1.0$) and two extended barrels at both sides to cover $0.8 < |\eta| < 1.7$. It consists of plastic scintillating tiles as the active material, together with steel absorbers. The scintillating light produced by the shower particles is transmitted to photomultiplier tubes via wavelength shifting fibres.

HECs are placed behind the end-cap electromagnetic calorimeters and share the same cryostat to reduce dead material in the detector. They cover a range of $1.5 < |\eta| < 3.2$ and are equipped with LAr, as an active material, and copper absorbers. The FCal covers the region of $3.1 < |\eta| < 4.9$ and is placed in the forward region of the detector. The first layer with copper as absorber is used for electromagnetic calorimetry, while the other two layers with tungsten as an absorber are used for hadronic calorimetry. An additional benefit of the forward detector is the shielding of the muon spectrometer against non-muon particles in that $\eta$ region.

### 4.2.5. Muon spectrometer

Muons and neutrinos (apart from other particles predicted by BSM theories) are the only particles that escape the sections of the ATLAS detector previously discussed. Since neutrinos cannot be detected by ATLAS, muons are the only particles that can be further investigated after leaving the hadronic calorimeter. The ATLAS muon spectrometer (shown in Figure 4.12) detects muons and measures the properties of their tracks bent in the toroidal magnet field, using high-precision tracking chambers. The measurement of the track bending is used to determine the transverse momentum and charge of the muons. The toroidal field is generated in a way that the magnetic field lines are mostly orthogonal to the muon tracks. The magnetic field of the large barrel toroid covers the pseudorapidity range of $|\eta| < 1.4$ and the magnetic field of the end-cap toroids cover the range of $1.6 < |\eta| < 2.7$. In the transition region between $1.4 < |\eta| < 1.6$, an overlap between both toroidal fields is used. The muon spectrometer also consist of trigger chambers, see Section 4.2.6. For muons with a transverse momentum of $p_T > 1$ TeV, the momentum resolution of the muon spectrometer is independent of the inner detector and approximately $\sigma(p_T)/p_T = 10\%$.

The muon spectrometer is made out of three layers of muon chambers in all detector regions. In the pseudorapidity range $|\eta| < 2.7$ for the outermost and mid layers and $|\eta| < 2.0$ for the innermost layer, muon tracks are measured using *Monitored Drift Tubes* (MDTs). MDTs are drift tubes pressurised with an $Ar/CO_2$ gas mixture using a tungsten-rhenium wire as an anode. For the range of $2.0 < |\eta| < 2.7$ in the innermost layer, multi-wire proportional chambers with the cathodes segmented into strips, called *Cathode Strip Chambers* (CSCs) are used. The CSCs have a higher granularity than the MDTs in order to cope with the higher expected background in that region.

The part of the muon spectrometer belonging to the trigger system consists of *Resistive Plate Chambers* (RPCs) and *Thin Gap Chambers* (TGCs) in the end-cap regions. RPCs are pairs of resistive plates with an electric field between them, allowing avalanches to form once a muon ionises the material between the two plates. TGCs are a variant of multi-wire proportional chambers. The RPCs and TGCs are installed in a range of $|\eta| < 2.4$ and are used for the identification of bunch-crossings and $p_T$ thresholds, see Section 4.2.6. They are also used for muon track measurements, complementary to the measurements done by the MDTs and CSCs.

During LS1, an upgrade with additional *small-diameter Muons Drift Tubes* (sMDTs) was started to improve the muon spectrometer performance [188]. These modules have half of the drift tube diameter of MDTs and fit in regions where the MDTs do not fit. Over the course of Run-II, the upgrade continues, improving the momentum resolution and also allowing more space for trigger chambers by replacing old MDTs with sMDTs.

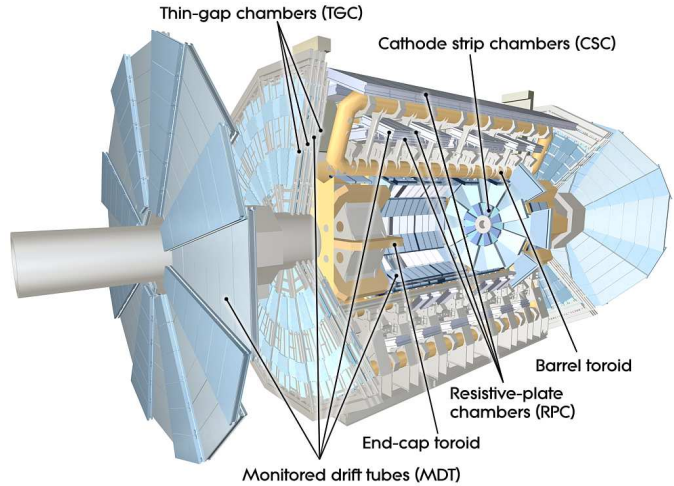

Figure 4.12.: Cutaway view of the ATLAS muon spectrometer © CERN.

## 4.2.6. The ATLAS trigger system

During the LS1, the ATLAS trigger system [174, 189, 190] was revised to be able to cope with the increased centre-of-mass energy, luminosity and pileup values (see Section 6.1). With a bunch crossing of 25 ns, the collision rate at the LHC is 40 MHz. The ATLAS trigger system selects data in a way that relevant events are stored at a frequency of approximately 1 kHz. The ATLAS trigger system consists of the hardware-based *level-1 trigger* (L1) and the software-based *high-level trigger* (HLT).

The L1 trigger reduces the event rate to approximately 100 kHz, using information from muons with high transverse momentum and objects depositing energy in the calorimeter system. It has a latency of ∼ 25 ns. For muons, information from dedicated trigger chambers (see Section 4.2.5) in the muon spectrometer is used (*L1Muon*). For other objects, calorimeter information with artificially reduced granularity is used (*L1Calo*). The L1 trigger defines *Regions-of-Interest* (RoIs), which define the locations in $\eta$ and $\phi$ where relevant features are identified. The RoIs also store additional information about the identified trigger features and the type of trigger requirements which are passed. The third trigger component, called *L1Topo*, was introduced for Run-II. It performs an additional selection based on kinematic and geometric relations between the trigger objects identified with L1Muon and L1Calo. Hardware-based track reconstruction information for each event accepted by the L1 is provided by the *Fast Tracker* (FTK) [191], which was introduced in 2015.

The trigger decision, RoI and FTK information from the L1 trigger is passed to the HLT. The HLT runs on a computing farm of around 40,000 processor cores, where an event recording rate of around 1 kHz is achieved. For the best performance in terms of physics and system capabilities, a so-called *seeded* and *stepwise* reconstruction approach is performed. The HLT makes use of information passed ("seeded") from the L1 trigger. Objects are rejected "stepwise" as soon as they fail one trigger requirement, instead of checking all requirements before rejecting the object, to save processing time. The information is processed in 2500 independent trigger chains, making use of RoIs seeded from the L1 trigger, but providing higher precision reconstruction. Reconstructed tracks from the FTK are used, since this approach is much faster than reconstructing tracks offline using CPU systems. The HLT is capable of reconstructing full events. Full event information will be written out as data streams for physics analyses.

Object Reconstruction and Selection

In the analysis presented in this thesis, electrons, muons, jets, $b$-jets and missing transverse momentum are considered as physics objects. Their definitions are presented in this chapter. Tau leptons are not considered as reconstructed objects but their decay products from the leptonic and hadronic decay modes are reconstructed as electrons, muons, jets and missing transverse momentum on their own.

Leptons from hadronic processes can be mistakenly reconstructed as isolated leptons from the initial event. These leptons are called *fake leptons from hadronic processes*. To determine their contribution using the matrix method, *loose* and *tight* electron and muon definitions are introduced. Section 6.3 discusses fake leptons and the matrix method in more detail. For the overlap removal between single final state objects shown in Section 5.7, leptons defined using the loose selection are used. Leptons selected with the tight selection will be called *tight leptons* and leptons selected by the loose selection will be called *loose leptons*, respectively.

## 5.1. Track and primary vertex reconstruction

The tracks of charged particles in the inner detector need to be reconstructed from hits in the Pixel detector[1] and the SCT, so that they can be used for the object definitions described below. This is done in several steps [192]. First, clusters of hits in the sensors are formed to identify three-dimensional space-points potentially caused by the trajectories of charged particles. In dense environments, a single cluster may contain hits from multiple charged particles, called a *merged cluster*, see Figure 5.1. Two different classes of merged clusters exist: those that are *identified as merged* clusters, and so-called *shared clusters*. The latter cannot be distinguished from single-particle clusters.

---

1. If not mentioned otherwise, this includes the IBL.

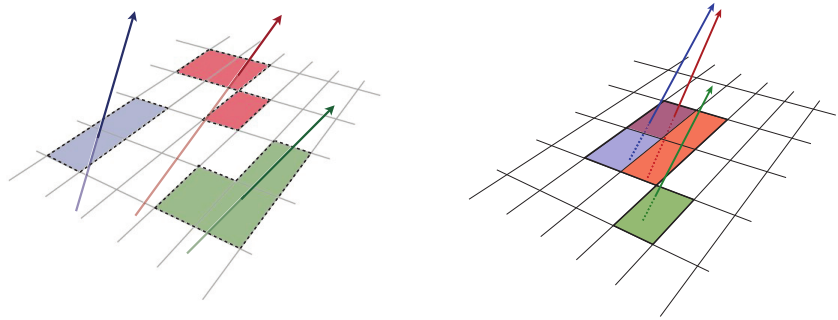

Figure 5.1.: Left: Single particle pixel clusters. Right: Merged pixel clusters due to highly collimated charged particles in the inner detector. Different particles (trajectories shown as arrows) and their corresponding energy deposits in the sensor are represented by different colours © CERN.

In the next step, *track seeds* are formed from groups of three space points and their momentum and impact parameter (displacement of the track with respect to the primary vertex, see below) values are determined. The track seeds are extrapolated into track candidates, based on their $p_T$, impact parameters and cluster locations, as well as on information of the additional cluster positions. Next, multiple reconstructed track candidates with shared particle clusters need to be removed based on several track quality requirements, such as the number of assigned clusters, number of holes (spots without a cluster where the track should have caused a cluster), the $\chi^2$ of the track fit and the track $p_T$. When two track candidates have a shared cluster, only the candidate with the highest track quality is kept. Merged clusters are identified using a Neural Network. For tracks selected with this method, a high-resolution fit is performed afterwards, taking into account more detailed information.

The position of the hard scattering process, called the *primary vertex*, needs to be precisely known, for instance for *b*-tagging and pileup[2] suppression (see below). For the reconstruction of the primary vertex [193,194], tracks reconstructed in the inner detector are used. These tracks are required to have a transverse momentum of $p_T > 400$ MeV, at least nine hits in the Pixel detector and the SCT for $|\eta| < 1.65$ and at least 11 hits for $|\eta| > 1.65$. The tracks need to have at least one hit in the first two pixel layers, at most one shared pixel hit or at most two shared SCT hits, exactly zero pixel holes and at most one SCT hole. Tracks fulfilling these requirements are assigned a seed position for a vertex candidate. The seed and the tracks are fitted to find the best vertex position. Once the fit has determined the optimal vertex position for vertices with at least two tracks, all tracks incompatible with the determined vertex position are removed. Afterwards, the fit is repeated with the remaining tracks, until no further tracks need to be removed or no vertex can be identified. The final vertex is associated with the primary vertex via the JVT method (see Section 5.4) and *b*-tagging techniques, see Section 5.5.

---

2. Pileup is discussed in detail in Section 6.1.

## 5.2. Electrons

Electrons are defined by clusters of deposited energy in the electromagnetic calorimeter associated to tracks in the inner detector [195]. These clusters are reconstructed within fixed-sized rectangles using the so-called *sliding-window algorithm* [196]. First, the deposited energy is divided by $\phi$ and $\eta$ into a grid of *tower energies*. Pre-clusters are formed by sliding a window of fixed size over the grid of towers. The window position that yields a local energy maximum from the towers inside that window is chosen to include the pre-cluster. Finally, the position and energy of the EM clusters are determined by iterating over the calorimeter layers within the pre-clusters.

The identification uses a likelihood-based selection, taking into account several parameters from the inner detector and the electromagnetic calorimeter [197,198]. A cut on the likelihood reduces the amount of fake leptons. For the tight selection, the cut is chosen to reject more fakes than for the loose selection. The prompt lepton efficiency is $\sim 80\%$ for the tight selection and $\sim 95\%$ for the loose selection for a transverse electron energy of $\sim 40$ GeV [199].

The electron isolation for the tight lepton definition requires the sum of the calorimeter transverse energies inside a cone of $\Delta R \equiv \sqrt{(\Delta\eta)^2 + (\Delta\phi)^2} = 0.2$ around the electron candidate to be less than 6% of the electron candidate's transverse momentum ($p_{\mathrm{T}}^{\mathrm{can}}$). It also requires the sum of track transverse momenta around the candidate inside a cone of $\Delta R = \min(10 \text{ GeV}/p_{\mathrm{T}}^{\mathrm{can}}, 0.2)$ to be less than 6% of $p_{\mathrm{T}}^{\mathrm{can}}$. These two isolation requirements are found to be most effective in order to suppress fake leptons and are dropped for the loose selection.

In addition to the identification and isolation criteria, the absolute value of the pseudorapidity of the calorimeter energy deposit $|\eta_{\text{cluster}}|$ is required to be less than 2.47. Candidates in the transition area between the barrel and the end-cap part of the electromagnetic calorimeter at $1.37 < |\eta_{\text{cluster}}| < 1.52$ are excluded. The electrons are also required to have a minimum transverse momentum of $p_{\text{T}} > 7$ GeV.

Electrons also need to fulfil the recommended impact parameter cuts of $|d_0|/\sigma(d_0) < 5$, $z_0 \sin(\theta) < 0.5$ mm, where $d_0$ is the transverse impact parameter of a track (and $\sigma(d_0)$ its resolution), defined as the distance between the primary vertex and the point of closest approach in the r-$\phi$ plane to a track. The parameter $z_0 \sin(\theta)$ is the longitudinal impact parameter of a track, defined as the distance of a track to the point of closest approach in the r-$\phi$ plane.

The electron charge can be misidentified if the electron emits a photon via bremsstrahlung, which subsequently decays into an electron-positron pair, and the charge of the resulting positron is misidentified as the charge of the initial electron. This effect is called a *charge-flip*. To reduce the number of charge-flips, the output of a boosted decision tree, trained to discriminate against these objects using electron cluster and track properties, is used. A cut on the output is chosen in a way that 97% of electrons with the proper charge are kept and that the rejection factor for electrons with a wrong charge assignment is between 7 and 8.

## 5.3. Muons

Muons are reconstructed from track segment information in the layers of the muon spectrometer and tracks in the inner detector. For the muon identification, track quality requirements are imposed to veto against fakes from pion and kaon decays. Only muons with $|\eta| < 2.5$ are selected, which are reconstructed independently for the muon spectrometer and the inner detector and are then combined in a global refit of hits from both sub-detectors. This is done in a way that a muon efficiency of 96.1% for muons with $p_{\text{T}} > 20$ GeV is achieved, with a fake muon efficiency of 0.17% in the same momentum regime [200].

In addition, muons are required to have transverse momenta of $p_{\text{T}} > 7$ GeV, as well as impact parameters of $|d_0|/\sigma(d_0) < 3$ and $z_0 \sin(\theta) < 0.5$ mm. For the tight selection cut, the sum of the track transverse momenta within the cone of $\Delta R = \min(10 \text{ GeV}/p_{\text{T}}^{\text{can}}, 0.3)$ around the muon candidate cannot be more than 6% of the muon $p_{\text{T}}$. For the loose selection, this requirement is dropped.

## 5.4. Jets

Due to QCD confinement, the production of quarks and gluons at the ATLAS detector causes a cascade of hadronisation processes, resulting in a stream of particles, called a *jet*, see Section 2.1.4. Jets in the analysis presented in this thesis are reconstructed using the anti-$k_T$ jet clustering algorithm [201], starting with topological clusters in the calorimeters [202].

Topological clustering algorithms merge together neighbouring calorimeter cells as long as the signal in those cells is significant compared to noise. Cluster seeds are chosen from calorimeter cells where the ratio $\varsigma$ between the deposited energy and the energy from noise is $\varsigma > 4$. Neighbouring cells with $\varsigma > 2$ are merged into these clusters. The results are three dimensional clusters with cell cores containing highly significant signals. Another result of this technique is the suppression of calorimeter noise which also partially suppresses pileup, see Section 6.1.

The anti-$k_T$ algorithm defines the following distance measure for two objects $i$ and $j$ (such as clusters) that potentially belong to a jet:

$$d_{ij} = \min\left(\frac{1}{p_{Ti}^2}, \frac{1}{p_{Tj}^2}\right) \frac{(\eta_i - \eta_j)^2 + (\phi_i - \phi_j)^2}{R^2}, \qquad (5.1)$$

where $p_{Ti}$ is the transverse momentum of the $i^{\text{th}}$ particle, $\eta_i$ and $\phi_i$ its pseudorapidity and azimuthal angle and $R$ is the radius parameter. The algorithm identifies the minimum distance value $d_{\min}$ of all $d_{ij}$. If $d_{\min}$ is below a certain threshold $d_{\text{cut}}$, particles $i$ and $j$ are combined into a new particle called *pseudojet*. This step is repeated until there are no cases left where $d_{\min}$ is below $d_{\text{cut}}$. All remaining particles are then considered to be jets by the algorithm. For the analysis presented in this thesis, the radius parameter is chosen to be $R = 0.4$. An example of jet clustering using the anti-$k_T$ algorithm is visualised in Figure 5.2. A benefit of the anti-$k_T$ algorithm is that it is safe against infrared and ultraviolet divergences of jets from Monte Carlo events, since the number of hard anti-$k_T$ jets is unaffected by soft gluon emissions and collinear splitting.

For further pileup suppression, the *Jet vertex tagger* [203, 204] (JVT) is used. This tagger uses jet track information to differentiate between jets from the hard scattering process and pileup. Two track variables are included in the JVT. The first variable is the *corrected jet vertex fraction* (corrJVF). It is the fraction of the sum of $p_T$ from tracks in a jet associated with the primary vertex over the total momentum of that jet, corrected to the average scalar $p_T$ sum of pileup jets dependent on the number of vertices. The second track variable is the ratio $R_{pT}$ between the sum of $p_T$ from tracks in a jet associated with the primary vertex and the fully calibrated jet $p_T$ including pileup subtraction. Both corrJVF and $R_{pT}$ are utilised in a two-dimensional likelihood which defines the JVT. A cut on the JVT output is chosen in a way that a 92% efficiency for jets from the hard scattering event is achieved.

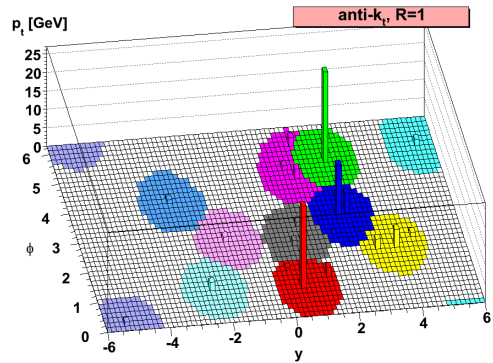

Figure 5.2.: Example of jet clustering using the anti-$k_T$ algorithm. The coloured areas show the clustered jets from that algorithm [201].

The energy of a reconstructed jet is calibrated in terms of the properties of the energy depositions in the calorimeter system, called *jet energy scale* (JES) calibration. This is done by applying $p_T$ and $\eta$ dependent corrections, derived from Monte Carlo simulations, after removing the pileup contributions. Additional calibrations from in-situ measurements are taking into account the difference in the jet response between data and Monte Carlo [205, 206]. The jet energy resolution is determined in in-situ measurements of $Z \to ee/\mu\mu$+jets and $\gamma \to ee/\mu\mu$+jets, see Section 8.3.4. A minimum transverse momentum and a maximum pseudorapidity of $p_T^{\text{jet}} > 25$ GeV and $\eta^{\text{jet}} < 2.5$ are required.

## 5.5. *b*-tagging

Identifying *b-jets*, which are jets from hadrons containing bottom quarks[3], is crucial for analyses with top quarks, since the top quark decays in almost 100% of all cases into a $W$ boson and a bottom quark. For this identification, the relatively long lifetime of *b*-hadrons of $\sim 1.5$ ps is exploited. These particles travel a distance of several hundred $\mu$m in the beam pipe before decaying, depending on their momenta (see Figure 5.3). This allows to tag their decay vertex as a *secondary vertex* of the process. Information from the inner detector, mainly from the Pixel detector, is used to distinguish the secondary vertices from the primary vertex.

---

3. These hadrons are called *b-hadrons*.

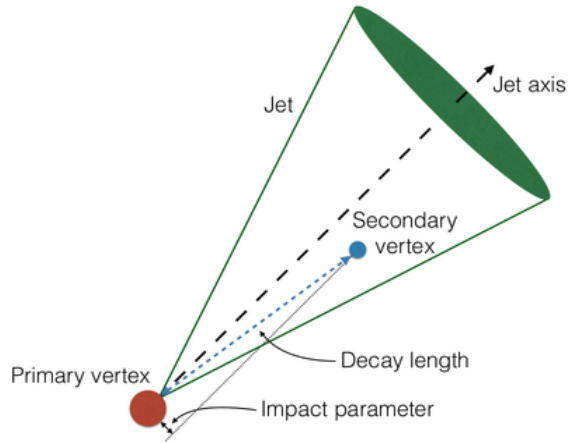

Figure 5.3.: Secondary vertex displacement of a *b*-jet © CERN.

For the analysis presented in this thesis, the multivariate-based algorithm `MV2c10` [168, 207] is used, which utilises the output of several *b*-tagging algorithms to train a Boosted Decision Tree (BDT): two impact-parameter based algorithms (`IP2D` and `IP3D`) [208], a secondary vertex finding algorithm [208] and a decay-chain multi-vertex algorithm called `JetFitter` [209]. In addition, the variables jet $p_T$ and $\eta$ are used in the training.

The algorithms `IP2D` and `IP3D` make use of the impact parameters (see Section 5.2) and the hit patterns of charged particles in the Pixel detector and the SCT. The `IP2D` algorithm uses the transverse impact parameters $d_0$ of tracks. Therefore, the `IP2D` algorithm performs *b*-tagging in a two-dimensional plane. The `IP3D` algorithm takes into account $d_0$, but also the longitudinal impact parameters $z_0 \sin(\theta)$ of the tracks. Consequently, the `IP3D` algorithm performs *b*-tagging in the three-dimensional space. In both cases, Monte Carlo events are used to determine likelihoods in order to differentiate between *b*-jets and non-*b*-jets.

The secondary vertex algorithm reconstructs displaced secondary vertices within jets. All jets are checked if two tracks could potentially form a secondary vertex, which does not come from a long-lived particle such as $K_s$ or $\Lambda$. If the requirement is fulfilled together with some additional quality requirements, a secondary vertex is reconstructed from additional tracks close to the two-track vertex.

The decay-chain multi-vertex algorithm `JetFitter` uses the topological structure of electroweak decays of *c*- and *b*-hadrons inside a jet to reconstruct the complete *b*-hadron decay chain.

For the training of the BDT, the background samples are composed of 10% *c*-flavour jets (thus the "c10" in `MV2c10`) and of 90% light-flavour jets. A cut on the `MV2c10` output to yield a 77% efficiency of tagging *b*-jets from $t\bar{t}$ is found to be the best fitting working point. For that cut, a *c*-jet rejection factor of 6, a light-jet rejection factor of 134 and a rejection factor for jets from tau leptons of 22 is achieved. A tighter requirement would have removed too many possible signal events while a looser cut would have resulted in too much background contamination. Tagged *b*-jets from Monte Carlo events are calibrated to match the expectation in data.

## 5.6. Missing transverse momentum

Missing transverse momentum $E_{\mathrm{T}}^{\mathrm{miss}}$ is defined as the magnitude of the transverse momentum vector which quantifies the transverse momentum imbalance of all detectable momenta. Since neutrinos will pass the ATLAS detector without interacting with any of its sub-detectors, events containing neutrinos will contain $E_{\mathrm{T}}^{\mathrm{miss}}$ due to energy-momentum conservation. Particles hitting a part of the ATLAS detector which is not covered by any sub-detector, or badly reconstructed objects, can also cause missing transverse momentum. In addition, several theories for BSM physics predict particles that do not interact with the detector, which is not taken into account for this analysis. The missing transverse momentum is calculated using the *track-based soft term* (TST) approach, which is the current ATLAS recommendation [210]. The TST $E_{\mathrm{T}}^{\mathrm{miss}}$ is calculated as the absolute value of the sum of the transverse momentum vectors of all reconstructed objects, taking into account calorimeter information. An additional soft term is added, which is defined as the sum of all transverse momenta of tracks that are not already matched to any physics object but are associated to the primary vertex [211]. Alternative $E_{\mathrm{T}}^{\mathrm{miss}}$ definitions are the *calorimeter-based soft term* (CST, primary $E_{\mathrm{T}}^{\mathrm{miss}}$ definition during Run-I) approach and *Track* $E_{\mathrm{T}}^{\mathrm{miss}}$. The CST $E_{\mathrm{T}}^{\mathrm{miss}}$ is similar to the TST $E_{\mathrm{T}}^{\mathrm{miss}}$ but uses calorimeter information for the soft term instead of tracking information. It is very sensitive to pileup. The Track $E_{\mathrm{T}}^{\mathrm{miss}}$ however uses only momentum information from the inner detector. This method is very stable against pileup but is insensitive to electrically neutral particles which leave no track in the inner detector. The TST $E_{\mathrm{T}}^{\mathrm{miss}}$ is a combination of both the CST and the Track $E_{\mathrm{T}}^{\mathrm{miss}}$, combining their advantages and cancelling out their disadvantages [210].

The $E_{\mathrm{T}}^{\mathrm{miss}}$ scale for the TST $E_{\mathrm{T}}^{\mathrm{miss}}$ is determined from $Z \to \mu\mu$ events. A measure of this scale is the mean value $\langle \vec{E}_{\mathrm{T}}^{\mathrm{miss}} \cdot \vec{A}_Z \rangle$ of the direction of the reconstructed missing transverse momentum $\vec{E}_{\mathrm{T}}^{\mathrm{miss}}$ projected onto the momentum axis of the $Z$ boson $\vec{A}_Z$.

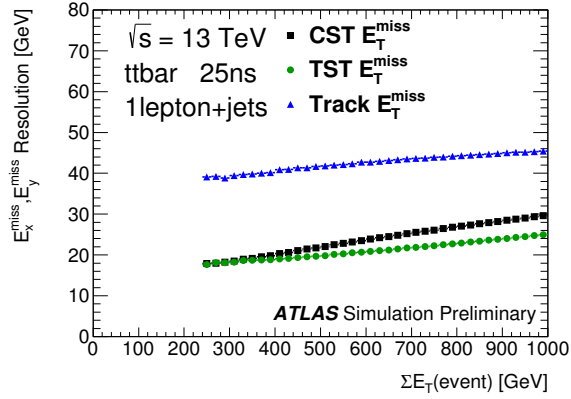

Figure 5.4.: Comparison of the $E_\mathrm{T}^\mathrm{miss}$ resolution (RMS) between TST $E_\mathrm{T}^\mathrm{miss}$, CST $E_\mathrm{T}^\mathrm{miss}$ and Track $E_\mathrm{T}^\mathrm{miss}$ in $t\bar{t}$ Monte Carlo events. The resolution is shown as a function of the scalar sum of transverse momenta, including the soft term, using the CST method [210].

This component is sensitive to biases in the detector response. Ideally, this would be zero since no neutrinos are expected in this process. Possible reasons of non-zero values are the contributions from soft neutral particles and the limited acceptance in the inner detector [211]. The $E_\mathrm{T}^\mathrm{miss}$ resolution is determined from $Z \to \mu\mu$, $W \to \mu\nu$ and $t\bar{t}$ events from Monte Carlo simulations. The transverse momentum using the TST has the best resolution, which is shown for example in Figure 5.4 [210].

## 5.7. Overlap removal

Double-counting between the reconstructed objects, as defined above, can occur, for example by using the same energy deposits in the electromagnetic calorimeter to reconstruct both a jet and an isolated electron. Therefore, an overlap removal procedure between the reconstructed objects is performed. Leptons with the loose object definition are used for this task. The result is extrapolated to the tight object definition.

**Electron-jet overlap removal** A jet candidate is removed if an electron candidate is reconstructed at a distance of $\Delta R < 0.2$. In case there are multiple jets within that distance, only the closest jet candidate is removed. If the distance between a jet candidate and an electron candidate is $0.2 < \Delta R < 0.4$, the electron candidate is removed.

*5. Object reconstruction and selection*

**Muon-jet overlap removal**   If a jet candidate shares two or more tracks in the inner detector with a muon candidate and the distance to the muon is $\Delta R < 0.4$, the muon is dropped. If a jet candidate at a distance of $\Delta R < 0.4$ shares one track or fewer with a muon candidate, the jet candidate is removed.

**Lepton-lepton overlap removal**   An electron candidate is removed if it shares a track with a muon candidate in the inner detector.

## Measured Data, Signal and Background Samples

This analysis uses data taken with the ATLAS detector during the years 2015 and 2016, corresponding to an integrated luminosity of 36.1 fb$^{-1}$, which is further discussed in Section 6.1. In order to model most of the processes contributing to the signal and control regions, defined in Chapter 7, Monte Carlo simulations are used, see Section 6.2. For the fake lepton estimation, the data driven matrix method and the semi data driven fake factor method are used, as shown in Section 6.3.

## 6.1. ATLAS dataset of 2015 and 2016

For this analysis, data from proton-proton collisions with stable beams taken with the ATLAS detector at a centre-of-mass energy of 13 TeV are used. The data were taken between June and November 2015, as well as between April and November 2016, and are classified into runs corresponding to the data taking periods of the ATLAS data acquisition system. These runs are further divided into *luminosity blocks* (LBs) corresponding to a few minutes of data taking each. The quality of the LBs is ensured by so-called *good run lists* (GRLs) which contain information on whether or not a LB is suitable for a certain type of analysis. For instance, some analyses might not rely on a certain part of the detector, so that such analyses can use LBs in which that part of the detector is switched off. The analysis presented in this thesis, however, needs the full capability of the ATLAS detector, so only LBs with all sub-detectors and the magnet system switched on are used. This also requires the IBL to be included in the data taking runs, since the analysis heavily relies on *b*-tagging. LBs in which the LAr and tile calorimeters are in an error state or the SCT is in recovery mode are removed. Only runs for which the bunch crossing frequency is 25 ns are taken into account. Consequently, data taken during proton-proton runs with a bunch crossing frequency of 50 ns are not used for this analysis.

Using the GRL requirements discussed above, the dataset used for this analysis corresponds to $3.2 \pm 0.07$ fb$^{-1}$ of data taken during 2015 and $32.9 \pm 0.72$ fb$^{-1}$ of data taken during 2016. In 2015 and 2016, the luminosity and its uncertainty was determined with the *LUCID-2* system [212]. The key elements of this system are Cherenkov detectors placed around the beam-pipe on both forward ends of the ATLAS detector, which detect hits from particles travelling with a small angle with respect to the beam pipe [173]. The integrated luminosity for 2015 and 2016 as a function of time is shown in Figure 6.1. The instantaneous peak luminosity per beam fill as a function of time is shown in Figure 6.2. Over the course of the two years, the performance of the data taking in terms of luminosity has increased significantly, so that during 2016, approximately ten times as much data was obtained as for 2015.

A major challenge when working with hadron collisions at a high frequency are additional detector hits from *pileup*. Pileup consists of additional proton-proton collisions during the same bunch-crossing or bunch-crossings shortly before and after it (*in-time* and *out-time* pileup), of neutrons and photons from the gas inside the cavern (*cavern background*), of collisions from protons with a collimator (*beam halo events*) and of collisions between a proton bunch and residual gas inside the beam-pipe (*beam gas events*) [213]. For this analysis, in-time and out-time pileup are the main contributors to this effect. Therefore, the term "pileup" is used synonymously for "number of interactions per bunch crossing" in this thesis. The pileup profile of collisions from 2015 and 2016, displayed in Figure 6.3, shows that the amount of pileup in 2016 has increased significantly with respect to 2015. The pileup contribution due to jets is reduced using the jet vertex tagger (JVT), see also Section 5.4.

All events are required to have a primary vertex, as described in Section 5.1. Events with non-collision background and cosmic events, leading to falsely reconstructed jets called *fake jets*, or events containing jets falsely reconstructed from calorimeter signals caused by noise are removed using the JVT tool.

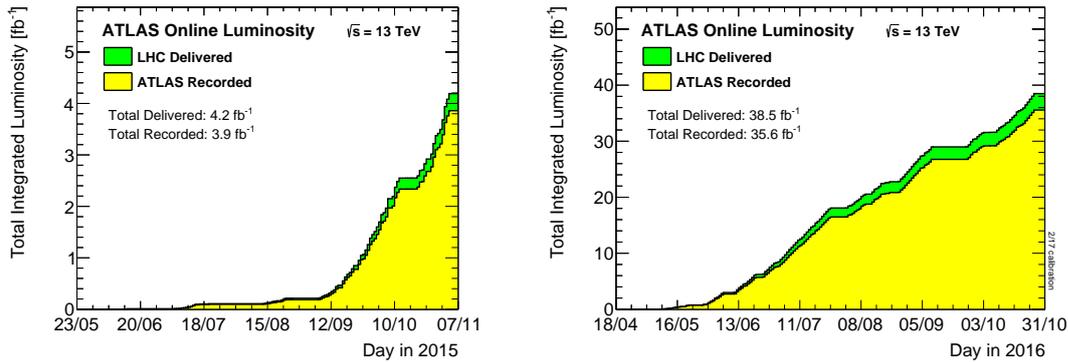

Figure 6.1.: Integrated luminosity from proton-proton collisions at a centre-of-mass energy of 13 TeV at the ATLAS detector as a function of time for the year 2015 (left) and the year 2016 (right). The sum of the green and yellow parts of the histograms show the delivered luminosity from the LHC while the yellow parts shows the fraction that was actually recorded by ATLAS. *These figures are provided by the ATLAS Luminosity Working Group.*

## 6.2. Monte Carlo samples

Since elementary particle interactions follow the rules of quantum mechanics, their occurrence and properties are based on probabilities. This can be best modelled using computational algorithms relying on random event generation, called *Monte Carlo generators*. They play a crucial role for most of the physics analyses using the ATLAS detector and give predictions about the event yields, decay signatures and kinematic distributions of physics processes. Monte Carlo samples are used to determine the optimal cuts on a dataset in order to select as many signal events as possible while rejecting many background events, see Chapter 7. Also, they are used for fits to data to determine production cross sections, see Chapter 11.

### 6.2.1. Production of Monte Carlo samples for ATLAS

The generation of the Monte Carlo samples used in this analysis is done in several steps as shown in this section. The Monte Carlo normalisation and event weighting is done within the analysis and is not a part of the event simulation, which is done in central productions for all ATLAS analyses.

**Parton distribution functions**   Since protons are composite particles, the gluons and quarks inside the protons are the particles actually that interact with each other in hadron collisions. Therefore, the momenta of the colliding partons from the *pp* collision have to be determined. This is done with so called *parton distribution functions* (PDFs). PDFs define the probability density of a parton to carry a certain fraction of the proton momentum. These distributions heavily rely on the centre-of-mass energy of the collision. Typical PDFs used for Monte Carlo samples at ATLAS for 13 TeV analyses are `NNPDF3.0NLO` [158], `CTEQ6L1` [214] and `CT10` [215]. The PDFs must be considered when calculating the hard scattering process.

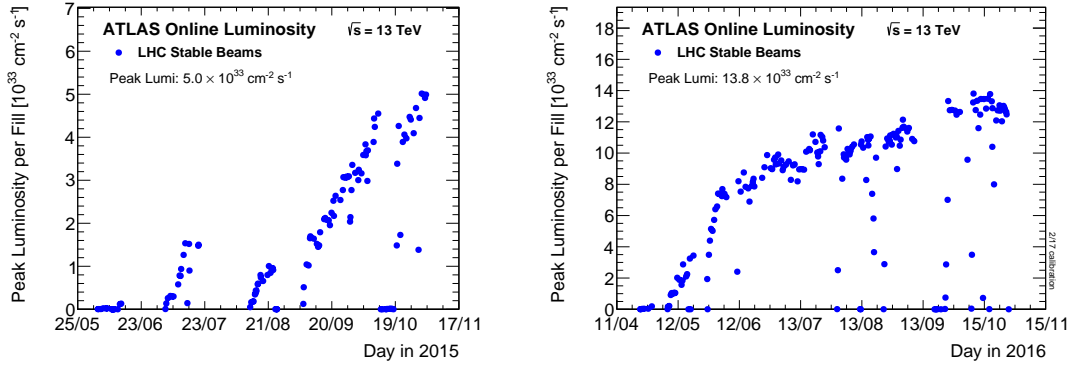

Figure 6.2.: Peak luminosity per fill from stable proton-proton collisions at a centre-of-mass energy of 13 TeV at the ATLAS detector as a function of time for the year 2015 (left) and the year 2016 (right). *These figures are provided by the ATLAS Luminosity Working Group.*

**Hard scattering process**  In this step, the matrix element of the hard scattering process is calculated from the Feynman diagrams up to a certain order in perturbation theory. According to the amplitude corresponding to this calculation, random events are generated. These random events contain elementary particles, such as charged leptons, gauge and Higgs bosons, neutrinos and quarks, together with their corresponding properties. Typical Monte Carlo generators for the hard scattering process are MAD-GRAPH5_aMC@NLO [156] and SHERPA [216].

**Parton showering and hadronisation**  Due to the high energy regime of the simulated processes, the produced partons are repeatedly radiating off gluons due to QCD bremsstrahlung, causing a cascade of strongly interacting particles. Modelling this process is called *parton showering*. Typical Monte Carlo generators used for parton showering are PYTHIA6 [217] and PYTHIA8 [157]. Due to QCD confinement, the produced quarks and gluons are not allowed to exist as free particles and are always bound into hadrons, see Section 2.1.4. The process of forming hadrons from the previously produced quarks and gluons is called *hadronisation*. There are several models which describe hadronisation, such as the *string model* [59] used by PYTHIA and the *cluster model* [218] used by HERWIG [219]. Unstable hadrons produced during hadronisation further decay into other particles until stable particles are formed. The result of showering and hadronisation is a stream of particles, called a *jet*, see also Section 5.4.

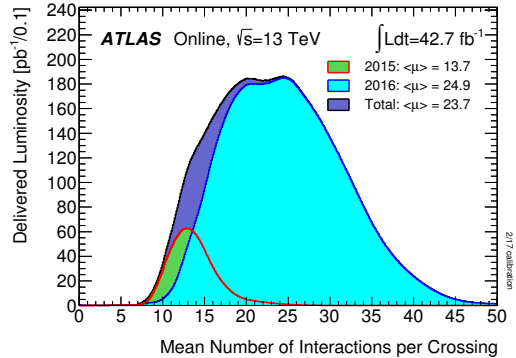

Figure 6.3.: Distribution of the mean number of interactions per bunch crossing for stable beams during proton-proton collisions with a centre-of-mass energy of 13 TeV at the ATLAS detector for 2015 and 2016. The distributions for 2015 and 2016 are weighted by the integrated luminosity of the corresponding dataset. *This figure is provided by the ATLAS Luminosity Working Group.*

**Monte Carlo tuning**  Various low-energy QCD interactions like multiple parton interactions, initial- and final-state radiation, as well as fragmentation processes are hard to predict in theory due to the increase of the strong coupling constant at low energies. Therefore, there are plenty of free parameters in the Monte Carlo production, e.g. in the parton showering, that need to be tuned with respect to data. Common Monte Carlo tunes are A14 [159] and Perugia2012 [220].

**Detector simulation**  The interaction of the generated stable particles with all parts of the ATLAS detector needs to be simulated. This includes the hits in active parts of the detector, as well as in passive material such as cables and steel beams. The ATLAS detector is usually simulated using the software package GEANT4 [221, 222]. The calorimeter response is either also simulated using GEANT4 or using a faster parametrisation in some cases, called *ATLAS fast simulation II* (AFII), which simplifies the longitudinal and lateral energy profiles of showers [223].

**Monte Carlo normalisation and event weights**  The number of generated Monte Carlo events depends on the restrictions from the computing infrastructure and the relevance of the corresponding sample. Therefore, the number of those Monte Carlo events does not reflect the expected yields. The Monte Carlo yields are matched to the expectation by scaling the simulated events according to the corresponding cross section, cut efficiency and the integrated luminosity of the dataset. The Monte Carlo events are also reweighted to data with respect to the object identification, object reconstruction, trigger efficiencies, energy scales and energy resolutions. The generated pileup profile also needs to be reweighted in terms of Monte Carlo events, according to the actual pileup profile in data.

| Sample | Generator | ME PDF | Shower | Normalisation | Cross section [pb] |
|---|---|---|---|---|---|
| $t\bar{t}Z$ $(Z \to \ell^+\ell^-)$ | aMC@NLO | NNPDF3.0NLO | Pythia8 | NLO | 0.12 |
| $t\bar{t}W$ | aMC@NLO | NNPDF3.0NLO | Pythia8 | NLO | 0.60 |
| $WZ$ $\to \ell\ell\ell\nu + \text{jets}$ | Sherpa2.1 | CT10 | Sherpa2.1 | NLO | 4.57 |
| $ZZ$ $\to \ell\ell\ell\ell + \text{jets}$ | Sherpa2.1 | CT10 | Sherpa2.1 | NLO | 1.05 |
| $tZ$ | MadGraph | CTEQ6L1 | Pythia6 | LO | $\sim$0.24 |
| $tWZ$ | aMC@NLO | NNPDF3.0NLO | Pythia8 | NLO | $\sim$0.015 |
| $t\bar{t}H$ | aMC@NLO | NNPDF3.0NLO | Pythia8 | NLO | $\sim$0.51 |

Table 6.1.: Details of the most relevant Monte Carlo samples.

### 6.2.2. Monte Carlo samples used in this analysis

This section lists the processes simulated by Monte Carlo generators. They are sorted as they are shown in the yield tables in subsequent chapters. Information concerning the Monte Carlo generators is provided. The details of the most relevant Monte Carlo samples are also listed in Table 6.1. The top quark mass is set to 172.5 GeV and the Higgs boson mass is set to 125 GeV in all samples. Pileup is simulated using Pythia8 and the MSTW2008LO [224] PDF set which is then superimposed to the hard scattering events. The decay of hadrons with heavy flavour quarks is modelled using the EvtGen program [225], except for the processes simulated with the Sherpa generator.

**Monte Carlo samples for the $t\bar{t}Z$ and $t\bar{t}W$ processes**   For the $t\bar{t}Z$ modelling, only samples with the $Z$ boson decaying into two electrons, muons or tau leptons are considered since only these decay modes are contributing to the channel presented in this thesis. All $t\bar{t}$ decay modes are considered. The $t\bar{t}W$ sample includes all possible decay modes of the $W$ boson and the top quark pair. The hard process is generated at next-to-leading-order (NLO) using MadGraph5_aMC@NLO with the NNPDF3.0NLO parton distribution function. Partons are showered using Pythia8, and EvtGen is used for heavy flavour decays. To tune the Monte Carlo to data, the A14 tunes are applied. The $t\bar{t}Z$ and $t\bar{t}W$ samples are normalised to the NLO QCD and electroweak predictions of $\sigma_{t\bar{t}Z,\mathrm{lep}} = 0.12$ pb and $\sigma_{t\bar{t}W} = 0.60$ pb, respectively [155]. In the $t\bar{t}Z$ sample, off-shell $Z$ bosons, photons and their interference are included with a minimum mass requirement of 5 GeV to avoid divergences.

**Monte Carlo samples for the diboson processes $WZ$ and $ZZ$** The process of a pair of a $W$ and a $Z$ bosons decaying into three changed leptons and a neutrino ($\ell\ell\ell\nu$) is the most important background that is described using Monte Carlo events. The Monte Carlo samples also contain events with additional jets. It will be referred to from here on as the $WZ$ process. The decay of a $Z$ boson pair into four charged leptons ($\ell\ell\ell\ell$) together with additional jets will further be referred to as the $ZZ$ process. The matrix element, containing all Feynman diagrams with four electroweak vertices, is calculated using Sherpa2.1, which is also used for the simulation of the showering. The PDF used for these samples is CT10. A dedicated parton shower tuning is provided by the Sherpa authors.

The Matrix element of events with up to one additional parton (in addition to the ones listed above) are generated at NLO and events with up to three additional partons are simulated at leading order (LO). For these matrix element calculations, Comix [226] and OpenLoops [227] are used. They are matched with the Sherpa showering using the ME+PS@NLO prescription [228]. For parton shower matching, see also Section 8.5.1. The $WZ$ and $ZZ$ events are scaled by their NLO QCD calculations of $\sigma_{WZ\to\ell\ell\ell\nu+\mathrm{jets}} = 4.57$ pb and $\sigma_{ZZ\to\ell\ell\ell\ell+\mathrm{jets}} = 1.05$ pb, respectively [229].

**Monte Carlo samples for the $tZ$ process** The matrix element of the $t$-channel single top production in association with a $Z$ boson ($tZ$) is calculated at leading order in the four-flavour scheme (see Section 2.2.2) using MadGraph [230]. For the PDF, CTEQ6L1 is used and the showering is performed using Pythia6. The Monte Carlo tuning is performed using the Perugia2012 tunes. The events are normalised to a cross section of $\sim 0.24$ pb, which is calculated at leading order QCD.

**Monte Carlo samples for the $tWZ$ process** The matrix element for the $Wt$ channel single top production in association with a $Z$ boson (called $tWZ$) is calculated at NLO using MadGraph5_aMC@NLO. The PDF NNPDF3.0NLO is used. The showering is simulated using PYTHIA8 and EVTGEN is used for heavy flavour decays. Monte Carlo tuning is performed using the `A14` tunes. The events are normalised to the NLO QCD cross section prediction of $\sim 0.015$ pb.

**Monte Carlo samples for the $t\bar{t}H$ process**  The $t\bar{t}H$ samples include all $t\bar{t}$ and Higgs boson decay channels. The matrix element is calculated at NLO using the MAD-GRAPH5_aMC@NLO generator with the `NNPDF3.0NLO` PDF. The showering is simulated using PYTHIA8 and EVTGEN is used for heavy flavour decays. For Monte Carlo tuning, the `A14` tunes are used. The process is normalised to its NLO QCD prediction of $\sim 0.51$ pb.

**Monte Carlo samples for processes with smaller contributions**  Other samples with minor background contribution to this analysis are the following ones:

- The production of a top quark pair in association with two $W$ bosons ($t\bar{t}WW$) and the production of three or four top quarks are generated with MADGRAPH, using the `NNPDF2.3LO` PDF [231], and showered with PYTHIA8.

- The production of three heavy gauge bosons in any combination of $W$ and $Z$ bosons are generated using SHERPA2.1 and the `CT10` PDF.

- The Higgs boson production via the radiation from a $W$ or $Z$ boson is generated using PYTHIA8 for both the hard scattering process and the showering, together with the `NNPDF2.3LO` PDF.

- The Higgs boson production via gluon-gluon fusion and its subsequent decay into two $Z$ bosons, decaying into charged leptons, is generated using POWHEG-BOX v2 [232], with the `CT10` PDF, and showered using PYTHIA8.

## 6.3. Estimation of the fake lepton background

For some contributions, Monte Carlo samples do not describe the respective physics processes sufficiently. This is the case for events containing *fake leptons* from hadronic processes (also called *hadronic fake leptons*), which are one of the main backgrounds for this analysis. These leptons are the decay products of heavy flavour hadrons (containing charm or bottom quarks), pions or kaons inside a jet, and are mistakenly isolated from the jet. Jets misidentified as an electron also contribute to these hadronic fake leptons. They all have in common that they are misidentified as a lepton from the hard scattering process, also called a *real* or *prompt* lepton. Fake electrons can also occur from photon conversion, which is discussed below. In the trilepton channel, fake lepton background mostly comes from the $Z$+jets process.

For the estimation of events containing hadronic fake leptons in the signal regions (see Section 7.3), the data driven *matrix method* is used, which is common in high energy physics, e.g. in [166] and [167]. However, due to computational limitations, the matrix method is not applied for events with no *b*-tags. Therefore, Monte Carlo events are used to describe fake leptons in the control regions, using the so-called *fake factor method*, see Section 6.3.2 and 7.4. Fake factors are also used to describe another source of electron fakes, called *photon conversions*. They are caused by the production of an electron-positron pair from a photon in the detector. The photon conversion fakes arise by mistakenly identifying the resulting electron or positron as a prompt electron or positron.

The estimation of the fake lepton background via the matrix method and fake factor method is done as a group effort in the analysis group. It is performed for all analysis channels simultaneously.

## 6.3.1. Fake lepton background description using the matrix method

The matrix method fully relies on measured data. A tight and a loose lepton definition is used, as described in Section 5.2 and 5.3. The method is based on the idea that events with prompt leptons passing the loose selection have a certain *real lepton efficiency* $r$ to turn up in the tight selection, while events with fake leptons passing the loose selection have a *fake lepton efficiency* $f$ passing also the tight selection according to

$$
\begin{aligned}
N^{\text{loose}} &= N^{\text{loose}}_{\text{real}} + N^{\text{loose}}_{\text{fake}} \\
N^{\text{tight}} &= r \cdot N^{\text{loose}}_{\text{real}} + f \cdot N^{\text{loose}}_{\text{fake}},
\end{aligned} \tag{6.1}
$$

where $N^{\text{loose}}$ and $N^{\text{tight}}$ are the total numbers of events in the loose and the tight selection. The number of events with prompt leptons passing the loose selection is described by $N^{\text{loose}}_{\text{real}}$ and the number of events with fake leptons passing the loose selection is described by $N^{\text{loose}}_{\text{fake}}$. The interesting value is $f \cdot N^{\text{loose}}_{\text{fake}}$ since it describes the actual fake yield in the tight selection. The efficiencies $r$ and $f$ are derived from a fit in five control regions orthogonal to all multilepton analysis regions of the overall analysis, with exactly two leptons and different flavour combinations: two regions enriched with prompt leptons ($OS_r$) and three regions enriched with fake leptons ($SS_f$), see Table 6.2. The $OS_r$ regions require the two leptons to have opposite electric charge and same lepton flavour, at least one jet and at least one *b*-jet. The $SS_f$ requires the two leptons to have the same electric charge, at least two jets, at least two *b*-jets and a missing transverse momentum of at least 20 GeV. Events with prompt leptons only are subtracted from data using Monte Carlo estimations. Events with two fake leptons are assumed to be zero in the fit. Because the fake lepton contribution is also estimated for other channels in the overall analysis (see Chapter 9), a veto on signal regions of the $2\ell$SS channel (see Section 9.3) is applied for the $SS_f$ regions.

| Variable | $OS_r$ | $SS_f$ |
|---|---|---|
| Leptons | exactly 2 | |
| Leading lepton $p_T$ | $> 27$ GeV | |
| Second lepton $p_T$ | $> 20$ GeV | |
| Dilepton invariant mass | $> 15$ GeV | |
| Sum of lepton charges | 0 | $\pm 2e$ |
| Lepton flavour | $ee,\mu\mu$ | $ee,\mu\mu,e\mu$ |
| $n_{b-\text{jets}}$ | $\geq 1$ | $\geq 1$ |
| $n_{\text{jets}}$ | $\geq 1$ | $\geq 2$ |
| $E_T^{\text{miss}}$ | - | $\geq 20$ GeV |
| Veto $2\ell$SS signal regions | no | yes |

Table 6.2.: Control regions used in the fit for the $r$ and $f$ lepton efficiencies. The $2\ell$SS signal region definition is discussed in Section 9.3.

The fit is performed in bins of leading and second leading $p_T$ for events with exactly one $b$-jet or at least two $b$-jets by maximising the following likelihood function with respect to the efficiencies $r_i$, $f_i$, $r_j$ and $f_j$:

$$L = \prod_{i,j} \prod_{k \in CR} P\left( (N_{ij}^k \,|\, M_{ij}^{kl}(r_i, f_i, r_j, f_j) \, n_{ij}^l \right) , \qquad (6.2)$$

where $i$ and $j$ are the indices in terms of the leading and second leading lepton $p_T$ and $k$ is the index running over the five control regions. The observed data yields without the events containing only real leptons is denoted by $N_{ij}^k$. The real and fake lepton yields are denoted by $n_{ij}^l$, where the index $l$ stands for either "real" or "fake". The matrix $M_{ij}^{kl}(r_i, f_i, r_j, f_j)$ relates the measured and estimated yields and $P(x|y)$ is a Poisson distribution. The resulting real and fake efficiencies are shown in Figure 6.4. These efficiencies are used to scale data events with leptons passing the loose object definition in order to estimate the fake lepton events passing the tight selection.

To avoid overlap with fake lepton events from Monte Carlo samples, events with at least one fake lepton out of the three leptons with the highest transverse momentum are removed from the Monte Carlo, which is called *truth matching*. This is achieved by accessing additional information on the lepton origin, stored in the Monte Carlo files, and identifying leptons with non-prompt origin as fakes. Events containing fake electrons from photon conversion are also kept since the Monte Carlo events are used for the $\gamma + X$ fake background estimation using the fake factor method. The results of the fake lepton estimation in the $SS_f$ regions, compared to data, are shown in Figure 6.5.

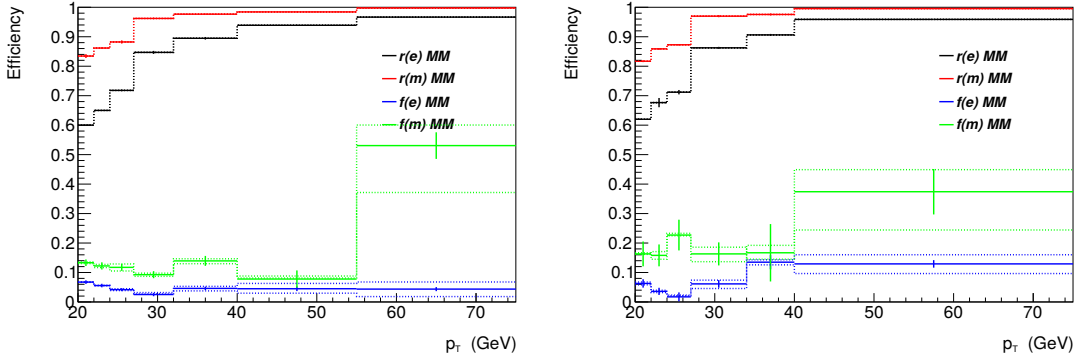

Figure 6.4.: Fake efficiencies ($f$) and real efficiencies ($r$) for electrons ($e$) and muons ($m$), depending on the combined lepton $p_T$. The figure on the left hand side shows the efficiencies for events with exactly one $b$-tag and the figure on the right hand side shows the efficiencies for events with at least two $b$-tags [3].

### 6.3.2. Fake lepton background description using the fake factor method

The background due to fake electrons from photon conversion, called $\gamma + X$, is described by Monte Carlo events like $Z + \gamma$ and $t\bar{t}\gamma$ which are reweighted according to *fake factors*, derived from fits to data. Therefore, in contrast to the matrix method which is fully data driven, the fake factor method is partly data driven. Overlap between the other Monte Carlo samples is avoided by removing $\gamma + X$ events from those files using truth matching in case the matrix method is used to evaluate fake leptons from hadronic processes. Fake leptons from hadronic processes are also modelled with the fake factor method for the control regions, since the matrix method is not applied for events with exactly zero $b$-tags (see Table 6.2). In these cases, the $\gamma + X$ contribution is evaluated together with the hadronic fake lepton contribution. For the photon conversion background, Monte Carlo samples for the following contributions are taken into account:

- $W + \gamma$ and $Z + \gamma$ events are generated with SHERPA2.1 using the `CT10` PDF.

- $Z$+jets events with the $Z$ boson decaying into either electrons or muons are generated with SHERPA2.2.1 and using the `NNPDF3.0NLO` PDF.

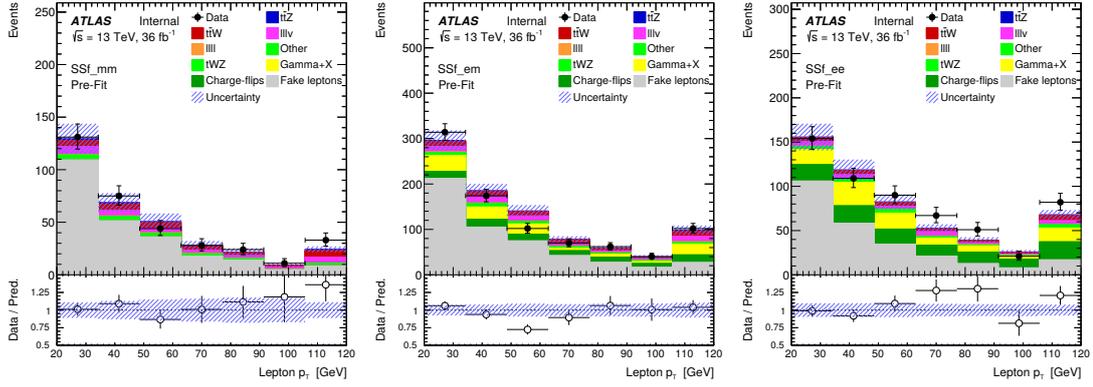

Figure 6.5.: Matrix method prediction for the fake lepton background in the $SS_f$ regions, together with other backgrounds, in comparison to data. The combined lepton $p_T$ is shown for the dimuon (left), $e\mu$ (center) and dielectron regions (right). The shaded error bands include the statistical uncertainty, the uncertainty of the luminosity and the uncertainties of the matrix method [3].

- $t\bar{t}\gamma$ events are generated with MADGRAPH interfaced to PYTHIA8 with the NNPDF2.3LO PDF.

- $ZZ$ events with one $Z$ decaying hadronically and the other one into charged leptons are generated with SHERPA2.2.1 using the NNPDF3.0NLO PDF.

For the fake lepton background from hadronic processes, all processes mentioned before in this chapter are used. Leptons are required to fulfil the tight lepton selection criteria. The fake factor method is performed for multiple channels of the overall analysis. Four fake factors are derived:

- $F_{heavy}^e$: for fake electrons from heavy flavour hadronic processes

- $F_{other}^e$: for fake electrons from other sources (mainly photon conversion)

- $F_{heavy}^\mu$: for fake muons from heavy flavour hadronic processes

- $F_{other}^\mu$: for fake muons from other sources.

The two sets of fake factors ($F_{heavy}^e$, $F_{other}^e$) and ($F_{heavy}^\mu$, $F_{other}^\mu$) are derived via separate simultaneous fits in two control regions for fake electrons and muons, respectively. Two control regions are enriched with $t\bar{t}$ events and two region are enriched with $Z$+jets events, both with an additional fake lepton (electron or muon), see Table 6.3. These regions are defined to be orthogonal to all other signal and control regions used in the overall analysis. In the $Z$+jets control region, a fake lepton is defined as the lepton that does not come from the $Z$ boson, and therefore is not part of the pair with the opposite charge and same flavour (OSSF). In the $t\bar{t}$ control region, the lepton that shares its charge with another lepton and has the lowest transverse momentum is defined as the fake lepton. The results of the fits to the fake lepton yields in data are

$$
\begin{aligned}
F_\mathrm{heavy}^{e} &= 0.90 \pm 0.14 \\
F_\mathrm{other}^{e} &= 1.84 \pm 0.27 \\
F_\mathrm{heavy}^{\mu} &= 1.07 \pm 0.09 \\
F_\mathrm{other}^{\mu} &= 1.00 \pm 0.50 \,.
\end{aligned}
\tag{6.3}
$$

|                        | $Z$+jets CR     | $t\bar{t}$ CR              |
|------------------------|-----------------|---------------------------|
| Number of leptons      | exactly 3       | exactly 3                 |
| Lepton pair            | 1 OSSF pair     | no OSSF pair, one OS pair |
| $E_\mathrm{T}^\mathrm{miss}$ | < 50 GeV  | —                         |
| $m_\mathrm{T}$         | < 50 GeV        | —                         |
| Number of jets         | —               | $\geq 1$                  |
| $p_\mathrm{T}^\mathrm{leading\ jet}$ | —    | > 30 GeV                  |

Table 6.3.: Control region definition for the fake factor determination. The missing transverse mass is referred to as $m_\mathrm{T}$. The control regions are further divided into regions for fake electrons or muons, respectively.

Note that the number of fakes for muons coming from the "other" processes is very small. Therefore, the fit has only low sensitivity to $F_\mathrm{other}^{\mu}$, which is consequently fixed to one with a 50% uncertainty. Figure 6.6 and 6.7 show the Monte Carlo yields compared to data before and after the extraction of the fake factors for electrons and muons, respectively.

The fake factors are applied by checking the Monte Carlo information for the lepton origins. For each fake lepton, the event is scaled by the fake factor. For the regions using the matrix method to determine fake leptons from hadronic processes, only the $\gamma + X$ processes are scaled by fake factors. For the other regions, all events are scaled by fake factors.

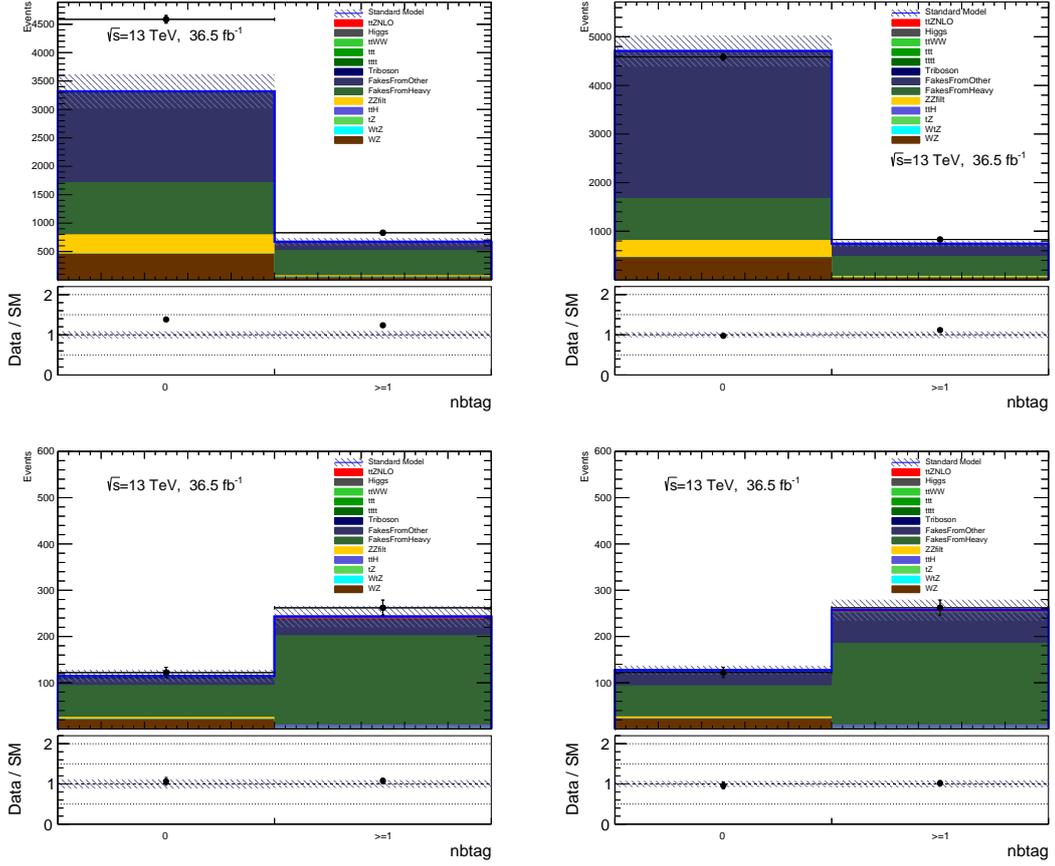

Figure 6.6.: Distribution of the $b$-jet multiplicity in the $Z$+jets control region (top row) and $t\bar{t}$ control region (bottom row) for fake electrons, before (left column) and after (right column) the extraction of the fake factors $F_{\text{heavy}}^e$ and $F_{\text{other}}^e$ [3].

Figure 6.7.: Distribution of the $b$-jet multiplicity in the $Z$+jets control region (top row) and $t\bar{t}$ control region (bottom row) for fake muons, before (left column) and after (right column) the extraction of the fake factors $F^{\mu}_{\text{heavy}}$ and $F^{\mu}_{\text{other}}$ [3].

Event Selection

This chapter discusses the selection of signal, control and validation regions for the *trilepton channel* sensitive to the $t\bar{t}Z$ process. The name of this channel already implies that the basic selection requires exactly three charged leptons (electrons or muons[1]). The three different types of regions serve different purposes in the analysis. The validation regions are used to check the agreement between data and the background modelling without being biased by signal events. The signal and control regions are actually used for a profile likelihood fit, see Chapter 10. The concept is to fit the yields of each region. The signal regions are selected in a way to be sensitive to the $t\bar{t}Z$ process, and the control regions are chosen to be sensitive to important background processes[2] in the analysis. In the fit, the $t\bar{t}Z$ signal strength and the normalisation of the main background processes, described by Monte Carlo events, are chosen as free fit parameters. Therefore, the fit in the trilepton channel will be performed in six bins (four signal and two control regions) with three free fit parameters. A combination with other analysis channels is also performed. A trilepton channel sensitive to the $t\bar{t}W$ process is also defined for a different part of the overall analysis which is discussed in Chapter 9.

---

1. Tau leptons are not considered as reconstructed objects but their decay products from the leptonic and hadronic decays are reconstructed as electrons, muons, jets and missing transverse momentum on their own. Therefore, the electrons and muons from tau lepton decays can end up in the trilepton selection without explicitly requiring tau leptons.
2. As it is discussed later, $WZ$+jets and $ZZ$+jets are the processes for which the control regions are optimised. While the $ZZ$ process is not a dominant process in the trilepton channel, it is an important background for the tetralepton channel.

Two validation regions for the trilepton selection serve to check for good agreement between data and Monte Carlo in terms of event yields and variable distributions. Three trilepton signal regions are defined to be sensitive to on-shell $Z$ bosons and one trilepton signal region is defined to be sensitive to off-shell $Z$ bosons. Two control regions are defined. One region is sensitive to fully leptonic $WZ$ decays with additional light and heavy flavoured jets, using a trilepton selection. The other region is sensitive to the decay of a $Z$ boson pair into four charged leptons with additional light and heavy flavoured jets and selects at least four charged leptons. The diboson processes in the control regions are called $WZ$ and $ZZ$ for simplicity. The control regions are used to constrain the $WZ$ and $ZZ$ cross section normalisation in the fit described in Chapter 10 to reduce systematic uncertainties.

To avoid overlap with the other channels of the overall analysis, which are defined by different lepton multiplicities, events with additional leptons passing the loose requirements (see Chapter 5) are vetoed for the trilepton regions. Otherwise, for example, a trilepton event with an additional loose lepton might simultaneously be accounted for in an event with four leptons in another part of the overall analysis. The $ZZ$ control region however only requires at least four charged leptons defined by the tight object selection and any number of additional loose leptons, see Section 7.4.1.

## 7.1. Event preselection

In addition to the requirements for data, presented in Section 6.1, both data and Monte Carlo events need to fulfil trigger requirements to pass the selection of this analysis. This analysis uses triggers that require at least one lepton (electron or muon) per event. All events need to fulfil the trigger requirements of at least one of these triggers. Different triggers are used for data taken during 2015 [189] and 2016 [233], respectively. Therefore, Monte Carlo events are semi-randomly assigned[3] to 2015 and 2016 data to match the two different trigger sets. This separation is also done to match the Monte Carlo events to the different pileup profiles for the two data taking periods. The trigger sets used for both years are listed in Table 7.1 together with the HLT trigger requirements. The HLT trigger requirements are also seeded with L1 triggers which are not further discussed here.

---

3. The semi-random assessment divides the Monte Carlo samples according to the integrated luminosities of the corresponding datasets to keep the Monte Carlo statistical uncertainties low. The ratio between the raw Monte Carlo events assigned to 2015 and 2016 approximately corresponds to the ratio of the integrated luminosities of both datasets.

| Period | Trigger type | Trigger | $p_T$ threshold | Identification | Isolation |
|---|---|---|---|---|---|
| 2015 | Single elec. | `HLT_e24_lhmedium_L1EM20VH` | 24 GeV | Medium LH ID | — |
| | | `HLT_e60_lhmedium` | 60 GeV | Medium LH ID | — |
| | | `HLT_e120_lhloose` | 120 GeV | Loose LH ID | — |
| | Single muon | `HLT_mu20_iloose_L1MU15` | 20 GeV | Comb. muons | Loose iso |
| | | `HLT_mu50` | 50 GeV | Comb. muons | — |
| 2016 | Single elec. | `HLT_e26_lhtight_nod0_ivarloose` | 26 GeV | Tight LH ID | Loose var iso |
| | | `HLT_e60_lhmedium_nod0` | 60 GeV | Medium LH ID | — |
| | | `HLT_e140_lhloose_nod0` | 140 GeV | Loose LH ID | — |
| | Single muon | `HLT_mu26_ivarmedium` | 26 GeV | Comb. muons | Medium var iso |
| | | `HLT_mu50` | 50 GeV | Comb. muons | — |

Table 7.1.: HLT requirements for the single lepton triggers used in the analysis for 2015 and 2016 data. The identification and isolation criteria are further discussed in Section 7.1.

For the HLT, electrons are reconstructed in the electromagnetic calorimeter from deposited energy clusters, which are then matched to tracks in the inner detector. A multivariate technique is used to produce a likelihood discriminant which is then cut on in order to get the *medium LH* and *loose LH* selections shown in the Table 7.1. Electrons are not required to be isolated by the HLT except for the `HLT_e26_lhtight_nod0_ivarloose` trigger, where the isolation track parameter is used (see below).

Muons are reconstructed in the HLT using a combined muon approach, which is similar to the one described in Section 5.3, but comparatively simpler. For the HLT part of the `HLT_mu20_iloose_L1MU15` trigger, the scalar sum of the track $p_T$ in a cone of $\Delta R = 0.2$ around the muon candidate is required to be less than 12% than the muon candidate's transverse momentum. For the `HLT_mu50` triggers in both 2015 and 2016, no muon isolation is required.

For the 2016 triggers denoted by the "var iso" isolation in Table 7.1, a new HLT online isolation requirement is used. It is based on an isolation track parameter, instead of a fixed cut on track $p_T$ within a certain cone of $\Delta R$ around the lepton candidate in the inner detector.

## 7.2. Validation regions

To check the modelling of the Monte Carlo samples, and the validity of the matrix and fake factor methods for the trilepton channel, two validation regions are defined. The general validity of the fake factor method and the matrix method is already checked while deriving the fake estimations. For other channels of the overall analysis, similar tests are performed.

| Variable | 3ℓ-Z-2j-VR | 3ℓ-1b-VR |
|---|---|---|
| Number of leptons | =3 | =3 |
| Leading lepton $p_T$ | > 27 GeV | > 27 GeV |
| Second and third lepton $p_T$ | > 20 GeV | > 20 GeV |
| One OSSF lepton pair | Required | Not required |
| $Z$-window cut | Required | Not required |
| $n_{\mathrm{jets}}$ | $\geq 2$ | 2 or 3 |
| $n_{b-\mathrm{jets}}$ | $\geq 0$ | $= 1$ |

Table 7.2.: Summary of the validation region definitions.

Only after good agreement between data and Monte Carlo in the validation regions is ensured, can data be studied in the signal regions. The definition of the validation regions is shown in Table 7.2. The cut on the transverse momentum of the lepton with the highest transverse momentum (called the *leading lepton*) of $p_\mathrm{T} > 27$ GeV is chosen to match the single lepton trigger $p_\mathrm{T}$ requirements. The other lepton $p_\mathrm{T}$ cuts are chosen to match the ones used for the signal regions, see Section 7.3.1. For the 3ℓ-Z-2j-VR region, at least one lepton pair with the same lepton flavour and opposite electric charge (*opposite-sign-same-flavour*, *OSSF*) is required. A cut on the OSSF lepton pair invariant mass $M_{\ell\ell}^Z$ closest to the $Z$ boson mass of $|M_{\ell\ell}^Z - 91.2 \text{ GeV}| < 10$ GeV is required for the 3ℓ-Z-2j-VR region to enhance sensitivity to the $Z$ decay into two charged leptons from the $WZ$ process. This range of $M_{\ell\ell}^Z$ is referred to as the *Z-window*. Both regions are chosen to be sensitive to the most prominent backgrounds in the trilepton channel, which are the fully leptonic decay of $WZ$ with additional jets (called $WZ$ for simplicity) and lepton fakes, both derived using the matrix method (for the signal regions) and the fake factor method (for the control regions). In addition, the signal contribution needs to be small to avoid any bias on the $t\bar{t}Z$ cross section measurement. Due to its sensitivity to events with exactly zero $b$-jets, the 3ℓ-Z-2j-VR region serves to give additional cross-checks for the validation of the fake factor method. For additional cross-checks of the validation of the matrix method, the 3ℓ-1b-VR is used.

Table 7.3 shows the yields in the two validation regions. The 3ℓ-Z-2j-VR region is mostly sensitive to the $WZ$ process, while allowing also other backgrounds like fake leptons derived using the fake factor method and $ZZ$ events to contribute. The 3ℓ-1b-VR region is mostly sensitive to $WZ$ and events with fake leptons estimated using the matrix method. The $t\bar{t}Z$ contribution is $\sim 10\%$ with respect to the total Monte Carlo yields. In terms of the yields, good agreement between data and Monte Carlo is achieved within the statistical and systematic uncertainties. Figure 7.1 shows the distributions of the missing transverse momentum, the electron multiplicity, the leading lepton transverse momentum and the jet multiplicity for the 3ℓ-Z-2j-VR region. An overall good agreement between data and Monte Carlo within the statistical and systematic uncertainties (see Chapter 8) is achieved. Only a slight mismodelling in the electron multiplicity is observed.

|  | 3ℓ-Z-2j-VR | 3ℓ-1b-VR |
|---|---|---|
| $t\bar{t}Z$ | $179.36 \pm 66.40$ | $43.57 \pm 5.18$ |
| $t\bar{t}W$ | $5.78 \pm 7.38$ | $16.81 \pm 2.38$ |
| $WZ$ | $1405.79 \pm 93.24$ | $139.68 \pm 12.05$ |
| $ZZ$ | $135.50 \pm 13.02$ | $13.80 \pm 1.41$ |
| $tZ$ | $49.87 \pm 15.10$ | $29.87 \pm 9.04$ |
| $tWZ$ | $34.37 \pm 6.69$ | $12.99 \pm 2.66$ |
| $t\bar{t}H$ | $5.00 \pm 0.47$ | $5.97 \pm 0.65$ |
| Other | $18.77 \pm 9.69$ | $2.75 \pm 1.60$ |
| DD fakes | — | $124.50 \pm 14.40$ |
| $\gamma + X$ fakes | — | $20.85 \pm 5.66$ |
| MC fakes | $60.80 \pm 18.97$ | — |
| Total | $1895.23 \pm 141.04$ | $410.79 \pm 22.97$ |
| Observed | 1850 | 418 |

Table 7.3.: The expected and observed event yields in the 3ℓ-Z-2j-VR and 3ℓ-1b-VR validation regions. Statistical and systematic uncertainties are included as described in Section 8. Contributions not considered for the selection are denoted with a solid line. The hadronic fake lepton contribution determined via the matrix method is denoted by "DD fakes", while the hadronic fake lepton contribution determined via the fake factor method is denoted by "MC fakes".

Figure 7.2 shows the distributions of the missing transverse momentum, the electron multiplicity, the leading lepton transverse momentum and the jet multiplicity for the 3ℓ-1b-VR region. Again, good agreement between data and Monte Carlo is found within the statistical and systematic uncertainties. The only exception is again the number of electrons.

The slight mismodelling in the electron multiplicity is not considered to be worrisome since it is not an important variable in the trilepton channel. An important influence on this mismodelling seems to be the $WZ$ background and the fake lepton estimate from the fake factor method. However, fake leptons estimated using the fake factor method do not play a huge role in the analysis. Looking at the other, more important variables like jet multiplicity and lepton $p_T$, the modelling looks much better.

Figure 7.3 shows the distribution of $M_{\ell\ell}^Z$ for the 3$\ell$-1b-VR region. Also in this case, the distribution shows good agreement between data and Monte Carlo. In addition, it shows that the fake lepton background (both from hadronic sources and gamma conversions) is almost flatly distributed in $M_{\ell\ell}^Z$ while the rest of the background peaks around the $Z$ boson mass.

## 7.3. Signal regions

The signal regions in the trilepton channel are required to be sensitive to the $t\bar{t}Z$ signal process. Four of them are defined with different background contributions to allow more freedom for the fit described in Chapter 10. One of the four signal regions is sensitive to off-shell $Z$ bosons, while the other three are sensitive to on-shell $Z$ bosons, see Table 7.4. Since the interference of the $Z$ boson with an off-shell photon ($\gamma^*$) is included in the $t\bar{t}Z$ samples, it is expected to have a contribution in the latter region, but cannot be shown separately.

### 7.3.1. Identifying signal region cuts

For the leading lepton $p_T$, the minimum cut is required to be slightly higher than for the single lepton trigger requirement (see Section 7.1). Therefore, a minimum leading lepton $p_T$ of 27 GeV is chosen. To avoid background from sources of soft leptons, such as fake leptons, the other two leptons are required to have a minimum $p_T$ cut of 20 GeV. In the trilepton channel, the $Z$ boson is expected to decay into two electrons or two muons. Therefore, an OSSF lepton pair is expected and the invariant mass distribution of this lepton pair is expected to peak at the $Z$ boson mass of $\approx 91.2$ GeV. The top quark pair is assumed to decay in the lepton+jets channel (see Section 2.2). Therefore, one charged lepton, missing transverse momentum, two light jets and two $b$-jets are expected from the $t\bar{t}$ decay.

This expectation has to be taken into account when selecting the signal regions. The chosen $b$-tagging working point corresponds to a $b$-tagging efficiency of 77%. Therefore, a fraction of 23% of the $b$-jets is tagged as non-$b$-jets. To keep more signal events, events with exactly one $b$-tagged jet are also accepted. The sum of the electric charges of the three leptons must be $\pm 1\,e$, where $e$ is the elementary charge. It is also possible that the jet reconstruction fails to identify some jets. Therefore, also events with three jets are taken into account. Figure 7.4 shows the jet and $b$-jet multiplicities for this selection. It is obvious that the background composition is different for the bins in these two distributions. The $t\bar{t}W$ process, which is treated as a background in this channel, populates the bins with lower jet multiplicities. This is because for the trileptonic $t\bar{t}W$ decay, the $t\bar{t}$ pair decays in the dilepton channel, hence yielding only two $b$-tagged jets and no light jets from the initial process. The diboson $WZ$ and $ZZ$ processes mostly populate the bins with only one $b$-tag. Events containing fake leptons also prefer lower $b$-tag multiplicities.

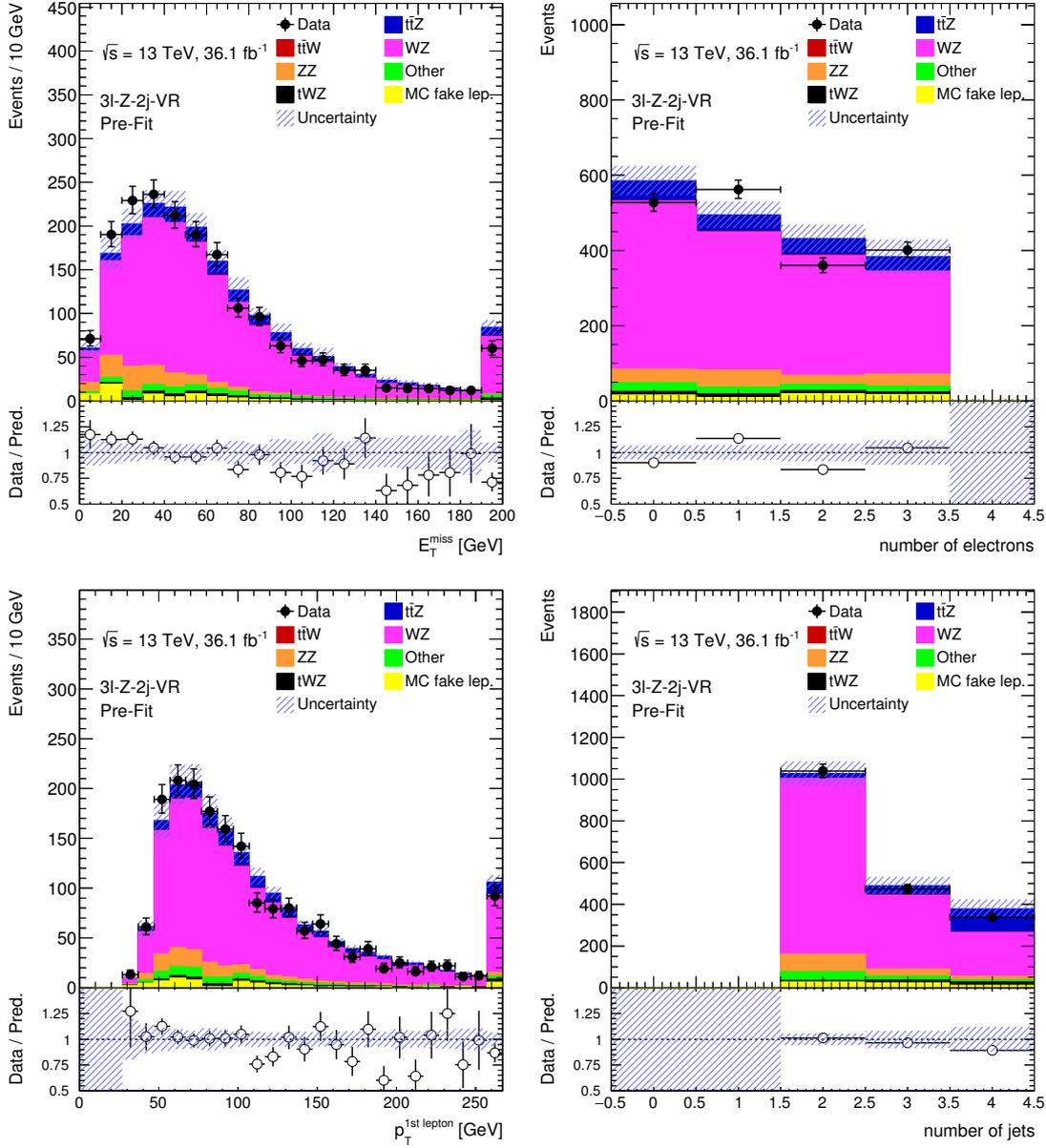

Figure 7.1.: The missing transverse momentum, electron multiplicity, jet multiplicity and leading lepton $p_T$ (clockwise from the top left) for data and Monte Carlo events in the 3ℓ-Z-2j-VR validation region. Statistical and systematic uncertainties are included as described in Section 8. Background events with fake lepton contributions, estimated via the fake factor method, are denoted by "MC fake lep.".

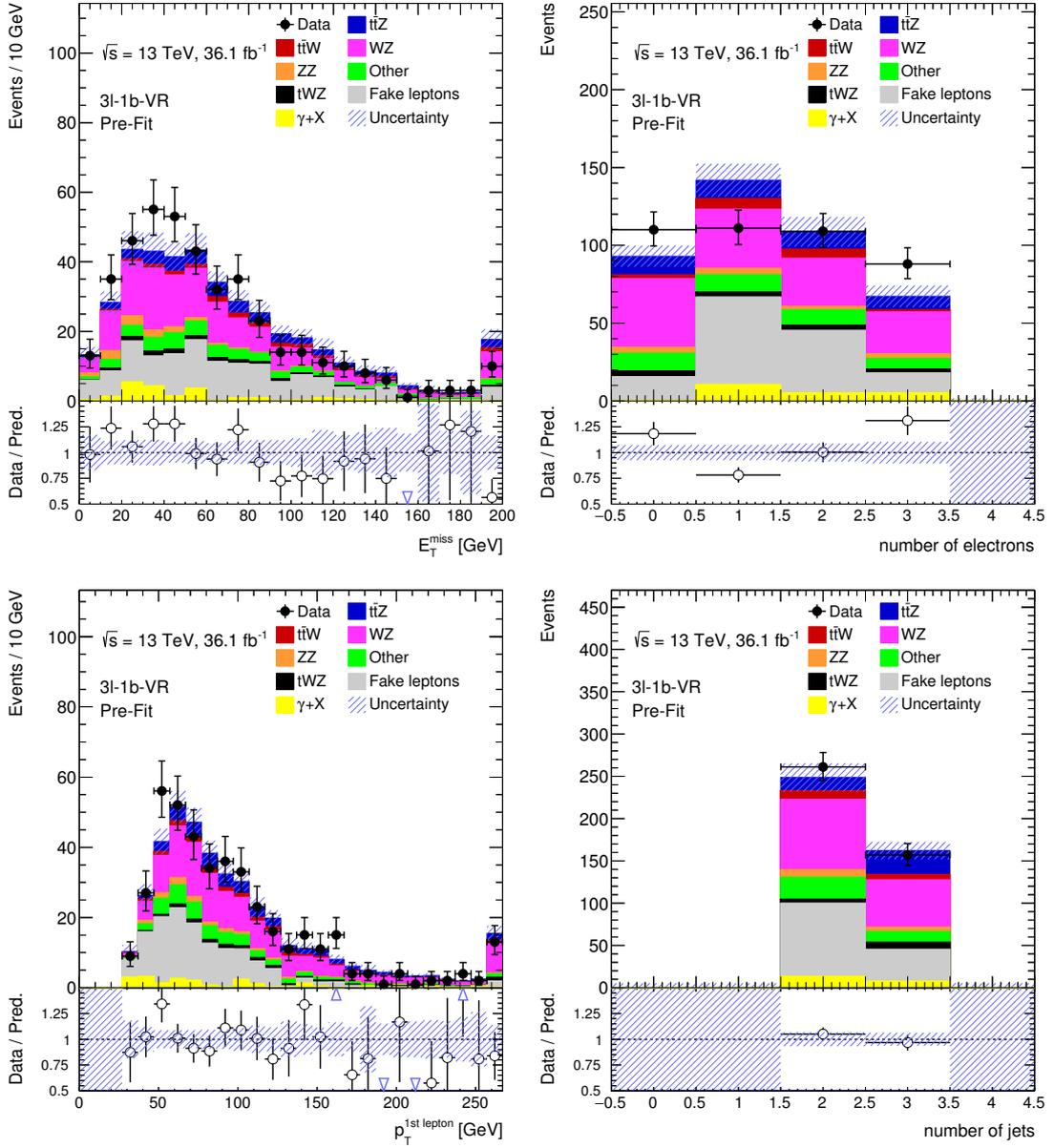

Figure 7.2.: The missing transverse momentum, electron multiplicity, jet multiplicity and leading lepton $p_T$ (clockwise from the top left) for data and Monte Carlo events in the $3\ell$-1b-VR validation region. Statistical and systematic uncertainties are included as described in Section 8. Background events with hadronic fake lepton contributions, estimated via the matrix method, are denoted by "Fake leptons".

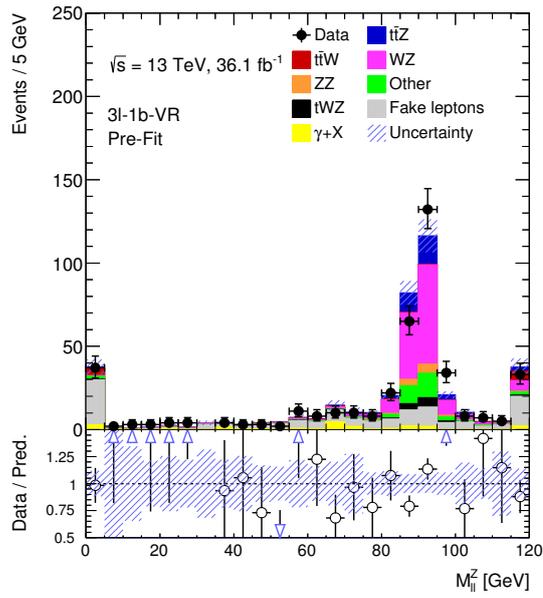

Figure 7.3.: The OSSF lepton pair invariant mass closest to the $Z$ boson mass for data and Monte Carlo events in the $3\ell$-1b-VR validation region. Statistical and systematic uncertainties are included as described in Section 8. The fake lepton background has an almost flat distribution in $M_{\ell\ell}^Z$ while the rest of the background peaks around the $Z$ boson mass. Background events with hadronic fake lepton contributions, estimated via the matrix method, are denoted by "Fake leptons".

| Variable | 3ℓ-Z-1b4j | 3ℓ-Z-2b4j | 3ℓ-Z-2b3j | 3ℓ-noZ-2b4j |
|---|---|---|---|---|
| Number of leptons | | =3 | | |
| Leading lepton $p_T$ | | > 27 GeV | | |
| $2^{nd}$ and $3^{rd}$ lepton $p_T$ | | > 20 GeV | | |
| One OSSF lepton pair | | Required | | |
| Sum of lepton charges | | $\pm 1\,e$ | | |
| $Z$-window | Required | Required | Required | Vetoed |
| $H_T^{\text{jets}}$ | > 200 GeV, < 450 GeV | — | — | — |
| $n_{\text{jets}}$ | $\geq 4$ | $\geq 4$ | $= 3$ | $\geq 4$ |
| $n_{b-\text{jets}}$ | $= 1$ | $\geq 2$ | $\geq 2$ | $\geq 2$ |

Table 7.4.: Summary of the event selection in the trilepton signal regions. The sum of the four highest jet $p_\mathrm{T}$ is called $H_\mathrm{T}^{\text{jets}}$ in this table.

To allow the fit to work in signal regions with different signal-to-background ratios and different diboson contributions (which are also free parameters in the fit), different signal regions are defined by the jet and $b$-jet multiplicities.

Figure 7.5 shows $M_{\ell\ell}^Z$ for different jet and $b$-jet multiplicities. The peak at the $Z$ boson mass of $\sim 91.2$ GeV is clearly visible. A lot of background, especially from lepton fakes from hadronic processes and from photon conversion can be suppressed by imposing a $Z$-window cut.

Figure 7.6 shows that the sum of the four highest jet $p_\mathrm{T}$, called $H_\mathrm{T}^{\text{jets}}$, can also be used to remove further $WZ$ background from the 3ℓ-Z-1b4j region. Therefore, a cut of $200$ GeV $< H_\mathrm{T}^{\text{jets}} < 450$ GeV is chosen for this region. Table 7.4 shows the definitions of the three regions defined this way: 3ℓ-Z-1b4j, 3ℓ-Z-2b4j and 3ℓ-Z-2b3j. An additional region called 3ℓ-noZ-2b4j is defined which vetoes the $Z$-region to allow access to off-shell $Z$ bosons. This region has a tight cut on the jet multiplicity to avoid contamination from $t\bar{t}W$ which is expected for less than four jets.

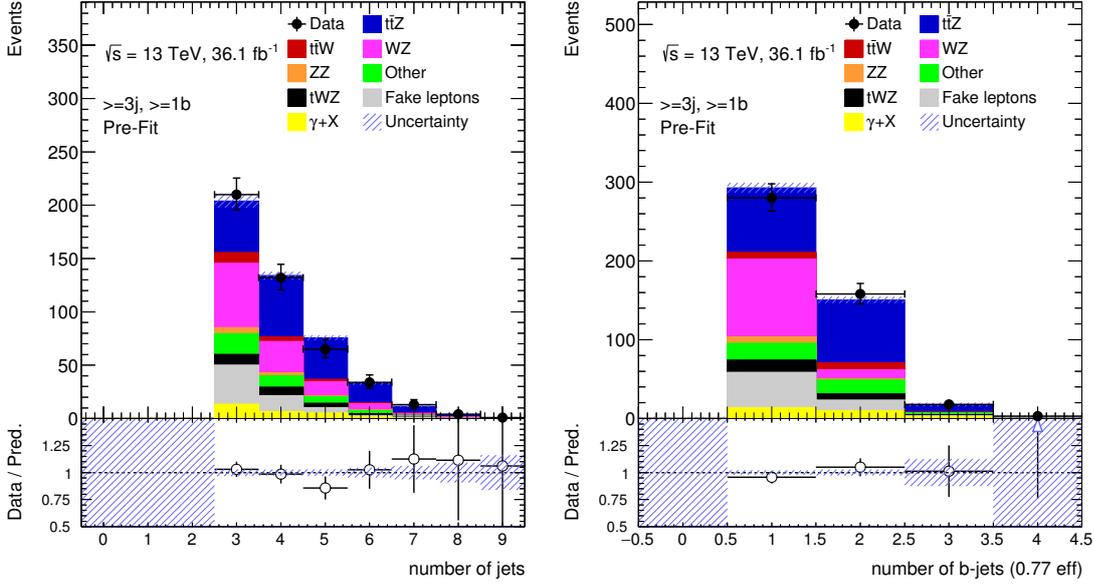

Figure 7.4.: Jet multiplicity (left) and *b*-jet multiplicity (right) for the following basic selection: exactly three leptons with one OSSF pair, at least three jets, at least one *b*-jet, a leading lepton transverse momentum of $p_T > 27$ GeV and $p_T > 20$ GeV for the other leptons. The shaded bands show the statistical uncertainty of the Monte Carlo events. Background events with hadronic fake lepton contributions, estimated via the matrix method, are denoted by "Fake leptons".

### 7.3.2. Yields in the signal regions

Table 7.5 shows the expected yields together with the data in the four signal regions. All systematic uncertainties, as defined in Chapter 8, are included. Approximately half of the expected yields come from signal. Dominating backgrounds for the regions accepting the *Z*-window cut are *WZ*, especially for the 3ℓ-*Z*-1b4j region due to the low *b*-jet multiplicity, *tZ*, *tWZ* and fake leptons from hadronic processes. The fake lepton events in these signal regions mostly come from the *Z*+jets process. In the 3ℓ-no*Z*-2b4j region, the main backgrounds are *t̄tW*, *t̄tH*, and fake leptons from hadronic processes and photon conversion. An overall good agreement between data and Monte Carlo is achieved within the Monte Carlo statistical and systematic uncertainties and the data statistical uncertainties (the latter ones are not shown in the table but can be seen in Figures 7.7 to 7.11). However, there is a slight enhancement of data events in the 3ℓ-*Z*-2b3j region which will be discussed in Section 11.1.

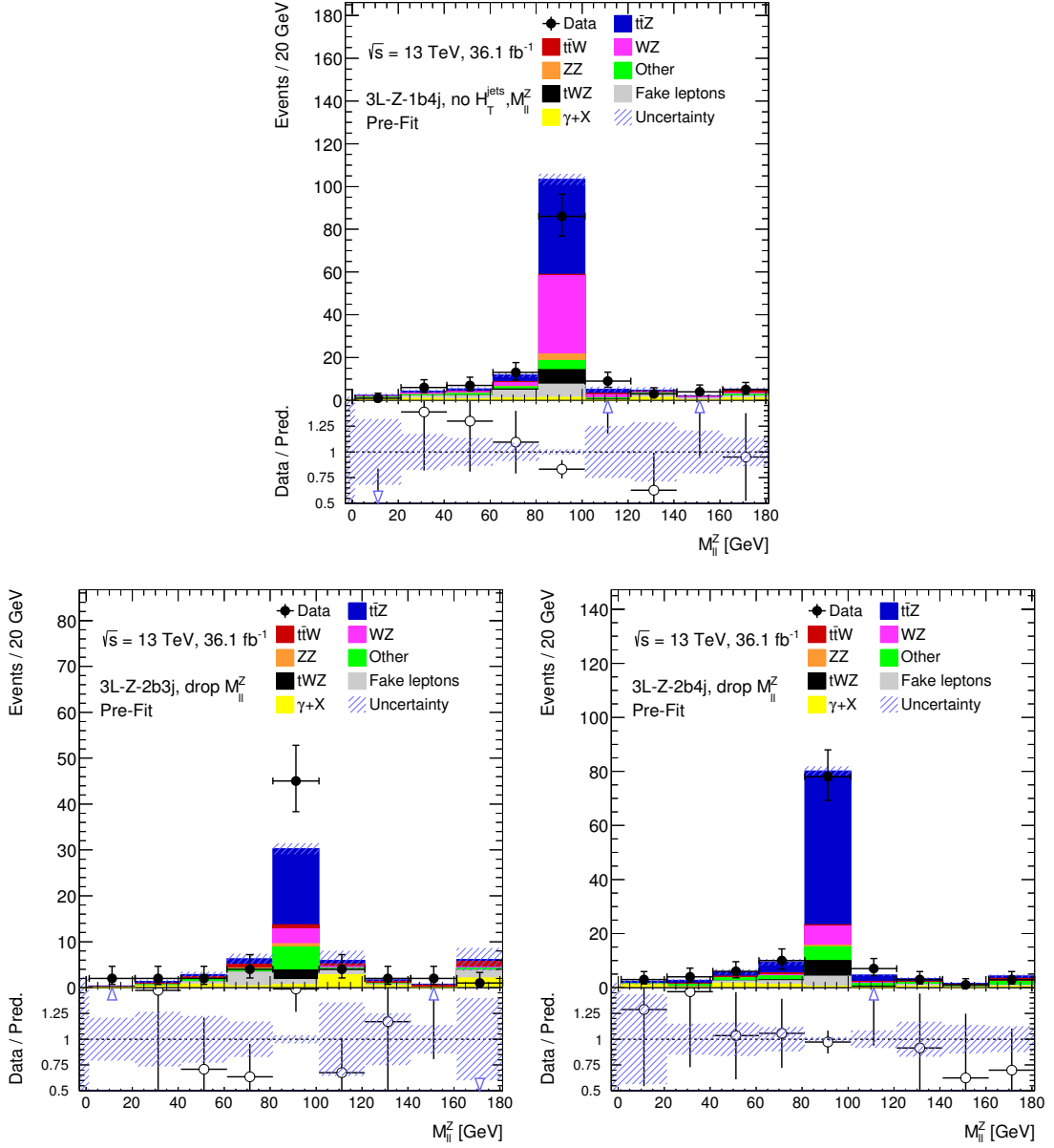

Figure 7.5.: OSSF lepton pair invariant mass $M_{\ell\ell}^Z$ closest to the $Z$ boson mass for (clockwise from the top) the $3\ell$-$Z$-1b4j, $3\ell$-$Z$-2b4j and $3\ell$-$Z$-2b3j signal regions with the $Z$-window requirement and the cut on $H_T^{jets}$ dropped. Note that $3\ell$-$Z$-2b4j and $3\ell$-no$Z$-2b4j are identical without the $Z$-window cut. Only statistical uncertainties are shown. Background events with hadronic fake lepton contributions, estimated via the matrix method, are denoted by "Fake leptons".

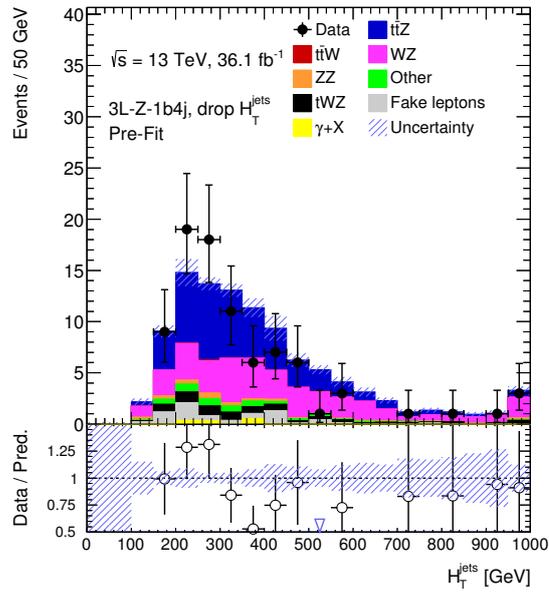

Figure 7.6.: Sum of the four highest jet transverse momenta $H_T^{\text{jets}}$ in the $3\ell$-$Z$-1b4j signal region with the cut on $H_T^{\text{jets}}$ dropped. Only statistical uncertainties are shown. Background events with hadronic fake lepton contributions, estimated via the matrix method, are denoted by "Fake leptons".

|          | 3ℓ-Z-1b4j        | 3ℓ-Z-2b4j        | 3ℓ-Z-2b3j        | 3ℓ-noZ-2b4j      |
|----------|------------------|------------------|------------------|------------------|
| $t\bar{t}Z$ | 29.88 ± 2.23  | 57.00 ± 6.69     | 16.63 ± 3.56     | 12.71 ± 1.48     |
| $t\bar{t}W$ | 0.35 ± 0.18   | 0.52 ± 0.23      | 0.83 ± 0.22      | 3.67 ± 1.05      |
| $WZ$     | 17.84 ± 5.89     | 7.05 ± 3.76      | 3.32 ± 1.59      | 1.05 ± 0.53      |
| $ZZ$     | 1.70 ± 0.40      | 0.53 ± 0.08      | 0.68 ± 0.23      | 0.33 ± 0.16      |
| $tZ$     | 1.95 ± 0.64      | 3.41 ± 1.11      | 3.66 ± 1.17      | 0.32 ± 0.13      |
| $tWZ$    | 4.03 ± 1.77      | 5.77 ± 2.15      | 2.07 ± 0.51      | 0.67 ± 0.29      |
| $t\bar{t}H$ | 0.85 ± 0.12   | 1.42 ± 0.19      | 0.51 ± 0.08      | 4.87 ± 0.63      |
| Other    | 0.14 ± 0.08      | 0.37 ± 0.37      | 0.87 ± 0.84      | 2.13 ± 1.08      |
| DD fakes | 4.39 ± 1.80      | 4.01 ± 1.60      | 1.17 ± 0.82      | 3.16 ± 1.44      |
| $\gamma + X$ | 1.31 ± 0.99  | 0.49 ± 0.42      | 0.62 ± 0.81      | 4.88 ± 1.98      |
| Total    | 62.43 ± 7.55     | 80.57 ± 9.29     | 30.36 ± 4.71     | 33.78 ± 3.90     |
| Observed | 61               | 78               | 45               | 37               |

Table 7.5.: The expected and observed event yields in the trilepton channel signal regions sensitive to the $t\bar{t}Z$ process. Statistical and systematic uncertainties are included as described in Chapter 8. The hadronic fake lepton contribution determined via the matrix method is denoted by "DD fakes".

Figures 7.7 to 7.11 show various kinematic distributions for the four signal regions. The Monte Carlo describes the data well. Only the slight enhancement in the 3ℓ-Z-2b3j signal region hints at a fit result that will be larger than one for the $t\bar{t}Z$ signal strength.

## 7.4. Control regions

For estimating the event yields of the $WZ$ process, which is the most dominating background from Monte Carlo in the trilepton channel, a control region is defined. For the tetralepton channel in the overall analysis, a control region is defined for the $ZZ$ process, which is the dominant background in that channel. Both control regions are added to the final fit, described in Chapter 10, to be able to fit the $WZ$ and $ZZ$ normalisations as additional free parameters. The control region definitions are shown in Table 7.6.

### 7.4.1. Identifying $WZ$ and $ZZ$ control region cuts

The fully leptonic decay of $WZ$ with additional jets is the most dominant Monte Carlo driven background in the trilepton channel. Due to its importance, it is constrained in the final fit using an additional control region, defined orthogonally to the signal regions. Exactly three leptons are required, with the leading one fulfilling the recommendation of $p_{\mathrm{T}} > 27$ GeV due to the single lepton trigger requirements. The other two leptons need to fulfil the minimum transverse momentum requirement of $p_{\mathrm{T}} > 20$ GeV. At least two of the leptons need to form an OSSF pair. A $Z$-window cut is required which takes into account the $Z$ boson decay in the diboson signature. The OSSF requirement also implies that the sum of all lepton charges has to be $\pm 1\,e$. Due to the required orthogonality to the signal regions, exactly three jets and exactly zero $b$-jets are required. The matrix method (see Section 6.3.1) does not determine the hadronic fake lepton background for events with exactly zero $b$-jets, so hadronic fake leptons derived via the fake factor method have to be used (see Section 6.3.2). Figure 7.12 shows the distribution of missing transverse momentum for this selection. For $E_{\mathrm{T}}^{\mathrm{miss}} < 40$ GeV, the contribution from the fully leptonic $ZZ$ decay in association with jets (in the case of one lepton failing the reconstruction criteria), as well as the contribution from events containing fake leptons from hadronic processes, increases. Therefore, this region of $E_{\mathrm{T}}^{\mathrm{miss}}$ is excluded.

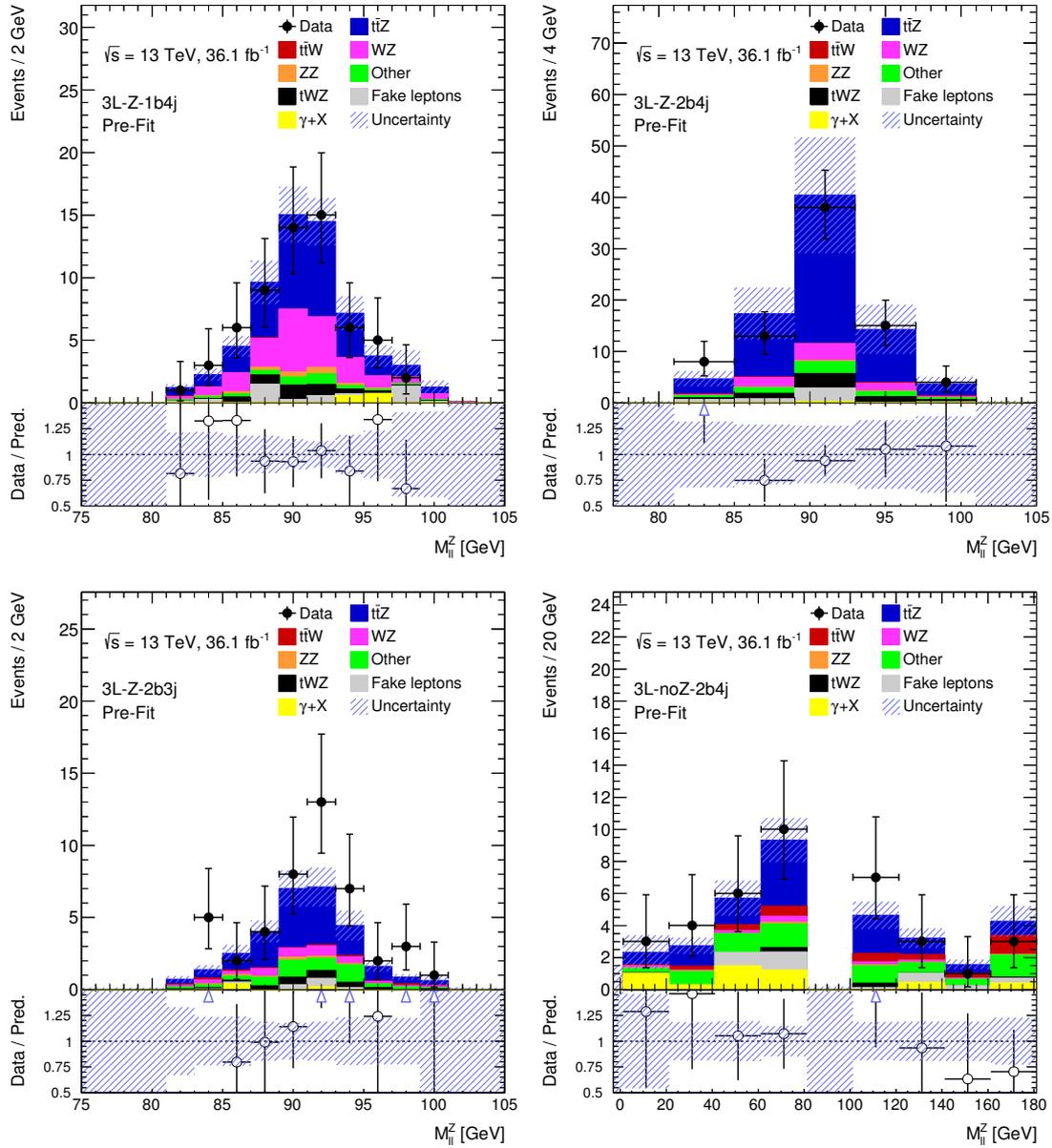

Figure 7.7.: Invariant mass of the OSSF lepton pair for data compared to the expectation from Monte Carlo events in (clockwise from the top left) the 3ℓ-Z-1b4j, 3ℓ-Z-2b4j, 3ℓ-noZ-2b4j and 3ℓ-Z-2b3j signal regions. The shaded bands show statistical and systematic uncertainties as defined in Chapter 8. Background events with hadronic fake lepton contributions, estimated via the matrix method, are denoted by "Fake leptons".

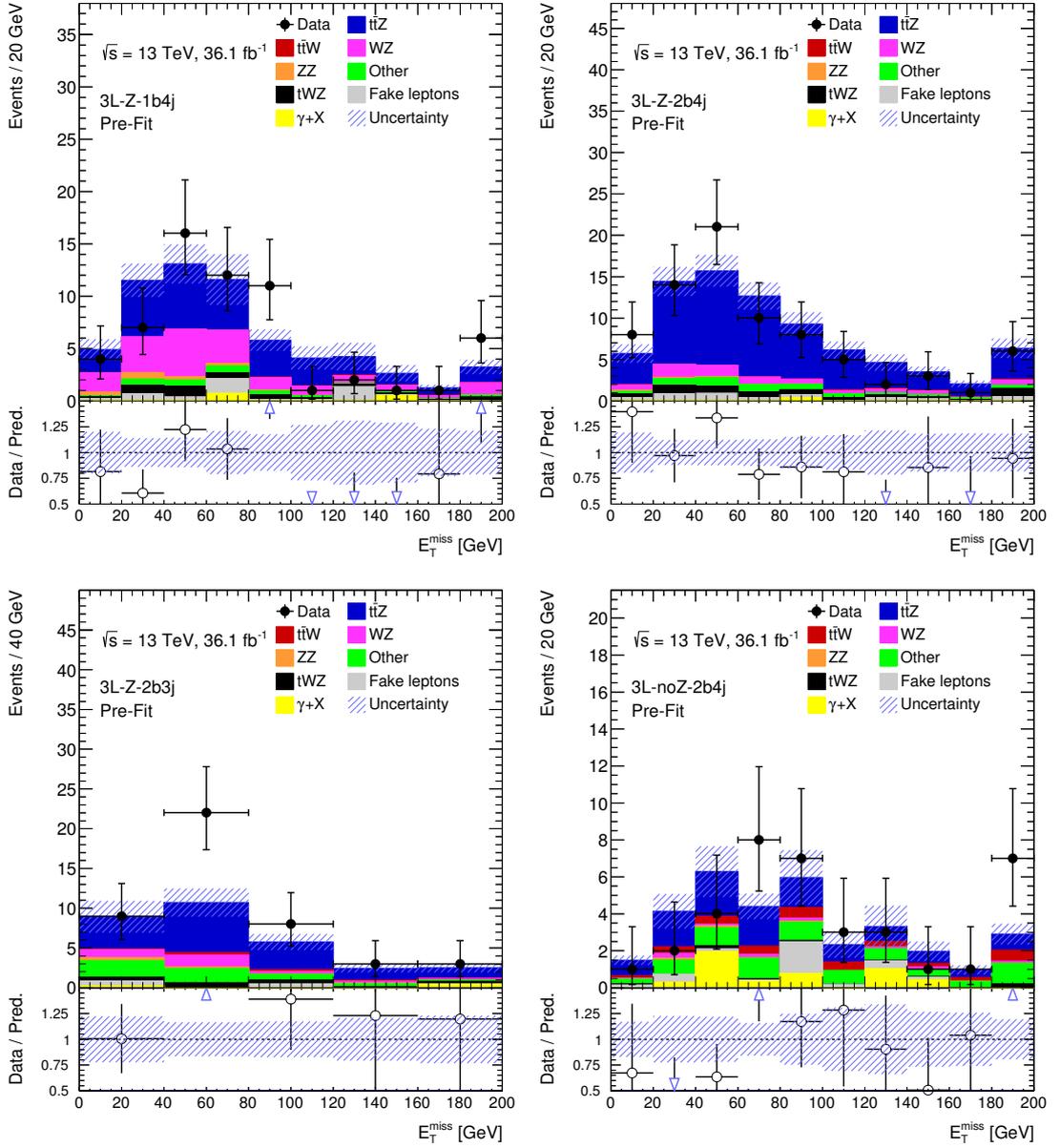

Figure 7.8.: Missing transverse momentum for data compared to the expectation from Monte Carlo events in (clockwise from the top left) the $3\ell\text{-}Z\text{-}1\text{b}4\text{j}$, $3\ell\text{-}Z\text{-}2\text{b}4\text{j}$, $3\ell\text{-no}Z\text{-}2\text{b}4\text{j}$ and $3\ell\text{-}Z\text{-}2\text{b}3\text{j}$ signal regions. The shaded bands show statistical and systematic uncertainties as defined in Chapter 8. Background events with hadronic fake lepton contributions, estimated via the matrix method, are denoted by "Fake leptons".

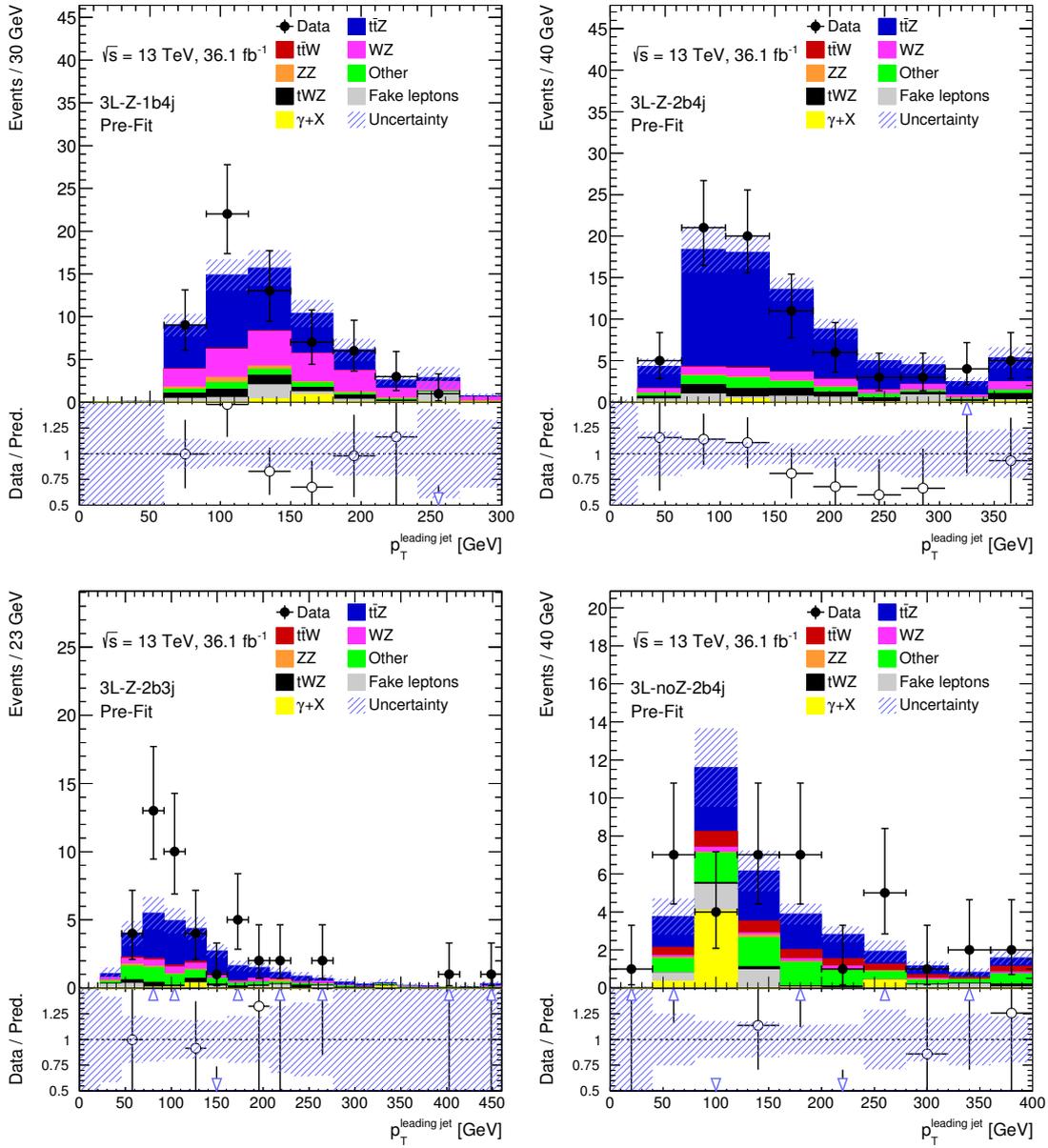

Figure 7.9.: Leading jet transverse momentum for data compared to the expectation from Monte Carlo events in (clockwise from the top left) the 3ℓ-Z-1b4j, 3ℓ-Z-2b4j, 3ℓ-noZ-2b4j and 3ℓ-Z-2b3j signal regions. The shaded bands show statistical and systematic uncertainties as defined in Chapter 8. Background events with hadronic fake lepton contributions, estimated via the matrix method, are denoted by "Fake leptons".

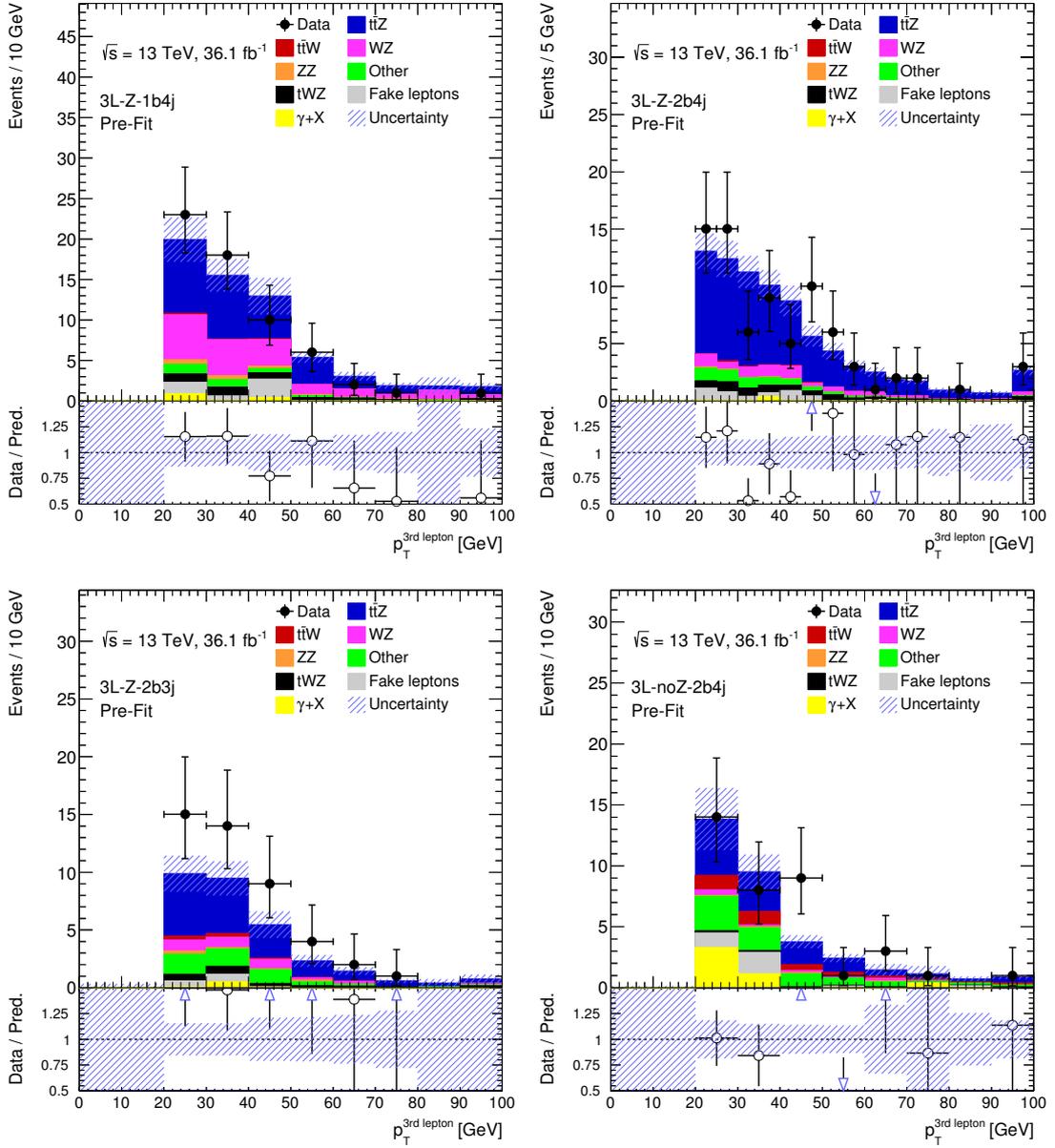

Figure 7.10.: Lowest lepton transverse momentum for data compared to the expectation from Monte Carlo events in (clockwise from the top left) the 3ℓ-*Z*-1b4j, 3ℓ-*Z*-2b4j, 3ℓ-no*Z*-2b4j and 3ℓ-*Z*-2b3j signal regions. The shaded bands show statistical and systematic uncertainties as defined in Chapter 8. Background events with hadronic fake lepton contributions, estimated via the matrix method, are denoted by "Fake leptons".

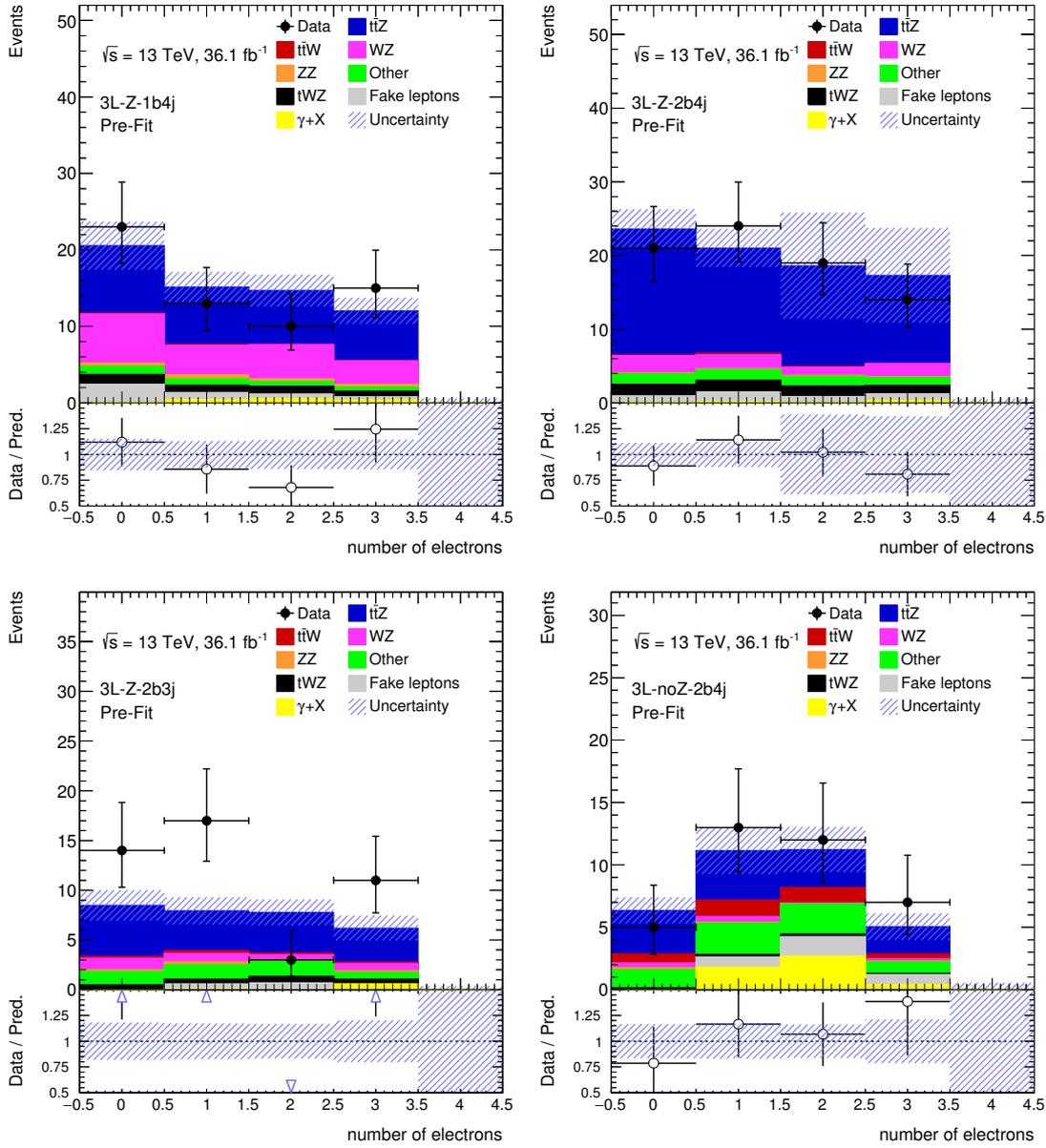

Figure 7.11.: Electron multiplicity for data compared to the expectation from Monte Carlo events in (clockwise from the top left) the $3\ell$-$Z$-1b4j, $3\ell$-$Z$-2b4j, $3\ell$-no$Z$-2b4j and $3\ell$-$Z$-2b3j signal regions. The shaded bands show statistical and systematic uncertainties as defined in Chapter 8. Background events with hadronic fake lepton contributions, estimated via the matrix method, are denoted by "Fake leptons".

| Variable | $3\ell\text{-}WZ$-CR | $4\ell\text{-}ZZ$-CR |
|---|---|---|
| Lepton definition | Loose and tight | Tight |
| Number of leptons | =3 | 4 electrons or<br>4 muons or<br>2 electrons and 2 muons |
| Leading lepton $p_T$ | > 27 GeV | > 27 GeV |
| Non-leading lepton $p_T$ | > 20 GeV | > 7 GeV |
| One OSSF lepton pair | Required | Required |
| Second OSSF lepton pair | — | Required |
| Sum of lepton charges | $\pm 1\,e$ | $0\,e$ |
| $Z$-window | Required | Required for both OSSF pairs |
| $E_{\mathrm{T}}^{\mathrm{miss}}$ | > 40 GeV | < 40 GeV |
| $n_{\mathrm{jets}}$ | = 3 | $\geq 0$ |
| $n_{b-\mathrm{jets}}$ | = 0 | $\geq 0$ |

Table 7.6.: Definitions of the control regions for the $WZ$ and $ZZ$ background processes.

An additional control region for the decay of a $Z$ boson pair into four charged leptons with additional jets ($4\ell\text{-}ZZ$-CR) is defined. The $ZZ$ process has a small contribution to the trilepton channel but is an important background for the $t\bar{t}Z$ analysis in the tetralepton channel. Therefore, it will be included in the combined fit of all $t\bar{t}Z$ analysis channels. As a test for the full capabilities of the fit in the trilepton channel, the $4\ell\text{-}ZZ$-CR will also be included in the analysis presented in this thesis.

|  | 3ℓ-$WZ$-CR | 4ℓ-$ZZ$-CR |
|---|---|---|
| $t\bar{t}Z$ | $5.10 \pm 1.15$ | $0.18 \pm 0.04$ |
| $t\bar{t}W$ | $0.18 \pm 0.09$ | — |
| $WZ$ | $210.77 \pm 22.07$ | — |
| $ZZ$ | $11.49 \pm 1.96$ | $371.78 \pm 19.28$ |
| $tZ$ | $1.42 \pm 0.50$ | — |
| $tWZ$ | $2.17 \pm 0.72$ | $0.06 \pm 0.07$ |
| $t\bar{t}H$ | $0.11 \pm 0.03$ | — |
| Other | $1.53 \pm 1.10$ | $0.57 \pm 0.46$ |
| MC fakes | $5.03 \pm 2.67$ | $0.32 \pm 0.26$ |
| Total | $237.80 \pm 22.67$ | $372.91 \pm 19.24$ |
| Observed | 211 | 435 |

Table 7.7.: The expected pre-fit event yields and observed event yields in the 3ℓ-$WZ$-CR and 4ℓ-$ZZ$-CR control regions. Statistical and systematic uncertainties are included as described in Chapter 8. Processes with a contribution of less than 0.01 events are dropped and are denoted by solid lines. The hadronic fake lepton contribution determined via the fake factor method is denoted by "MC fakes".

The 4ℓ-$ZZ$-CR needs to be orthogonal to the other tetralepton signal regions. Exactly four electrons or muons or exactly two electrons and muons are required for this channel. These leptons only need to fulfil the tight selection criteria since the tetralepton channel is the channel with the highest lepton multiplicity. Therefore, leptons passing the loose selection but not the tight one do not cause an overlap with other channels. Due to the requirement on the number of electrons and muons, more than four tight leptons can occur in the channel. The mandatory leading lepton $p_{\mathrm{T}}$ cut of $> 27$ GeV is applied. Because fake leptons are far less dominant in the tetralepton channel than in the trilepton channel, the other leptons only have to fulfil the minimum transverse momentum requirement from the object reconstruction of $p_{\mathrm{T}} > 7$ GeV, see Chapter 5. Two OSSF lepton pairs need to be reconstructed, of which each pair needs to pass a $Z$ region cut to allow sensitivity to both $Z$ bosons. Missing transverse momentum of $E_{\mathrm{T}}^{\mathrm{miss}} < 40$ GeV is required. No cuts on the jet or $b$-jet multiplicities are required. Due to events with no $b$-jets in the 4ℓ-$ZZ$-CR region, fake leptons from hadronic sources need to be derived using the fake factor method, see Section 6.3.2. Table 7.6 shows the full definition of the 3ℓ-$WZ$-CR and 4ℓ-$ZZ$-CR control regions.

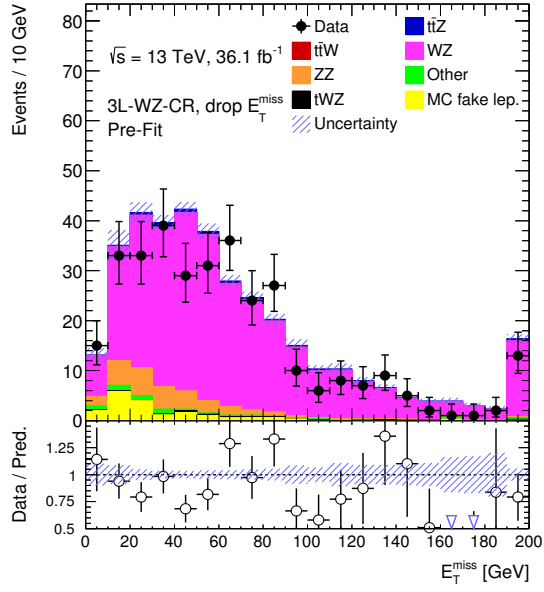

Figure 7.12.: Distribution of the missing transverse momentum $E_T^{miss}$ using the selection for the $3\ell$-$WZ$-CR region from Table 7.6 but with the $E_T^{miss}$ requirement dropped to show the contamination from other background sources in this channel for $E_T^{miss} \leq 40$ GeV. Only statistical uncertainties are shown. Background events with fake lepton contributions, estimated via the fake factor method, are denoted by "MC fake lep.".

### 7.4.2. Yields in the $WZ$ and $ZZ$ control regions

Table 7.7 shows the expected yields from Monte Carlo events for both the $3\ell$-$WZ$-CR and $4\ell$-$ZZ$-CR control regions compared to data. Note that for the $4\ell$-$ZZ$-CR region, Monte Carlo contributions with less than 0.01 expected events have been dropped to avoid issues from samples with low Monte Carlo statistics. The $4\ell$-$ZZ$-CR region is almost exclusively populated with events from the $ZZ$ process. The $3\ell$-$WZ$-CR region is also very clean with some contamination from $t\bar{t}Z$, $ZZ$ and fake leptons.

Figure 7.13 shows the expectation of the first and third highest lepton $p_T$, as well as the electron multiplicity and leading jet $p_T$ in the $3\ell$-$WZ$-CR region compared to data. The agreement between data and Monte Carlo is good within the systematic and statistical uncertainties. Figure 7.14 shows the same distributions for the $4\ell$-$ZZ$-CR region. For this region, the agreement between data and Monte Carlo is not optimal. Validation tests conducted within the tetralepton sub-analysis show that the slight disagreement between data and Monte Carlo is not due to missing Monte Carlo samples or due to bad modelling of other processes. Therefore, the fit of the $ZZ$ normalisation will derive the correct contribution of this process.

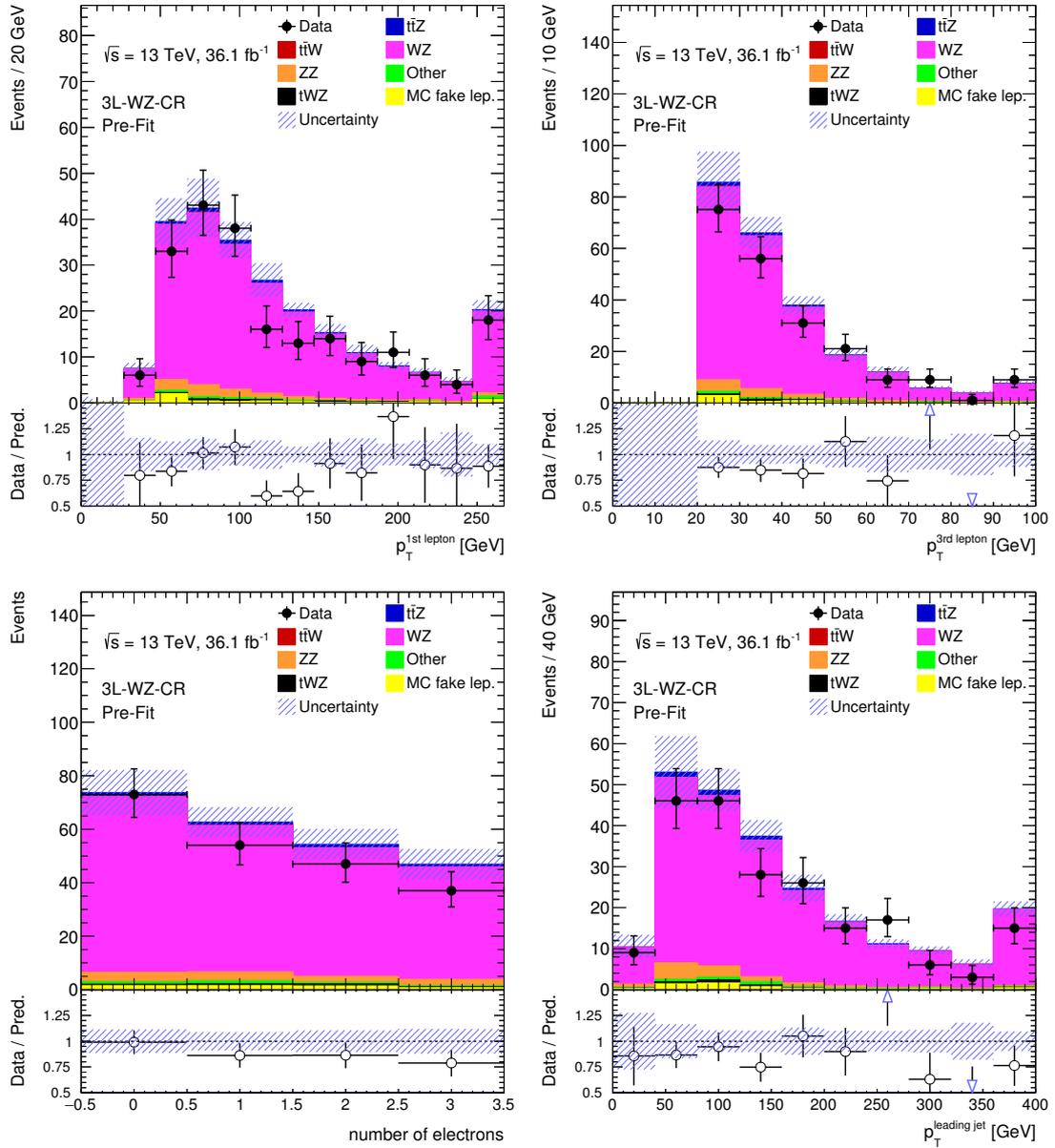

Figure 7.13.: Expectation of leading and third lepton $p_T$, leading jet $p_T$ and electron multiplicity (clockwise from the top left) in the $3\ell$-$WZ$-CR region, compared to data. The shaded bands show statistical and systematic uncertainties as defined in Chapter 8. Background events with fake lepton contributions, estimated via the fake factor method, are denoted by "MC fake lep.".

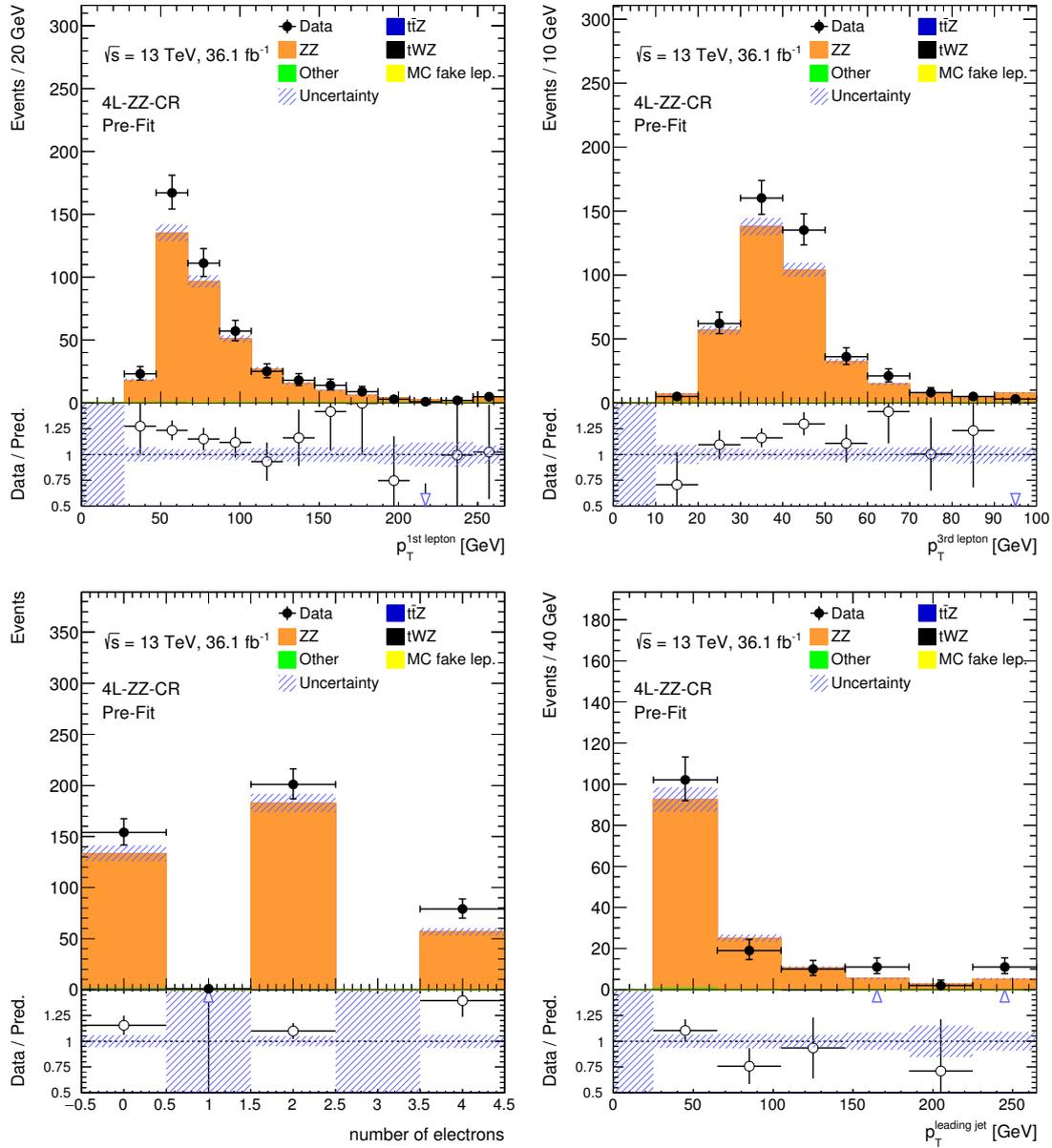

Figure 7.14.: Expectation of leading and third lepton $p_\mathrm{T}$, leading jet $p_\mathrm{T}$ and electron multiplicity (clockwise from the top left) in the $4\ell$-$ZZ$-CR region, compared to data. The shaded bands show statistical and systematic uncertainties as defined in Chapter 8. Background events with fake lepton contributions, estimated via the fake factor method, are denoted by "MC fake lep.".

## Systematic Uncertainties

Systematic uncertainties play a crucial role in the analysis presented in this thesis since they will be included in the profile likelihood fit as nuisance parameters. These uncertainties are applied to the Monte Carlo events and to the data driven background[1]. In this chapter, the systematic uncertainties are categorised into five groups. The lepton systematics (Section 8.2) and jet systematics (Section 8.3) are related to all systematic uncertainties of charged leptons and jets, respectively. The systematic uncertainties of the data driven fake lepton background estimation are also discussed Section 8.2. Uncertainties related to the different Monte Carlo signal and background processes are discussed in Section 8.5. The uncertainties for the luminosity (Section 8.1) and the missing transverse momentum (Section 8.4) do not fit into these categories and thus form their own groups of systematic uncertainties for this chapter. These categories are created just for the structure of this chapter and do not play a role in how the systematic uncertainties are treated in the analysis. While some systematic uncertainties are applied globally, others have different impacts on the different signal and control regions because they are dependent on the distribution of a certain value or because they are defined separately for each region.

The determination of the systematic uncertainties is either performed centrally for the whole ATLAS collaboration or by the analysis team, depending on the uncertainty.

---

1. Only the systematic uncertainties of the matrix method and the statistical uncertainty of the sample are applied to the data driven background. All other systematic uncertainties are only applied to Monte Carlo events.

Therefore, estimating the systematic uncertainties is a group effort. Table 8.1 lists all different systematic variations, considered for this analysis, and refers to the section in this chapter where they are discussed.

## 8.1. Luminosity

All Monte Carlo samples are normalised to data using the luminosity of the corresponding data taking period. Therefore, the uncertainty of this parameter needs to be included in the measurement. The luminosity and its uncertainty is determined as explained in Section 6.1, see also [173]. The uncertainty is $\pm 2.1\%$ for data taken during 2015 and 2016. It is applied as normalisation uncertainty to the Monte Carlo samples.

## 8.2. Systematic uncertainties related to leptons

This section lists the uncertainties for leptons considered for the trilepton channel. This includes also the uncertainties from the fake lepton estimation methods. Since tau leptons are not considered as reconstructed particles and neutrinos cannot be detected by the ATLAS detector, the term "leptons" refers to electrons and muons in this analysis.

### 8.2.1. Lepton selection efficiencies

The analysis presented in this paper heavily relies on the exact knowledge of the lepton selection. Therefore, the selection uncertainties need to be carefully investigated. The discrepancies for reconstruction, identification, isolation and trigger selection efficiencies between data and Monte Carlo must be taken into account for electrons and muons.

Separate scale factors are derived to match the lepton efficiencies in Monte Carlo in terms of these four selection requirements. For electrons, this is done using the tag-and-probe method (see below) in $Z \to ee$, $W \to e\nu$ and $J/\Psi \to ee$ events, for example as shown in [199]. The corresponding uncertainties are taken as shape uncertainties. The muon scale factor uncertainties are derived from $Z \to \mu\mu$ and $J/\Psi \to \mu\mu$ tag-and-probe experiments [200] and are divided into statistical and systematic contributions. The efficiencies and their uncertainties are determined, for both electrons and muons, by applying the corresponding reconstruction, identification, isolation or trigger requirements to the probe lepton.

The tag-and-probe method makes use of well known processes such as $Z \to \ell^+\ell^-$ and $J/\Psi \to \ell^+\ell^-$. A lepton pair is required to originate from the corresponding process.

| Systematic uncertainty | Components | Section |
|---|---|---|
| Luminosity | 1 | 8.1 |
| Electron reco, ID, iso and trigger | 4 | 8.2.1 |
| Muon reco, ID, iso and trigger | 8 | |
| Electron resolution and scale | 2 | 8.2.2 |
| Muon resolution and scale | 5 | |
| Matrix method fakes | 2 | 8.2.3 |
| Fake factors | 4 | |
| Pileup reweighting | 1 | 8.3.1 |
| JVT scale factor | 1 | 8.3.2 |
| Jet energy scale | 20 | 8.3.3 |
| Jet energy resolution | 1 | 8.3.4 |
| $b$-tagging | 7 | 8.3.5 |
| $c$-tagging | 4 | |
| light-jet tagging | 12 | |
| $E_\mathrm{T}^\mathrm{miss}$ scale | 1 | 8.4 |
| $E_\mathrm{T}^\mathrm{miss}$ resolution | 2 | |
| $t\bar{t}Z$ `A14` | 1 | 8.5.2 |
| $t\bar{t}W$ `A14` | 1 | |
| $t\bar{t}Z$ generator | 1 | |
| $t\bar{t}W$ generator | 1 | |
| $t\bar{t}Z$ $\mu_R$ and $\mu_F$ scale choice | 3 | |
| $t\bar{t}W$ $\mu_R$ and $\mu_F$ scale choice | 3 | |
| $t\bar{t}Z$ PDF | 6 | |
| $t\bar{t}W$ PDF | 4 | |
| $t\bar{t}W$ NLO QCD normalisation | 1 | |
| $WZ$ theory | 4 | 8.5.3 |
| $tWZ$ modelling | 1 | 8.5.4 |
| $tWZ$ shower | 1 | |
| $t\bar{t}H$ xsec QCD scale | 1 | 8.5.5 |
| $t\bar{t}H$ xsec PDF uncertainty | 1 | |
| $tZ$ normalisation | 1 | |
| Other normalisation | 1 | |
| MC statistics | 5 | 8.5.6 |

Table 8.1.: Systematic uncertainties considered in this analysis, sorted by the sections where they are discussed in detail. The number of separate variations that these systematics are composed of is also shown.

From this pair, the *tag* lepton is selected using tight requirements and the *probe* lepton is selected using very loose requirements. The motivation of this method is that if the tag lepton is a prompt one from a well known process, such as $Z \to \ell\ell$, the probe lepton is assumed to be real. Therefore, efficiencies for prompt leptons can be tested using the probe lepton. The systematic uncertainties are applied by scaling separate events in the Monte Carlo samples accordingly.

### 8.2.2. Lepton momentum scale and resolution

For the event selection in the trilepton channels, the electron energies need to be well known, for example for the $Z$-window and lepton $p_{\mathrm{T}}$ requirements. Therefore, the uncertainties for the lepton momentum calibration techniques and their resolutions need to be considered. The uncertainties of the lepton momentum scales and resolutions in Monte Carlo are determined by investigating reconstructed $Z \to \ell^+\ell^-$ and $J/\Psi \to \ell^+\ell^-$ mass distributions [200, 234]. An additional Monte Carlo scale factor uncertainty for electrons stems from the determination of the ratio between the deposited energy in the electromagnetic calorimeter and the momentum in the inner detector, determined in $W \to e\nu$ events. Lepton momentum scales and resolutions are treated as separate uncertainties for each lepton. For muons, the uncertainties are both evaluated in the inner detector and the muon spectrometer. The uncertainties are applied by varying the lepton momenta in the Monte Carlo samples accordingly.

### 8.2.3. Uncertainties from the determination of the fake lepton background

Fake leptons from hadronic processes or from photon conversions, that are misreconstructed as prompt leptons, are one of the most important background sources in the trilepton channel. Therefore, the uncertainties from the fake lepton estimation methods are important parameters to consider in this analysis. Hadronic fakes in the $3\ell$-$WZ$-CR and $4\ell$-$ZZ$-CR control regions and the $3\ell$-Z-2j-VR validation region, as well as photon conversion fakes in all regions, are estimated using the fake factor method (see Section 6.3.2). For the estimation of hadronic fakes in the $3\ell$-1b-VR region and in the signal regions, the data driven matrix method (see Section 6.3.1) is used.

For the fake factor method, the uncertainties of the fits, shown in Section 6.3.2, are used. Separate Monte Carlo events, containing fake leptons, are scaled up and down, according to the uncertainties for the corresponding fake leptons. Since four different fake factors are derived for electrons and muons from light and heavy flavour hadronic processes, four corresponding systematic variations are used.

For the matrix method, the subtraction of Monte Carlo real lepton events from the data in the control regions (see Section 6.3.1) is varied up and down by a factor of 30%. This variation is applied as the systematic uncertainty for this method. It is derived for the electrons and muons separately. The resulting systematic variations for electrons and muons are used to shift the data driven fake estimate accordingly.

## 8.3. Systematic uncertainties related to jets

Jets, $b$-tagging and pileup play important roles in this measurement and therefore need to be treated with care. This section includes all systematic variations for Monte Carlo, related to jet measurements. This includes the uncertainties related to the pileup reweighting, the efficiency of the jet vertex tagger algorithm, jet energy scale and resolution, as well as to $b$-tagging.

### 8.3.1. Pileup reweighting

The pileup profile needs to be properly modelled for Monte Carlo samples in order to match the profile in data. However, the correct pileup profile of the ATLAS data taking run that these samples should describe, is often not yet well known at the time when the Monte Carlo samples are generated. Therefore, the Monte Carlo pileup distribution needs to be reweighted, which is called *pileup reweighting*. The uncertainty for the pileup reweighting scale for Monte Carlo events is represented by one shape variation. It includes the uncertainty of the reconstructed jet $p_{\mathrm{T}}$, depending on the mean number of inelastic $pp$ interactions per bunch crossing $\langle \mu \rangle$ and the number of reconstructed primary vertices $N_{\mathrm{PV}}$. Two methods are used to determine this uncertainty by comparing data and Monte Carlo events. The first one uses jets reconstructed from tracking information (called *track-jets*) and the other one uses the $p_{\mathrm{T}}$ imbalance between a reconstructed jet and a $Z$ boson [235]. The systematic uncertainty is applied by assigning weights to separate events in the Monte Carlo samples.

### 8.3.2. Jet vertex tagger efficiency

For a successful $t\bar{t}Z$ cross section measurement, it is necessary to distinguish between pileup and jets from the hard scattering process. Therefore, the efficiency of the jet vertex tagger (JVT) needs to be precisely known. For the uncertainty related to the Monte Carlo scale factors of the JVT algorithm (see Section 5.4), three contributions are merged into one systematic variation. The performance of the algorithm is probed by running on $Z$+jets events from different Monte Carlo generators. An uncertainty for the residual pileup contamination after applying the JVT algorithm is included. Lastly, statistical uncertainties for the JVT scale factor determination are applied [203, 204]. This systematic uncertainty is also applied by assigning weights to separate events in the Monte Carlo samples.

### 8.3.3. Jet energy scale

The precise measurement of jet energies is very important in ATLAS analyses. The calibration of the jet energy is therefore a crucial task. The jet energy is determined from energy depositions in the calorimeter system using the *jet energy scale* (JES) calibration, see Section 5.4. For the JES uncertainty, 20 separate systematic variations are included. The JES is determined in test beam data, data from LHC collisions and Monte Carlo simulations. The uncertainties of these methods propagate into the JES systematics. Uncertainties on the $Z$+jets, $\gamma$+jets and multijet in-situ calibrations, as well as on pileup are part of these systematics [205, 206]. The systematic uncertainties are applied by varying the individual jet energies of the Monte Carlo events.

### 8.3.4. Jet energy resolution

Similar to the JES, the *jet energy resolution* (JER, see Section 5.4) needs to be precisely known for the $t\bar{t}Z$ analysis which relies on jet information. The JER is defined as

$$\frac{\sigma(p_{\mathrm{T}})}{p_{\mathrm{T}}} = \frac{N}{p_{\mathrm{T}}} \oplus \frac{S}{\sqrt{p_{\mathrm{T}}}} \oplus C \; , \tag{8.1}$$

where $N$ describes the effect of pileup and electronic noise at low $p_{\mathrm{T}}$, $S$ describes the stochastic effect from the sampling structure of the calorimeters and $C$ is a constant term in $p_{\mathrm{T}}$. These JER parameters are determined in in-situ calibrations. Events from the processes $Z \rightarrow ee/\mu\mu$+jets and $\gamma \rightarrow ee/\mu\mu$+jets are used to measure the imbalance between the reconstructed gauge bosons and the jets to determine the detector resolution. Di-jet events are used to determine the resolution in the higher $p_{\mathrm{T}}$ and $|\eta|$ ranges [205, 206, 236]. The JER uncertainty is expressed by one systematic variation, shifting the individual jet energies of the Monte Carlo events.

### 8.3.5. Jet flavour tagging uncertainties

Since top quarks decay in almost 100% of all cases into bottom quarks and $W$ bosons, $b$-tagging is a necessary tool to identify the $t\bar{t}Z$ signal process and to reject many background processes. The uncertainties on the flavour tagging efficiencies for the `MV2c10` tagger are parametrised by 23 different systematic variations, derived from studies using Monte Carlo samples, similar to the methods described in [237,238]. They are separated into several $p_T$ bins for the $b$-tagging (seven variations), $c$-tagging (four variations) and light jet tagging efficiencies, respectively. The light jet tagging efficiency systematics are further divided into bins of $\eta$, which yields a total of 12 light jet variations. The uncertainties for the $b$-tagging efficiency are derived from calibration studies using top quark dilepton Monte Carlo samples. Monte Carlo samples of $D^*$ meson decays are used for the $c$-tagging efficiency uncertainties. For the light-jet tagging efficiency uncertainties, Monte Carlo multijet events are used. The uncertainties are evaluated for a `MV2c10` working point corresponding to a nominal $b$-tagging efficiency of 77% [207]. The $c$- and light-jet tagging efficiencies are also referred to as mistag efficiencies. The systematic uncertainties are applied by assigning weights to separate events in the Monte Carlo samples, according to the jet flavours, jet $p_T$ and light jet $\eta$ in these events.

## 8.4. Uncertainties on the missing transverse momentum

The definitions of the $3\ell$-$WZ$-CR and $4\ell$-$ZZ$-CR control regions rely on cuts on the missing transverse momentum. Therefore, the $E_T^{\text{miss}}$ uncertainty also has to be taken into account. One systematic variation on the $E_T^{\text{miss}}$ scale and two variations on the $E_T^{\text{miss}}$ resolution are considered.

Data and Monte Carlo events of the $Z \to \mu\mu$ process in association with additional jets are used to study the $E_T^{\text{miss}}$ properties. Since no $E_T^{\text{miss}}$ from neutrinos is expected in these events, they are used to check the momentum imbalance between all reconstructed objects and the soft term to determine the $E_T^{\text{miss}}$ scale and resolution as well as their corresponding uncertainties [211], see also Section 5.6.

## 8.5. Uncertainties for the different Monte Carlo processes

In this section, the systematic uncertainties related to the different Monte Carlo signal and background processes are discussed. This includes uncertainties on the overall Monte Carlo normalisation and uncertainties depending on kinematic distributions. The uncertainties of the most important processes, including $t\bar{t}Z$, $WZ$ and $tWZ$, are discussed in detail. A short overview of renormalisation, factorisation, resummation and shower matching scales is given. This section also discusses the statistical uncertainties of the Monte Carlo samples.

### 8.5.1. Monte Carlo scales as sources of systematic uncertainties

For the discussion of the systematic uncertainties related to the Monte Carlo samples, some concepts need to be explained in advance. In this section, the renormalisation, factorisation, resummation and shower matching scales are discussed. Their influence is determined by varying their values within a certain range and checking the normalisation and shape variations in Monte Carlo.

#### Renormalisation and factorisation scale

In perturbative QCD, ultraviolet (UV) and infrared (IR) divergences can occur because the strong coupling $\alpha_s(\mu^2)$ is scale dependent. Therefore, dedicated scales need to be considered in QCD calculations for the energies at which the theory is probed. The *renormalisation scale* $\mu_R$ is chosen in a way to avoid the UV divergences and the *factorisation scale* $\mu_F$ is chosen to avoid the IR divergences. For Monte Carlo samples in ATLAS analyses, the scales $\mu_R$ and $\mu_F$ are usually set to the same value. For analyses using data taken at a centre-of-mass energy of $\sqrt{s} = 13$ TeV, both scales are set to e.g. $\mu_R = \mu_F = H_{\mathrm{T}}/2$. In this case, $H_{\mathrm{T}}$ is defined as the scalar sum of the transverse masses $\sqrt{p_{\mathrm{T}}^2 + m^2}$ of all final state particles. Usually, the uncertainties due to these scales are determined by varying $\mu_R$ and $\mu_F$ by factors of 2 and 0.5 respectively and comparing the effect on the Monte Carlo samples.

#### Resummation scale (QSF parameter)

In the parton shower generation of the Monte Carlo samples, the emission of real and virtual soft gluons can cause IR divergences. These IR divergences cancel out, but large logarithmic terms in some regions of the phase space still remain. Those logarithms are resummed at the *resummation scale* (also called QSF parameter). The systematic uncertainty due to the choice of this scale is determined by varying it by factors of 2 and 0.5 respectively and comparing the effect on the Monte Carlo samples.

|            | 3ℓ-$Z$-1b4j | 3ℓ-$Z$-2b4j | 3ℓ-$Z$-2b3j | 3ℓ-no$Z$-2b4j | 3ℓ-$WZ$-CR | 4ℓ-$ZZ$-CR |
|------------|-------------|-------------|-------------|---------------|------------|------------|
| $t\bar{t}Z$ | +1.44%      | +1.19%      | +0.68%      | +1.27%        | +2.38%     | +0.79%     |
|            | −0.52%      | −1.59%      | −1.06%      | −0.50%        | −2.29%     | −0.62%     |
| $t\bar{t}W$ | +7.12%      | +7.13%      | +7.22%      | +7.30%        | —          | —          |
|            | −7.08%      | −7.09%      | −7.16%      | −7.24%        |            |            |

Table 8.2.: PDF uncertainties for the $t\bar{t}Z$ and $t\bar{t}W$ Monte Carlo samples for each region. The $t\bar{t}W$ PDF uncertainties are not applied in the $WZ$ and $ZZ$ control regions.

**Shower matching scale (CKKW matching)**

Some Monte Carlo backgrounds are modelled by multi-leg generators, which already generate multiple additional quarks and gluons at the matrix element level. For those samples, double counting of jet configurations between jets from matrix element quarks or gluons and parton shower jets needs to be avoided. A resolution parameter, called the *shower matching scale* is defined to separate between the jets from the matrix element and from parton showering. This is done via the Catani-Krauss-Kuhn-Webber (CKKW) method [239]. The uncertainty from the shower matching scale is calculated by varying its value by a certain amount and checking the effect on the Monte Carlo yields, see Section 8.5.3.

## 8.5.2. Uncertainties on $t\bar{t}W$ and $t\bar{t}Z$

The uncertainties due to the signal modelling need to be well understood. The influence of varying the renormalisation and factorisation scales by factors of 2 and 0.5 is determined by reweighting the nominal $t\bar{t}Z$ and $t\bar{t}W$ samples according to these scale choices. Three systematic variations are applied for both $t\bar{t}Z$ and $t\bar{t}W$, respectively: one variation for each of the separate $\mu_R$ or $\mu_F$ variations, while keeping the remaining one fixed at the nominal value, and one variation for $\mu_R$ and $\mu_F$ varied up and down simultaneously.

The choice of the parton distributions function (PDF) can cause different results for the $t\bar{t}W$ and $t\bar{t}Z$ yields. Different PDFs use different input datasets and parametrisations. For this analysis, the uncertainty of the PDF choice is determined from the envelope of the $t\bar{t}Z$ and $t\bar{t}W$ yields in the separate regions using different PDFs. This is done using the `CT14NLO` [240], `MMHT2014NLO` [241] and `NNPDF3.0NLO` PDF sets. The corresponding up and down variations for each region are shown in Table 8.2.

The uncertainties from the `A14` Monte Carlo tunes corresponding to the parton shower-
ing are amongst the most important systematics in this analysis, especially for the $t\bar{t}Z$
samples. Monte Carlo samples are generated with the parameters of the `A14` shower
tunes varied up and down. For the detector simulation of these samples, AFII is used
(see Section 6.2). The shape uncertainties from the `A14` tunes on the $t\bar{t}W$ and $t\bar{t}Z$ sam-
ples are estimated by comparing these samples with an AFII sample using the nominal
tunes.

For the uncertainty of the $t\bar{t}W$ and $t\bar{t}Z$ generator choice, samples generated with MAD-
GRAPH5_aMC@NLO (at NLO) and with SHERPA (at leading order) are compared. These
samples use the AFII detector simulation and do not have truth matching applied. Ex-
cept for these two differences, the MADGRAPH5_aMC@NLO samples are identical to the
nominal samples. The SHERPA samples use the `NNPDF3.0NNLO` PDF and a dedicated
SHERPA parton shower tune. The difference between the MADGRAPH5_aMC@NLO and
SHERPA samples are taken as the shape uncertainty for the $t\bar{t}Z$ and $t\bar{t}W$ samples, re-
spectively.

For the NLO QCD and electroweak cross section calculation, used for the normalisation
of the $t\bar{t}W$ sample, an uncertainty of 13% is assigned. This uncertainty takes into ac-
count the PDF and scale uncertainties. It is rounded and symmetrised with respect to
the reference [155].

### 8.5.3. $WZ$ background with additional jets

The process of $WZ \to \ell\ell\nu$ in association with additional heavy and light flavour jets
is the most relevant background in the trilepton channel that is estimated from Monte
Carlo. Therefore, its systematic uncertainties have to be carefully evaluated. The sys-
tematic uncertainties on the $WZ$ Monte Carlo yields are evaluated for each signal region
separately. They are estimated using Monte Carlo samples with variations in the renor-
malisation, factorisation and resummation scales (see Section 8.5.1) of a factor 2 and 0.5,
respectively. In addition, variations in the shower matching scale are taken into account
by using samples with matching scales of 15 GeV and 30 GeV (the nominal value is
20 GeV).

All samples are assumed to have the same cross sections as the corresponding nominal
$WZ$ samples. The yields in all four trilepton signal regions, as well as in the $3\ell\text{-}WZ\text{-CR}$
control region (see Chapter 7), are evaluated for all variations. Next, *transfer factors*
are calculated for each variation and each signal bin:

$$f_{var,SR} = \frac{N_{\text{var,SR}}}{N_{\text{var},3\ell\text{-}WZ\text{-CR}}} \tag{8.2}$$

| $3\ell$-$Z$-1b4j | $3\ell$-$Z$-2b4j | $3\ell$-$Z$-2b3j | $3\ell$-no$Z$-2b4j |
|---|---|---|---|
| $\pm 30\%$ | $\pm 50\%$ | $\pm 45\%$ | $\pm 42\%$ |

Table 8.3.: Systematic uncertainty for the $WZ$ background in the different signal regions.

where $N_{\text{var,SR}}$ is the event yield in the signal region SR for the systematic variation "var" and $N_{\text{var,}3\ell\text{-}WZ\text{-CR}}$ is the corresponding event yield in the $3\ell$-$WZ$-CR region. The total uncertainty for each signal region is calculated by summing all up and down variations from the transfer functions in quadrature, including statistical uncertainties. The resulting uncertainties for the different signal regions are shown in Table 8.3.

### 8.5.4. Uncertainties on the $tWZ$ background

The single top production in the $Wt$ channel in association with an additional $Z$ boson is an important background for the $t\bar{t}Z$ trilepton channel. Therefore, uncertainties concerning the normalisation and the shape of the Monte Carlo samples have to be carefully evaluated. One systematic variation is assigned for the normalisation and shape uncertainties, respectively.

The up and down variations of the normalisation are dominated by different uncertainties. For the down variation of the $tWZ$ normalisation, the overlap removal procedure between $tWZ$ at NLO and $t\bar{t}Z$ at LO is the biggest source of systematic uncertainty. To estimate it, the difference between the yields using the nominal diagram removal technique and an alternative one [242, 243] is evaluated. The alternative overlap removal technique yields 28% less $tWZ$ events. This variation is therefore set as the down variation of the normalisation uncertainty. The up variation of the normalisation uncertainty is determined by NLO QCD calculations [244]. Since no dedicated uncertainty estimation is available for $tWZ$ or $tZ$ for $\sqrt{s} = 13$ TeV, the uncertainty on the $t$-channel $tZ$ process for $\sqrt{s} = 8$ TeV is chosen to be equal to the $tWZ$ normalisation uncertainty for $\sqrt{s} = 13$ TeV. Therefore, the up variation is chosen to be 10% for the $tWZ$ Monte Carlo events. These up and down variations are applied to all regions of the trilepton channel. Another uncertainty originates from the choices of the renormalisation $\mu_R$ and factorisation $\mu_F$ scales. The renormalisation and factorisation scales are varied by factors of 2 and 0.5 from the nominal value of $\mu_R = \mu_F = H_T/2$, respectively. The resulting envelope of the variations is taken as the $tWZ$ shape uncertainty. The systematic uncertainties related to showering are determined comparing HERWIG++ to the nominal shower generators PYTHIA8 and EVTGEN.

### 8.5.5. Normalisation uncertainties on other backgrounds

The $t\bar{t}H$ NLO QCD normalisation uncertainty due to the choice of normalisation and factorisation scales is $+5.8\%$ and $-9.2\%$, according to the latest ATLAS recommendations [155]. The corresponding $t\bar{t}H$ PDF uncertainty is $\pm 3.6\%$. The uncertainty on the $tZ$ process is considered to be $\pm 30\%$. This number is estimated by varying the QCD scales of the LO sample. For all other Monte Carlo backgrounds, a normalisation uncertainty of $50\%$ is applied.

### 8.5.6. Monte Carlo statistical uncertainties and statistical uncertainties of the data driven fake background

For the profile likelihood fit (see Section 10.1), the total Monte Carlo statistical uncertainties from all processes are treated as nuisance parameters for each separate region. In addition to Monte Carlo, these uncertainties also include the statistical uncertainties of the data driven fake lepton background in the signal regions. For simplicity, they will only be referred to as "Monte Carlo statistical uncertainties". These uncertainties are evaluated for each region separately. The Poissonian uncertainties are symmetrised so they can be used as the $1\sigma$ standard deviations of Gaussian distributions. They are applied in all signal regions and the $3\ell$-$WZ$-CR control region. Since the total Monte Carlo statistical uncertainty in the $4\ell$-$ZZ$-CR region is smaller than $1\%$, it is not considered as a nuisance parameter in this region.

# Other Channels of the $t\bar{t}Z$ and $t\bar{t}W$ Analysis

The analysis presented in this thesis is part of a comprehensive one, targeting both the measurement of the $t\bar{t}Z$ and the $t\bar{t}W$ cross sections. The separate results of all channels contribute to a global fit to data. While the trilepton channel is said to be the "golden channel" for $t\bar{t}Z$, due to the high expected sensitivity, other channels can help to add sensitivity. The $t\bar{t}W$ process is usually measured together with the $t\bar{t}Z$ process since it has a similar decay signature and can therefore be treated similarly in terms of background and systematic uncertainties. The overall analysis that measures the $t\bar{t}Z$ and $t\bar{t}W$ cross sections has three different channels sensitive to the $t\bar{t}Z$ process: the dilepton opposite-sign same-same flavour channel (short $2\ell$OSSF), the trilepton channel which is extensively discussed in this thesis and the tetralepton channel. Two channels are sensitive to the $t\bar{t}W$ process: the dilepton same-sign channel (short $2\ell$SS) and a separate trilepton channel. Only the signal regions will be presented in this chapter. The decay signatures of the top quark pair and the vector boson for the different analysis channels are summarised in Table 9.1. The work on the channels presented in this chapter was done by other people from the analysis group.

| Process | $t\bar{t}$ decay | Boson decay | Channel |
|---------|------------------|-------------|---------|
| $t\bar{t}W$ | $(\ell^{\pm}\nu b)(q\bar{q}b)$ | $\ell^{\pm}\nu$ | $2\ell$SS |
| | $(\ell^{\pm}\nu b)(\ell^{\mp}\nu b)$ | $\ell^{\pm}\nu$ | $t\bar{t}W$-trilepton |
| $t\bar{t}Z$ | $(q\bar{q}b)(q\bar{q}b)$ | $\ell^{+}\ell^{-}$ | $2\ell$OSSF |
| | $\boldsymbol{(\ell^{\pm}\nu b)(q\bar{q}b)}$ | $\boldsymbol{\ell^{+}\ell^{-}}$ | $\boldsymbol{t\bar{t}Z\text{-trilepton}}$ |
| | $(\ell^{\pm}\nu b)(\ell^{\mp}\nu b)$ | $\ell^{+}\ell^{-}$ | Tetralepton |

Table 9.1.: Decay signatures of the top quark pair and the vector boson for the different analysis channels. The trilepton channel sensitive to the $t\bar{t}Z$ process, which is the main topic of this thesis, is highlighted.

## 9.1. The dilepton opposite-sign same-flavour channel

The dilepton opposite-sign same-flavour ($2\ell$OSSF) channel is sensitive to the $t\bar{t}Z$ process with the top quark pair decaying fully hadronically and the $Z$ boson decaying into two charged leptons. To ensure sensitivity to this signature, two leptons with opposite electric charge and same lepton flavour are required. The lepton pair is also required to have an invariant mass inside the $Z$-window (see Chapter 7). To avoid overlap with other regions, events with additional loose leptons are vetoed. The requirements on the transverse momenta are $p_{\mathrm{T}} > 30$ GeV for the highest $p_{\mathrm{T}}$ lepton and $p_{\mathrm{T}} > 15$ GeV for the other lepton. Due to the large background contamination from $t\bar{t}$+jets and $Z$+jets in this channel (see below), a Boosted Decision Tree (BDT) is trained to differentiate $t\bar{t}Z$ from all backgrounds. The BDT discriminant (see Figure 9.1) is used to impose further cuts on the signal regions to reduce background. Three different signal regions are defined based on jet and $b$-jet multiplicities, as well as on the BDT output values. The selection criteria of the signal regions are shown in Table 9.2.

The signal and background contributions for the $2\ell$OSSF channel are shown in Table 9.3, compared to data. One main background comes from $Z$+jets events with heavy flavour contributions. To deal with this background, three dedicated control regions are used to fit the normalisation factors of $Z$+jets with exactly one or at least two heavy flavour jets.

The other background, which comes from $t\bar{t}$ events with additional jets, is highly mis-modelled in Monte Carlo and would lead to huge systematic uncertainties. Therefore, this background needs to be derived using a data driven technique. Three $t\bar{t}$ enriched validation regions are derived, using similar cuts as for the signal regions but with an opposite-flavour requirement for the leptons and no BDT cut. To cross check if $t\bar{t}$ events can be derived from these regions, the BDT output distributions are compared between the validation regions and the corresponding signal regions, using $t\bar{t}$ events from Monte Carlo. After subtracting the non-$t\bar{t}$ events from the validation regions using the Monte Carlo predictions of the corresponding samples, the extracted results are extrapolated to the signal regions by scaling them according to the factor of $N_{t\bar{t}}^{SR}/N_{t\bar{t}}^{VR}$, where $N_{t\bar{t}}$ is the number of $t\bar{t}$ events in the signal and validation regions respectively, estimated from Monte Carlo.

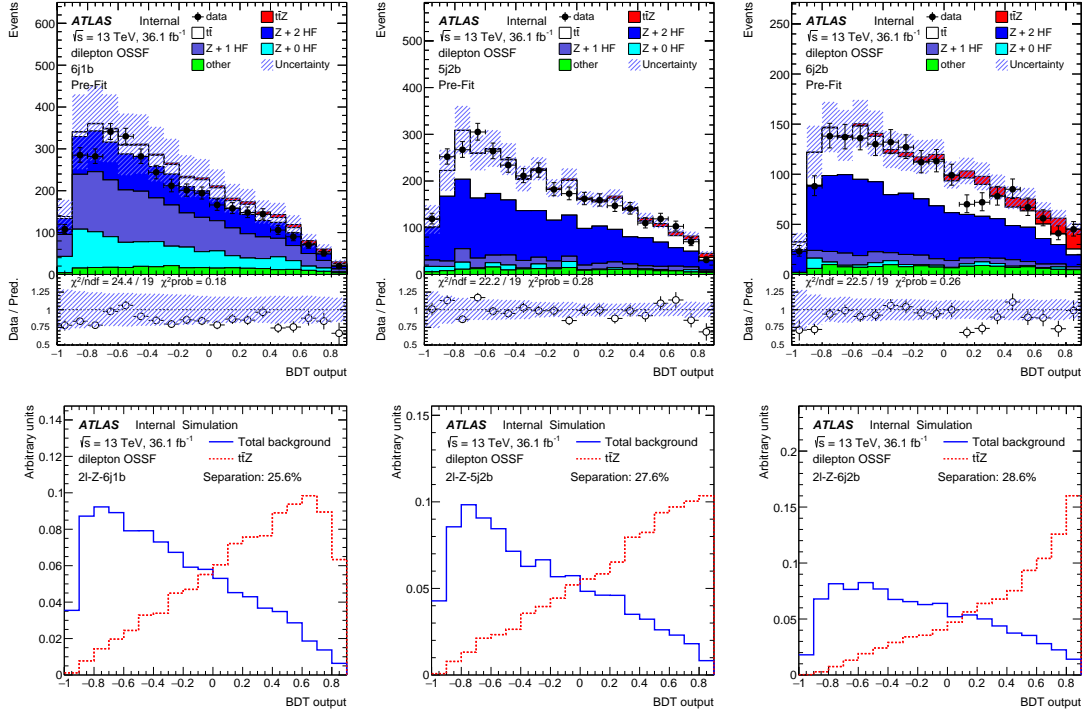

Figure 9.1.: BDT discriminants for the three signal regions of the 2ℓOSSF channel. Pre-fit plots are shown in the top row and separation plots between signal and background are shown in the bottom row. The shaded bands show statistical and systematic uncertainties [3].

| Variable | 2ℓ-Z-6j1b-SR | 2ℓ-Z-5j2b-SR | 2ℓ-Z-6j2b-SR |
|---|---|---|---|
| Number of leptons | | = 2 | |
| Lepton flavour | | Same | |
| Lepton charge | | Opposite | |
| $Z$-window cut | | Required | |
| $p_{\mathrm{T}}$ (1st lepton) | | > 30 GeV | |
| $p_{\mathrm{T}}$ (2nd lepton) | | > 15 GeV | |
| $n_{b-\mathrm{jets}}$ | =1 | ≥ 2 | ≥ 2 |
| $n_{jets}$ | ≥ 6 | =5 | ≥ 6 |

Table 9.2.: Summary of the event selection for the 2ℓOSSF signal regions. In addition, different cuts on the BDT output are applied for each region.

| | 2ℓ-Z-6j1b-SR | 2ℓ-Z-5j2b-SR | 2ℓ-Z-6j2b-SR |
|---|---|---|---|
| $t\bar{t}Z$ | 35.0 ± 6.40 | 36.8 ± 2.74 | 100 ± 13.1 |
| DD $t\bar{t}$ | 35.8 ± 5.99 | 108 ± 10.8 | 202 ± 15.3 |
| $Z + 2$ HF | 111 ± 25.5 | 206 ± 30.7 | 297 ± 64.7 |
| $Z + 1$ HF | 135 ± 31.0 | 31.6 ± 8.79 | 42.7 ± 13.7 |
| $Z + 0$ HF | 72.3 ± 33.4 | 11.6 ± 10.2 | 15.2 ± 9.11 |
| Other | 43.9 ± 16.2 | 30.7 ± 8.20 | 64.0 ± 16.8 |
| Total | 434 ± 86.0 | 425 ± 41.8 | 706 ± 91.5 |
| Observed | 338 | 368 | 613 |

Table 9.3.: The expected event yields in the 2ℓOSSF signal regions compared to data. Monte Carlo statistical and systematic uncertainties are shown.

For the training of the BDT however, $t\bar{t}$ events from Monte Carlo samples are used. Otherwise, the BDT would suffer from overtraining due to the low statistics in the data driven $t\bar{t}$ estimate.

| Variable | $4\ell$-SF-1b | $4\ell$-SF-2b | $4\ell$-DF-1b | $4\ell$-DF-2b |
|---|---|---|---|---|
| Number of leptons | | $= 4$ | | |
| Leading lepton $p_T$ | | $> 27$ GeV | | |
| Other lepton $p_T$ | | $> 7$ GeV | | |
| $Z_2$ leptons | $e^\pm e^\mp,\ \mu^\pm\mu^\mp$ | | $e^\pm\mu^\mp$ | |
| $p_{T4}$ | $> 7$ GeV | $> 10$ GeV | $> 7$ GeV | $> 10$ GeV |
| $p_{T34}$ | $> 25$ GeV | — | $> 35$ GeV | — |
| $E_T^{\mathrm{miss}}$ outside $Z$-window | $> 40$GeV | — | — | — |
| $E_T^{\mathrm{miss}}$ inside $Z$-window | $> 80$GeV | $> 40$GeV | — | — |
| $N_{b\text{-jets}}$ | 1 | $\geq 2$ | 1 | $\geq 2$ |

Table 9.4.: Summary of the event selection for the tetralepton signal regions. The regions with same flavour $Z_2$ leptons are denoted by *SF* and the regions with different flavour $Z_2$ leptons are denoted by *DF*.

## 9.2. The tetralepton channel

The channel with four charged leptons, called the *tetralepton channel*, is sensitive to the $t\bar{t}Z$ process in which the top quark pair decays fully leptonically and the $Z$ boson decays into two charged leptons. Two lepton pairs with opposite electric charge are required with one of them being an OSSF lepton pair. The OSSF lepton pair with the invariant mass closest to the $Z$ boson mass is called the $Z_1$ lepton pair and the other lepton pair is called the $Z_2$ lepton pair. Four signal regions are defined by the $b$-jet multiplicity and the lepton flavours of the $Z_2$ lepton pair (same flavour and opposite flavour). Additional cuts on the missing transverse momentum, the scalar sum $p_{T34}$ of the two lowest transverse momenta and on the lowest lepton transverse momentum $p_{T4}$ are required to reduce the background from fake leptons. The definition of the signal regions in the tetralepton channel is shown in Table 9.4.

Table 9.5 shows the expected event yields in the four tetralepton signal regions compared to data. All regions have a high signal purity but low event yields compared to the other multilepton channels. This is a characteristic property of this channel due to the low decay branching ratio of the top quark pair in the dilepton channel. Depending on the region, $ZZ$, $tWZ$ and fake leptons are the dominant backgrounds. The contribution from fake leptons is determined via the fake factor method, see Section 6.3.2. Pre-fit distributions for all tetralepton signal regions combined are shown in Figure 9.2.

|         | $4\ell$-SF-1b     | $4\ell$-SF-2b    | $4\ell$-DF-1b     | $4\ell$-DF-2b    |
|---------|-------------------|------------------|-------------------|------------------|
| $t\bar{t}Z$  | $6.56 \pm 0.39$ | $6.14 \pm 0.56$ | $7.38 \pm 0.42$ | $5.99 \pm 0.74$ |
| $ZZ$    | $2.27 \pm 0.99$   | $1.06 \pm 0.48$  | $0.19 \pm 0.06$   | $0 \pm 0$        |
| $tWZ$   | $1.60 \pm 0.53$   | $0.55 \pm 0.27$  | $1.57 \pm 0.40$   | $0.51 \pm 0.27$  |
| $t\bar{t}H$  | $0.58 \pm 0.07$ | $0.62 \pm 0.09$ | $0.68 \pm 0.08$ | $0.57 \pm 0.08$ |
| Other   | $0.18 \pm 0.05$   | $0.15 \pm 0.07$  | $0.22 \pm 0.06$   | $0.10 \pm 0.02$  |
| MC fakes| $1.83 \pm 0.82$   | $1.23 \pm 0.62$  | $0.93 \pm 0.16$   | $0.39 \pm 0.11$  |
| Total   | $13.02 \pm 1.49$  | $9.75 \pm 1.16$  | $10.96 \pm 0.70$  | $7.57 \pm 0.89$  |
| Observed| 18                | 14               | 11                | 5                |

Table 9.5.: The expected event yields in the tetralepton channel signal regions compared to data. Statistical and systematic uncertainties are shown.

## 9.3. The $t\bar{t}W$ channels

To optimise the performance of the analysis, a total of 16 different signal regions with two or three charged leptons, sensitive to the $t\bar{t}W$ process, are defined. To avoid overlap with other regions, events with additional loose leptons are vetoed. Four of those regions are trilepton signal regions, vetoing $Z$-like OSSF lepton pairs with invariant mass inside the $Z$-window and requiring two or three jets. These four regions are called *$t\bar{t}W$-trilepton regions*. They are sensitive to the leptonic $W$ decay and the dileptonic $t\bar{t}$ decay from the $t\bar{t}W$ process. The $t\bar{t}W$-trilepton regions are divided by sum of the lepton charges and by the $b$-jet multiplicities (one or at least two $b$-jets). A cut on the total sum of transverse momentum of $H_T > 240$ GeV is applied for the regions with exactly one $b$-jet. A minimum transverse momentum of $p_T > 27$ GeV is required for all three charged leptons. The dominant backgrounds in the $t\bar{t}W$-trilepton regions are $t\bar{t}Z$, $t\bar{t}H$, $WZ$, $tZ$ and fake leptons, depending on the region. Figure 9.3 shows the second highest lepton $p_T$ distributions for the $t\bar{t}W$-trilepton regions.

The other twelve signal regions are sensitive to the $t\bar{t}W$ process with the leptonic decay of the $W$ boson and the lepton+jets decay channel of the top quark pair. To avoid background from $t\bar{t}Z$ events and overlap with the $2\ell$OSSF channel, the two leptons from the $t\bar{t}W$ decay signature are required to have the same electric charge. Therefore, this channel is also called the dilepton same-sign ($2\ell$SS) channel. The $2\ell$SS signal regions are divided by lepton flavour (electron-electron, muon-muon or electron-muon), $b$-jet multiplicity (one or at least two $b$-jets), and the lepton charges (both positive or both negative). The latter separation is made to take into account the imbalance between the $t\bar{t}W^+$ and $t\bar{t}W^-$ production. Missing transverse momentum of $E_T^{\text{miss}} > 20$ GeV is required for the two di-muon channels with at least two $b$-jets and $E_{\mathrm{T}}^{\mathrm{miss}} > 40$ GeV for the other regions. Both leptons are required to have a transverse momentum of $p_{\mathrm{T}} > 27$ GeV. At least four jets[1] are required for the electron-muon and di-electron channels, as well as for the di-muon channels with exactly one $b$-jet. An additional veto on the invariant mass $m_{ll}$ of same flavour lepton pairs of $|m_{ll} - m_Z| > 10$ GeV is required to reject $t\bar{t}Z$ and $Z$+jets backgrounds with misidentified charge (for example from charge flips, see Section 5.2). The main backgrounds in the $2\ell$SS regions are lepton fakes from hadronic processes (mostly from $t\bar{t}$ events), followed by fake lepton events from photon conversions and electrons with misidentified charge. Figure 9.4 shows the second highest lepton $p_{\mathrm{T}}$ distributions for the $2\ell$SS regions.

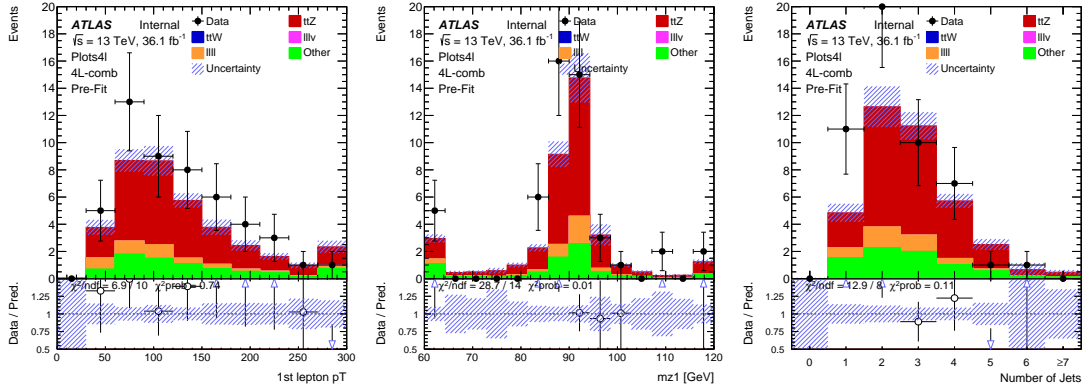

Figure 9.2.: Pre-fit distributions of the leading lepton $p_{\mathrm{T}}$, the invariant mass of the $Z_1$ lepton pair and the jet multiplicity (from left to right) with all tetralepton signal regions combined. The shaded bands show statistical and systematic uncertainties [3].

In both the $t\bar{t}W$-trilepton and $2\ell$SS regions, leptons are defined slightly differently with respect to the $t\bar{t}Z$ channels, in order to deal with the more dominant background from fake leptons. The likelihood cut for the electron identification is tightened (see Section 5.2) and for the lepton isolation, the so-called *prompt lepton isolation* method is used. This method was developed for the most recent $t\bar{t}H$ analysis in multilepton final states [88]. The prompt lepton isolation method makes use of a Boosted Decision Tree to distinguish between prompt and non-prompt leptons by using $b$-tagging variables, secondary vertex information, lepton and jet track variables, lepton calorimeter and track isolation information, as well as the track jet multiplicity. For all $t\bar{t}W$ regions, lepton fakes from hadronic sources are estimated using the matrix method, see Section 6.3.1.

---

1. For the di-muon channel with at least two $b$-tags, at least two jets are required due to the $b$-jet requirement.

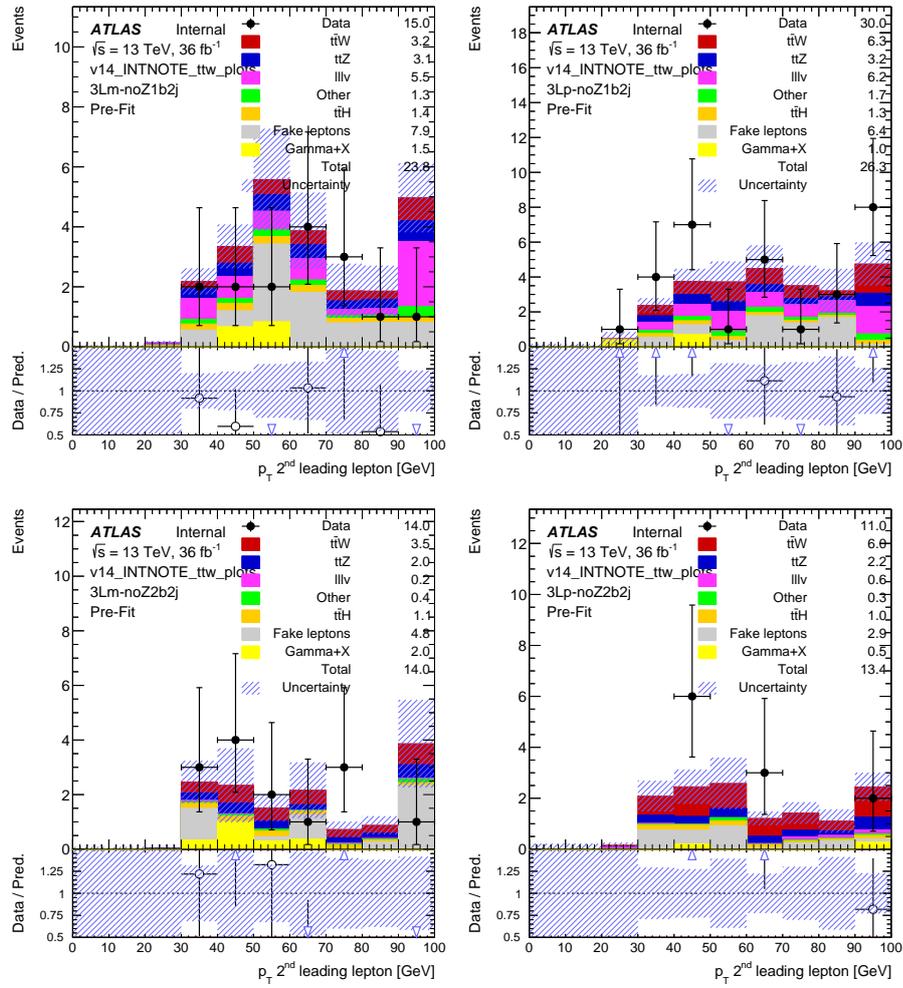

Figure 9.3.: Plots for the second highest lepton $p_T$ in the $t\bar{t}W$-trilepton regions. The shaded bands show statistical and systematic uncertainties [3].

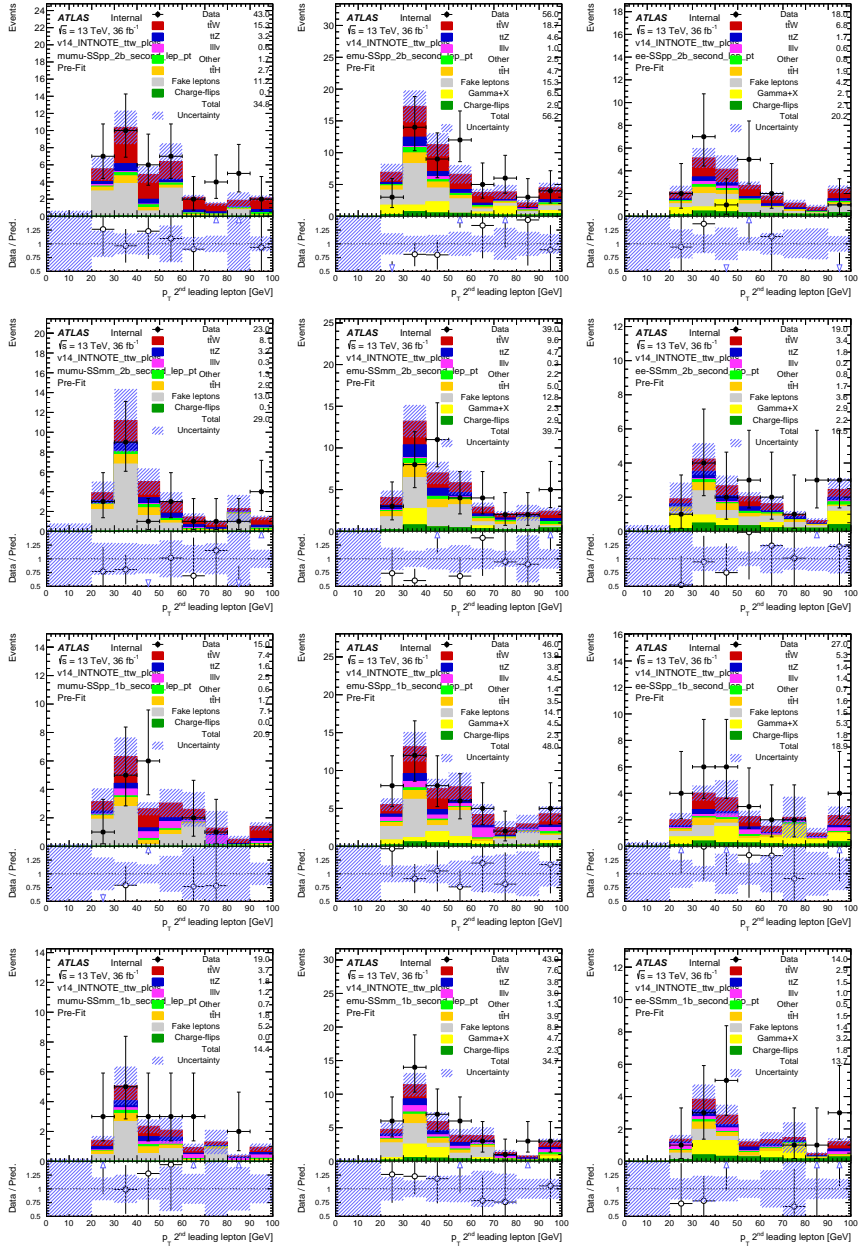

Figure 9.4.: Plots for the second highest lepton $p_{\mathrm{T}}$ in the 2$\ell$SS regions. The shaded bands show statistical and systematic uncertainties [3].

This chapter describes the setup that will be used in the final fit to data shown in Chapter 11. The basic idea of a profile likelihood fit is discussed, as well as the strategy of the fit. An Asimov fit is employed to test its performance.

For the fit, the `TRExFitter` software framework is used, which includes the `HistFactory` tool [245]. This framework follows a similar approach as the `HistFitter` software framework [246]. Both are fitting tools developed for high energy physics, incorporating the idea of signal, control and validation regions.

## 10.1. Profile likelihood fit

The profile likelihood method implemented in the `TRExFitter` software framework includes the signal strength and background normalisation as free fit parameters. Additional fit parameters are the impacts of the systematic uncertainties which are treated as nuisance parameters. The fit is performed by maximising the likelihood function

$$
\begin{aligned}
L(\boldsymbol{n}, \boldsymbol{\theta}^0 | \mu_{\text{sig}}, \boldsymbol{b}, \boldsymbol{\theta}) \quad = \quad & \prod_{i \in SR} P(n_i | \lambda_i(\mu_{\text{sig}}, \boldsymbol{b}, \boldsymbol{\theta})) \\
& \times \prod_{i \in CR} P(n_i | \lambda_i(\mu_{\text{sig}}, \boldsymbol{b}, \boldsymbol{\theta})) \\
& \times C_{\text{syst}}(\boldsymbol{\theta}^0, \boldsymbol{\theta}) \,,
\end{aligned}
\tag{10.1}
$$

where $P(n_i|\lambda_i(\mu_{\text{sig}}, \boldsymbol{b}, \boldsymbol{\theta}))$ are the Poisson distributions of the number of observed events $n_i$ in each signal and control region (denoted by $SR$ and $CR$). The functions $\lambda_i(\mu_{\text{sig}}, \boldsymbol{b}, \boldsymbol{\theta})$ for each signal and control region depend on the signal strength $\mu_{\text{sig}}$, the background normalisation factors $\boldsymbol{b}$ (in the case of this analysis a two-dimensional vector for the $WZ$ and $ZZ$ normalisations) and the nuisance parameters $\boldsymbol{\theta}$ (vectors are written in boldface). The signal strength $\mu_{\text{sig}}$ is the ratio between the observed signal cross section and the theory expectation. A signal strength of $\mu_{\text{sig}} = 1$ means that the measured cross section exactly fits the theoretical prediction and a signal strength of $\mu_{\text{sig}} = 0$ means that signal is completely absent. The background normalisation factors $b_{WZ}$ and $b_{ZZ}$ describe the deviations from the $WZ$ and $ZZ$ background expectations, respectively.

The systematic uncertainties are treated as uncorrelated nuisance parameters in

$$C_{\text{syst}}(\boldsymbol{\theta}^0, \boldsymbol{\theta}) \quad = \quad \prod_{j \in S} G(\theta_j^0 - \theta_j) \,. \tag{10.2}$$

The nominal value of the systematic uncertainty $j$ within the full set of all systematics $S$ is denoted by $\theta_j^0$, which is set to zero for most of the nuisance parameters[1] and its deviation is denoted by $\theta_j$ within the Gaussian function $G(\theta_j^0 - \theta_j)$ with unit width [245, 246]. Therefore, varying $\theta_j$ by $\pm 1$ corresponds to a shift of the corresponding systematic of $\pm 1\sigma$. In this model, systematic uncertainties are assumed to be uncorrelated in the likelihood. However, a correlation is determined during the fit by evaluating which systematic uncertainties have similar effects on the result.

The benefit of the profile likelihood fit is that the nuisance parameters can be used to optimise the fit sensitivity in the multidimensional fit. This way, systematic uncertainties can be constrained or pulled towards higher or lower values. However, these side effects need to be understood and physically motivated in order to achieve a valid fit result.

## 10.2. Fit strategy

The strategy of this analysis is to determine the signal strength parameter $\mu_{t\bar{t}Z}$ of the $t\bar{t}Z$ process while including systematic uncertainties as nuisance parameters in the profile likelihood fit as described in Section 10.1. Four different signal regions, as defined in Section 7.3, with different amounts of signal and background contributions, are used to determine $\mu_{t\bar{t}Z}$. As discussed in Section 7.4, a control region is defined to determine the normalisation of the $WZ$ background process, which is the most dominant one modelled using Monte Carlo. An additional control region for the $ZZ$ background process is defined because it is an important background for the tetralepton channel in the overall analysis. Both, the signal strength and the two diboson normalisation factors are fitted as free parameters in all six regions simultaneously. The contributions of all other backgrounds are set to their expected values and are allowed to vary within their systematic uncertainties, which are included as nuisance parameters in the fit. For the cross section of $t\bar{t}Z$, where the $Z$ boson decays into two charged leptons ($e^+e^-$, $\mu^+\mu^-$ or $\tau^+\tau^-$), $\sigma_{t\bar{t}Z,\text{lep}} = 123.7$ fb is assumed as the nominal value [155] for the fit, which corresponds to a signal strength of $\mu = 1$. The $t\bar{t}W$ inclusive cross section is fixed at a value of $\sigma_{t\bar{t}W} = 600.8$ fb with an additional scale and PDF uncertainty of 13%, included as a nuisance parameter for the fit [155]. These cross sections are determined in NLO QCD and electroweak calculations. For the cross sections of $WZ$ and $ZZ$, nominal values of $\sigma_{WZ\rightarrow\ell\ell\ell\nu+\text{jets}} = 4570$ fb and $\sigma_{ZZ\rightarrow\ell\ell\ell\ell+\text{jets}} = 1053$ fb, calculated in NLO QCD, are chosen [229]. This takes into account decays into three and four charged leptons, respectively, with associated production of light and heavy flavour jets.

---

1. The statistical uncertainties from Monte Carlo and the data driven fake estimate are fixed at one, since they refer to the raw number of events without any weights applied.

In addition to the fit in the trilepton channel, a combined fit in the 2$\ell$OSSF, trilepton and tetralepton channels is performed. In the tetralepton channel, the $ZZ$ normalisation factor via the $4\ell$-$ZZ$-CR region is determined. In the 2$\ell$OSSF channel, three control regions are used to determine the background contribution of the processes with one $Z$ boson in association with one or at least two heavy flavour jets.

The $t\bar{t}W$ signal strength is determined in a fit in the 2$\ell$SS and a dedicated trilepton region, see Section 9.3. In this case, $t\bar{t}Z$ is fixed to its theoretical expectation with an additional scale and PDF uncertainty of 12%, included as a nuisance parameter for the fit [155].

## 10.3. Expected results using an Asimov fit

In order to test the capabilities of the fit setup described above, an Asimov fit is performed. The name of this method is derived from a short story written by Isaac Asimov called "Franchise". The topic of that story is that one single person is asked a number of questions and a computer then makes political decisions based on the assumption that the opinions of an average person are equivalent to the opinions of all possible voters. This method of course does not work for society but can be used as cross checks of statistical analyses.

In the case of this analysis, the term "Asimov fit" refers to the fit of the Monte Carlo prediction to itself. This means that the total Monte Carlo yields in each region are assumed to be the data yields that will be fitted to later, as it is done in Chapter 11. Of course, the result of the signal strengths and background normalisations will always be one in these cases. However, the uncertainty of the fit results will indicate the expected uncertainties of the fit to data. On the other hand, the effect of each systematic uncertainty on the result can be studied. It is also important to know if the profile likelihood fit already constrains certain systematic uncertainties when fitting the Monte Carlo to itself.

The Asimov test shown in this section is used prior to the unblinding of the signal regions. When setting up an analysis, it is a common practice that the signal regions are *blinded*. This means that one should not look at data in these regions. Otherwise, the choices of the signal region selections and the Monte Carlo samples, as well as the fake estimations might be done unintentionally in a way that good agreement between data and Monte Carlo is achieved artificially in the signal regions. That would be called a *bias*. However, having an unbiased analysis is the key to a scientifically correct estimation of the signal strength. Only this way, is it assured that a possible deviation from the Standard Model expectation can be determined, if present. In addition to the Asimov fit result, agreement between data and Monte Carlo in the validation and control regions (see Sections 7.2 and 7.4) is used to check the validity of the selection prior to unblinding, which is extensively discussed in these sections. The unblinded fit to data is shown in Chapter 11.

The result of the Asimov fit of the $t\bar{t}Z$ signal strength in the trilepton regions, including the $WZ$ and $ZZ$ normalisation as free parameters, is

$$
\begin{aligned}
\Delta\mu(t\bar{t}Z, \text{ up}) &= 19\% \\
\Delta\mu(t\bar{t}Z, \text{ down}) &= 17\% \\
\Delta b(WZ) &= 13\% \\
\Delta b(ZZ) &= 7\% \,.
\end{aligned}
\tag{10.3}
$$

The multidimensional profile likelihood fit aims at obtaining the best possible fit value for each nuisance parameter and also on their post-fit uncertainty. No deviations from the nominal nuisance parameter values are expected.

Figure 10.1 shows the post-fit uncertainty of all nuisance parameters compared to the pre-fit input. The green area shows the pre-fit $1\sigma$ area of the pre-fit Gaussian distribution of all systematics, normalised to the same width for the plot, and the yellow area shows the $2\sigma$ area, respectively. The black bars indicate the $1\sigma$ area of the post-fit Gaussian distributions of all nuisance parameters. In most cases, the post-fit Gaussian fits approx-imately the pre-fit one. There are some exceptions where the Asimov fit already slightly optimises the systematic uncertainties. This is the case for the `A14` tunes corresponding to the parton showering for the $t\bar{t}Z$ generator, as well as the $t\bar{t}Z$ generator choice, the $WZ$ theory uncertainty in the $3\ell$-$Z$-1b4j signal region and one of the $b$-tagging systematics. These differences are reasonably small. A much larger difference would hint at an overestimation of systematics and/or issues with Monte Carlo statistics, which is not the case. However, broader post-fit distributions than the pre-fit ones would hint towards a conceptual issue in the fit, since the fit is supposed to improve the knowledge about the systematic uncertainties by constraining them.

Although no assumptions of correlations are made for the pre-fit likelihood, the fit determines the correlation between nuisance parameters with a similar impact on the fit result. Figure 10.2 shows the correlation between the different nuisance parameters as well as the signal strength and the diboson normalisations for the highest correlated systematics. The highest correlations are between the pileup reweighting uncertainty and the normalisation factor of $ZZ$+jets, as well as between a jet energy scale nuisance parameter and the normalisation factor of $WZ$+jets.

In order to determine which systematic uncertainty has the highest influence on the result, a set of four auxiliary fits is performed for each nuisance parameter: one fit with the nominal value shifted up by $1\sigma$ and one with a $1\sigma$ shift down in the fitted Monte Carlo samples. This is done with the pre-fit and post-fit uncertainties, respectively. The Asimov dataset is kept identical. The 15 systematics with the highest impact on the result are shown in Figure 10.3. Their impact on the fit result is represented by the blue bars, both before and after the nuisance parameters underwent the optimisation via the profile likelihood fit. The most important systematic uncertainties are for the choice of the $t\bar{t}Z$ generator, $b$-tagging, $WZ$ and $tWZ$ theory uncertainties, jet energy scale, the `A14` tunes for the $t\bar{t}Z$ samples, and the luminosity. The post-fit impact of the most relevant systematics decrease after the optimisation, as expected due to the limited statistics of the Asimov set. The impact of the $tWZ$ modelling is asymmetric, which is expected because the systematic uncertainty is defined asymmetrically. The black bars are equivalent to the ones shown in Figure 10.1.

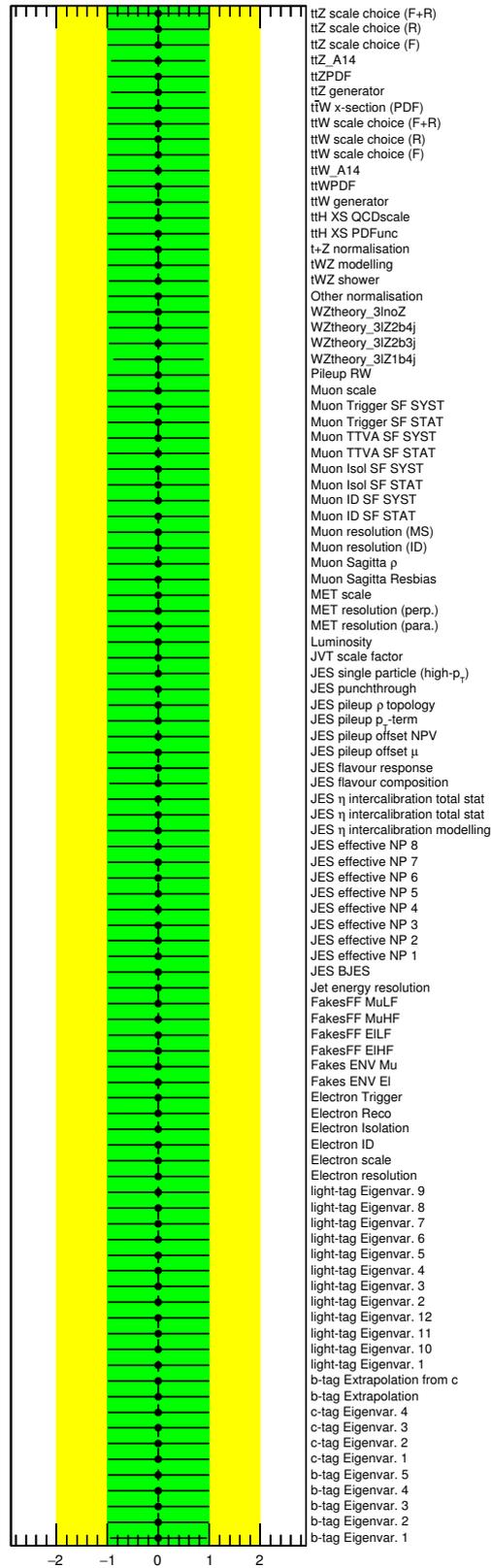

Figure 10.1.: Constraints and pulls of all nuisance parameters for the Asimov fit in the $t\bar{t}Z$ trilepton channel.

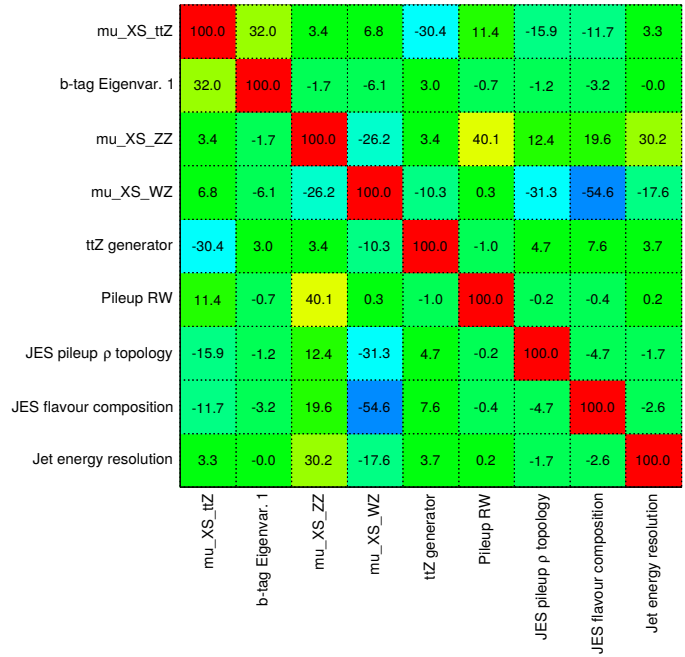

Figure 10.2.: Matrix showing the highest correlated nuisance parameters for the Asimov fit.

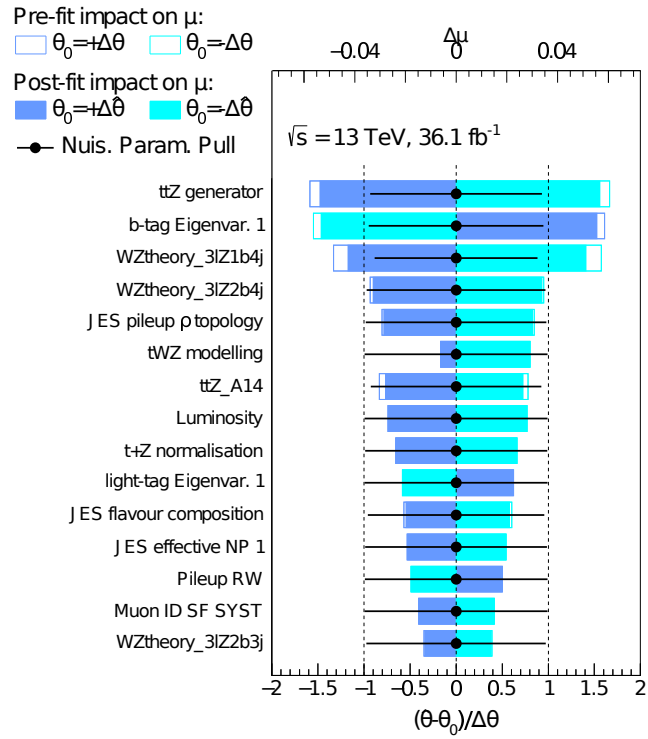

Figure 10.3.: Nuisance parameters with the highest impacts to the fit result. The blue bars show the pre-fit and post-fit impacts and the black bars show the pulls and constraints for these parameters.

The results of the fit to data are shown in this chapter. The fit procedure is explained in Chapter 10 and the signal and control regions used for this fit are defined in Chapter 7. The fit results in the trilepton channel are discussed in Section 11.1. Section 11.2 shows the result of the combined fit with the other channels of the overall analysis. Section 11.3 compares the result with other recent $t\bar{t}Z$ measurements and the theory prediction from NLO QCD and electroweak calculations [155].

## 11.1. Fit results for the $t\bar{t}Z$ cross section in the trilepton channel

For the fit to data in the trilepton channel, the same setup as described in Chapter 10 is used. First, the influence of the fit on the signal and background yields is investigated. As a second step, the fit results of the nuisance parameters are discussed. Finally, the resulting $t\bar{t}Z$ signal strength and diboson background normalisations are discussed and extrapolated to the assumed theory predictions of the corresponding cross sections.

### 11.1.1. Post-fit yields

The post-fit yields for the $t\bar{t}Z$ signal regions are shown in Table 11.1. Figure 11.1 shows the comparison between the pre-fit (left) and post-fit (right) yields in these regions.

|              | 3ℓ-Z-1b4j        | 3ℓ-Z-2b4j        | 3ℓ-Z-2b3j        | 3ℓ-noZ-2b4j      |
|--------------|------------------|------------------|------------------|------------------|
| $t\bar{t}Z$  | $32.53 \pm 5.62$ | $62.76 \pm 8.38$ | $22.86 \pm 4.44$ | $13.78 \pm 2.47$ |
| $t\bar{t}W$  | $0.34 \pm 0.18$  | $0.49 \pm 0.22$  | $0.82 \pm 0.22$  | $3.57 \pm 1.00$  |
| $WZ$         | $15.07 \pm 4.65$ | $5.28 \pm 2.81$  | $3.46 \pm 1.63$  | $0.98 \pm 0.50$  |
| $ZZ$         | $1.81 \pm 0.46$  | $0.61 \pm 0.10$  | $0.77 \pm 0.26$  | $0.38 \pm 0.19$  |
| $tZ$         | $1.90 \pm 0.61$  | $3.39 \pm 1.09$  | $3.88 \pm 1.21$  | $0.32 \pm 0.13$  |
| $tWZ$        | $3.80 \pm 1.65$  | $5.71 \pm 2.09$  | $2.08 \pm 0.51$  | $0.63 \pm 0.28$  |
| $t\bar{t}H$  | $0.84 \pm 0.12$  | $1.41 \pm 0.18$  | $0.54 \pm 0.08$  | $4.77 \pm 0.60$  |
| Other        | $0.15 \pm 0.07$  | $0.24 \pm 0.22$  | $1.15 \pm 0.97$  | $2.27 \pm 1.09$  |
| DD fakes     | $4.37 \pm 1.79$  | $3.98 \pm 1.59$  | $1.17 \pm 0.82$  | $3.19 \pm 1.44$  |
| $\gamma + X$ | $1.37 \pm 1.00$  | $0.50 \pm 0.42$  | $0.75 \pm 0.92$  | $4.94 \pm 1.97$  |
| Total        | $62.18 \pm 6.17$ | $84.37 \pm 7.92$ | $37.50 \pm 4.71$ | $34.85 \pm 3.68$ |
| Observed     | 61               | 78               | 45               | 37               |

Table 11.1.: The post-fit event yields in the trilepton signal regions for a fit in those regions together with the $ZZ$ and $WZ$ control regions. Statistical and systematic uncertainties are included.

Compared to the pre-fit yields in Table 7.5, the profile likelihood fit reduces the systematic uncertainties in all regions. The $t\bar{t}Z$ signal and the $WZ$ and $ZZ$ background yields are the result of the free floating signal strength and normalisation factors. The other backgrounds are allowed to vary within their systematic uncertainties and are therefore also modified with respect to the pre-fit yields. The fit partially compensates the slightly lower number of pre-fit Monte Carlo events compared to data in the 3ℓ-Z-2b3j region. Therefore, the post-fit Monte Carlo yields in the signal regions all agree with data within the statistical and systematic uncertainties.

The post-fit yields in the 3ℓ-$WZ$-CR and 4ℓ-$ZZ$-CR control regions are shown in Table 11.2 (see Table 7.7 for the pre-fit yields). Figure 11.2 shows the comparison between the pre-fit and post-fit yields in these regions. Due to the high purity of $ZZ$ events in the 4ℓ-$ZZ$-CR region and the low $ZZ$ contributions in other channels, the $ZZ$ normalisation is fitted in a way that the Monte Carlo yields match exactly the data. For the $WZ$ process, the contribution to the signal regions and the contamination of the 3ℓ-$WZ$-CR control region with other backgrounds is higher. Still, the fit achieves almost perfect agreement between data and Monte Carlo predictions in the 3ℓ-$WZ$-CR region. For both control regions, the systematic uncertainties are reduced by the profile likelihood fit, relative to the nominal Monte Carlo yields.

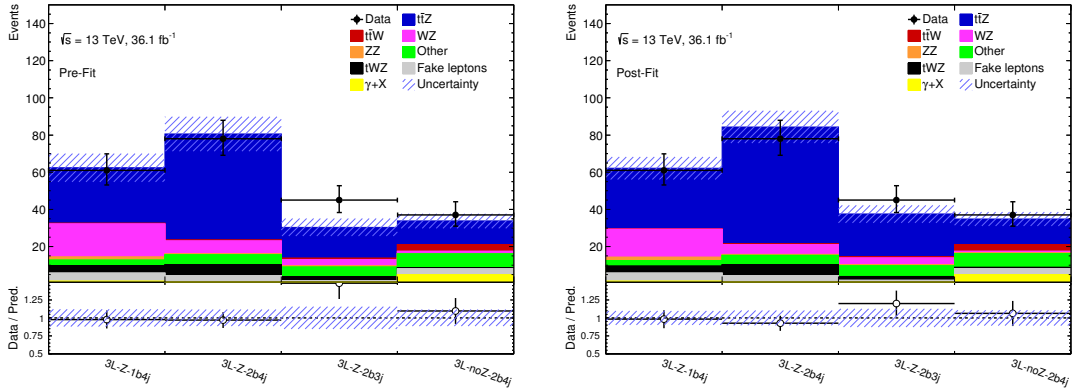

Figure 11.1.: Pre-fit (left) and post-fit (right) yields in the four trilepton $t\bar{t}Z$ signal regions.

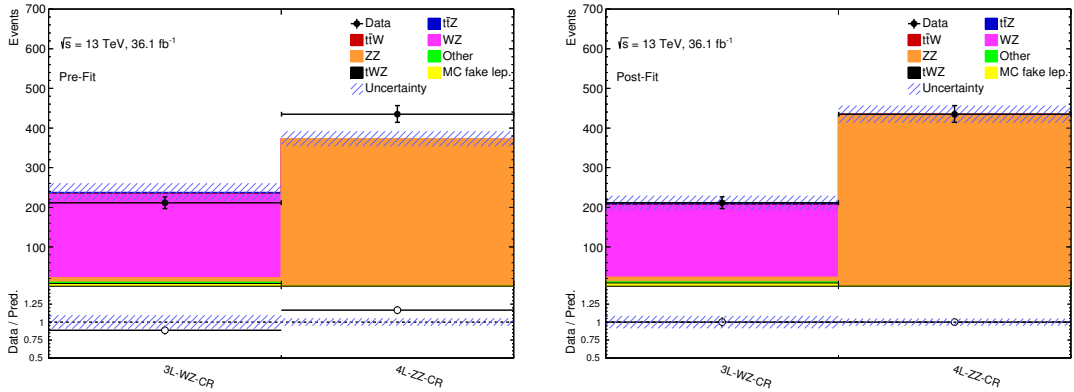

Figure 11.2.: Pre-fit (left) and post-fit (right) yields in the $3\ell$-$WZ$-CR and $4\ell$-$ZZ$-CR control regions.

## 11.1.2. Nuisance parameters

One benefit of using a profile likelihood fit is that systematic uncertainties are treated as nuisance parameters in the fit. In cases where the size of a systematic uncertainty is not reflected by the statistical uncertainty of the data sample, the corresponding nuisance parameter can be *constrained* by the fit. This helps to reduce systematic uncertainties. The other feature of including nuisance parameters is that they can be *pulled* when fitted to data. This means that the maximised likelihood favours a value of this nuisance parameter that deviates from the nominal value assumed in the Gaussian prior of this parameter. These effects can be helpful to obtain an optimised fitting result. However, they have to be treated with care, since too many and unjustified constrained and pulled nuisance parameters will cause drastic deviations from the usually well validated input systematic uncertainties. This would be a hint of a suboptimal fit model. Therefore, pulls and constraints need to be well understood and motivated.

|  | 3ℓ-*WZ*-CR | 4ℓ-*ZZ*-CR |
|---|---|---|
| $t\bar{t}Z$ | 6.31 ± 1.54 | 0.22 ± 0.06 |
| $t\bar{t}W$ | 0.18 ± 0.09 | — |
| $WZ$ | 181.80 ± 17.03 | — |
| $ZZ$ | 11.87 ± 2.42 | 433.80 ± 20.62 |
| $tZ$ | 1.41 ± 0.48 | — |
| $tWZ$ | 2.23 ± 0.73 | 0.07 ± 0.07 |
| $t\bar{t}H$ | 0.11 ± 0.03 | — |
| Other | 1.60 ± 1.11 | 0.62 ± 0.46 |
| MC fakes | 5.74 ± 2.93 | 0.30 ± 0.24 |
| Total | 211.25 ± 16.47 | 435.00 ± 20.62 |
| Observed | 211 | 435 |

Table 11.2.: The post-fit event yields in the 3ℓ-*WZ*-CR and 4ℓ-*ZZ*-CR control regions for a fit together with the trilepton $t\bar{t}Z$ signal regions. Statistical and systematic uncertainties are included.

Figure 11.3 shows the pulls and constraints of all nuisance parameters for the fit in the trilepton channel. No extreme nuisance parameter constraints can be seen. Some parameters are pulled in different directions, mostly the systematic uncertainty of the `A14` tunes for parton showering of the $t\bar{t}Z$ generator. This pull will be studied later in this section. Other nuisance parameters with larger pulls are the uncertainty for the $t\bar{t}Z$ generator choice, as well as some $WZ$ theory, jet energy scale and *b*-tagging systematic uncertainties. Some of these nuisance parameters are among the systematic uncertainties with the highest expected impact on the fit result, see Section 10.3.

The highest correlations between the nuisance parameters are shown in Figure 11.4. Compared with Figure 10.2 from the Asimov fit, no big differences can be seen. The jet energy scale nuisance parameter for the "JES pileup $\rho$ topology" however is dropped from the plot of the correlation matrix since the absolute value of its anti-correlation with the $WZ$ background normalisation is below 30% for the fit to data, which is the threshold for correlated variables to be displayed.

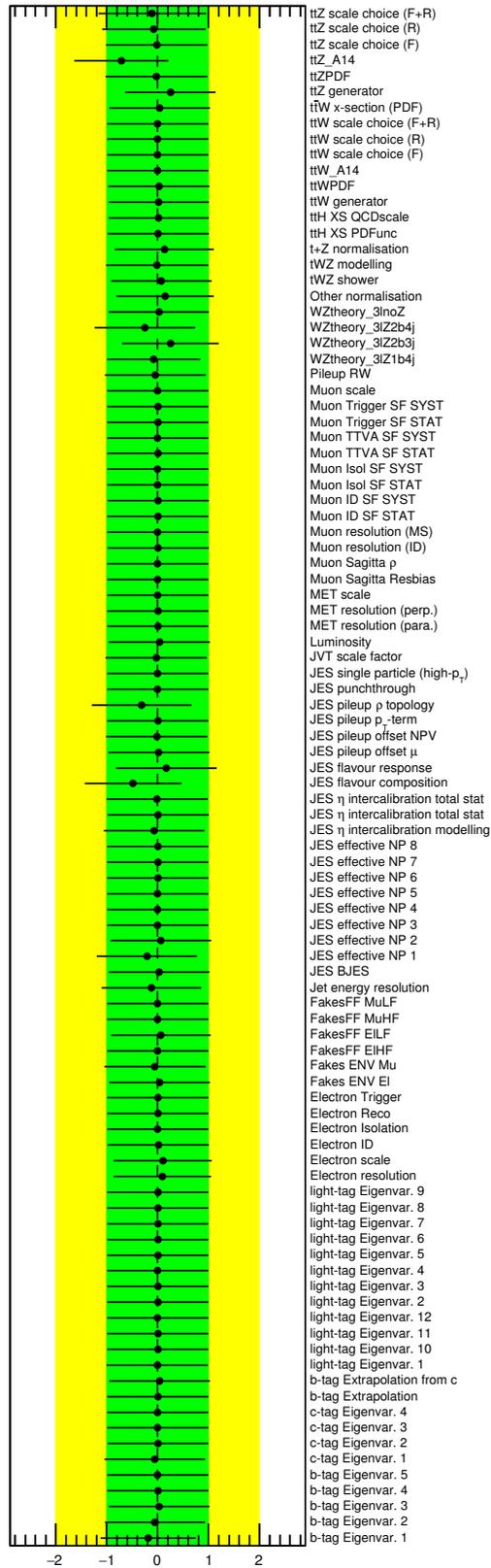

Figure 11.3.: Constraints and pulls of all nuisance parameters for the fit to data in the $t\bar{t}Z$ trilepton channel.

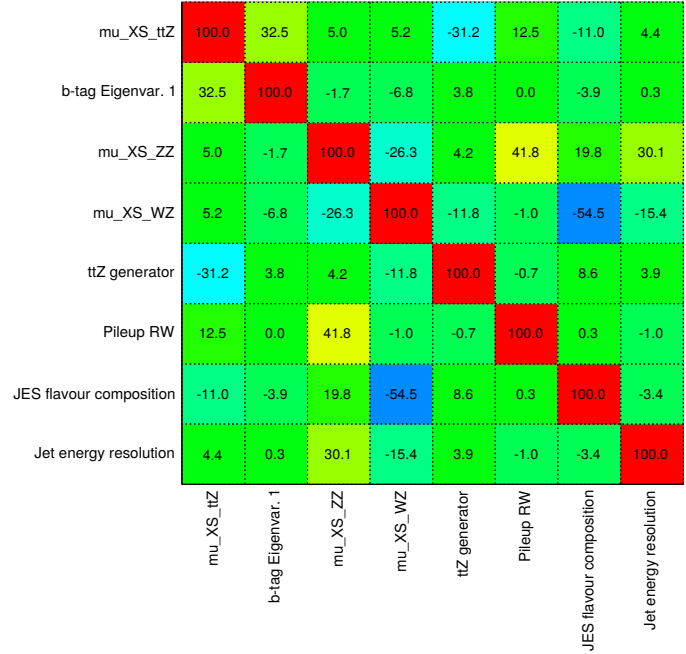

Figure 11.4.: Matrix showing the highest correlated nuisance parameters for the fit to data.

The systematic uncertainties with the highest impacts on the fit result are shown in Figure 11.5. Compared to the ranking from the Asimov fit shown in Figure 10.3, *b*-tagging is now the most important systematic uncertainty, followed by the $t\bar{t}Z$ generator choice. The uncertainties belonging to the `A14` tunes in the $t\bar{t}Z$ samples and the luminosity are now placed higher in the ranking compared to the Asimov result.

In this list of the 15 most important uncertainties, the statistical uncertainty of the Monte Carlo samples and the data driven background in the $3\ell$-$Z$-1b4j region is listed as "$\gamma$ *(trilepZ1b4j [...])*". Table A.1 in Appendix A shows the number of raw Monte Carlo events contributing to each channel for each process. Comparing these numbers with Table A.2, which shows the pre-fit yields with statistical uncertainties only in the $3\ell$-$Z$-1b4j region, it becomes apparent that this uncertainty comes mostly from the statistical limitations of the $WZ$ and data driven fake backgrounds in the $3\ell$-$Z$-1b4j region. Note that, unlike for the other nuisance parameters, the default value for the $\gamma$ parameter is 1, since it refers to the raw number of Monte Carlo and data driven events without any weights applied. The width for the $\gamma$ nuisance parameter is set to the statistical uncertainty and therefore the error bar is very small compared to the other nuisance parameters.

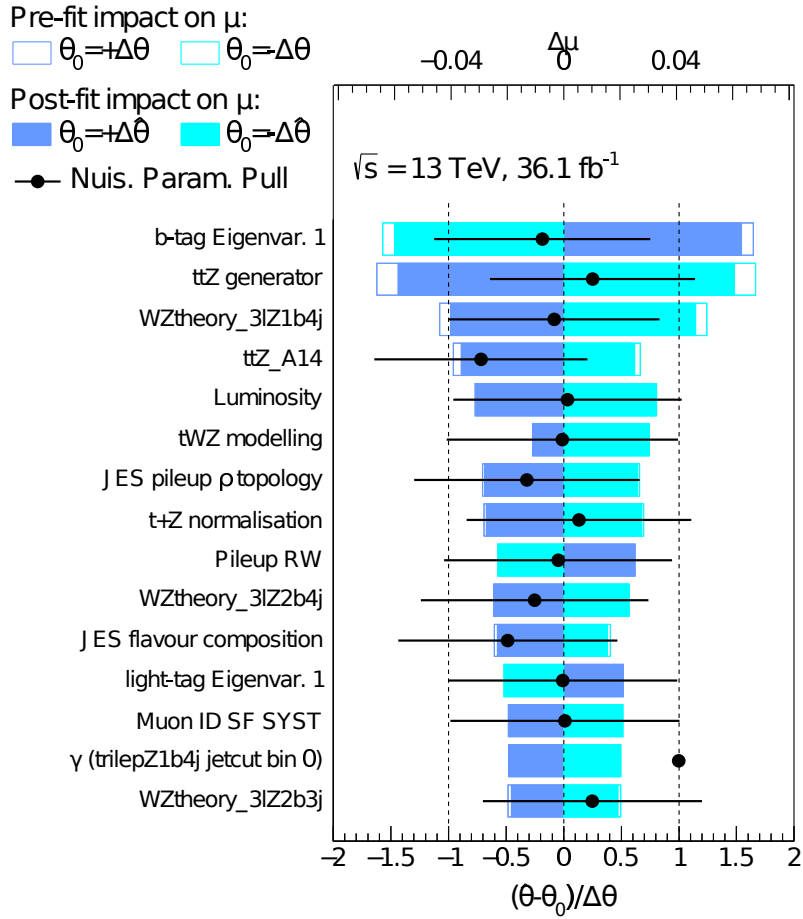

Figure 11.5.: Nuisance parameters with the highest impacts to the fit result. The blue bars show the pre-fit and post-fit impacts and the black bars show the pulls and constraints for these parameters.

The origin of the strongest pull among the nuisance parameters with the highest impacts on the fit result is the systematic uncertainty from the A14 tunes for the parton showering of the $t\bar{t}Z$ Monte Carlo samples (called *A14 systematic uncertainty* from now on). This nuisance parameter needs to be studied in more detail. It is important to know which region causes the pull to be that large. For this purpose, the nuisance parameter for the A14 systematic uncertainty is split and decorrelated for each region and the fit is repeated.

Figure 11.6 shows the ranking of the decorrelated A14 systematic uncertainties in all regions. The impact on the $3\ell$-$WZ$-CR and $4\ell$-$ZZ$-CR control regions is negligible because the $t\bar{t}Z$ contribution is small there. The regions sensitive to on-shell $Z$ bosons with at least two $b$-jets are the ones in which the `A14` systematic uncertainties have the highest impact. While the largest impact is in the $3\ell$-$Z$-2b4j region, the largest pull comes from the $3\ell$-$Z$-2b3j region.

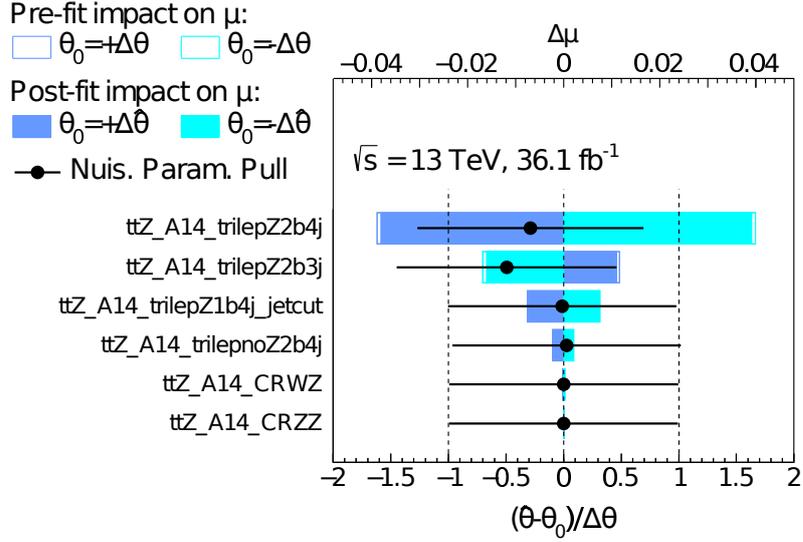

Figure 11.6.: Pre- and post-fit impacts, as well as the pulls and constraints for the `A14` tune uncertainties on the $t\bar{t}Z$ Monte Carlo samples. This nuicance parameter has been split and decorrelated for each separate region in the fit.

Figure 11.7 shows the systematic uncertainty from the `A14` tunes for the $t\bar{t}Z$ yield in the $3\ell$-$Z$-2b4j (left) and the $3\ell$-$Z$-2b3j (right) region. The symmetrisation is performed because the fit setup requires symmetrised nuisance parameters. The impact of the `A14` systematic uncertainty on the $t\bar{t}Z$ yields is higher in the $3\ell$-$Z$-2b3j region (11.6% uncertainty) than in the $3\ell$-$Z$-2b4j region (5.8% uncertainty). However, the overall impact of this nuisance parameter on the result is much higher in the $3\ell$-$Z$-2b4j region because the signal ($t\bar{t}Z$) to background ratio is much higher in this region than in the $3\ell$-$Z$-2b3j region.

To conclude the findings, the $t\bar{t}Z$ pre-fit (Table 7.5) and post-fit yields (Table 11.1) need to be compared. The pre-fit Monte Carlo yields in the $3\ell$-$Z$-2b3j region are slightly less than the observed event yields. For the other signal regions, the pre-fit agreement between data and Monte Carlo predictions is much better. While the $t\bar{t}Z$ yield is scaled by the factor of $\sim 1.4$ in the $3\ell$-$Z$-2b3j region, the signal yields in the other regions are scaled by a factor of $\sim 1.1$. Considering the different $t\bar{t}Z$ scaling in the signal regions, it seems like the profile likelihood fit uses nuisance parameters with large uncertainties on $t\bar{t}Z$ in the $3\ell$-$Z$-2b3j region, like the `A14` systematic uncertainty, to compensate for the slightly lower number of Monte Carlo events. This also explains the strong pull for the

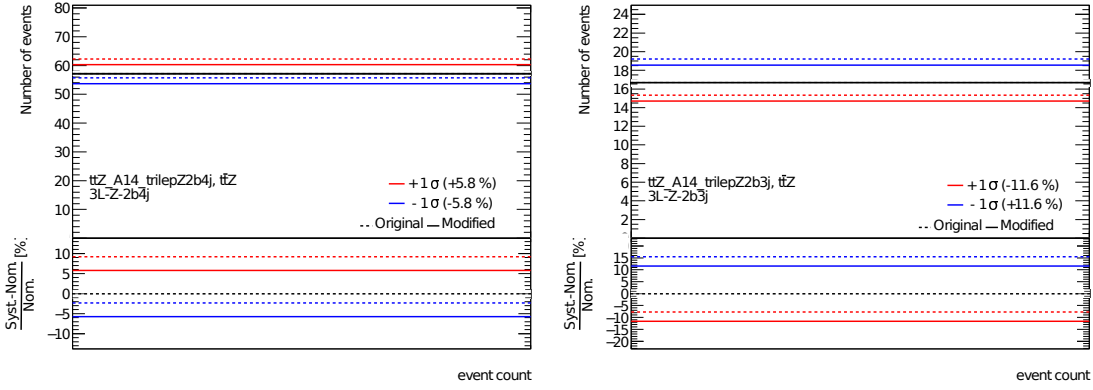

Figure 11.7.: Decorrelated systematic uncertainties for the `A14` tunes for parton show-
ering of the $t\bar{t}Z$ samples for the 3ℓ-Z-2b4j (left) and the 3ℓ-Z-2b3j (right)
regions. The black dotted line shows the nominal value of the $t\bar{t}Z$ yields.
The red and blue dotted lines show the original input uncertainty and the
solid lines show the symmetrised uncertainty.

`A14` nuisance parameter in this region. Due to the different amount of scaling needed
in order to achieve good agreement between data and Monte Carlo predictions, a global
scaling of $\mu_{t\bar{t}Z}$ without any pulls would not have achieved the same results if the nuisance
parameter for the `A14` tunes would not be pulled.

### 11.1.3. Resulting fit parameter values and cross section extrapolations

The $t\bar{t}Z$ signal strength $\mu_{t\bar{t}Z}$ is derived from the fit to data in the trilepton channel, as dis-
cussed above. Additional results are the background normalisation factors $b_{WZ\to\ell\ell\nu+\text{jets}}$
and $b_{ZZ\to\ell\ell\ell\ell+\text{jets}}$ for the diboson processes $WZ \to \ell\ell\ell\nu$ and $ZZ \to \ell\ell\ell\ell$ in association
with additional light and heavy flavoured jets. These results are

$$
\begin{aligned}
\mu_{t\bar{t}Z}^{3\ell} &= 1.15^{+0.14}_{-0.12}(\text{stat.}) \pm 0.14(\text{syst.}) &= 1.15^{+0.19}_{-0.18} \\
b_{WZ\to\ell\ell\nu+\text{jets}}^{3\ell} &= 0.92 \pm 0.07(\text{stat.}) \pm 0.10(\text{syst.}) &= 0.92 \pm 0.12 \\
b_{ZZ\to\ell\ell\ell\ell+\text{jets}}^{3\ell} &= 1.14 \pm 0.06(\text{stat.}) \pm 0.06(\text{syst.}) &= 1.14 \pm 0.08 \,.
\end{aligned}
$$

The signal strength, as well as the background normalisation factors, scale the expected
event yields to the measured ones. The "3ℓ" in the superscript indicates that this result
is only obtained by the fit in the trilepton channel. The uncertainties represent the $1\sigma$
confidence level intervals. They are similar to the total expected uncertainties obtained
by the Asimov fit in Section 10.3. The statistical uncertainties from data are calculated by repeating the fit with all nuisance parameters fixed to the previously fitted values. Since no correlation between statistical and systematic uncertainties are expected, the contributions of the systematic uncertainties are determined by subtracting the statistical uncertainties from the total uncertainties in quadrature.

The measured value of the $t\bar{t}Z$ signal strength is in good agreement with the Standard Model prediction of $\mu_{t\bar{t}Z} = 1$. The observed (expected) significance is $7.2\sigma$ ($6.4\sigma$), corresponding to a deviation from the background-only hypothesis of $\mu_{t\bar{t}Z} = 0$. The expected significance is obtained from the pre-fit yields and their uncertainties. The observed significance is obtained from the fitted result.

The result of $\mu_{t\bar{t}Z}^{3\ell} > 1$ is expected from the pre-fit agreement between data and Monte Carlo predictions in the signal regions, see Table 7.5 and Figure 11.1 (left). This is mostly driven by the slight enhancement of data events in the 3$\ell$-$Z$-2b3j region, although the nuisance parameters of the fit are covering a part of it. The value of $b_{WZ \to \ell\ell\ell\nu + \text{jets}}^{3\ell} < 1$ is also expected due to the fact that there are slightly more Monte Carlo events than data events in the 3$\ell$-$WZ$-CR control region (see Table 7.7 and Figure 11.2, left) for the pre-fit yields. Also, the signal regions which have slightly more Monte Carlo events, compared to data, (3$\ell$-$Z$-1b4j and 3$\ell$-$Z$-2b4j) are the ones where the $WZ$ background is most dominant, while the two signal regions with slightly less events from Monte Carlo predictions than from data (3$\ell$-$Z$-2b3j and 3$\ell$-no$Z$-2b4j) have a much lower $WZ$ contribution. The result of $b_{ZZ \to \ell\ell\ell\ell + \text{jets}}^{3\ell} > 1$ is obvious since the $ZZ$ Monte Carlo events only have a significant contribution to the 4$\ell$-$ZZ$-CR control region, where slightly more data is observed than expected from the pre-fit Monte Carlo yields.

These values can be extrapolated to determine the cross section values of the corresponding processes. For the fit, $t\bar{t}Z$ samples with all decay channels of the top quark pair and the decays of the $Z$ boson into a pair of electrons, muons or tau leptons are used. Of these signatures, only the lepton+jets decay of the top quark pair is taken into account for the trilepton channel. Extrapolating to the inclusive $t\bar{t}Z$ decay[1], the NLO (in QCD and electroweak calculations) cross section of $\sigma_{t\bar{t}Z} = 839.3$ fb is used [155]. Scaling this cross section with the signal strength obtained from the fit yields the following result:

$$
\begin{aligned}
\sigma_{t\bar{t}Z}^{3\ell} &= 966_{-102}^{+114}(\text{stat.})_{-114}^{+115}(\text{syst.}) \,\text{fb} \\
&= 966 \pm 162 \,\text{fb} \,.
\end{aligned}
$$

---

1. This means also taking into account the decays of the $Z$ boson into neutrinos and quarks, as well as all $t\bar{t}$ decay channels.

In the second line, both uncertainties are added in quadrature and symmetrised to the largest value. For the diboson background, the QCD NLO cross sections assumed for the fit are $\sigma_{WZ\to\ell\ell\ell\nu+\text{jets}} = 4571$ fb and $\sigma_{ZZ\to\ell\ell\ell\ell+\text{jets}} = 1053$ fb [229]. Therefore, the resulting cross sections for these processes are

$$
\begin{aligned}
\sigma^{3\ell}_{WZ\to\ell\ell\ell\nu+\text{jets}} &= 4213 \pm 338(\text{stat.}) \pm 444(\text{syst.})\,\text{fb} &= 4213 \pm 558\,\text{fb} \\
\sigma^{3\ell}_{ZZ\to\ell\ell\ell\ell+\text{jets}} &= 1206 \pm 58(\text{stat.}) \pm 60(\text{syst.})\,\text{fb} &= 1206 \pm 84\,\text{fb} \,.
\end{aligned}
$$

These cross sections refer to the phase space described by the Monte Carlo samples and are not inclusive diboson cross sections.

## 11.2. Combination with other multilepton channels

A higher precision of the $t\bar{t}Z$ cross section measurement can be achieved by performing a combined fit in all signal and control regions of the three multilepton channels ($2\ell$OSSF, trilepton and tetralepton, see Chapter 9) sensitive to this process. The $t\bar{t}W$ cross section is determined by a fit in the $2\ell$SS channel and the dedicated $t\bar{t}W$ trilepton channel. Both processes are fitted independently of each other, by fixing either the $t\bar{t}Z$ or $t\bar{t}W$ normalisation to its Standard Model expectation and keeping the other one as a free fit parameter. This combination is performed by another member of the overall analysis group.

The results of the combined fit in the $2\ell$OSSF, trilepton and tetralepton regions are

$$
\begin{aligned}
\mu^{\text{comb, 1D}}_{t\bar{t}Z} &= 1.04^{+0.10}_{-0.09}(\text{stat.}) \pm 0.11(\text{syst.}) &= 1.04^{+0.15}_{-0.14} \\
b^{\text{comb, 1D}}_{WZ\to\ell\ell\ell\nu+\text{jets}} &= 0.88 \pm 0.07(\text{stat.}) \pm 0.07(\text{syst.}) &= 0.88 \pm 0.10 \\
b^{\text{comb, 1D}}_{ZZ\to\ell\ell\ell\ell+\text{jets}} &= 1.16 \pm 0.06(\text{stat.}) \pm 0.07(\text{syst.}) &= 1.16 \pm 0.09 \\
b^{\text{comb, 1D}}_{Z+1\text{HF}} &= 1.01 \pm 0.05(\text{stat.})^{+0.23}_{-0.20}(\text{syst.}) &= 1.01^{+0.24}_{-0.21} \\
b^{\text{comb, 1D}}_{Z+2\text{HF}} &= 0.92 \pm 0.03(\text{stat.})^{+0.12}_{-0.11}(\text{syst.}) &= 0.92^{+0.12}_{-0.11} \,,
\end{aligned}
$$

with an observed (expected) significance of $9.0\sigma$ ($8.5\sigma$). The measured value of the $t\bar{t}Z$ signal strength is in good agreement with the Standard Model prediction of $\mu_{t\bar{t}Z} = 1$. The normalisation factors for the $Z$ boson production in association with exactly one or at least two heavy flavour jets are denoted by $b_{Z+1\text{HF}}$ and $b_{Z+2\text{HF}}$, respectively. They are derived for the $2\ell$OSSF channel in three dedicated control regions.

| Fitted signal | Channel | $\mu$ of fitted signal | Expected significance | Observed significance |
|---|---|---|---|---|
| $t\bar{t}Z$ | 2$\ell$OSSF | $0.69^{+0.18}_{-0.17}$(stat.)$^{+0.18}_{-0.17}$(syst.) | $3.8\sigma$ | $3.0\sigma$ |
| | **Trilepton** | $\mathbf{1.15^{+0.14}_{-0.12}(stat.) \pm 0.14(syst.)}$ | $\mathbf{6.4\sigma}$ | $\mathbf{7.2\sigma}$ |
| | Tetralepton | $1.16^{+0.28}_{-0.26}$(stat.)$^{+0.13}_{-0.10}$(syst.) | $5.0\sigma$ | $5.3\sigma$ |
| $t\bar{t}Z$ | Combination | $1.04^{+0.10}_{-0.09}$(stat.) $\pm 0.11$(syst.) | $8.5\sigma$ | $9.0\sigma$ |
| $t\bar{t}W$ | 2$\ell$SS + $t\bar{t}W$ 3$\ell$ | $1.17^{+0.17}_{-0.16}$(stat.)$^{+0.22}_{-0.20}$(syst.) | $4.2\sigma$ | $4.8\sigma$ |

Table 11.3.: Comparison between the measured signal strengths, expected and observed significances in the different channels of the multilepton analysis. The trilepton channel sensitive to the $t\bar{t}Z$ process, which is the main topic of this thesis, is highlighted.

The fit result of the $t\bar{t}W$ signal strength in the 2$\ell$SS and $t\bar{t}W$ trilepton signal regions is

$$\mu_{t\bar{t}W}^{1D} = 1.17^{+0.17}_{-0.16}(\text{stat.})^{+0.22}_{-0.20}(\text{syst.}) = 1.17^{+0.28}_{-0.26},$$

which is in good agreement with the Standard Model expectation of $\mu_{t\bar{t}W} = 1$ and corresponds to an observed (expected) significance of $4.8\sigma$ ($4.2\sigma$).

The results of the individual fits, as well as of the combined fit in all $t\bar{t}Z$ channels, are shown in Table 11.3. The result of the $t\bar{t}W$ fit in the 2$\ell$SS channel and the dedicated trilepton channel for $t\bar{t}W$ are also included, see Section 9.3. For the $t\bar{t}W$ fit, the $t\bar{t}Z$ contribution is fixed to the Standard Model expectation with an additional scale and PDF uncertainty of 12% [155], while the scale and PDF uncertainty on the $t\bar{t}W$ normalisation is removed.

The largest sources of systematic uncertainties for the combined fit of the $t\bar{t}Z$ signal strength come from $b$-tagging, the $t\bar{t}Z$ `A14` tunes and the normalisation of the $WZ$ background. The trilepton channel, as the "golden channel" of the overall analysis, has the highest significance contributing to the fit. It has the smallest statistical uncertainty and a smaller systematic uncertainty than the 2$\ell$OSSF channel. The 2$\ell$OSSF channel has the highest total systematic uncertainty due to statistical uncertainties from the Monte Carlo samples and the modelling of the $Z$+jets background. This is expected because of the large background contribution from $Z$+jets events in this channel. The tetralepton channel, being the purest channel in terms of signal and background, has the lowest total systematic uncertainty with main contributions from $t\bar{t}Z$ and $tWZ$ modelling, as well as from $b$-tagging. However, it has the highest statistical uncertainty due to the small event yields in this channel.

Using the same method as in Section 11.1.3, the resulting $t\bar{t}Z$ cross section is

$$\sigma_{t\bar{t}Z}^{\text{comb, 1D}} = 873^{+84}_{-76}(\text{stat.})^{+94}_{-90}(\text{syst.})\,\text{fb}$$
$$= 873 \pm 126\,\text{fb}\,.$$

This method is also used to extract the $t\bar{t}W$ cross section from the signal strength. As the nominal inclusive $t\bar{t}W$ cross section in the fit, $\sigma_{t\bar{t}W} = 600.8$ fb (NLO QCD and electroweak prediction) is used [155]. The resulting cross section, extrapolated from the signal strength, is

$$\sigma_{t\bar{t}W}^{\text{1D}} = 703^{+102}_{-96}(\text{stat.})^{+134}_{-123}(\text{syst.})\,\text{fb}$$
$$= 703 \pm 168\,\text{fb}\,.$$

The corresponding diboson cross sections are

$$\sigma_{WZ\to\ell\ell\nu+\text{jets}}^{\text{comb, 1D}} = 4022 \pm 320(\text{stat.}) \pm 326(\text{syst.})\,\text{fb} = 4022 \pm 457\,\text{fb}$$
$$\sigma_{ZZ\to\ell\ell\ell\ell+\text{jets}}^{\text{comb, 1D}} = 1222 \pm 63(\text{stat.}) \pm 71(\text{syst.})\,\text{fb} = 1222 \pm 95\,\text{fb}\,,$$

for the phase space taken into account by the diboson samples.

## 11.3. Comparison with other $t\bar{t}Z$ measurements

For a centre-of-mass energy of 13 TeV, four measurements of the $t\bar{t}Z$ and $t\bar{t}W$ cross sections have been performed, two by the ATLAS collaboration and two by the CMS collaboration, respectively, see Section 3.2. The results of the most recent ATLAS analysis, using 36.1 fb$^{-1}$ of data, are the ones presented in Section 11.2.

Table 11.4 shows the results of the four measurements, using separate fits for the $t\bar{t}Z$ and $t\bar{t}W$ cross sections, respectively. They are compared to the cross section predictions used for signal normalisation in the analysis presented in this thesis [155]. The main reason for the improvement of the results presented in Section 11.2 with respect to the ATLAS measurement using 3.2 fb$^{-1}$ of data is the increase in data statistics. Another reason for the improved uncertainty on the $t\bar{t}Z$ cross section is the inclusion of the $2\ell$OSSF region. It must be stated that the $t\bar{t}Z$ cross section measurement in the trilepton channel alone has a far smaller statistical uncertainty than the 3.2 fb$^{-1}$ measurement with all combined channels. The $t\bar{t}W$ cross section result has improved even more with respect to the 3.2 fb$^{-1}$ ATLAS analysis. The reasons for this improvement are the increase in data statistics but also the inclusion of the electron-electron and electron-muon channels, as well as the implementation of prompt lepton isolation. Another reason for the improved $t\bar{t}W$ result is the careful re-evaluation of the signal region selection.

| Experiment | $\int \mathcal{L} dt$ [fb$^{-1}$] | $\sigma_{t\bar{t}Z}$ [pb] | $\sigma_{t\bar{t}W}$ [pb] |
|---|---|---|---|
| *Current analysis* | | | |
| **ATLAS trilep.** | **36.1** | **$0.97^{+0.11}_{-0.10}$(stat.)$^{+0.12}_{-0.11}$(syst.)** | — |
| ATLAS comb. | 36.1 | $0.87 \pm 0.08$(stat.) $\pm 0.09$(syst.) | $0.70 \pm 0.10$(stat.)$^{+0.13}_{-0.12}$(syst.) |
| *Other results* | | | |
| ATLAS [134] | 3.2 | $0.92 \pm 0.29$(stat.) $\pm 0.10$(syst.) | $1.50 \pm 0.72$(stat.) $\pm 0.33$(syst.) |
| CMS [135] | 12.9 | $0.70^{+0.16}_{-0.15}$(stat.)$^{+0.14}_{-0.12}$(syst.) | $0.98^{+0.23}_{-0.22}$(stat.)$^{+0.22}_{-0.18}$(syst.) |
| CMS [97] | 35.9 | $0.99^{+0.09}_{-0.08}$(stat.)$^{+0.12}_{-0.10}$(syst.) | $0.77^{+0.12}_{-0.11}$(stat.)$^{+0.13}_{-0.12}$(syst.) |
| NLO QCD +EWK [155] | — | $0.84 \pm 0.10$ | $0.60 \pm 0.08$ |

Table 11.4.: Results of the recent $t\bar{t}Z$ and $t\bar{t}W$ measurements conducted at the LHC at a centre-of-mass energy of $\sqrt{s} = 13$ TeV compared to the theoretical predictions (NLO QCD and electroweak) used for the Monte Carlo normalisation in this analysis. The trilepton channel sensitive to the $t\bar{t}Z$ process, which is contributing to the latest ATLAS result and is the main topic of this thesis, is highlighted.

The full ATLAS measurement presented in this thesis clearly outperforms the CMS measurement with an integrated luminosity of 12.9 fb$^{-1}$ due to data statistics. Even the separate fit in the trilepton channel yields a better result for the $t\bar{t}Z$ cross section than the CMS measurement. The results of the CMS analysis with an integrated luminosity of 35.9 fb$^{-1}$ are comparable to the results of the combined ATLAS measurement presented here. It should be mentioned that the smaller systematic uncertainties of the ATLAS result (both absolute and relative to the nominal value) can be the outcome of a different evaluation of the systematic uncertainties between the ATLAS and CMS measurements. Although a multivariate approach is chosen for the $t\bar{t}W$ signal region definition in the CMS analysis, the ATLAS measurement of the same process has almost the same uncertainty on the $t\bar{t}W$ cross section, with a slightly smaller statistical uncertainty on the dataset. All measurements compared here show good agreement with the Standard Model cross section predictions (NLO in QCD and electroweak calculations) that are used for the signal normalisation in this analysis [155]. Summarising, the current ATLAS measurement, that the analysis presented in this thesis is a part of, outperforms the previous ATLAS measurement and has a similar performance as the latest CMS measurement which uses a similar amount of data.

Further Studies to Improve the Sensitivity of the Measurement

The trilepton channel is the most sensitive channel of the $t\bar{t}Z$ analysis, see Chapter 3. However, some techniques can be implemented to optimise the analysis in this channel. Two of those techniques are shown in this chapter. One of them is based on a more efficient usage of the $b$-tagging information and the other makes use of $t\bar{t}Z$ reconstruction information.

The idea behind these two approaches is to find variables that are sensitive to the $t\bar{t}Z$ signature and allow this signature to be distinguished from the most prominent backgrounds. These studies are feasibility studies and do not contribute to the final result shown in Chapter 11. They should serve as an inspiration for future $t\bar{t}Z$ measurements, as well as bachelor, master or PhD projects to come.

## 12.1. Continuous $b$-tagging

The analysis presented in this thesis uses the number of $b$-tagged jets to define the signal and control regions. For this purpose, the `MV2c10` tagger is used. It provides a BDT discriminant which will be referred to as "*b-tag weight*" in this chapter. A $b$-jet is defined as a jet that has a $b$-tag weight higher than a certain value, in a way that the efficiency of correctly identifying an actual $b$-jet is around 77%, see Section 5.5.

A different approach will be tested in this section. Instead of using the number of $b$-tagged jets for cuts, the $b$-tag weight itself will be used directly to study the separa-tion of the $t\bar{t}Z$ signal from the background. This technique is called *pseudo continuous b-tagging*. In this case, *continuous* stands for the use of the whole *b*-tag weight by itself. *Pseudo* refers to the technical limitation that Monte Carlo event weights for the *b*-tagging calibration of this distribution are still derived in five fixed bins, corresponding to different *b*-tagging efficiencies. Therefore, the binning of the *b*-tag weight distribution relies on the binning of these event weights, as shown in Table 12.1.

| Bin # | *b*-tag efficiency range |
|-------|--------------------------|
| 1 | $> 85\%$ |
| 2 | $77\% - 85\%$ |
| 3 | $70\% - 77\%$ |
| 4 | $60\% - 70\%$ |
| 5 | $< 60\%$ |

Table 12.1.: Definition of the five bins of the pseudo continuous *b*-tag weight distribution.

For the $t\bar{t}Z$ signal, the *b*-tag weight of two jets is expected to be high due to the two *b*-jets expected from the $t\bar{t}$ decay. For the two main backgrounds, lepton fakes and $WZ \rightarrow \ell\ell\ell\nu$ with additional jets, the highest two *b*-tag weights are expected to be smaller. Therefore, the *b*-tag weight is a promising variable for distinguishing between signal and background.

To study the pseudo continuous *b*-tagging, the same object definition and preselection criteria as described in Chapter 5 and Section 7.1 (except for the definition of *b*-jets) are used. Exactly three charged leptons are required with at least one OSSF pair fulfilling the *Z*-window requirement (see Chapter 7). The requirements on the lepton transverse momenta are $p_T > 27$ GeV for the highest $p_T$ lepton and $p_T > 20$ GeV for the other two leptons. Regions with exactly three jets (*3jex*) and at least four jets (*4jin*) are studied separately. Because this study takes into account events with exactly zero *b*-jets according to the 77% efficiency criterion, the fake lepton background is modelled using the fake factor method. To be able to study the contribution from heavy flavour jets in the diboson background, the $WZ$ process is split into $WZ$ with light flavour jets ($WZ\_$LF) and $WZ$ with at least one *b*-jet ($WZ\_$HF). This split is done according to generator level information.

Figure 12.1 shows the highest (left) and second highest (right) *b*-tag weights in the 3jex region. For the leading *b*-tag weight, the highest bin (corresponding to a *b*-tagging efficiency of less than 60% for fixed cuts) is dominated by $t\bar{t}Z$, $WZ\_$HF and $tWZ$ events. The lower bins are dominated by $WZ$ in association with light flavour jets, $ZZ$ with additional jets and lepton fakes. For the second highest *b*-tag weight, most of the events populate the lowest bin, corresponding to a *b*-tagging efficiency of at least 85% in terms of fixed cuts. However, the $WZ$ background mostly populates this bin, while the $t\bar{t}Z$ process also populates higher bins. Therefore, cuts on the *b*-tag weights in the 3jex region can mostly be used to reduce the background from $WZ$ with additional light flavour jets, $ZZ$ with additional jets and lepton fakes.

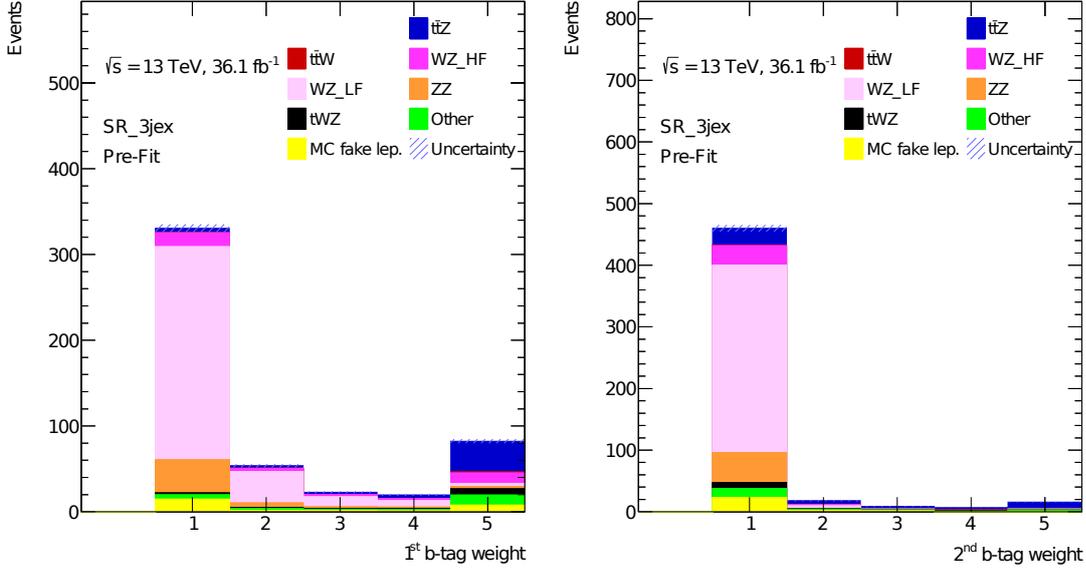

Figure 12.1.: The highest (left) and second highest (right) *b*-tag weights in the 3jex region. Only statistical uncertainties are shown.

Figure 12.2 shows the highest and second highest *b*-tag weights in the 4jin region. The highest bin in the distribution of the highest *b*-tag weight is populated even more by $t\bar{t}Z$ events than in the 3jex case. The background is distributed in a similar way as it is in the 3jex region. For the second highest *b*-tag weight, the $t\bar{t}Z$ events are almost uniformly distributed over all five bins, while most of the background populates the lowest bin. As for the 3jex region, the *b*-tag weights can be used to discriminate against $WZ$ with additional light flavour jets, $ZZ$ with additional jets, and lepton fakes.

The separation power of the *b*-tag weights can be used to define signal and control regions, based on the signal-to-background ratio (S/B). Figure 12.3 shows the S/B, depending on the highest and second highest *b*-tag weights for the 3jex (left) and 4jin (right) regions. For this study, one control region (CR_3jin) and five signal regions are defined, based on the jet multiplicity and S/B in the two-dimensional *b*-tag weight distribution, see

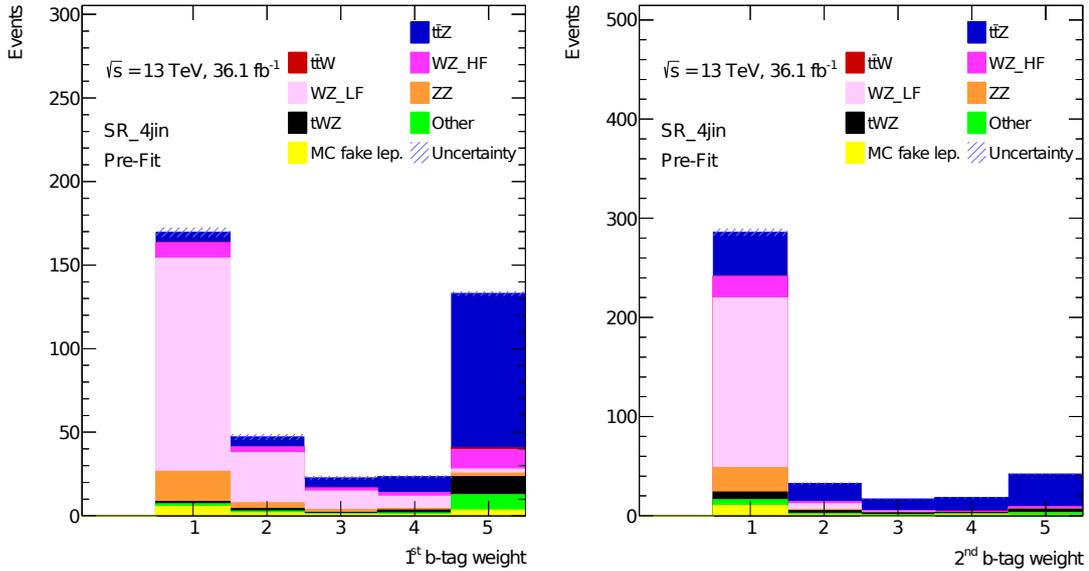

Figure 12.2.: The highest (left) and second highest (right) *b*-tag weights in the 4jin region. Only statistical uncertainties are shown.

Table 12.2. Regions are created from bins that have a similar S/B, which allows more flexibility for a fit. Three distinctive clusters of S/B are identified for the 3jex selection and four clusters for the 4jex selection. The clusters with the lowest S/B are merged into the CR_3jin control region and the remaining clusters are defined as the signal regions. Figure 12.4 shows the first, second and third leading lepton transverse momenta, as well as the electron multiplicity in the CR_3jin control region. Only statistical uncertainties are shown. The control region is contaminated with only a few $t\bar{t}Z$ signal events, while the dominating process is $WZ$ with additional light flavour jets. Other contributing background processes are $ZZ$ with additional jets, lepton fakes, $WZ$ with heavy flavour jets and $tWZ$. Good agreement between data and Monte Carlo is achieved in all four distributions within the statistical uncertainties.

Figure 12.5 shows the expected yields from Monte Carlo events in the five signal regions. Only statistical uncertainties are shown. All signal regions are heavily populated by the $t\bar{t}Z$ signal, while the dominating backgrounds are $WZ$ with heavy flavour jets and $tWZ$. The process of $WZ$ in association with light flavour jets, which dominates in the control region, is only relevant in the SR_4jin_low signal region. This means that for this approach, the $WZ$ background cannot be split into light and heavy (*b*-jet) flavour contributions since the selection criterion is based on exactly this property. Heavy and light flavour contributions have to be treated together to be able to define a useful $WZ$ control region using pseudo continuous $b$-tagging. With the same Monte Carlo samples, the nominal trilepton analysis presented in this thesis already uses a combined $WZ$ background from heavy and light flavour jets. This is done because the heavy flavour contributions are expected to be modelled well enough.

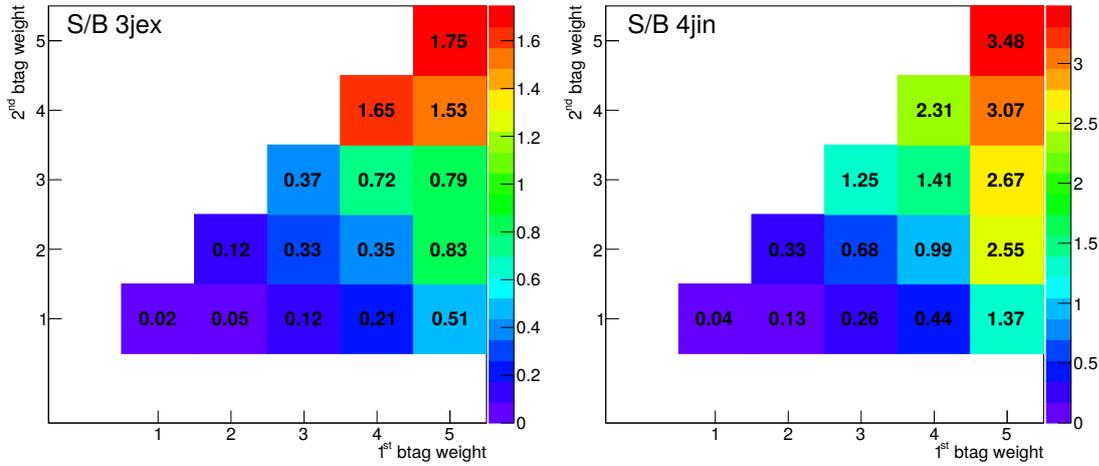

Figure 12.3.: S/B depending on the highest and second highest $b$-tag weights for the 3jex (left) and 4jin (right) regions.

In conclusion, it is possible to create signal and control regions by cutting on the BDT output of the `MC2c10` tagger. The selected signal regions are highly populated with $t\bar{t}Z$ signal events, while the control region has a low $t\bar{t}Z$ contamination. The light and heavy flavour components of the $WZ$+jets background should be treated together in order to obtain valid control regions. This study can be seen as a starting point for further investigations of pseudo continuous $b$-tagging in the $t\bar{t}Z$ analysis.

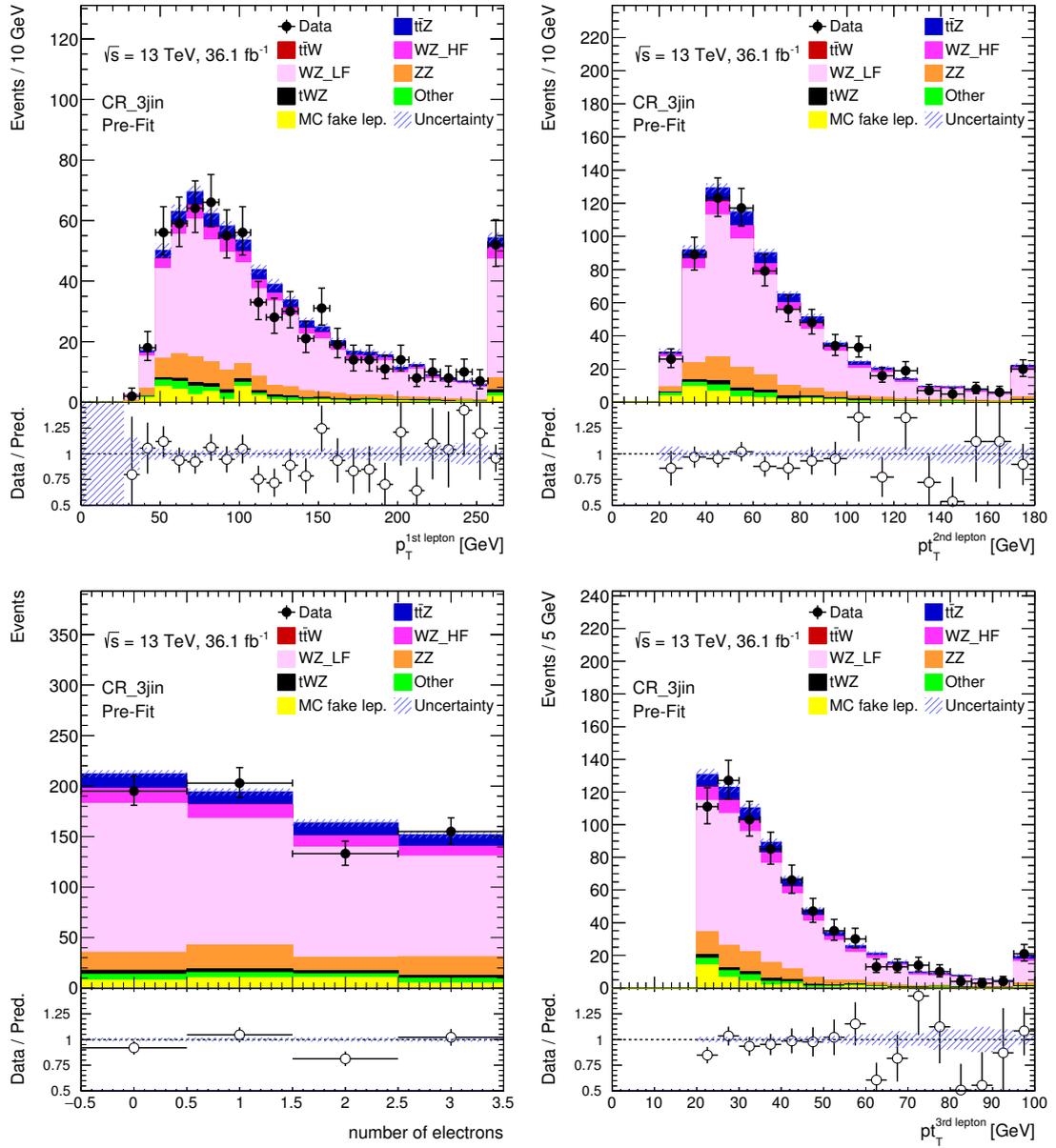

Figure 12.4.: First, second and third leading lepton transverse momenta, as well as the electron multiplicity in the CR_3jin control region (clockwise from the top left) for the continuous *b*-tagging approach. Only statistical uncertainties are shown.

| Region | # jets | S/B range |
|--------|--------|-----------|
| SR_3jex_low | =3 | [0.6 , 1.0] |
| SR_3jex_high | =3 | > 1.0 |
| SR_4jin_low | ≥ 4 | [0.6 , 1.5] |
| SR_4jin_mid | ≥ 4 | [1.5 , 3.0] |
| SR_4jin_high | ≥ 4 | > 3.0 |
| CR_3jin | ≥ 3 | < 0.6 |

Table 12.2.: Definition of signal and control regions based on two-dimensional *b*-tag distributions.

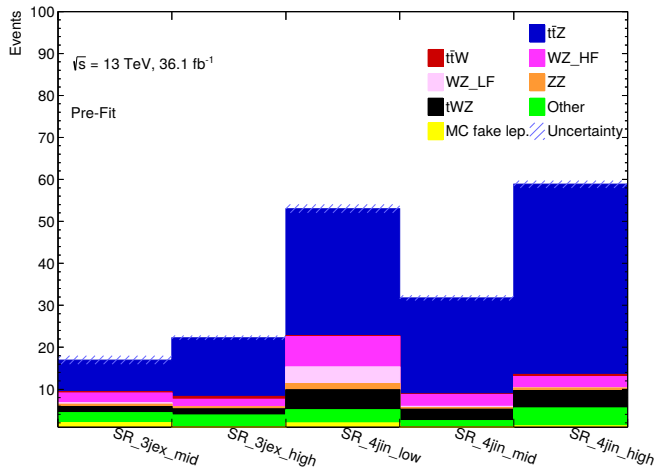

Figure 12.5.: Yields in the five signal regions for the continuous *b*-tagging approach. Only statistical uncertainties are shown.

## 12.2. Reconstruction of the $t\bar{t}Z$ process using the KLFitter framework

A second study refers to the reconstruction of the $t\bar{t}Z$ trilepton final state using the KLFitter tool [4], based on the Bayesian Analysis Toolkit (BAT) [247]. KLFitter uses a likelihood-based reconstruction algorithm to find the best match between reconstructed jets and leptons, and the corresponding particles from the hard scattering process (called *parton level*). For this study, it is assumed that the four parton-level quarks from the semileptonic decay of the top quark pair end up as four reconstructed jets. Ignoring the commutation of the two jets from the hadronically decaying $W$ boson, there are twelve possible jet reconstruction permutations. This number can be further reduced by using *b*-tagging information. Additional jets from pileup and gluon radiation are also taken into account.

For the reconstruction of a semileptonically decaying top quark pair, the corresponding likelihood is maximised [4]:

$$
\begin{aligned}
L_{t\bar{t}} \;=\;& B(m_{q_1 q_2 q_3}|m_{\text{top}}, \Gamma_{\text{top}}) \cdot B(m_{q_1 q_2}|m_W, \Gamma_W) \\
& \times B(m_{q_4 \ell \nu}|m_{\text{top}}, \Gamma_{\text{top}}) \cdot B(m_{\ell \nu}|m_W, \Gamma_W) \\
& \times \prod_{i=1}^{4} W_{\text{jet}}(E_{\text{jet},i}^{\text{meas}}|E_{\text{jet},i}) \cdot W_{\ell}(E_{\ell}^{\text{meas}}|E_{\ell}) \\
& \times W_{\text{miss}}(E_x^{\text{miss}}|p_x^{\nu}) \cdot W_{\text{miss}}(E_y^{\text{miss}}|p_y^{\nu}) \,.
\end{aligned}
\tag{12.1}
$$

The Breit-Wigner functions

$$
B(m_{\text{a,b,c}}|m_{\text{initial}}, \Gamma_{\text{initial}}) \propto \frac{1}{(m_{\text{a,b,c}}^2 - m_{\text{initial}}^2)^2 + m_{\text{initial}}^2 \Gamma_{\text{initial}}^2}
\tag{12.2}
$$

are distributions for the invariant mass $m_{\text{a,b,c}}$ of the reconstructed particles $a$, $b$ and $c$ (charged leptons, neutrinos and quarks), assigned to the *initial* particle (top quark or $W$ boson). These distributions peak at the initial particle's mass $m_{\text{initial}}$ and have widths corresponding to the decay width of the initial particle, $\Gamma_{\text{initial}}$. The terms $W(E^{\text{meas}}|E)$ are so-called *transfer functions*. They constrain the particle energies $E$ of particle-level objects based on the measured energies $E^{\text{meas}}$. In the case of neutrinos, the transfer functions constrain the $x$ and $y$ components of the neutrino momentum $p_{x/y}^{\nu}$ to the measured missing transverse momentum components $E_{x/y}^{\text{miss}}$. The transfer functions represent the detector resolution in terms of the energy and transverse momentum.

For the $t\bar{t}Z$ process in the trilepton channel, the top quark pair is assumed to decay semileptonically. Therefore, the likelihood in Equation (12.1) is expanded by an additional likelihood part for the $Z$ boson [248]:

$$
\begin{aligned}
L_{t\bar{t}Z} \;=\;& L_{t\bar{t}} \times L_Z \\
=\;& L_{t\bar{t}} \times W_{\ell}(E_{\ell_2}^{\text{meas}}|E_{\ell_2}) \times W_{\ell}(E_{\ell_3}^{\text{meas}}|E_{\ell_3}) \\
& \times \left[ f \cdot B(m_{\ell_2 \ell_3}|m_Z, \Gamma_Z) \,+\, (1-f) \cdot \frac{c_{\text{norm}}}{m_{\ell_2 \ell_3}^2} \right] .
\end{aligned}
\tag{12.3}
$$

This modification includes a transfer function for each of the two additional leptons, as well a Breit-Wigner function for assigning the two leptons to the $Z$ boson. Due to the interference between $Z$ bosons and photons in the production of the additional lepton pair, a weighting factor $f$ is introduced between the contributions of these two processes. The photon interference term is taken into account by the factor $c_{\text{norm}} m_{\ell_2 \ell_3}^{-2}$, where $c_{\text{norm}}$ is a normalisation parameter.

For the study presented in this section, the likelihood from Equation (12.3) is used. However, this likelihood only takes into account events with at least four jets. Therefore, events with three jets or fewer have to be vetoed. However, recent studies [249] have shown that a KLFitter likelihood can also be introduced for $t\bar{t}Z$ events with only three reconstructed jets.

In this section, the distribution of the highest $t\bar{t}Z$ likelihood value for all permutations of leptons and jets, determined by the KLFitter reconstruction, is studied. High likelihood values are expected for the signal process, since the KLFitter likelihood aims to reconstruct $t\bar{t}Z$ events. The $t\bar{t}Z$ likelihood in Equation (12.3) is used for these studies. Therefore, only events with at least four jets can be taken into account. In addition, exactly three charged leptons with at least one OSSF pair with an invariant mass within the $Z$-window (see Chapter 7) is required. The requirements on the lepton transverse momenta are $p_{\text{T}} > 27$ GeV for the highest $p_{\text{T}}$ lepton and $p_{\text{T}} > 20$ GeV for the other two leptons. The object selections discussed in Chapter 5 and the event preselection discussed in Section 7.1 are used.

For these KLFitter studies, the fake lepton background cannot be considered and Monte Carlo events with fake leptons are not subtracted from the background samples. A fraction of lepton fake events are still included in the background samples, especially in "others". However, this does not represent the full fake lepton background and fake factors are not applied. It is expected that the method shown in this chapter provides good discrimination against lepton fakes, because the KLFitter likelihood for the $t\bar{t}Z$ reconstruction will most probably yield low values for lepton fakes.

Figure 12.6 shows the distribution of the highest KLFitter logarithmic likelihood values per event for the selection discussed above. Only statistical uncertainties are shown. The separation shown in the figure is defined as

$$ S = \frac{1}{2} \sum_{i \in \text{bins}} \frac{(s_i - b_i)^2}{s_i + b_i} , \qquad (12.4) $$

where $s_i$ and $b_i$ are the number of entries in the i$^{\text{th}}$ bin of the signal and total background distributions, which are both normalised to one. The index $i$ runs over all bins.

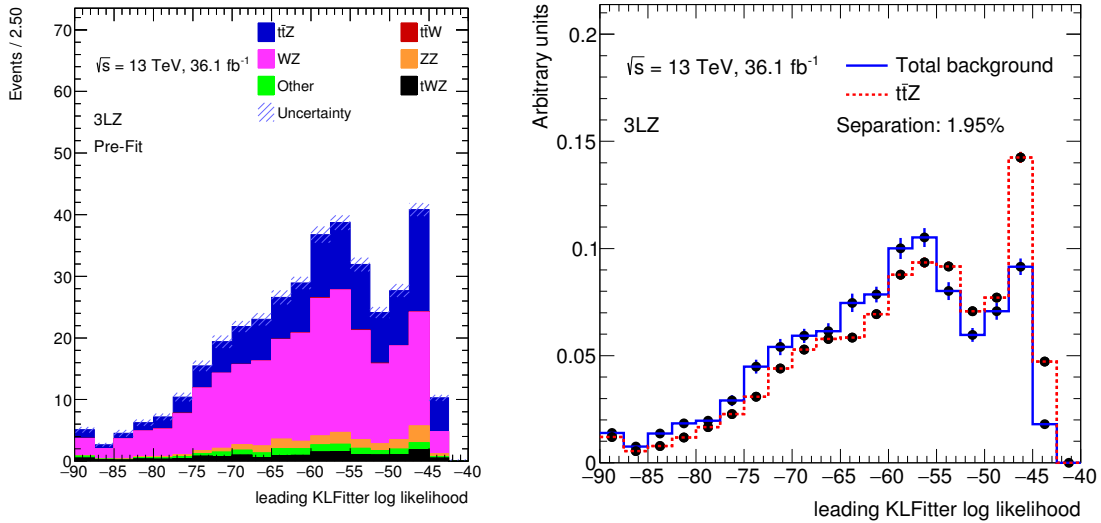

Figure 12.6.: Highest KLFitter logarithmic likelihood distribution for a basic trilepton selection. Only statistical uncertainties are shown. The distributions of the different signal and background contributions are shown on the left hand side. The separation between signal and background is shown on the right hand side.

It is not possible to construct signal and background regions based on cuts with this variable, as it is done with the pseudo continuous $b$-tagging in Section 12.1. However, the separation plot between signal and background reveals some separation power. A peak at a logarithmic likelihood value of $\sim -47$ is much more pronounced for $t\bar{t}Z$ than for the background.

Studying the KLFitter likelihood in the signal regions defined in Chapter 7 can give a further insight in the separating power of this variable. Due to the missing background from fake leptons, data is not shown in the signal regions for this study. Figure 12.7 and 12.8 show the highest KLFitter logarithmic likelihoods in the $3\ell$-$Z$-1b4j and $3\ell$-$Z$-2b4j signal regions, respectively. Although the statistical uncertainty of the background has increased with respect to the looser selection shown in Figure 12.6, a separation between signal and background is clearly visible. Furthermore, the peak at $\sim -47$ is well pronounced for the signal. In the $3\ell$-$Z$-1b4j region, the signal is clearly shifted towards higher logarithmic likelihood values with respect to the background. In the $3\ell$-$Z$-2b4j region, the separation is well pronounced above logarithmic likelihood values of $-55$. Since the fake lepton background contribution is relatively small in these two regions (see Table 7.5), these statements are expected to be still valid after including this process in these regions.

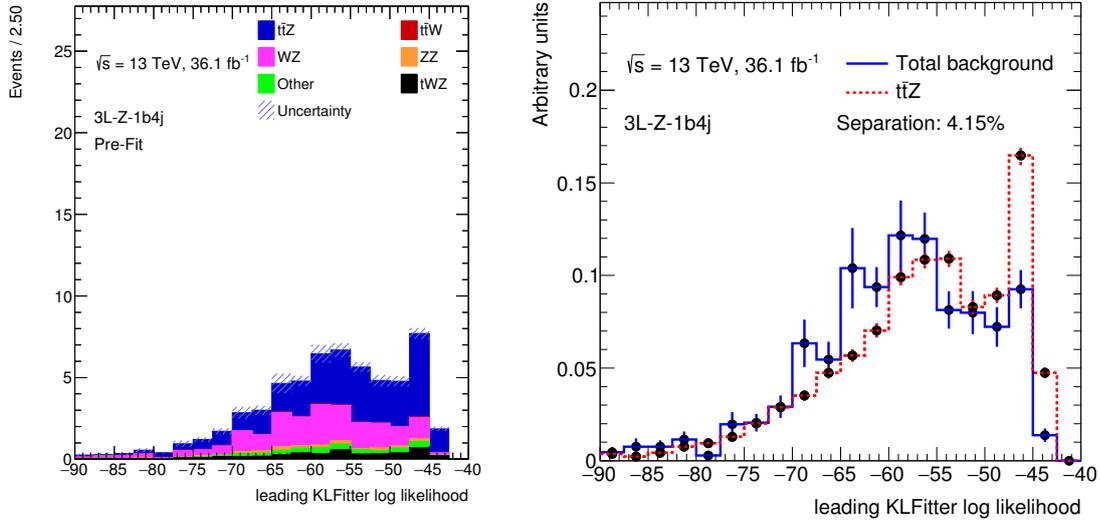

Figure 12.7.: Highest KLFitter logarithmic likelihood distribution in the 3ℓ-Z-1b4j region. Only statistical uncertainties are shown. The distributions of the different signal and background contributions are shown on the left hand side. The separation between signal and background is shown on the right hand side.

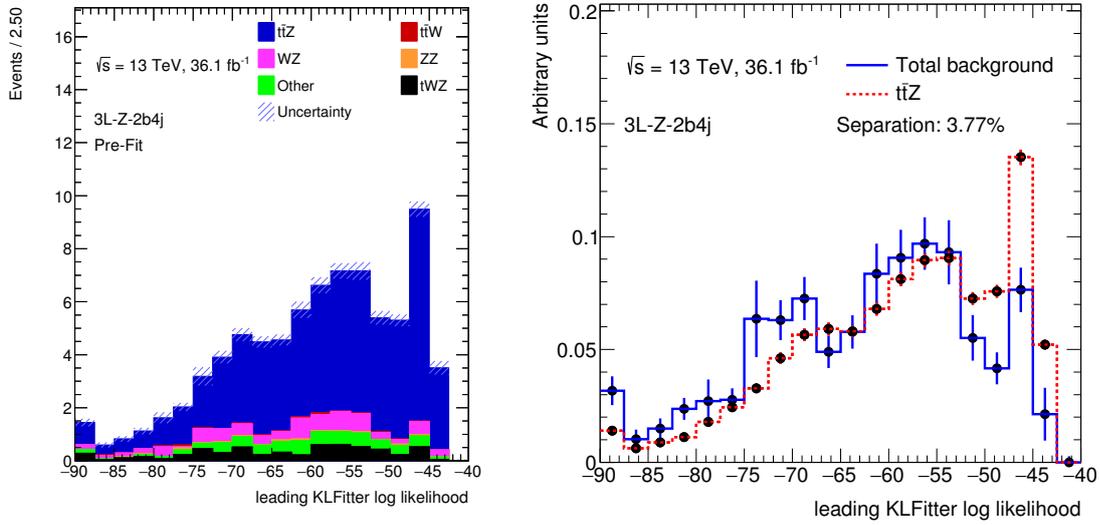

Figure 12.8.: Highest KLFitter logarithmic likelihood distribution in the 3ℓ-Z-2b4j region. Only statistical uncertainties are shown. The distributions of the different signal and background contributions are shown on the left hand side. The separation between signal and background is shown on the right hand side.

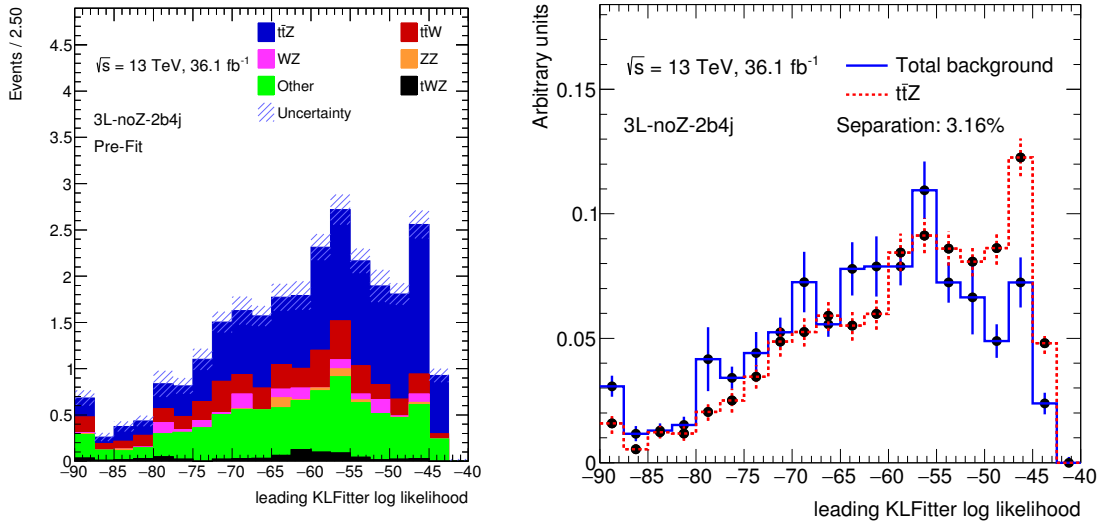

Figure 12.9.: Highest KLFitter logarithmic likelihood distribution in the 3ℓ-noZ-2b4j region. Only statistical uncertainties are shown. The distributions of the different signal and background contributions are shown on the left hand side. The separation between signal and background is shown on the right hand side.

Figure 12.9 shows the same distribution in the 3ℓ-noZ-2b4j region. This region is sensitive to off-shell $Z$ bosons. The Monte Carlo statistical uncertainty is comparable to the 3ℓ-$Z$-1b4j and 3ℓ-$Z$-2b4j region. The contribution from lepton fakes is much stronger in this region, see Table 7.5. This has to be kept in mind when interpreting the separation in this region. Although the $t\bar{t}Z$ events contributing to this channel are events where the $Z$ boson is produced off-shell, the signal and background distributions show an adequate separation.

## 12.3. Conclusion for the additional studies

Further optimisation of the signal and control region selection can be beneficial for a future $t\bar{t}Z$ analysis. Two studies for possible improvements of the $t\bar{t}Z$ analysis in the trilepton channel are presented in this chapter. The pseudo continuous $b$-tagging approach discussed in Section 12.1 shows very good results in terms of separation between signal and background. Using the KLFitter likelihood as an additional criterion for separating signal and background, as shown in Section 12.2, is by far not as powerful as the continuous $b$-tagging approach. However, the variable shows reasonable separation and can therefore possibly be used as an input variable for multivariate analyses. Furthermore, a template fit of the KLFitter likelihood distribution is imaginable.

CHAPTER 13

---

Summary, Conclusion and Outlook

---

This thesis presents the measurement of the $t\bar{t}Z$ production cross section in the channel with exactly three charged leptons (called the *trilepton channel*), using a dataset corresponding to an integrated luminosity of 36.1 fb$^{-1}$. This dataset was produced during 2015 and 2016 in proton-proton collisions at the ATLAS detector at a centre-of-mass energy of $\sqrt{s} = 13$ TeV. The $t\bar{t}Z$ process describes the top quark pair production in association with a $Z$ boson from both initial and final state radiation. Measuring the $t\bar{t}Z$ cross section is an interesting task, since this cross section depends on the coupling of the $Z$ boson to the top quark in final state $Z$ radiation and therefore also depends on the third component of the weak isospin $T_t^3$ of the top quark. Deviations from the expectation of the $t\bar{t}Z$ cross section can hint at new physics at the $t\bar{t}Z$ vertex, for instance a deviation from the Standard Model expectation of $T_t^3 = 1/2$. In addition, the $t\bar{t}Z$ process is an important background for $t\bar{t}H$ analyses and new physics searches.

The analysis presented in this thesis is part of an analysis conducted by the ATLAS collaboration to measure the $t\bar{t}Z$ and $t\bar{t}W$ cross sections [3]. The $t\bar{t}W$ process is the production of a top quark pair in association with a $W$ boson, radiated from an initial state quark. In addition to the trilepton channel, other multilepton channels are used for the overall ATLAS analysis to determine the $t\bar{t}Z$ and $t\bar{t}W$ cross sections.

Due to the low expected $t\bar{t}Z$ production cross section of $\sim 0.8$ pb, the statistical uncertainty on the dataset is still a considerable factor when performing this measurement. The trilepton channel, being sensitive to the decay of the leptonically decaying $Z$ boson and the lepton+jets decay mode of the top quark pair, has the highest expected significance for the $t\bar{t}Z$ process of all analysis channels used for the current iteration of this analysis. The signature with exactly three leptons also allows for sensitivity to the $t\bar{t}W$ process, which is not part of the analysis channel presented in this thesis but is considered in another part of the overall analysis. A main contribution to the background in this channel comes from $WZ$ events with additional jets, with the $WZ$ pair decaying into three charged leptons and a neutrino. The other important background contribution comes from events containing leptons from secondary processes, misidentified as prompt leptons from the hard scattering process.

For the trilepton channel, a profile likelihood fit of the yields in four signal regions and two background control regions is performed. Three of these signal regions are sensitive to $t\bar{t}Z$ with on-shell $Z$ bosons, and one signal region is sensitive to off-shell $Z$ bosons. The on-shell and off-shell sensitivity is achieved by requirements on the flavours, charges and invariant masses of lepton pairs. The main differences between the three on-shell signal regions are the jet and $b$-jet multiplicities. The signal strength of the $t\bar{t}Z$ process is fitted as a free parameter. The $WZ$+jets background is controlled by treating its normalisation as an additional free fit parameter with a dedicated control region. For a better comparison with the other $t\bar{t}Z$ channels, a control region for $ZZ$+jets, where the $ZZ$ pair decays into four charged leptons, is added. The $ZZ$+jets process is the main background in the tetralepton channel (see below). The normalisation of this process is also treated as a free fit parameter. The advantage of the profile likelihood fit is that it includes the systematic uncertainties of the measurement as nuisance parameters. Therefore, the fit helps to constrain systematic uncertainties and uses them to modify the scaling of the fit parameters.

The resulting $t\bar{t}Z$ cross section from the profile likelihood fit in the trilepton channel is

$$
\begin{aligned}
\sigma_{t\bar{t}Z}^{3\ell} &= 966_{-102}^{+114}(\text{stat.})_{-114}^{+115}(\text{syst.})\,\text{fb} \\
&= 966 \pm 162\,\text{fb} \;,
\end{aligned}
$$

where the uncertainty in the second row is determined by adding both statistical and systematic uncertainties quadratically and symmetrising the result. The most important sources of systematic uncertainties come from $b$-tagging, the $t\bar{t}Z$ generator choice, the normalisation of the $WZ$+jets process, the luminosity of the dataset and the Monte Carlo tunes for the parton showering of the $t\bar{t}Z$ samples. The nuisance parameters of the profile likelihood fit are well understood and do not show any unexpected behaviour.

The result is in good agreement with the NLO QCD and electroweak Standard Model prediction of $\sigma_{t\bar{t}Z}^{\text{NLO}} = 839.3 \pm 101$ fb [155]. The observed (expected) sensitivity of this result is $7.2\sigma$ ($6.4\sigma$), corresponding to a deviation from the background-only hypothesis. According to the fit result, approximately 130 $t\bar{t}Z$ events are produced in the signal regions of the trilepton channel. The result of the $t\bar{t}Z$ cross section measurement in the trilepton channel alone outperforms the previous combined result of the ATLAS collaboration [134] of $\sigma_{t\bar{t}Z} = 0.92 \pm 0.29(\text{stat.}) \pm 0.10(\text{syst.})$ pb in terms of the statistical uncertainty of the dataset. The previous ATLAS measurement uses $\int \mathcal{L} dt = 3.2$ fb$^{-1}$ of data from 2015, taken in proton-proton collisions at a centre-of-mass energy of 13 TeV.

Two additional analysis channels are used for measuring the $t\bar{t}Z$ cross section in different charged lepton multiplicities in a combined fit[1]. All of the signal and control regions of the three channels are used. One of these two analysis channels is the $2\ell$OSSF channel, which requires exactly two charged leptons with the same flavour and opposite electric charge. It is sensitive to the $Z$ boson decay into charged leptons and the $t\bar{t}$ decay in the all-jets channel. While having a large branching fraction, this channel has to deal with large backgrounds from $t\bar{t}$+jets and $Z$+jets. To differentiate between $t\bar{t}Z$ and these backgrounds, a Boosted Decision Tree is trained. The normalisation of the $Z$+jets background is fitted in two control regions. A separate fit in the $2\ell$OSSF channel yields an observed (expected) significance of $3.0\sigma$ ($3.8\sigma$).

The second additional analysis channel is the tetralepton channel, which requires exactly four charged leptons. It is sensitive to the decay of the $Z$ boson into two charged leptons and the dileptonic $t\bar{t}$ decay. While this channel has a high signal purity, the number of $t\bar{t}Z$ events is low. Therefore, this channel struggles with a high statistical uncertainty on the dataset. The main background of this channel is the $ZZ$+jets process, where both $Z$ bosons decay into charged leptons. The normalisation of this background is fitted as a free parameter in a dedicated control region. A separate fit in the tetralepton channel yields an observed (expected) significance of $5.3\sigma$ ($5.0\sigma$).

The result of a combined fit of all signal and control regions of all three $t\bar{t}Z$ channels is $\sigma_{t\bar{t}Z}^{\text{comb}} = 873^{+84}_{-76}(\text{stat.})^{+94}_{-90}(\text{syst.})$ fb, which is in good agreement with the Standard Model prediction. The observed (expected) significance of the combined fit result is $9.0\sigma$ ($8.5\sigma$). The trilepton channel, which is the main topic of this thesis, contributes to this result with the highest observed significance. This result is compatible with the $t\bar{t}Z$ cross section obtained from the most recent CMS analysis of

---

1. The combined fit and the separate analyses in all channels other than the trilepton channel are not part of this PhD project.

$\sigma_{t\bar{t}Z}^{\text{CMS}} = 0.99_{-0.08}^{+0.09}(\text{stat.})_{-0.10}^{+0.12}(\text{syst.})$ pb [97]. The CMS measurement uses a dataset from proton-proton collisions at the same centre-of-mass energy and with a similar integrated luminosity. The sizes of the statistical and systematic uncertainties of both the ATLAS and CMS measurements are similar. The overall ATLAS measurement also performs a measurement of the $t\bar{t}W$ cross section in a similar fashion. Regions with exactly two or three leptons are used as $t\bar{t}W$ signal regions, vetoing $t\bar{t}Z$ events. The result is in agreement with the Standard Model prediction.

In conclusion, the trilepton channel is proven to be the most relevant channel for the overall $t\bar{t}Z$ cross section measurement. The precision of this measurement improves with respect to the previous ATLAS measurement and is compatible with the latest CMS result and the Standard Model prediction. With an increasing Run-II dataset, statistical uncertainties become less dominant and optimising the analysis channels, for instance to reduce the effects of systematic uncertainties, becomes necessary. Additional studies, performed in this thesis for the trilepton channel, already show possible modifications for further analyses. Continuous $b$-tagging can be used to separate the signal from background containing light flavour jets. A dedicated $t\bar{t}Z$ KLFitter likelihood can potentially be used as a variable to further optimise the analysis in this channel.

Further possibilities for future $t\bar{t}Z$ analyses are manifold. The result of the $t\bar{t}Z$ cross section can be used to test the Standard Model hypothesis of the weak isospin $T_t^3 = 1/2$ of the top quark or on an alternative hypothesis like $T_t^3 = -1/2$, which would mean that the top quark is a down-type quark of a higher quark mass generation. This can already be done with the result of the current ATLAS analysis. However, Monte Carlo samples with alternative values of $T_t^3$ are needed which do not exist yet.

Two auxiliary studies [3], conducted in parallel to the overall $t\bar{t}Z$ and $t\bar{t}W$ cross section measurements, already point to new directions for the next $t\bar{t}Z$ analyses. One study uses an effective field theory to probe higher dimensional operators at the $t\bar{t}Z$ vertex, see also [165]. The second study is a fiducial measurement in a fraction of the total phase space. This can minimise effects from extrapolating cross sections into detector regions with little to no sensitivity. Fiducial measurements are often used for setting limits on physics models beyond the Standard Model. Both methods can be studied in further detail in future $t\bar{t}Z$ analyses.

With an increasing Run-II dataset at the LHC, statistical uncertainties are further decreasing. Therefore, an attempt can be made to measure differential cross sections in the future. In addition, several observables sensitive to the $t\bar{t}Z$ vertex can be studied, for example the angular separation of the charged leptons from the $Z$ boson and their invariant mass, as well as the transverse momentum of the $Z$ boson and the invariant mass of the top quark pair. Therefore, the $t\bar{t}Z$ vertex can be probed in more detail, for example to measure the third component of the weak isospin of the top quark.

Another possible expansion of the $t\bar{t}Z$ analysis would be the extension into multilepton channels in which the $Z$ boson decays hadronically or into neutrinos. Possibilities would be a zero-lepton channel, where the top quark pair decays in the all-jets channel, or the single-lepton channel, in which the $t\bar{t}$ pair decays in the lepton+jets channel. First studies in these channels are already ongoing, for instance in the single-lepton channel with hadronic $Z$ boson decays [250].

# Acknowledgements

Doing a PhD in elementary particle physics means working together with a lot of awesome people. No matter if it is during analysis meetings, conferences, workshops, in the ATLAS control room or even during the daily coffee break after lunch: you will always meet people that will inspire you. Therefore, I would first like to express my gratefulness for being able to work in the diverse and energetic environment of the ATLAS collaboration.

All of this was made possible by my supervisor and first referee Prof. Dr. Arnulf Quadt, for which I am deeply thankful. Being able to do my PhD in his group helped me to expand my horizon in particle physics. He was a great advisor to me, with a lot of experience and good ideas. I also would like to thank Prof. Dr. Ariane Frey for being part of my PhD supervision committee, Prof. Dr. Stan Lai for being the second referee of this thesis, as well as Prof. Dr. Baida Achkar, Prof. Dr. Laura Covi and Prof. Dr. Steffen Schumann for being part of my PhD examination committee. All of these people ensure that the work I did during my three years of PhD fulfil the highest scientific standards.

I also would like to thank Dr. Elizaveta Shabalina for sharing her enormous experience in top quark physics with me. She has a lot of expertise working in ATLAS and I cannot remember an instance where her advice wasn't useful. I also thank Dr. Boris Lemmer for his useful advice in physics questions and beyond. Special thanks also go to Dr. Clara Nellist and Dr. Thomas Peiffer for being contact persons regarding questions of style and wording for this thesis.

I also would like to thank everyone from the $t\bar{t}Z$ analysis team for their collaboration. This thesis shows my contribution to the work of the group, but it represents only a part of the overall analysis that is determining the combined results of the $t\bar{t}Z$ and $t\bar{t}W$ cross sections in all analysis channels, which was a great group effort.

Doch nun möchte ich all denen danken, die mir diese Arbeit auf persönlicher Ebene ermöglicht haben, denn ohne den nötigen Rückenwind ist eine Promotion schwer schaffbar. Dazu gehören meine vielen Freunde, die mich in den letzten zwei Monaten nicht zu Gesicht bekommen haben. Sie werden demnächst wieder die Freude haben, sich mit mir über Physik zu unterhalten, oder einfach mal wieder eine Hopfenschorle zu genießen. Mein Dank richtet sich auch an meine ehemaligen Lehrer und Professoren, die mich dazu inspirierten, den Pfad bis hierhin zu verfolgen. Gerne erinnere ich mich noch an Exkursionen des Physik-Leistungskurses, die mir verdeutlichten, wie viel Spaß Physik machen kann.

Vielen Dank auch Dir, Ruth, für Dein liebevolles Verständnis während dieser Zeit. Du warst eine große Stütze und immer da, wenn ich Dich gebraucht habe. Ich hoffe, ich kann Dich ebenso unterstützen, wenn es bei Dir so weit ist. Und Euch, meinen lieben Eltern, möchte ich besonders danken, dass Ihr mich zwar stets habt meinen eigenen Weg gehen lassen, mich aber bis hierhin immer mit Worten und Taten begleitet habt. Es ist nicht selbstverständlich, so geduldige und weise Eltern zu haben, und es gibt mir stets Sicherheit zu wissen, dass Ihr an mich und meine Entscheidungen glaubt.

Wer nun allerdings behauptet, ich werde mich nach der Promotion, wie Herr Lehrer Lämpel bei Max und Moritz, zurücklehnen und mir sagen:

> *„Ach! [...] die größte Freud*
> *Ist doch die Zufriedenheit!"*,

der weiß nicht, wie Wilhelm Busch den vierten Streich weiter gedichtet hat. Es gilt nicht, sich auf seinen Lorbeeren auszuruhen, sondern stets weiter zu streben, neugierig zu bleiben und sich zu sagen „Dieses war der $n$-te Streich, doch der $n+1$-te folgt sogleich".

Hereafter, the papers and proceedings with my authorship or co-authorship, published during my time as a PhD student or close to publication, are listed together with their abstracts.

### EFTfitter—A tool for interpreting measurements in the context of effective field theories [251]

Abstract: *Over the past years, the interpretation of measurements in the context of effective field theories has attracted much attention in the field of particle physics. We present a tool for interpreting sets of measurements in such models using a Bayesian ansatz by calculating the posterior probabilities of the corresponding free parameters numerically. An example is given, in which top-quark measurements are used to constrain anomalous couplings at the $Wtb$-vertex.*

### Measurement of the $t\bar{t}Z$ and $t\bar{t}W$ production cross sections in multilepton final states using 3.2 fb$^{-1}$ of $pp$ collisions at $\sqrt{s} = 13$ TeV with the ATLAS detector [134]

Abstract: *A measurement of the $t\bar{t}Z$ and $t\bar{t}W$ production cross sections in final states with either two same-charge muons, or three or four leptons (electrons or muons) is presented. The analysis uses a data sample of proton-proton collisions at $\sqrt{s} = 13$ TeV recorded with the ATLAS detector at the Large Hadron Collider in 2015, corresponding to a total integrated luminosity of 3.2 fb$^{-1}$. The inclusive cross sections are extracted using likelihood fits to signal and control regions, resulting in $\sigma_{t\bar{t}Z} = 0.9 \pm 0.3$ pb and $\sigma_{t\bar{t}W} = 1.5 \pm 0.8$ pb, in agreement with the Standard Model predictions.*

**Top quark pair property measurements and $t\bar{t} + X$, using the ATLAS detector at the LHC [252]**  Abstract: *Measuring the properties of the top quark has been proven to be a reliable test of the Standard Model of particle physics (SM). At the Large Hadron Collider, top quark physics has entered the era of precision measurements which allows to measure new top quark properties and to repeat other top quark properties measurements with improved precision. This article presents the measurements of top quark spin observables, charge and CP asymmetries, W boson polarisation from $t\bar{t}$ events and the cross section of $t\bar{t}$ pairs produced in association with a W or Z boson. All these measurements are using data from proton-proton collisions taken with the ATLAS detector at the Large Hadron Collider. The spin correlation $C(n, n)$ is measured for the first time and direct CP violation is measured directly for the first time in the context of b-hadron decays. The measured value of the CP mixing asymmetry $A^b_{mix}$ presented here cannot disprove the deviation in the dimuon asymmetry seen by the DØ experiment. The W boson polarisation measurement from $t\bar{t}$ events is the most precise one to date and is used to constrain anomalous couplings at the Wtb vertex. All results are in agreement with the SM predictions.*

**Measurement of the $t\bar{t}Z$ and $t\bar{t}W$ production cross sections in multilepton final states using 36.1 fb$^{-1}$ of $pp$ collisions at 13 TeV at the LHC [3]**  Abstract: *A measurement of the $t\bar{t}Z$ and $t\bar{t}W$ production cross sections in final states with two same charge leptons, three or four isolated electrons or muons with the ATLAS detector at the LHC is presented. A measurement in the fiducial volume is also performed. The data sample used corresponds to an integrated luminosity of 36.1 fb$^{-1}$ of data collected in 2015 and 2016 during pp collisions at 13 TeV at the LHC.*

# Appendices

## Raw Event Yields

This appendix shows additional pre-fit tables to discuss the impact of the statistical uncertainties from Monte Carlo samples in Chapter 11.

| | 3ℓ-Z-1b4j | 3ℓ-Z-2b4j | 3ℓ-Z-2b3j | 3ℓ-noZ-2b4j | 3ℓ-WZ-CR | 4ℓ-ZZ-CR |
|---|---|---|---|---|---|---|
| $t\bar{t}Z$ | 35 268 | 62 385 | 12 590 | 13 921 | 5092 | 222 |
| $t\bar{t}W$ | 160 | 288 | 333 | 1575 | 68 | — |
| $WZ$ | 528 | 276 | 162 | 41 | 5263 | — |
| $ZZ$ | 256 | 116 | 118 | 41 | 1802 | |
| $tZ$ | 265 | 424 | 410 | 36 | 183 | — |
| $tWZ$ | 897 | 846 | 274 | 108 | 362 | 8 |
| $t\bar{t}H$ | 1444 | 2475 | 810 | 9590 | 238 | — |
| Other | 226 | 382 | 56 | 2119 | 2203 | 2851 |
| DD Fakes | 119 | 123 | 71 | 124 | — | — |
| $\gamma + X$ | 16 | 14 | 12 | 34 | — | — |
| MC Fakes | — | — | — | — | 176 | 45 |
| Total | 39 179 | 67 329 | 14 836 | 27 589 | 15 387 | 59 613 |

Table A.1.: Raw number of Monte Carlo and data driven events for the different signal and control regions.

|          | $3\ell$-$Z$-1b4j     |
|----------|----------------------|
| $t\bar{t}Z$  | $29.88 \pm 0.42$  |
| $t\bar{t}W$  | $0.35 \pm 0.05$   |
| $WZ$     | $17.84 \pm 1.16$     |
| $ZZ$     | $1.70 \pm 0.18$      |
| $tZ$     | $1.95 \pm 0.14$      |
| $tWZ$    | $4.03 \pm 0.31$      |
| $t\bar{t}H$  | $0.85 \pm 0.07$   |
| Other    | $0.14 \pm 0.02$      |
| DD Fakes | $4.39 \pm 1.69$      |
| $\gamma + X$ | $1.31 \pm 0.70$  |
| FakeFF   | —                    |
| Total    | $62.43 \pm 2.24$     |

Table A.2.: Pre-fit yields in the $3\ell$-$Z$-1b4j region, showing only statistical uncertainties.